\documentclass[aps,pre,reprint,conference]{ws-mpla}
\usepackage{graphicx,cite,multirow,hyperref,siunitx,listings,xspace,float,slashed}
\usepackage[utf8]{inputenc}
\hypersetup{colorlinks=true,linkcolor=blue,citecolor=magenta,filecolor=magenta,urlcolor=cyan}

\numberwithin{equation}{section}
\numberwithin{figure}{section}
\numberwithin{table}{section}
\DeclareSymbolFont{extraup}{U}{zavm}{m}{n}
\DeclareMathSymbol{\vardiamond}{\mathalpha}{extraup}{87}
\restylefloat{table}
\allowdisplaybreaks

\begin{document}

\catchline{}{}{}{}{}
\markboth{B.~Fuks {\it et al.}}{Proceedings of the first MadAnalysis 5 workshop on LHC recasting in Korea}
\title{PROCEEDINGS OF THE SECOND MADANALYSIS 5 WORKSHOP ON LHC RECASTING IN KOREA}

\author{Benjamin~Fuks$^{1,2}$, Pyungwon~Ko$^3$, Seung~J.~Lee$^4$ (editors);\\[.1cm]
  Jack~Y.~Araz$^5$, Eric~Conte$^{6,7}$, Robin~Ducrocq$^6$, Thomas~Flacke$^8$,
  Si Hyun Jeon$^9$, Taejeong~Kim$^{10}$, Richard~Ruiz$^{11,12}$, Dipan~Sengupta$^{13}$ (conveners);\\[.1cm]
  Sam~Bein$^{14}$, Jin~Choi$^9$, Luc~Darm\'e$^{15}$, Mark~D.~Goodsell$^1$, Ho~Jang$^{16}$, Adil~Jueid$^{17}$, Won~Jun$^9$,
  Yechan Kang$^{18}$,
  Jeongwoo~Kim$^{18}$, Jihun Kim$^9$, Jinheung~Kim$^{17}$, Jehyun~Lee$^9$, Joon-Bin~Lee$^9$, SooJin~Lee$^{17}$,
  Taegyu~Lee$^4$, Jongwon~Lim$^{10}$, Chih-Ting Lu$^3$, Ui~Min$^{19}$, Malte~Mrowietz$^{14}$,
  Kyungmin~Park$^{18}$, Jae-Hyeon~Park$^3$, Jiwon~Park$^{10}$, Jory~Sonneveld$^{14,20}$, Soohyun Yun$^{10}$\\[.2cm]}

\address{
  $^1$ Sorbonne Universit\'e, CNRS, Laboratoire de Physique Th\'eorique et Hautes \'Energies, LPTHE, F-75005 Paris, France\\[.1cm]
  $^2$ Institut Universitaire de France, 103 boulevard Saint-Michel, 75005 Paris, France\\[.1cm]
  $^3$ School of Physics, Korea Institute for Advanced Study, Seoul 02455, Korea\\[.1cm]
  $^4$ Department of Physics, Korea University, Seoul 136-713, Korea\\[.1cm]
  $^5$ Institute for Particle Physics Phenomenology, Durham University, South Road, Durham, DH1 3LE, UK\\[.1cm]
  $^6$ Institut Pluridisciplinaire Hubert Curien, D\'epartement Recherches Subatomiques, Universit\'e de
       Strasbourg/CNRS-IN2P3, 23 Rue du Loess, F-67037 Strasbourg, France \\[.1cm]
  $^7$ Universit\'e de Haute Alsace, Mulhouse, France\\[.1cm]
  $^8$ Center for Theoretical Physics of the Universe, Institute for Basic Science (IBS), Daejeon 34126, Korea\\[.1cm]
  $^9$ Department of Physics and Astronomy, Seoul National University, Gwanak-ro 1, Gwanak-gu, Seoul, 08826, South Korea\\[.1cm]
  $^{10}$ Department of Physics, Hanyang University, Seoul 04763, South Korea\\[.1cm]
  $^{11}$ Centre for Cosmology, Particle Physics and Phenomenology {\rm (CP3)},
       Universit\'e Catholique de Louvain, Chemin du Cyclotron, Louvain la Neuve, B-1348, Belgium\\[.1cm]
  $^{12}$ Institute of Nuclear Physics, Polish Academy of Sciences, ul.~Radzikowskiego, Cracow 31-342, Poland\\[.1cm]
  $^{13}$ Department of Physics and Astronomy, University of California, San Diego, 9500 Gilman Drive, La Jolla, USA\\[.1cm]
  $^{14}$ Institut f\"ur Experimental Physik, Universit\"at Hamburg, Luruper Chaussee 149, Hamburg, D-22671, Deutschland\\[.1cm]
  $^{15}$ Instituto Nazionale di Fisica Nucleare, Laboratori Nazionali di Frascati, C.P. 13, 00044 Frascati, Italy\\[.1cm]
  $^{16}$ Gwangju Institute of Science and Technology, 123, Cheomdangwagi-ro, Buk-gu, Gwangju, Republic of Korea\\[.1cm]
  $^{17}$ Department of Physics, Konkuk University, 120 Neungdong-ro, Gwangjin-gu, Seoul 05029, Korea \\[.1cm]
  $^{18}$ Department of Physics, University of Seoul, 163, Seoulsiripdae-ro, Dongdaemun-gu, Seoul, Republic of Korea\\[.1cm]
  $^{19}$ Department of Physics, Korea Advanced Institute of Science and Technology, 291 Daehak-ro, Yuseong-gu, Daejeon 34141, Republic of Korea\\[.1cm]
  $^{20}$ Nikhef, Science Park 105, 1098 XG Amsterdam, The Netherlands}

\maketitle


\begin{abstract}
We document the activities performed during the second {\sc MadAnalysis}~5 workshop on LHC recasting, that was organised in KIAS (Seoul, Korea) on February 12-20, 2020. We detail the implementation of 12 new ATLAS and CMS searches in the {\sc MadAnalysis}~5 Public Analysis Database, and the associated validation procedures. Those searches probe the production of extra gauge and scalar/pseudoscalar bosons, supersymmetry, seesaw models and deviations from the Standard Model in four-top production.
\end{abstract}

\clearpage

\markboth{B.~Fuks {\it et al.}}{Proceedings of the second MadAnalysis 5 workshop on LHC recasting in Korea}

\section{Introduction}
  \vspace*{-.1cm}\footnotesize{\hspace{.5cm}By Jack Y. Araz, Eric~Conte,
  Robin~Ducrocq, Thomas~Flacke, Benjamin Fuks, Si Hyun Jeon, Taejeong Kim, Pyungwon Ko, Seung J. Lee, Richard~Ruiz and Dipan Sengupta}
\vspace*{-.8cm}

\subsection*{}
Whereas the discovery of the Higgs boson almost a decade ago has accomplished
one of the main objectives of the LHC physics program, no significant deviation
beyond the Standard Model of particle physics has been found so far by the LHC
experiments. Therefore, the concrete mechanism triggering the breaking of the
electroweak symmetry remains unexplained, and no hint for any solution to the
issues and limitations (such as the hierarchy problem, neutrino masses, dark
matter, {\it etc.}) of the Standard Model has emerged from data. As new
physics must exist in some form, data therefore implies that either the new
states are too heavy and/or their interactions too feeble to leave any
observable signature within the present collider reach, or that new particles
lie in a configuration rendering their discovery challenging.

As a consequence of this non-observation of any new phenomenon at colliders, the
results of
the experimental searches are traditionally interpreted as constraints on
various theoretical models. These include popular scenarios like the Minimal
Supersymmetric Standard Model, as well as simplified models or effective field
theories. There is, however, a vast domain of new physics setups extending the
Standard Model, which all come with a variety of concrete realisations and
whose predictions should be compared with LHC data. It is therefore crucial to
develop a strategy allowing to exploit in the best possible way the current and
future results of the LHC, so that one could straightforwardly draw conclusive
statements on what physics beyond the Standard Model could or could not be.

Many groups have consequently developed and maintained public programs dedicated
to the re-interpretation of the results of the LHC~\cite{Khosa:2020zar,%
Drees:2013wra,Dumont:2014tja,Buckley:2010ar,Balazs:2017moi}. In practice, these
programs aim to predict the number of signal events that populate the different
signal regions of given LHC analyses, when one assumes a specific new physics
context.
From those predictions, considered together with information on data and on the
Standard Model expectation, it becomes possible to derive whether
the considered new physics scenario is excluded. {\sc MadAnalysis}~5 and its public analysis
database (PAD)\footnote{See the webpage
\href{http://madanalysis.irmp.ucl.ac.be/wiki/PublicAnalysisDatabase}
{http://madanalysis.irmp.ucl.ac.be/wiki/PublicAnalysisDatabase}.}
is one of these tools~\cite{Conte:2012fm,Conte:2014zja,Dumont:2014tja,Conte:2018vmg}.

The strategy that {\sc MadAnalysis}~5 follows relies on the generation of Monte
Carlo signal events representative of the signature(s) of a given model of
physics beyond the Standard Model. Those events are generated by relying on
calculations matching fixed order results with parton showers, and they are
further processed to include the modelling of hadronisation and multiple parton
interactions. Hadron-level events are handled to simulate the response of the
ATLAS or CMS detector, which can be either achieved with the {\sc Delphes}~3
software~\cite{deFavereau:2013fsa} or with the simplified SFS fast
detector simulation shipped with {\sc MadAnalysis}~5\cite{Araz:2020lnp}.
Next, the detector-level events are reconstructed and the code derives, by
employing validated and dedicated C++ recast codes, how those events populate
the signal regions of the different analyses of the PAD. From those predictions,
as above-mentioned, a statistical treatment allows for the derivation of
conclusions about whether the initially considered model of new physics is
excluded, and at which confidence level.

While implementing existing ATLAS and CMS searches in the {\sc MadAnalysis}~5
framework is not complicated {\it per se}, as this consists in transcribing in
C++ and in the code's internal data format a given search as described in the
experimental publications, doing so in a robust and trustable way is more
difficult\cite{Kraml:2012sg,Dumont:2014tja,Abdallah:2020pec}. This indeed
requires a careful validation of the implementation, which
can be achieved in several manners. For instance, one could derive cut-flows
for well-defined new physics scenarios, and compare them, on a cut-by-cut basis,
with the corresponding official results. One could also compare the shapes of
various differential distributions at different steps of the cut-flow with the
corresponding ATLAS or CMS curves, and finally, it is also possible to extract
exclusion contours for dedicated new physics frameworks, and confront them to
their official counterpart. In those comparison, we expect to obtain an
agreement at a satisfactory level, the precise definition of this level being
dependent on the analysis through (variable) quality of the associated
validation material released by the ATLAS and CMS collaborations.

In these proceedings, we report the activities that have been performed during
the second {\sc MadAnalysis}~5 workshop in Korea, that was held at the
Korea Institute for Advanced Study (KIAS) in
Seoul (South Korea), on February 12-20, 2020. Similar to its 2017
edition~\cite{Fuks:2018yku}, the workshop brought together an enthusiastic
group of students with post-doctoral, junior and senior researchers. Along with
the main theme of the workshop, namely the re-interpretation of the results of
the LHC searches for new physics, various lectures on collider physics, beyond
the standard model theories and LHC experimental aspects were organised,
together with dedicated sessions on Monte Carlo event generation and
{\sc MadAnalysis}~5.

The scope of the workshop consisted of a recasting exercise assigned
to the participants, who were tasked with implementing in the {\sc MadAnalysis}~5
framework several particular ATLAS and CMS searches for new physics. Moreover,
a careful validation of these implementations was required, so that they could
be shared with the community for dedicated physics studies without any concern.
These proceedings document those implementations and their validation. Twelve new
analyses have been added to the PAD as an outcome of the workshop. Their source
codes are available from the {\sc MadAnalysis}~5 dataverse\footnote{See the webpage
\href{https://dataverse.uclouvain.be/dataverse/madanalysis}
{https://dataverse.uclouvain.be/dataverse/madanalysis}.}, often with the Monte
Carlo configuration cards relevant for the validation of the different
implementations.
More details and validation information can be found in the following sections.

The list of analyses under consideration is provided below:
\begin{enumerate}
  \item ATLAS-EXOT-2018-30: an ATLAS search for $W'$-boson production and
    decay into a lepton-neutrino pair with 139~fb$^{-1}$ of LHC
    data~\cite{Aad:2019wvl}; see
    \url{https://doi.org/10.14428/DVN/GLWLTF}~\cite{GLWLTF_2020} and
    section~\ref{sec:wprime}.

  \item CMS-EXO-17-015: a CMS search for leptoquark pair-production, followed
    by decays into dark matter, one muon and one jet with 77.4~fb$^{-1}$ of LHC
    collisions~\cite{Sirunyan:2018xtm}; see
    \url{https://doi.org/10.14428/DVN/ICOXG9}\cite{ICOXG9_2020}
    and section~\ref{sec:lq}.

  \item CMS-EXO-17-030: a CMS search for the pair production of a new physics
    state decaying into three jets with 35.9~fb$^{-1}$ of LHC
    collisions~\cite{Sirunyan:2018duw}; see
    \url{https://doi.org/10.14428/DVN/GAZACQ}~\cite{GAZACQ_2020} and
    section~\ref{sec:ditrijet};

  \item CMS-EXO-19-002: a CMS search for new physics in final states containing
    multiple leptons with 137~fb$^{-1}$ of LHC
    collisions~\cite{Sirunyan:2019bgz}; see
    \url{https://doi.org/10.14428/DVN/DTYUUE}~\cite{DTYUUE_2021}
    and section~\ref{sec:multilep};

  \item CMS-HIG-18-011: a CMS search for exotic Higgs decays into a di-muon and
    di-$b$-jet system via two pseudo-scalars with 35.9~fb$^{-1}$ of LHC
    collisions~\cite{Sirunyan:2018mot}; see
    \url{https://doi.org/10.14428/DVN/UOH6BF}~\cite{UOH6BF_2020}
    and section~\ref{sec:nmssm};

  \item ATLAS-SUSY-2018-04: an ATLAS search for the pair production of staus
    that each decay into a tau lepton and missing transverse energy, with
    139~fb$^{-1}$ of LHC collisions~\cite{Aad:2019byo}; see
    \url{https://doi.org/10.14428/DVN/UN3NND}~\cite{UN3NND_2020} and
    section~\ref{sec:stau};

  \item ATLAS-SUSY-2018-06: an ATLAS search for electroweakino pair production
    and decay through Jigsaw variables, with 139~fb$^{-1}$ of LHC
    collisions~\cite{Aad:2019vvi}; see
    \url{https://doi.org/10.14428/DVN/LYQMUJ}~\cite{LYQMUJ_2020} and
    section~\ref{sec:jigsaw};

  \item ATLAS-SUSY-2018-31: an ATLAS search for sbottom pair production and
    decay in the multi-bottom plus missing transverse energy channel, with
    139~fb$^{-1}$ of LHC collisions~\cite{Aad:2019pfy}; see
    \url{https://doi.org/10.14428/DVN/IHALED}~\cite{IHALED_2020} and
    section~\ref{sec:sbottom};

  \item ATLAS-SUSY-2018-32: an ATLAS search for slepton or electroweakino
    pair production and decay in the di-lepton plus missing transverse energy
    channel, with 139~fb$^{-1}$ of LHC collisions~\cite{Aad:2019vnb}; see
    \url{https://doi.org/10.14428/DVN/EA4S4D}~\cite{EA4S4D_2020} and
    section~\ref{sec:sleptons};

  \item ATLAS-SUSY-2019-08: an ATLAS search for electroweakino
    pair production and decay with final states featuring a Higgs-boson decaying
    into a $b\bar b$ system, one lepton and missing transverse energy,
    with 139~fb$^{-1}$ of LHC collisions~\cite{Aad:2019vvf}; see
    \url{https://doi.org/10.14428/DVN/BUN2UX}~\cite{DVN/BUN2UX_2020} and
    section~\ref{sec:ewkinos_bbl};

  \item CMS-SUS-19-006: a CMS search for supersymmetry in events featuring a
    large hadronic activity and missing transverse energy, with 139~fb$^{-1}$ of
    LHC collisions~\cite{Sirunyan:2019ctn}; see
    \url{https://doi.org/10.14428/DVN/4DEJQM}~\cite{4DEJQM_2020} and
    section~\ref{sec:susyhad};

  \item CMS-TOP-18-003: a CMS search for the production of four top quarks in
    final states with a same-sign pair or more than three leptons, with
    137~fb$^{-1}$ of LHC collisions~\cite{Sirunyan:2019wxt}; see
    \url{https://doi.org/10.14428/DVN/OFAE1G}~\cite{OFAE1G_2020} and
    section~\ref{sec:4tops}.

\end{enumerate}

\subsection*{Acknowledgements}
We are especially indebted to JeongEun Yoon and Brad Kwon for their help with
local organisation, making this workshop a very live and stimulating event. We
are moreover greateful to the Asia Pacific Center for Theoretical Physics
(APCTP), KIAS, the Particle Theory Group of Korea University and the France
Korea Particle Physics and e-science Laboratory (FKPPL) of the CNRS for their
support. Resources have been provided by the supercomputing facilities of the
Universit\'e catholique de Louvain (CISM/UCL) and the Consortium des
\'Equipements de Calcul Intensif en F\'ed\'eration Wallonie Bruxelles (C\'ECI)
funded by the Fond de la Recherche Scientifique de Belgique (F.R.S.-FNRS) under
convention 2.5020.11 and by the Walloon Region.
\cleardoublepage
\markboth{Kyungmin Park, Ui Min, SooJin Lee and Won Jun}{Implementation of the ATLAS-EXOT-2018-30 analysis}

\section{Implementation of the ATLAS-EXOT-2018-30 analysis ($W'$ boson into a lepton and a neutrino; 139~fb$^{-1}$)}
  \vspace*{-.1cm}\footnotesize{\hspace{.5cm}By Kyungmin Park, Ui Min, SooJin Lee and Won Jun}
\label{sec:wprime}


\subsection{Introduction}
One of the testable models at the LHC is the Sequential Standard Model (SSM), where new heavy gauge bosons $W'$ and $Z'$ couple to the SM fermions with the same strength as the SM weak gauge bosons\cite{Altarelli:1989ff,Fuks:2017vtl}. In a simplified model approach, the SSM extends the SM gauge sector by an additional $SU(2)'$ symmetry, $SU(3)_c \times SU(2)_L \times U(1) \times SU(2)'$. Here, the new gauge bosons get their heavy masses after spontaneous symmetry breaking at the energy scale that is higher than the electroweak scale. We assume that any detail on the extended gauge symmetry breaking mechanism can be factored out and ignored at the LHC scale. For simplicity, we ignore interactions including $Z'$ and only consider those between $W'$ and the left-handed SM fermions. Its triple gauge couplings and couplings to Higgs are also neglected.

\begin{figure}[t]
  \centerline{\includegraphics[width=2.5in]{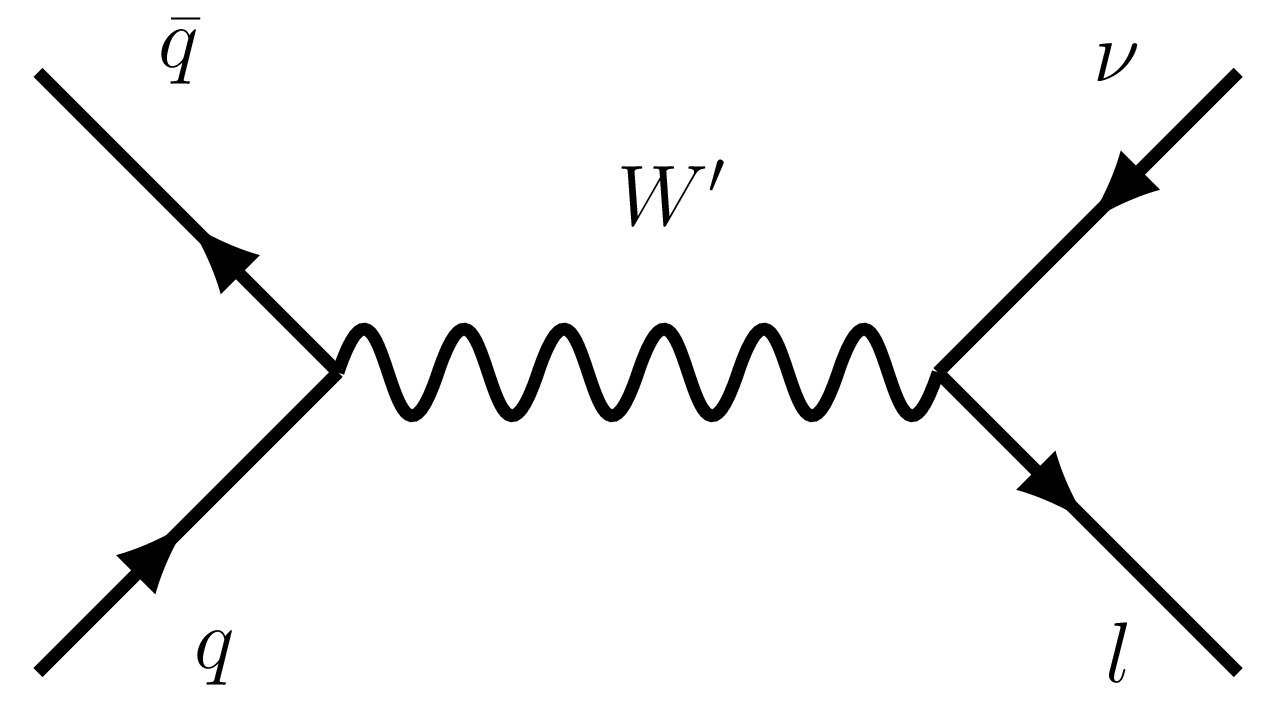}}
  \vspace*{8pt}
  \caption{$W'$ boson contribution to the production of a lepton and a neutrino in $q\bar{q}$ scattering. \protect\label{wp_fig1}}
\end{figure}

Under the {\sc \rm MadAnalysis 5}\cite{Conte:2012fm,Conte:2014zja,Dumont:2014tja,Conte:2018vmg} framework, we reimplement the ATLAS-EXOT-2018-30 analysis\cite{Aad:2019wvl}, a search for a $W' $ signal at the LHC using the ATLAS detector and $139 \, {\rm fb}^{-1}$ of proton-proton collisions, in the $ p \, p \rightarrow W' \rightarrow l \, \nu_l$  ($l \,  = \, e, \, \mu$) channel, as shown in Fig.~\ref{wp_fig1}. We then validate the reimplementation by comparing our signal predictions to those from the official ATLAS results, with $W'$ masses varying from 2 ${\rm TeV}$ to 6 ${\rm TeV}$.

In section~\ref{wp:sec2}, we define the objects such as electron, muon, jet and missing transverse energy, and we present how to select events for the electron and muon channels. In section~\ref{wp:sec3}, we describe processes of event generation for the decay channels  $ p \, p \rightarrow W' \rightarrow l \, \nu_l$  ($l \,  = \, e, \, \mu$) and compare the results with those of ATLAS analysis. We summarise our work in section~\ref{wp:sec4}.

\subsection{Description of the analysis}\label{wp:sec2}
The analysis targets a signature in which a heavy $W'$ boson decays into a single lepton and a neutrino. To extract the heavy charged gauge boson signal, events including high missing transverse energy ($ E\!\!\!/_T$) and a charged lepton with high transverse momentum ($p_T$) are selected.

\subsubsection{Object definitions}\label{sec:wp_obj}
As our main targets are the electron channel ($ p \, p \rightarrow W' \rightarrow e \, \nu$) and the muon channel ($ p \, p \rightarrow W' \rightarrow \mu \, \nu$), the analysis requires the reconstruction and identification of electrons and muons with high $p_T$, following the object selections defined in the considered ATLAS study\cite{Aad:2019wvl}.

For the electron candidates, they must have a transverse energy $E_T > 65 \, {\rm GeV}$ and a pseudo-rapidity $ \left| \eta \right| < 2.47$, where the barrel-endcap transition region $ 1.37 < \left| \eta \right| < 1.52$ is excluded. The candidates are required to satisfy the following isolation criteria based on both calorimeter and tracking measurements: $E^{cone20}_T/p_T < 0.06$ for calorimeter isolation and $p^{cone20}_T/p_T < 0.06$ for track isolation. Here, $p^{cone20}_T\left(E^{cone20}_T\right)$ is computed by summing the transverse momentum (energy) of all tracks (energy deposits) within a cone centered around the electron track, with a cone size of $\Delta R = 0.2$\cite{Aad:2019tso}. The reconstruction and identification efficiencies and the resolution of electrons are implemented in the {\sc \rm Delphes}~3\cite{deFavereau:2013fsa} card following Refs.~\cite{Aad:2019wvl,Aad:2019tso}. In the $p_T > 50 \, {\rm GeV}$ region, for example, this yields an electron reconstruction efficiency of $81.7\%$.

For the muon candidates, we require high-$p_T$ muons with $ p_T > 55 \, {\rm GeV} $ and $0.1 < \left| \eta \right| < 2.4$. Those with pseudo-rapidity in the range of $1.01 < \left| \eta \right| < 1.1$ are vetoed due to the significant drop in the efficiencies \cite{Aad:2016jkr}. The candidates must pass track-based isolation criteria, $p^{cone30}_T / p_{T} < 0.15$ where $p^{cone30}_T$ is defined as the scalar sum of the transverse momenta of all tracks with $p_T >1 \, {\rm GeV}$ in a cone size of $\Delta R = \min \left(10 \, {\rm GeV}/ p_T , 0.3 \right)$ around the muon transverse momentum $p_T$, excluding the muon track itself\cite{Aad:2016jkr}.
The reconstruction and identification efficiencies and resolution of muons are implemented in the {\sc \rm Delphes} card following Refs.~\cite{Aad:2019wvl,Aad:2016jkr}. For instance, this gives a muon efficiency of $53\% $ for $p_T > 3 \, {\rm TeV}$.
 
For the jet candidates, jet-reconstruction is achieved with the anti-$k_T$ algorithm~\cite{Cacciari:2008gp} as implemented in {\sc \rm FastJet}\cite{Cacciari:2011ma,Cacciari:2005hq} with a jet radius parameter $\Delta R = 0.4$. The kinematical region of interest is chosen by defining the jet candidates as those satisfying $p_{T} > 20 \, {\rm GeV}$ for $\left| \eta \right| < 2.4$ and $p_{T} > 30 \, {\rm GeV}$ for $2.4 < \left| \eta \right| < 2.5$. We enforce an overlap removal procedure with the electron collection, removing jets lying within a cone of $\Delta R(j,e) = 0.1$ of an electron.

The missing transverse energy $ E\!\!\!/_T$ is evaluated by the vector sum of the transverse momenta of the following components: leptons, photons\footnote{Photons are reconstructed as defined in the default ATLAS parameterization in {\sc Delphes}~3\cite{deFavereau:2013fsa}.}, and jets.
Table~\ref{wp_ta1} and Table~\ref{wp_ta2}  show the summary for these object selections and isolation criteria, respectively.

\begin{table}[t]
  \tbl{Object selections}
  {\begin{tabular}{@{}cccc@{}} \toprule
  Object & $E^e_{T}(p^{\mu,jet}_{T})$ & $\left| \eta \right|$ &  Identification \\
  \colrule
  $e$ \hphantom{0} & \hphantom{0} $> 65\, $ GeV & \hphantom{0} $<1.37, \, [1.52,2.47]$ & tight identification \\
  $\mu$ \hphantom{0} & \hphantom{0} $> 55\, $ GeV & \hphantom{0} $[0.1,1.01], \, [1.1,2.4]$ & high-$p_T$ identification \\
  jet \hphantom{0} & \hphantom{0} $> 20\, $ GeV & \hphantom{0} $< 2.4$ & - \\
  & \hphantom{0} $> 30\,$ GeV & \hphantom{0} $[2.4,2.5]$ & - \\
  \botrule
  \end{tabular}\label{wp_ta1} }
\end{table}

\begin{table}[t]
  \tbl{Isolation criteria}
  {\begin{tabular}{@{}cccc@{}} \toprule
  Object & Calorimeter isolation & Track isolation & $\Delta R$\ \\
  \colrule
  $e$ \hphantom{0} &  \hphantom{0} $E^{cone20}_T/p_T \, < \, 0.06 $ & \hphantom{0} $p^{cone20}_T/p_T \, < \, 0.06 $ & $0.2$\\
  $\mu$ \hphantom{0} & \hphantom{0} - & \hphantom{0}$ p^{cone30}_T/p_T \, < \, 0.15 $ & $\min (10\, GeV/p_T,0.3)$ for $p_T > 1 \, GeV$\\
  \botrule
  \end{tabular}\label{wp_ta2} }
\end{table}

\subsubsection{Event selection}
The missing transverse energy ($ E\!\!\!/_T$) and the transverse mass ($m_T$) observables are used to select events from the electron and muon channels. 
 Here, $m_T(l, E\!\!\!/_T)$ can be calculated by following formula,
 \begin{equation}
 m_T(l, E\!\!\!/_T)= \sqrt{2 \, p^l_T   E\!\!\!/_T \left(1-\cos \phi_{l \nu}\right)}~,
 \end{equation}
where $p^l_T$ is the lepton transverse momentum, and $\phi_{l \nu}$ refers the azimuthal angle difference between the lepton and missing energy momenta.

For the electron channel, each event must have exactly one electron satisfying the conditions stated in Section~\ref{sec:wp_obj}. Any events containing additional electrons or muons with $p_T > 20 \,{\rm GeV}$ are vetoed. Events are then required to satisfy $ E\!\!\!/_T >65 \, {\rm GeV}$ and $m_T(e, E\!\!\!/_T) > 130 \, {\rm GeV}$.
 
For the muon channel, there must be exactly one muon passing the selections listed in Section~\ref{sec:wp_obj}. Events are vetoed if they feature electrons that satisfy both $p_T > 20 \, {\rm GeV}$ and $\Delta R(e, \mu) > 0.1$. Events including any additional muons with $p_T > 20 \, {\rm GeV}$ are also vetoed. The missing transverse energy and the transverse mass must satisfy $ E\!\!\!/_T >55 \, {\rm GeV}$ and $m_T(\mu, E\!\!\!/_T) > 110 \, {\rm GeV}$.
 
\subsection{Validation}\label{wp:sec3}

\subsubsection{Event generation}

\begin{figure}[t]
  \centerline{\includegraphics[width=4in]{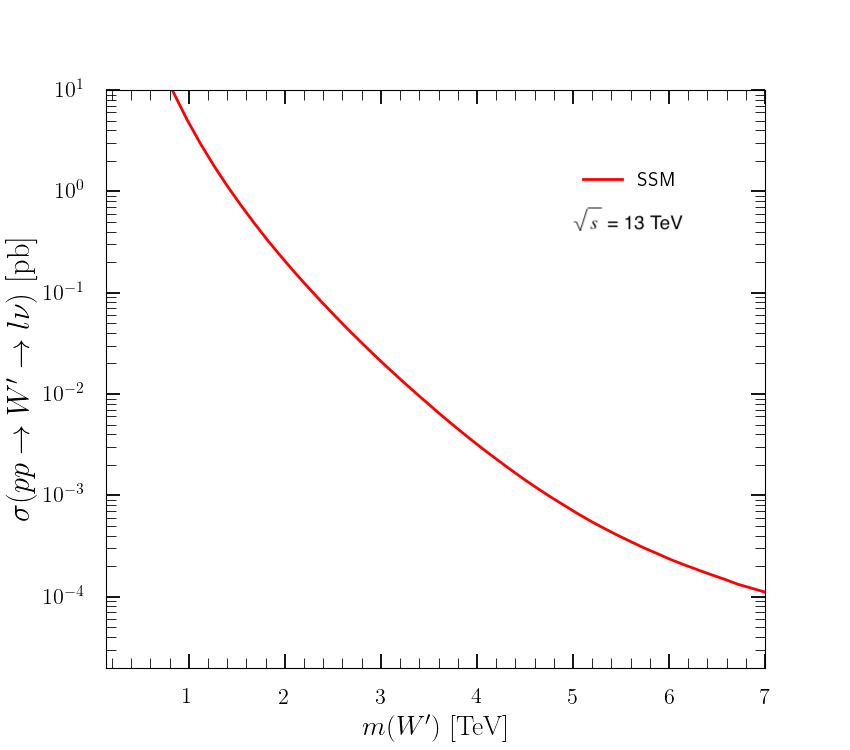}}
  \vspace*{8pt}
  \caption{
Cross-sections for $W'$ production and its decay into one lepton-neutrino pair, at leading order and at $\sqrt{s} = 13 \, {\rm TeV}$ as a function of $W'$ mass. 
\protect\label{wp_fig2}}
\end{figure}

The SSM with heavy gauge bosons has been implemented in the {\sc FeynRules} package\footnote{See the webpage \href{http://feynrules.irmp.ucl.ac.be/wiki/Wprime}{http://feynrules.irmp.ucl.ac.be/wiki/Wprime}.}\cite{Alloul:2013bka}, from which UFO model files have been generated. They are then imported into {\sc \rm MadGraph5\_aMC@NLO}\cite{Alwall:2014hca} to generate the signal samples relevant for the validation of our re-implementation. In the SSM simplified model set-up, we switch off all $W'$ couplings to right-handed SM fermions ($\kappa_R = 0$ in the model conventions), and set the couplings to the left-handed SM fermions to be the same as those of the SM $W$ boson ($\kappa_L = 1$ in the model conventions). The decay width of the $W'$ boson is finally automatically determined by its mass and couplings to fermions within {\sc \rm MadGraph5\_aMC@NLO} by means of {\sc MadSpin}\cite{Artoisenet:2012st} and {\sc MadWidth}\cite{Alwall:2014bza}.

Signal events describing the $pp \rightarrow W' \rightarrow l \, \nu_l$ ($l \,  = \, e, \, \mu$) process are generated\footnote{Events with hadronic taus in the final state are not generated.}. Both on-shell and off-shell heavy gauge boson contributions are included. The interference between the SM contributions and the SSM ones is, however, not considered, since the SM $W$ bosons are mostly produced almost on-shell and the mass gap between the $W$ and $W'$ bosons is much larger than their decay widths. Signal events with various $W'$ masses are generated by {\sc \rm MadGraph5\_aMC@NLO} v$2.6.7$\cite{Alwall:2014hca} at leading order (LO), with the LO set of NNPDF $2.3$ parton densities
with $\alpha_s (m_Z) = 0.130$~\cite{Ball:2012cx}, as obtained from LHAPDF6\cite{Buckley:2014ana}. We use {\sc \rm Pythia} 8.224\cite{Sjostrand:2014zea} for parton showering and hadronisation. 

The following commands were used to generate events in {\sc \rm MadGraph5\_aMC} {\sc \rm @NLO}.
\begin{equation}
\begin{split}
& \tt import \ model \ WEff \_ UFO
\\ & \tt define \ p \ = \ g \ u \ c \ d \ s \ u\sim \ c\sim \ d\sim \ s\sim
\\ & \tt define \ l+ \ = \ e+ \ mu+
\\ & \tt define \ l-  \ = \ e- \ mu-
\\ & \tt define \ vl \ = \ ve \ vm \ vt
\\ & \tt define \ vl\sim \ = \ ve\sim \ vm\sim \ vt\sim
\\ & \tt generate \ p \ p \ > \ wp- \ > \ l- \ vl\sim
\\ & \tt add \ process \ p \ p \ > \ wp+ \ > \ l+ \ vl
\\ & \tt output
\end{split}
\end{equation}
In the {\tt param\_card} file, {\tt kr} and {\tt kl} are set to $0$ and $1$ respectively, and the $W'$ mass {\tt MWP} varies from 2~TeV to 6~TeV with its decay width being automatically calculated. In the {\tt run\_card} file, both {\tt fixed\_ren\_scale} and {\tt fixed\_fac\_scale} are set to {\tt False}, and thus the QCD renormalization and collinear factorization scales are set to the averaged transverse mass of the final state particles. Half a million signal events are generated for each mass point.
The corresponding cross-sections estimated by {\sc \rm MadGraph5\_aMC@NLO} for various $W'$ masses are shown in Fig.~\ref{wp_fig2}. For a mass of $2(6) \, {\rm TeV}$ for example, we obtain the cross-sections of $195 (0.238) \, {\rm fb}$.
Overall, the cross-sections are in agreement with those from Fig.~2 in the considered ATLAS paper\cite{Aad:2019wvl}.

\begin{table}[t]
  \tbl{Comparison of {\sc \rm MadAnalysis 5} and ATLAS predictions for the $m_T$-spectrum in the electron channel ($W' \rightarrow e \,  \nu$). 
   The overflow bins are not accounted for. The relative differences ($\delta $) between our ratios ($R^{MA5}$) and those of ATLAS ($R^{ATLAS}$) are calculated by eq.~\eqref{eq:wp_delta}.
}
  {\begin{tabular}{@{}cccccccc@{}} \toprule
  \scriptsize$W'$ mass &\scriptsize $m_T$ range(${\rm GeV}$) & \scriptsize$130-400$ &\scriptsize$400-600$  &\scriptsize$600-1000$ &\scriptsize $1000-2000$ &\scriptsize $2000-3000$ & \scriptsize$3000-10000$  \\
  \colrule
  &\scriptsize $R^{ATLAS}$ &\scriptsize $0.0341 \pm 0.0019$ &\scriptsize  $0.0428 \pm 0.0029$ & \scriptsize$0.130 \pm 0.009$ & \scriptsize$0.725 \pm 0.046$ &\scriptsize $0.0672 \pm 0.0175$ & \scriptsize $0.000190 \pm 0.000017$   \\
  \scriptsize$ 2\,{\rm TeV}$ & \scriptsize $R^{MA5}$ &\scriptsize $0.0329$  &\scriptsize $0.0429$  & \scriptsize $0.134$  & \scriptsize$0.726$ & \scriptsize$0.0644$  & \scriptsize$0.000173$ \\ 
  & \scriptsize difference($\%$) &\scriptsize $3.50$ &\scriptsize$0.148$ & \scriptsize$2.75$  &\scriptsize$0.0468$  & \scriptsize $4.12$ & \scriptsize$9.25$ \\
    \colrule
  &\scriptsize $R^{ATLAS}$ & \scriptsize $0.0415 \pm 0.0017$&  \scriptsize $0.0355 \pm 0.0019$ &\scriptsize$0.0770 \pm 0.0049$ & \scriptsize $0.272 \pm 0.016$  & \scriptsize$0.522 \pm 0.029$ \scriptsize &$0.0528 \pm 0.0140$ \\
  \scriptsize$ 3\, {\rm TeV}$& $R^{MA5}$ & \scriptsize $0.0409$ & \scriptsize $0.0345$& \scriptsize $0.0779$ & \scriptsize $0.273$ & \scriptsize$0.521$ & \scriptsize $0.0530$  \\ 
  &\scriptsize difference($\%$) &\scriptsize$1.50$&\scriptsize$2.86$ &\scriptsize $1.15$ &\scriptsize$0.423$   &\scriptsize $0.114$&\scriptsize$0.377$ \\
    \colrule
  &\scriptsize $R^{ATLAS}$ & \scriptsize $0.0836 \pm 0.0028$& \scriptsize  $0.0563 \pm 0.0025$ &\scriptsize$0.0943 \pm 0.0052$ & \scriptsize $0.189 \pm 0.010$  & \scriptsize$0.210 \pm 0.011$  &\scriptsize$0.367 \pm 0.019$ \\
  \scriptsize $ 4\, {\rm TeV}$ & $R^{MA5}$ &\scriptsize $0.0812$ &\scriptsize  $0.0566$& \scriptsize $0.0875$ & \scriptsize $0.182$ &\scriptsize $0.212$ &\scriptsize  $0.381$  \\ 
  &\scriptsize difference($\%$) &\scriptsize$2.86$&\scriptsize$0.503$ & \scriptsize$7.24$ &\scriptsize$3.46$ & \scriptsize$0.786$&\scriptsize$3.77$ \\
    \colrule
 &\scriptsize $R^{ATLAS}$ & \scriptsize $0.159 \pm 0.005$& \scriptsize  $0.101 \pm 0.004$ &\scriptsize$0.147 \pm 0.008$ & \scriptsize $0.200 \pm 0.011$  & \scriptsize$0.118 \pm 0.006$  &\scriptsize$0.275 \pm 0.015$ \\
 \scriptsize $5\, {\rm TeV }$&$R^{MA5}$ & \scriptsize$0.159$ &\scriptsize  $0.0955$& \scriptsize $0.136$ &  \scriptsize$0.185$ & \scriptsize$0.119$ & \scriptsize $0.306$  \\ 
  &\scriptsize difference($\%$) &\scriptsize$0.0985$&\scriptsize$5.44$ &\scriptsize $7.69$ &\scriptsize$7.63$ &\scriptsize $1.25$&\scriptsize$11.1$ \\
    \colrule
  & \scriptsize $R^{ATLAS}$ &\scriptsize  $0.226 \pm 0.007$& \scriptsize  $0.141 \pm 0.005$ &\scriptsize$0.197 \pm 0.010$ & \scriptsize $0.230 \pm 0.012$  &\scriptsize $0.0848 \pm 0.0045$  &\scriptsize$0.121 \pm 0.006$ \\
  \scriptsize $6 {\rm TeV}$& $R^{MA5}$ & \scriptsize$0.230$ & \scriptsize $0.135$& \scriptsize $0.186$ & \scriptsize $0.213$ & \scriptsize$0.0844$ & \scriptsize $0.151$  \\ 
  &\scriptsize difference($\%$) &\scriptsize$1.76$&\scriptsize$4.26$ &\scriptsize $5.55$ &\scriptsize$7.22$ & \scriptsize$0.464$&\scriptsize$24.8$ \\
  \botrule
  \end{tabular}\label{wp_ta3} }
\end{table}
\begin{table}[t]
  \tbl{Same as in Table~\ref{wp_ta3}, but for the muon channel ($W' \rightarrow \mu \, \nu$)}
  {\begin{tabular}{@{}cccccccc@{}} \toprule
  \scriptsize $W'$mass & \scriptsize $m_T$ range(${\rm GeV}$) &\scriptsize $110-400$ &\scriptsize$400-600$  &\scriptsize$600-1000$ & \scriptsize$1000-2000$ &\scriptsize $2000-3000$ &\scriptsize $3000-10000$  \\
  \colrule
  & \scriptsize$R^{ATLAS}$ & \scriptsize$0.0443 \pm 0.0026$ & \scriptsize $0.0537 \pm 0.0036$ & \scriptsize$0.152 \pm 0.011$ &\scriptsize $0.606 \pm 0.054$ & \scriptsize$0.138 \pm 0.020$ &\scriptsize$0.00633 \pm 0.00453$   \\
 \scriptsize$2 \, {\rm TeV}$ &\scriptsize $R^{MA5}$ &\scriptsize $0.0416$  & \scriptsize$0.0547$  & \scriptsize $0.165$  & \scriptsize$0.608$ &\scriptsize $0.128$  &\scriptsize $0.00368$ \\ 
  &\scriptsize difference($\%$) &\scriptsize $6.21$ &\scriptsize$1.81$ &\scriptsize $8.47$  &\scriptsize$0.246$  & \scriptsize $7.23$ &\scriptsize $41.9$ \\
    \colrule
  &\scriptsize $R^{ATLAS}$ &\scriptsize $0.0544 \pm 0.0039$ & \scriptsize $0.0426 \pm 0.0033$ &\scriptsize $0.0955 \pm 0.0085$ &\scriptsize $0.302 \pm 0.035$ &\scriptsize $0.356 \pm 0.053$ &\scriptsize $0.150 \pm 0.039$   \\
  \scriptsize $3 \, {\rm TeV} $&\scriptsize $R^{MA5}$ &\scriptsize $0.0491$  &\scriptsize $0.0427$  & \scriptsize $0.104$  &\scriptsize $0.348$ &\scriptsize $0.328$  &\scriptsize $0.128$ \\ 
  &\scriptsize difference($\%$) & \scriptsize$9.84$ &\scriptsize$0.112$ &\scriptsize $9.05$  &\scriptsize$15.3$  & \scriptsize $7.745$ & \scriptsize$14.6$ \\
    \colrule
  &\scriptsize$R^{ATLAS}$&\scriptsize $0.107 \pm 0.010$ & \scriptsize $0.0634 \pm 0.0064$ & \scriptsize$0.105 \pm 0.011$ &\scriptsize $0.212 \pm 0.030$ &\scriptsize $0.224 \pm 0.039$ &\scriptsize$0.289 \pm 0.090$   \\
  \scriptsize $4\, {\rm TeV}$&\scriptsize$R^{MA5}$ &\scriptsize $0.0994$  &\scriptsize $0.0610$  & \scriptsize $0.109$  &\scriptsize $0.240$ &\scriptsize $0.241$  & \scriptsize$0.250$ \\ 
  &\scriptsize difference($\%$) &\scriptsize $7.21$ &\scriptsize$3.73$ &\scriptsize $3.83$  &\scriptsize$13.2$  & \scriptsize $7.78$ & \scriptsize$13.6$ \\
    \colrule
  &\scriptsize $R^{ATLAS}$ &\scriptsize $0.198 \pm 0.015$ & \scriptsize $0.108 \pm 0.009$ & \scriptsize$0.152 \pm 0.013$ &\scriptsize $0.204 \pm 0.023$ & \scriptsize$0.119 \pm 0.020$ & \scriptsize$0.219 \pm 0.072$   \\
  \scriptsize $5\, {\rm TeV}$ &\scriptsize $R^{MA5}$ & \scriptsize$0.184$  &\scriptsize $0.103$  & \scriptsize $0.150$  &\scriptsize $0.210$ &\scriptsize $0.140$  & $0.212$ \\ 
  &\scriptsize difference($\%$) &\scriptsize  $6.86$ & \scriptsize $4.45$ & \scriptsize $1.21$  & \scriptsize $3.01$  & \scriptsize $18.1$ &\scriptsize $3.38$ \\
    \colrule
  &\scriptsize $R^{ATLAS}$ & \scriptsize $0.272 \pm 0.012$ &\scriptsize  $0.146 \pm 0.008$ &\scriptsize $0.194 \pm 0.011$ &\scriptsize $0.215 \pm 0.017$ &\scriptsize $0.0771\pm 0.0109$ &\scriptsize $0.0954 \pm 0.309$   \\
  \scriptsize $6\, {\rm TeV} $&\scriptsize $R^{MA5}$ &\scriptsize $0.262$  &\scriptsize $0.145$  & \scriptsize $0.194$  & \scriptsize$0.217$ &\scriptsize $0.0803$  &\scriptsize $0.103$ \\ 
  &\scriptsize difference($\%$) &\scriptsize $3.67$ &\scriptsize$1.15$ &\scriptsize $0.419$  &\scriptsize$1.02$  & \scriptsize $4.14$ &\scriptsize $7.43$ \\
  \botrule
  \end{tabular}\label{wp_ta4} }
\end{table}

\begin{figure}[t]
 \centerline{\includegraphics[width=4.0in]{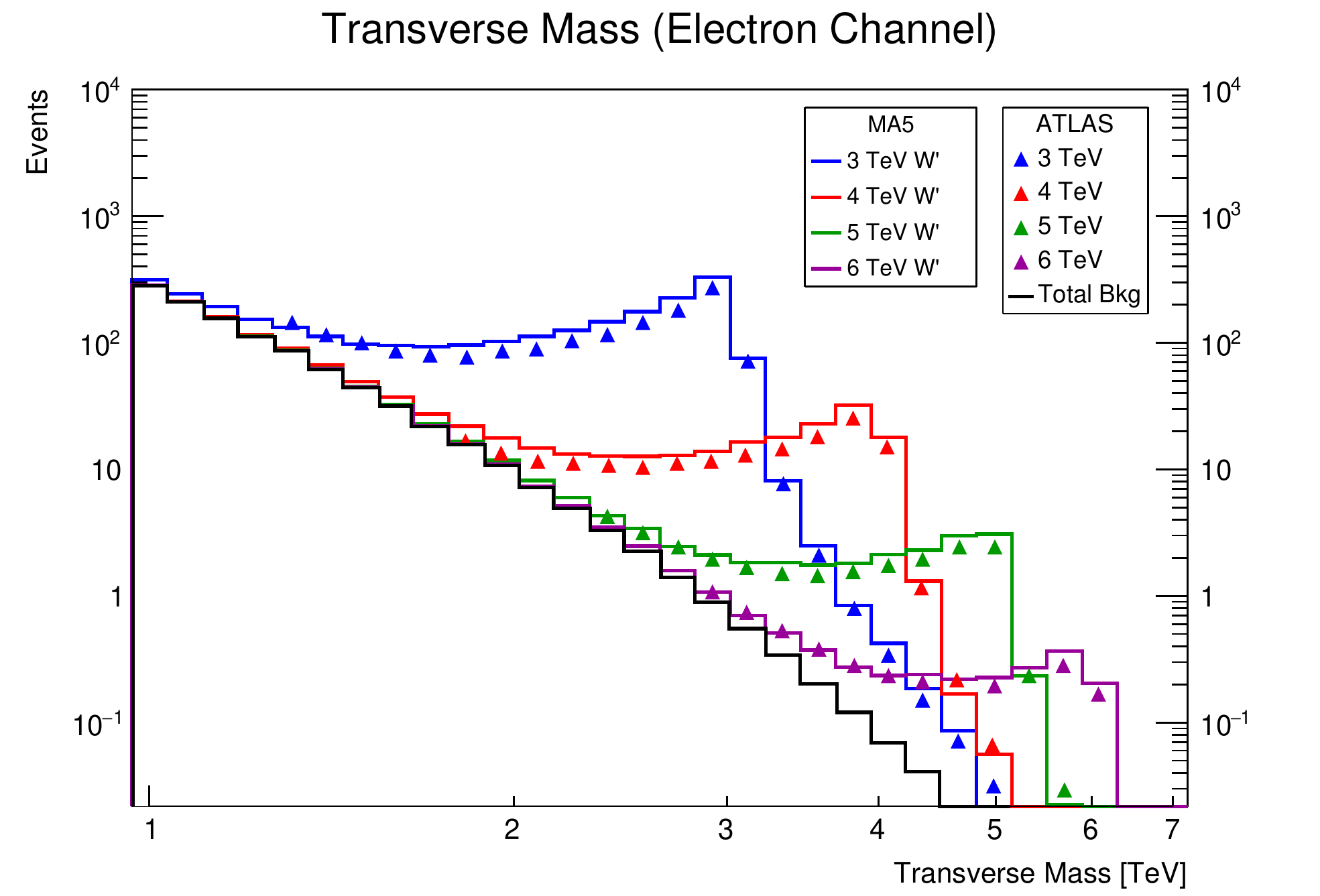}}
  \vspace*{8pt}
  \centerline{\includegraphics[width=4.0in]{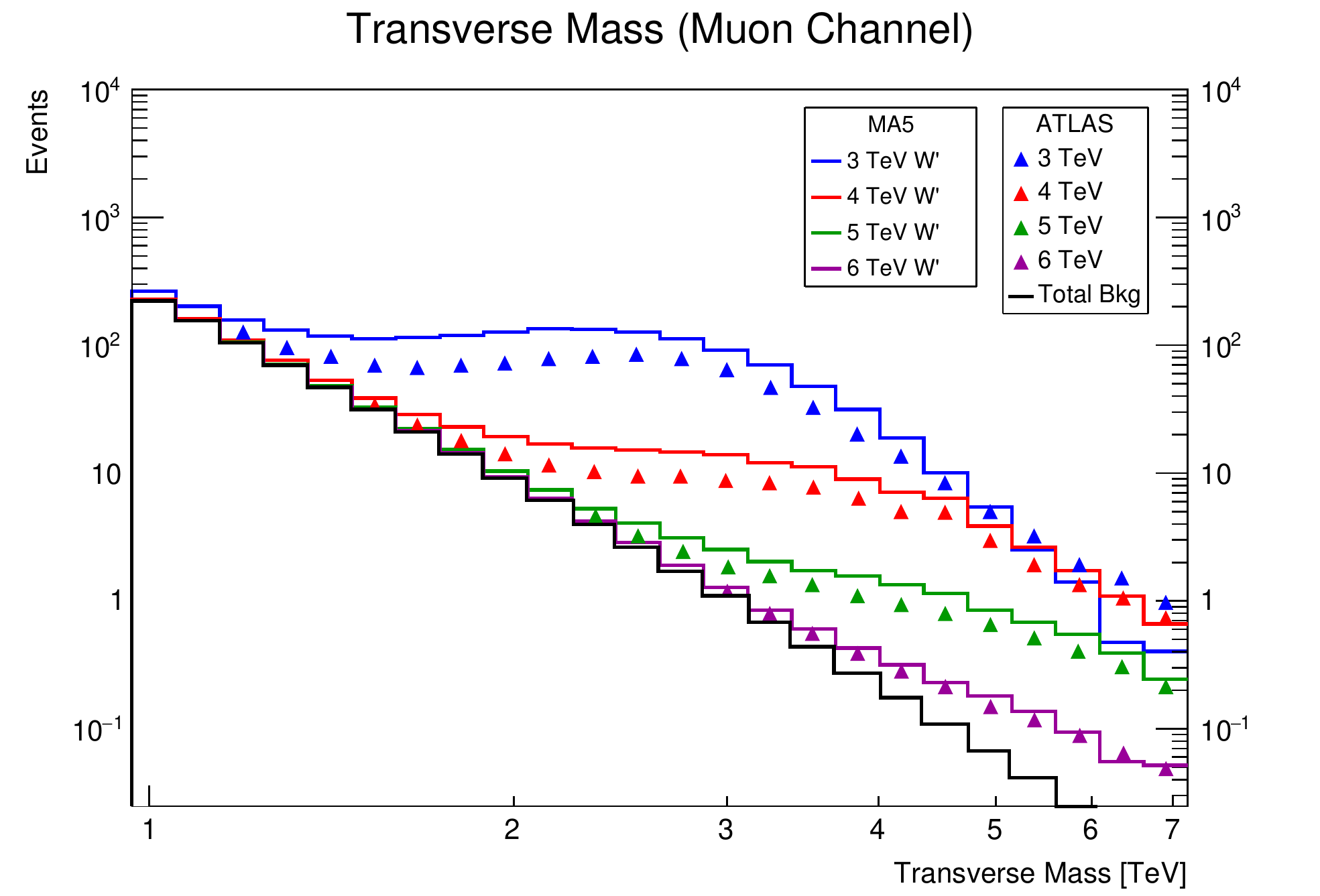}}
 \vspace*{8pt}
  \caption {$m_T$-distributions in the electron channel and muon channel. The solid lines represent our signal predictions for each $W'$ mass in MA5, and the triangle dots represent those of  ATLAS. \protect\label{fig3}}
\end{figure}

\subsubsection{Comparison with ATLAS result for a luminosity of $139 \, {\rm fb}^{-1}$}

In the absence of any official ATLAS cutflow in Ref.~\cite{Aad:2019wvl}, we decided to validate our implementation by comparing the $m_T$-distributions of our $W'$ signal events after all cuts with those of ATLAS. Here, $m_T$ refers to the transverse mass of the system comprising the signal lepton and the missing momentum. 

In Table~\ref{wp_ta3} (Table~\ref{wp_ta4}), we present the comparison of $m_T$-distributions for $m_T$ values ranging from $130$ $(110) \, {\rm GeV}$ to $10 \, {\rm TeV}$ in the electron (muon) channel between our {\sc \rm MadAnalysis 5} (MA5) results and the
ATLAS official results. For each $W'$ mass, there are six signal regions defined according to their $m_T$ ranges. We define the relative differences ($\delta$) between MA5 predictions and ATLAS official estimates as below,
\begin{equation}
 \delta = \frac{\left| R^{ATLAS} - R^{MA5} \right|}{R^{ATLAS}} \times 100 [\%]
\label{eq:wp_delta}\end{equation}
where $R^{MA5}$ and $R^{ATLAS}$ refer to the ratio of the number of events in each region over the total number of events for each $W'$ mass, for our analysis and for ATLAS study respectively. 
The relative differences $(\delta)$ are up to $20\%$ or within the uncertainty range given by ATLAS in most signal regions. In the electron channel, the differences are all below $10\%$ except for $24.8\%$ for the [3, 10] TeV bin. In the muon channel, for regions of $m_T$ below 1 TeV, the differences are all under $10\%$, while some differences reach up to around $15\%$ for those over 1 TeV. There is one $m_T$-region whose relative difference far exceeds $20\%$ --- when the transverse mass lies in the $[3,10]\, {\rm TeV}$ window for scenarios in which 2 ${\rm TeV}$ $W'$ boson decays into a muon-neutrino pair. However, this huge discrepancy can be well resolved when considering the large uncertainty associated with this region that is reported by the ATLAS collaboration. Therefore, we confirm that our reimplementation predictions are in good agreement with the official ATLAS results.

Fig.~\ref{fig3} shows the transverse mass distributions with $W'$ masses varying from 3 ${\rm TeV}$ to 6 ${\rm TeV}$. The signal predictions of our {\sc \rm MadAnalysis 5} implementation as well as those from ATLAS are stacked on top of the total background extracted from the official ATLAS results\cite{Aad:2019wvl}.
In both the electron and muon channels, we obtain a good agreement between the figures of our reimplementation and the original analysis\cite{Aad:2019wvl}.

\subsection{Conclusions}\label{wp:sec4}
We have presented the reimplementation of the heavy charged gauge boson search ATLAS-EXOT-2018-30~\cite{Aad:2019wvl} in the {\sc \rm MadAnalysis 5} framework. Samples of signal events describing the $ pp \rightarrow W' \rightarrow l \, \nu_l$ ($l$ = $e$ or $\mu$) process at $\sqrt{s} = 13\, {\rm TeV}$ in the sequential standard model have been generated with {\sc \rm MadGraph5\_aMC@NLO} at LO, and the simulation of the ATLAS detector has been achieved with {\sc \rm Delphes 3}. 
We have compared predictions made by {\sc \rm MadAnalysis 5} with the results provided by the ATLAS collaboration. We have considered various benchmark scenarios in both the electron and the muon channel, where a good agreement at the level of $m_T$ spectra is achieved between our reinterpretation and ATLAS results. Relative differences of at most 20\% have been observed, with the most extreme discrepancy being well explained by the large uncertainty populating the corresponding signal region.

The material that has been used for the validation of this implementation is available, together with the {\sc MadAnalysis}~5 C++ code, at the MA5 dataverse (\href{https://doi.org/10.14428/DVN/GLWLTF}{https://doi.org/10.14428/DVN/GLWLTF})~\cite{GLWLTF_2020}.

\subsection*{Acknowledgments}
We are very grateful to the ATLAS EXOT conveners for their help and all the additional information they provided, and to Magnar Bugge in particular who were invaluable in our validation process. We would also like to express our sincere gratitude to all the tutors at the second {\sc \rm MadAnalysis 5}  workshop on LHC recasting.

\cleardoublepage
\newcommand{\amc}{{\sc MadGraph5}\_a{\sc MC@NLO}}
\newcommand{\fr}{{\sc Feyn\-Rules}}
\newcommand{\fa}{{\sc Feyn\-Arts}}
\newcommand{\nloct}{{\sc NloCT}}
\newcommand{\mloop}{{\sc MadLoop}}
\newcommand{\mspin}{{\sc MadSpin}}
\newcommand{\mw}{{\sc MadWidth}}
\newcommand{\ma}{{\sc MadAnalysis~5}}
\newcommand{\py}{{\sc Pythia~8}}
\newcommand{\del}{{\sc Delphes~3}}
\newcommand{\fj}{{\sc FastJet}}
\newcommand{\gev}{\textrm{GeV}}
\newcommand{\MET}{$\vec{p}_{\textrm{miss}}$}

\newcommand{\be}{\begin{equation}}
\newcommand{\ee}{\end{equation}}
\def\bpm{\begin{pmatrix}}
\def\epm{\end{pmatrix}}
\def\bsp#1\esp{\begin{split}#1\end{split}}

\markboth{Benjamin Fuks and Adil Jueid}{Implementation of the CMS-EXO-17-015 analysis}

\section{Implementation of the CMS-EXO-17-015 analysis (leptoquark and dark matter with one muon,
  one jet and missing transverse energy; 77.4~fb$^{-1}$)}
  \vspace*{-.1cm}\footnotesize{\hspace{.5cm}By Benjamin Fuks and Adil Jueid}
\label{sec:lq}



\subsection{Introduction}
\label{sec:intro}
Since the discovery of the Higgs boson, the Standard Model (SM) of particle
physics is considered to be a good low energy approximation of a more complete,
yet undiscovered, theoretical framework. Such a theoretical framework may in
particular be able to address questions such as the nature of dark matter (DM)
in the universe, among many other interesting issues. Unfortunately, only little
is known about the true nature of DM, despite the extensive searches carried out
both in laboratories and astrophysical experiments.

At the LHC, one of the most known of and used strategies consists of looking for
the presence of a significant excess in the tail of the missing transverse
energy ($|$\MET$|$) distribution. A specific emphasis is put on a signature
comprised of dark matter particles recoiling against a visible hard
SM object like a photon, a jet, an electroweak gauge boson or even an SM Higgs
boson or a top quark~\cite{Feng:2005gj,Birkedal:2004xn,Bai:2010hh,Fox:2011fx,%
Bell:2012rg,Petrov:2013nia,Bai:2012xg,Andrea:2011ws}. Multiple associated searches have been
conducted by the ATLAS and the CMS collaborations, the most
recent ones analysing data recorded during the LHC Run~2~\cite{Aaboud:2016uro, Aaboud:2016tnv, Aaboud:2016qgg, Aaboud:2016obm,Aaboud:2017dor, Aaboud:2017uak, Aaboud:2017yqz, Aaboud:2017bja, Aaboud:2017rzf, Aaboud:2017phn, Aaboud:2018xdl, Sirunyan:2017hci, Sirunyan:2017hnk, Sirunyan:2017xgm, Sirunyan:2017qfc, Sirunyan:2017jix, Sirunyan:2018gka, Sirunyan:2018fpy, Sirunyan:2018dub, Sirunyan:2018gdw, Sirunyan:2019gfm, Sirunyan:2019zav, Sirunyan:2020fwm}. Consequently to the absence of direct evidence for
the existence of DM so far, these results have been used to severely constrain
the DM couplings and masses in large classes of new physics scenarios. In particular, the absence
of any DM signal at the LHC in the so-called thermal freeze-out mechanism has
called for either going beyond standard freeze-out, or investigating
alternative models.

One of the most attractive of those contexts
is the so-called co-annihilation paradigm in which DM is produced in
association with beyond-the-SM partners very close in mass.
In the framework developed in Ref.~\cite{Baker:2015qna}, the SM field content is extended by a
scalar leptoquark doublet $M_s$,
a weak doublet of Dirac fermions $X$ and a Majorana fermion
$\chi$ that plays the role of dark matter. These new states have the
following assignments under the SM gauge group $SU(3)_c \otimes SU(2)_L \otimes
U(1)_Y$,
\be
  M_s\equiv \bpm M^u_s\\M^d_s\epm: ({\bf 3}, {\bf 2}, 7/3) \qquad
  X \equiv \bpm X_u\\X_d\epm: ({\bf 3}, {\bf 2}, 7/3),
  \qquad \chi: ({\bf 1}, {\bf 1}, 0),
\ee
and the relevant interaction Lagrangian ${\cal L}_{\rm NP}$ can be written as
\begin{eqnarray}
  \mathcal{L}_{\rm NP} = - \left(y_D \overline{X} M_s \chi + y_{Q\ell} \overline{Q}_L M_s \ell_R + y_{L u} \overline{L}_L M_s^c u_R  + \textrm{h.c.} \right).
\label{eq:Lagrangian}
\end{eqnarray}

In parallel, the LHC collaborations developed search strategies dedicated to this
class of models. The CMS-EXO-17-015 analysis~\cite{Sirunyan:2018xtm} considered
in this proceedings contribution is one of these. In this analysis, the CMS
collaboration has focused on one of the benchmarks detailed in
Ref.~\cite{Baker:2015qna}. It assumes that
$y_{Lu} = 0$, and the other model parameters are chosen as
\begin{eqnarray}
y_{Q\ell} = \sqrt{2}/10,~ y_D = 0.1, ~\and~ \Delta = \frac{M_X - m_\chi}{m_\chi} = 0.1.
\label{eq:benchmark}
\end{eqnarray}
In this note, we report on the implementation of this CMS-EXO-17-015 analysis in
the \ma\ framework~\cite{Conte:2012fm, Conte:2014zja, Dumont:2014tja,
Conte:2018vmg, Araz:2019otb}. The relevant code for the \ma~ implementation can be found 
in \url{https://doi.org/10.14428/DVN/ICOXG9}. 
In Sec.~\ref{sec:analysis}, we describe the analysis that we implemented,
including a detailed description of the object definitions and  event selection
strategy. We discuss the validation of our implementation, focusing both on the
Monte Carlo event generation necessary for this task and on a comparison of the
\ma\ predictions with the official CMS results, in Sec.~\ref{sec:lq_validation}. We
summarise our work in Sec.~\ref{sec:conclusion}.

\subsection{Description of the analysis}
\label{sec:analysis}
\begin{figure}[t]
  \centerline{\includegraphics[width=0.6\textwidth]{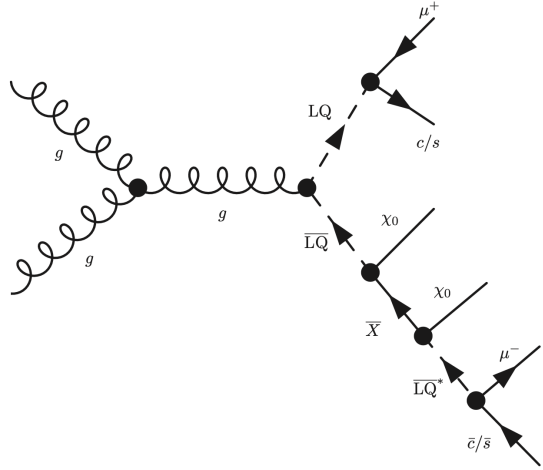}}
  \vspace*{8pt}
  \caption{Representative parton-level Feynman diagram illustrating the class of
    processes probed by the CMS-EXO-17-015 analysis. \label{fig:FD}}
  \end{figure}

In the considered theoretical framework, leptoquark (LQ) pair production and
decay lead to several signatures, their respective relevance depending on the LQ branching
ratios. In the CMS-EXO-17-015 analysis, the final state under consideration is
comprised of one isolated muon, one jet and a large amount of missing transverse
energy. This process is illustrated by the Feynman diagram shown in
Fig.~\ref{fig:FD}.

\subsubsection{Object definitions}
As above-mentioned, the considered analysis relies on the presence of hard
final-state jets and muons, as well as on the one of a large amount of missing
transverse energy. In addition, a veto is imposed on the presence of final-state
objects of a different nature.

Candidate muons (the leading one being assumed to originate from the decay of a
LQ, as illustrated by the Feynman diagram of
Fig.~\ref{fig:FD}) are required to satisfy tight selection
criteria~\cite{Sirunyan:2018fpa}. Moreover, their transverse momentum $p_T^\mu$
and pseudorapidity $\eta^\mu$ must fulfil
\begin{eqnarray}
p_T^\mu > 60~\textrm{GeV},~\textrm{and}~ |\eta^\mu| < 2.4.
\end{eqnarray}
In addition, those muons are enforced to be isolated to suppress any potential
contribution of muons arising from hadronic decays. This relies on an isolation
variable $I$ defined by 
\begin{equation}
   I \equiv \frac{1}{p_T^\mu} \sum_i p_{T,i},
\end{equation}
with the sum running over all photon, neutral hadron and charged hadron candidates
reconstructed within a distance, in the transverse plane, of $\Delta R \equiv
\sqrt{(\Delta \eta)^2 + (\Delta \phi)^2} = 0.4$ around the muon direction. This
isolation variable is required to satisfy  $I < 0.15$.
On the other hand, the analysis also makes use of \emph{loose muons} to reduce
the contribution of $Z$+jets background events (see below). Those are required
to satisfy $I < 0.25$ and $p_T^\mu > 10~\gev$. 

Reconstructed electron candidates are required to have a transverse momentum
$p_T^e > 15~\gev$ and a pseudorapidity $|\eta^e| < 2.5$. Moreover, they are
considered only if they satisfy loose identification
criteria~\cite{Khachatryan:2015hwa}. Hadronically decaying
tau leptons ($\tau_h$) are also identified through loose
criteria~\cite{Sirunyan:2018pgf}, their selection additionally enforcing
$p_T^\tau > 20~\gev~\and~ |\eta^\tau| < 2.3$. 

Jets are reconstructed by means of the anti-$k_T$ algorithm~\cite{Cacciari:2008gp},
with a radius parameter $R=0.4$\footnote{In the CMS-EXO-17-015 search, jet
clustering excludes the charged-particle tracks that are not associated with the
primary interaction vertex. This is irrelevant for our reimplementation as we
neglect any potential pile-up effects.}. The signal jet
collection is then comprised of all jets whose pseudorapidity satisfies
$|\eta^j|<2.4$. The Combined Secondary Vertex (CSVv2) algorithm is then used to
identify the jets originating from the fragmentation of a $b$-quark, the analysis
making use of its tight working point~\cite{Sirunyan:2017ezt}.
The corresponding $b$-tagging efficiency is given by
\begin{equation}
  \hspace{-.2cm} \mathcal{E}_{b|b}(p_T)   = \begin{cases}
   -0.033 + 0.0225~p_T - 3.5\!\cdot\! 10^{-4}~p_T^2 + 2.586\!\cdot\! 10^{-6}~p_T^3
- 9.096 \!\cdot\! 10^{-9}~p_T^4 \\ 
\quad + 1.212 \!\cdot\! 10^{-11}~ p_T^5 \quad \ \ \textrm{ for } ~20 ~\gev < p_T \leqslant 50~\gev\ , \\[0.5em] 
   0.169 + 0.013~p_T - 1.9\!\cdot\! 10^{-4}~p_T^2 + 1.373\!\cdot\! 10^{-6}~p_T^3 - 4.923\!\cdot\! 10^{-9}~p_T^4 \\ \quad + 6.87\!\cdot\! 10^{-12}~p_T^5 \quad \ \
   \textrm{ for } ~50~\gev < p_T \leqslant 160~\gev\ , \\[0.5em]
   0.62 - 8.3\!\cdot\! 10^{-4}~p_T + 4.3078\!\cdot\! 10^{-7}~p_T^2 \\ 
  \quad \ \ \textrm{ for } ~160~\gev < p_T \leqslant 1000~\gev ,
      \end{cases}
\end{equation}
while the associated mistagging probabilities of a charmed
jet ($\mathcal{E}_{c|b}$) and a light jet ($\mathcal{E}_{j|b}$) as a $b$-jet are
given by
\be\bsp
  \mathcal{E}_{c|b}(p_T) =&  \begin{cases}
     0.0234 - 8.417\!\cdot\! 10^{-5}~p_T + 1.24\!\cdot\! 10^{-6}~p_T^2 - 5.5\!\cdot\!10^{-9}~p_T^3 + 9.96 \!\cdot\! 10^{-12}~p_T^4 \\ 
- 6.32\!\cdot\! 10^{-15}~p_T^5 \quad \ \textrm{ for } ~20 ~\gev <  p_T \leqslant 65 ~\gev\ , \\[0.5em]
     0.0218 + 2.46 \!\cdot\! 10^{-5}~p_T - 2.021 \!\cdot\! 10^{-8}~p_T^2 \\ 
     \quad \ \textrm{ for } ~65 ~\gev <  p_T \leqslant 1000 ~\gev ,
  \end{cases}\\
  \mathcal{E}_{j|b}(p_T) =& \begin{cases}
     0.00284 - 8.63\!\cdot\!10^{-5}~p_T + 1.38\!\cdot\! 10^{-6} ~p_T^2
      - 9.69 \!\cdot\! 10^{-9}~p_T^3
      + 3.19 \!\cdot\! 10^{-11}~p_T^4 \\ 
      - 3.97 \!\cdot\! 10^{-14}~p_T^5 \quad 
      \  \textrm{ for } ~20 ~\gev <  p_T \leqslant 150 ~\gev, \\[0.5em]
        6.3\!\cdot\!10^{-4} + 4.51\!\cdot\!10^{-6}~p_T  + 2.83 \!\cdot\! 10^{-9}~p_T^2 \\ \quad
        \ \textrm{ for }  150  ~\gev < p_T \leqslant 1000 ~\gev.
  \end{cases}
\esp\ee

Finally, one defines the missing transverse energy as the magnitude of the
transverse momentum imbalance (\MET), which is computed as the opposite of the
vectorial sum of the transverse momentum of all reconstructed objects,
$$ 
\vec{p}_{\textrm{miss}} = - \left[\sum_{\textrm{electrons}} \vec{p}_T + \sum_{\textrm{muons}} \vec{p}_T + 
\sum_{\textrm{photons}} \vec{p}_T +
\sum_{\textrm{hadrons}} \vec{p}_T \right].
$$
In our simulation setup, we have implemented the above parametrisations in a
customized \del\ card that has then been used for the simulation of the
CMS detector response.

\subsubsection{Event selection}
The CMS-EXO-17-015 event selection strategy includes two stages, namely a preselection and the
definition of a signal region that we coin, in the following,
\texttt{SignalRegion}.

In the preselection procedure, events are first selected by requiring the
presence of at least one tightly isolated muon with $p_T > 60~\gev$ and $|\eta|
< 2.4$. The leading jet is then required to satisfy
$p_T > 100~\gev$ and to be separated from the leading muon in the transverse
plane by $\Delta R > 0.5$. Events satisfying those criteria are assumed to be
compatible with the production of a leptoquark that decays into those leading jet
and muon.

As a next step, several vetoes are applied to reduce the contamination of the
overwhelming
$t\bar{t}$, $W+$~jets and $Z+$~jets backgrounds. First, events are vetoed if at
least one $b$-tagged jet is present. Moreover, a veto on events featuring either a loose
electron candidate or a hadronic tau candidate is applied. These three vetoes are
necessary to jointly suppress the $t\bar{t}$ background, while the electron and
tau vetoes specifically suppress the  $W/Z+$~jets contributions.

Next, the transverse mass ($m_T$) of the
system comprised of the leading muon and the missing momentum is used to further
suppress the $W+$~jets background: One imposes that $m_T > 100~\gev$.
In addition, the contribution of the $Z+$~jets background is further reduced
by rejecting events that contain one extra loosely identified muon candidate with
an electric charge that is opposite to the one of the leading muon, and for which
\begin{equation}
|m_{\mu \mu} - M_Z| < 10~\textrm{GeV}.
\end{equation}
In this expression, $m_{\mu \mu}$ stands for the invariant mass of the system
comprised of this muon and the leading muon, such a system being thus
constrained to be incompatible with the decay of an on-shell $Z$-boson, if
present in the event final state.

Finally, the preselection ends by an extra requirement on the missing momentum
\MET\ that is enforced to be well separated in azimuth from the leading jet and
the leading muon. We require
\begin{equation}
\Delta\phi(\textrm{jet}, \vec{p}_\textrm{miss}) > 0.5 ~\and~ \Delta\phi(\mu, \vec{p}_\textrm{miss}) > 0.5,
\end{equation}
with $\Delta\phi(i,j) = |\phi_i - \phi_j|$.
Whereas these last requirements have very minor effects on the
considered signal, they allow in particular for the suppression of the multijet
background. For this
reason, while implemented in our recasting code, they will be absent from the
cut-flow tables presented in the next section.

After this preselection, the signal region is defined by a more stringent cut on
the $m_T$ variable,
\begin{equation}
m_T > 500~\textrm{GeV}.
\end{equation}
A summary of the full set of selection cuts is presented in table \ref{tab:selection}.

\begin{table}[t]
  \setlength\tabcolsep{10pt}
  \tbl{Selection cuts defining the unique CMS-EXO-17-015 signal region. The
    first column introduces our naming scheme for each cut, as used in the
    cut-flow tables presented in the next section.}
  {\begin{tabular}{@{}cccccc@{}} \toprule
    \multicolumn{6}{c}{\textbf{Basic requirements}} \\[.2cm]
    \texttt{SignalMuon} & \multicolumn{5}{|p{10cm}}{At least one isolated muon
      with $p_T > 60~\gev$ and $|\eta| < 2.4$.} \\[.1cm]
    \texttt{SignalJet} & \multicolumn{5}{|p{10cm}}{The leading jet should fulfil
      $p_T > 100~\gev$, $|\eta| < 2.4$ and be separated by $\Delta R > 0.5$ from
      the leading muon.} \\[.2cm]
    \colrule
    \multicolumn{6}{c}{\textbf{Vetoes}} \\[.2cm]
    $b$-\texttt{Veto} & \multicolumn{5}{|p{10cm}}{Veto of events featuring at
      least one $b$-jet with $p_T > 30~\gev$ and $|\eta| < 2.4$.} \\[.1cm]
    tau-\texttt{Veto} & \multicolumn{5}{|p{10cm}}{Veto of events featuring at
      least one hadronic tau with $p_T > 20~\gev$ and $|\eta| < 2.3$.} \\[.1cm]
    $e$-\texttt{Veto} & \multicolumn{5}{|p{10cm}}{Veto of events featuring at
      least one loosely reconstructed electron with $p_T > 15~\gev$ and $|\eta|<2.4$.} \\[.2cm]
    \colrule
    \multicolumn{6}{c}{\textbf{Further preselection requirements}} \\[.2cm]
    \texttt{ZMassWindow} & \multicolumn{5}{|p{10cm}}{No extra loose muon that
      could arise, together with the leading muon, from a
      $Z$-boson decay ({\it i.e.} if $|m_{\mu\mu} - M_Z| < 10~\gev$).} \\[.5cm]
    \MET-\texttt{threshold} & \multicolumn{5}{|p{10cm}}{$|\vec{p}_{\textrm{miss}}|  > 100~\gev$.} \\[.1cm]
    $m_T$-\texttt{threshold} & \multicolumn{5}{|p{10cm}}{The transverse mass of
      the muon-\MET\ system must fulfil $m_T > 100~\gev$.} \\[.2cm]
    \colrule
    \multicolumn{6}{c}{\textbf{Signal region}} \\[.2cm]
    \texttt{SignalRegion} & \multicolumn{5}{|p{10cm}}{Extra $m_T$ requirement: $m_T > 500~\gev$.} \\[.1cm]
  \botrule
  \end{tabular}\label{tab:selection} }
\end{table}

\subsection{Validation}
\label{sec:lq_validation}
\subsubsection{Event generation}\label{sec:MCgen}
For the validation of our implementation of the  CMS-EXO-17-015 analysis, we
generate events describing the dynamics of the signal of Fig.~\ref{fig:FD}
in the context of the model introduced in Sec.~\ref{sec:intro}. We use
\amc~\cite{Alwall:2014hca} to simulate hard-scattering events at the leading
order (LO) in the strong coupling, excluding the potentially relevant
$t$-channel leptonic exchange diagrams~\cite{Borschensky:2020hot}. In our
procedure, we convolute the LO matrix elements with the LO set of NNPDF~3.0
parton distribution functions in the four-flavour-number scheme, and with
$\alpha_s(M_Z) = 0.130$. Moreover, we set the renormalisation and factorisation
scales to the average transverse mass of the final-state objects.

We use \py~ (version 8.432)~\cite{Sjostrand:2014zea} to match the fixed-order
results with parton showers and to deal with the hadronisation of the resulting
partons, after ignoring multi-parton interactions. The response of the CMS
detector is then modeled by means of the fast detector simulation toolkit \del\
(version 3.4.2)~\cite{deFavereau:2013fsa}, that internally relies on \fj\ (version
3.3.0)~\cite{Cacciari:2011ma} for jet clustering. In this last step, we have
designed a customized \del~ parametrisation that accurately matches the actual
CMS performance working points of the analysis. This card is available, together
with our code, from the \ma\ Physics Analysis Database (PAD)\footnote{See the webpage
\url{http://madanalysis.irmp.ucl.ac.be/wiki/PublicAnalysisDatabase}.}.

\begin{table}[t]
  \setlength\tabcolsep{25pt}
  \tbl{Cut-flow charts associated with the CMS-EXO-17-015 analysis and the
    process depicted in Fig.~\ref{fig:FD} for the two
    benchmark scenarios \texttt{BP1} (upper panel) and \texttt{BP2} (lower
    panel). We show results obtained with
    \ma\ (second column) and those provided by the CMS collaboration (third
    column). The numbers inside brackets correspond to the selection efficiency
    of each cut and the ratio $\mathcal{R}_i$ depicting the differences between our
    predictions and the official CMS results is defined in
    Eq.~\eqref{eq:deviation}.}
{\begin{tabular}{@{}cccc@{}} \toprule
Cut &  \ma & CMS & $\mathcal{R}_i$  \\
\colrule
Initial events &   $99977~(100\%)$    &    $99977~(100\%)$   &     $0$ \\[.1cm]
\texttt{SignalMuon}   & $88583~(88.66\%)$ & $90104~(90.12\%)$ & $1.61 \times 10^{-2}$ \\[.1cm]
\texttt{SignalJet} &  $85594~(96.62\%)$   &    $88100~(97.77\%)$    &    $1.17 \times 10^{-2}$ \\[.1cm]
$b$-\texttt{Veto} &  $79367~(92.72\%)$    &    $84282~(95.66\%)$   &     $3.07 \times 10^{-2}$ \\[.1cm]
$\tau_h$-\texttt{Veto} &  $75572~(95.21\%)$ &       $83373~(98.92\%)$  &     $3.75 \times 10^{-2}$ \\[.1cm]
$e$-\texttt{Veto}  &   $75534~(99.94\%)$ & $83175~(99.7\%)$ & $2.41 \times 10^{-3}$ \\[.1cm]
\texttt{ZMassWindow}  &   $71795~(95.05\%)$ &       $81344~(97.79\%)$ & $2.80 \times 10^{-2}$ \\[.1cm]
\MET-\texttt{threshold} &   $69957~(97.44\%)$ & $78665~(96.71\%)$ & $1.71 \times 10^{-2}$ \\[.1cm]
$m_T$-\texttt{threshold} &   $65957~(94.29\%)$ & $74796~(95.08\%)$ & $8.30 \times 10^{-3}$ \\[.1cm]
\texttt{SignalRegion}  &   $51151~(77.75\%)$  & $54849~(73.33\%)$ & $6.02 \times 10^{-2}$ \\[.1cm]
\botrule\\[0.2cm]
Cut &  \ma & CMS & $\mathcal{R}_i$  \\
\colrule
Initial events &   $3996~(100\%)$    &    $3996~(100\%)$   &     $0$ \\[.1cm]
\texttt{SignalMuon}   & $3519~(88.07\%)$ & $3625~(90.71\%)$ & $2.90 \times 10^{-2}$ \\[.1cm]
\texttt{SignalJet} &  $3441~(97.80\%)$   &    $3586~(98.92\%)$    &    $1.12 \times 10^{-2}$ \\[.1cm]
$b$-\texttt{Veto} &  $3185~(92.57\%)$    &    $3433~(95.73\%)$   &     $3.33 \times 10^{-2}$ \\[.1cm]
$\tau_h$-\texttt{Veto} &  $3026~(95.01\%)$ &       $3401~(99.06\%)$  &     $4.08 \times 10^{-2}$ \\[.1cm]
$e$-\texttt{Veto} &   $3024~(99.93\%)$ & $3392~(99.73\%)$ & $2.06 \times 10^{-3}$ \\[.1cm]
\texttt{ZMassWindow} &   $2936~(97.11\%)$ &       $3327~(98.08\%)$ & $9.88 \times 10^{-3}$ \\[.1cm]
\MET-\texttt{threshold} &   $2897~(98.69\%)$ & $3277~(98.49\%)$ & $2.07 \times 10^{-3}$ \\[.1cm]
$m_T$-\texttt{threshold} &   $2678~(92.44\%)$ & $3160~(96.42\%)$ & $4.12 \times 10^{-2}$ \\[.1cm]
\texttt{SignalRegion} &   $2162~(80.74\%)$  & $2611~(82.62\%)$ & $2.27 \times 10^{-2}$ \\[.1cm]
\botrule
\end{tabular} \label{tab:cutflow-MLQ-1500}}
\end{table}

For the results presented in the rest of this contribution, we have generated
$200,000$ events for two benchmark points {\tt BP1} and {\tt BP2} defined by
\be\bsp
\texttt{BP1}:& \quad M_{\textrm{LQ}} = 1000~\gev,~\and~m_\chi = 400~\gev,\\
\texttt{BP2}:& \quad M_{\textrm{LQ}} = 1500~\gev,~\and~m_\chi = 600~\gev,
\esp\ee
with the other parameters fixed as in Eq.~\eqref{eq:benchmark}. About 102,326
(108,208) events pass all the selection criteria of the CMS analysis in the
framework of the \texttt{BP1} (\texttt{BP2}) scenario.

\subsubsection{Results}
In order to validate our implementation, we compare predictions obtained with
our implementation in \ma\ to the official results provided by the CMS
collaboration for the two benchmark scenarios {\tt BP1} and {\tt BP2} defined in
Sec.~\ref{sec:MCgen}. Our comparison is performed in two stages. First, we study the
resulting cut-flow tables. Next, we investigate the shape of the distributions of
several key observables.

To quantify the level of agreement between our results and the CMS ones at each
selection step of the cut-flow, we introduce a quantity ${\cal R}_i$ defined by
\begin{eqnarray}
  \mathcal{R} = \bigg|1 - \frac{\epsilon_{MA5}^i}{\epsilon_{CMS}^i}\bigg|,
\label{eq:deviation}
\end{eqnarray}
with $\epsilon^i$ being the selection efficiency of the $i^{\rm th}$ cut $i$,
\be
   \epsilon^i = \frac{n^i}{n^{i-1}}.
\ee
In this notation, $n_{i-1}$ events survive before the $i^{\rm th}$ cut, and
$n_i$ events survive after this cut. We present the results in the two panels of
table~\ref{tab:cutflow-MLQ-1500} for the {\tt BP1} and {\tt BP2} setup
respectively, after normalising our results to the same cross section as the
one used by the CMS collaboration in their analysis. We obtain an excellent level
of agreement, reaching $\mathcal{R}\sim 10^{-3}-10^{-2}$.

\begin{figure}[t]
  \includegraphics[width=0.49\textwidth]{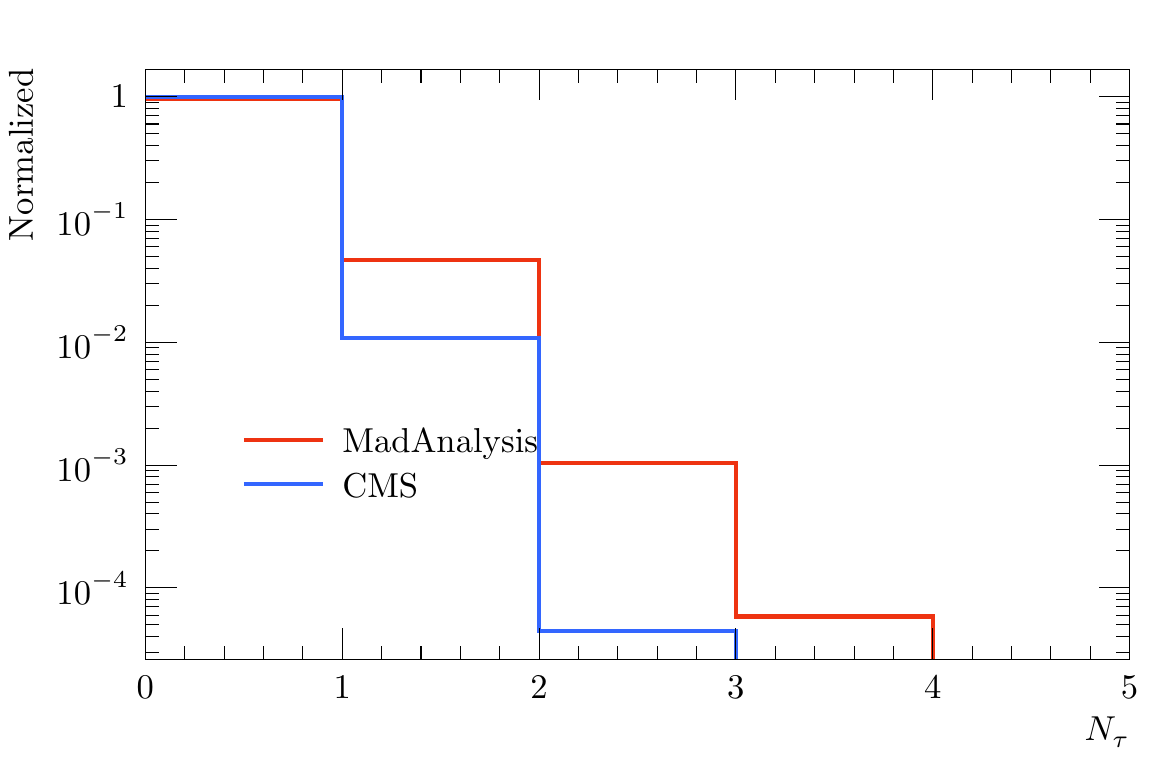}
  \includegraphics[width=0.49\textwidth]{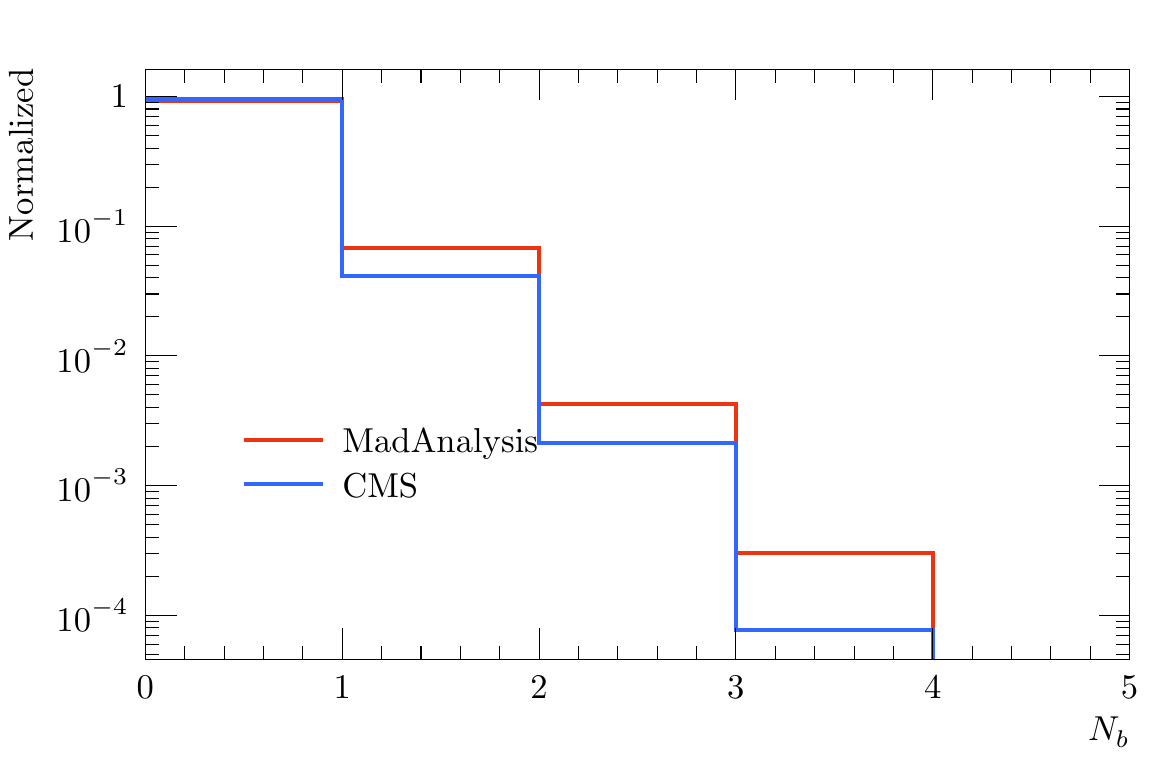}
  \includegraphics[width=0.49\textwidth]{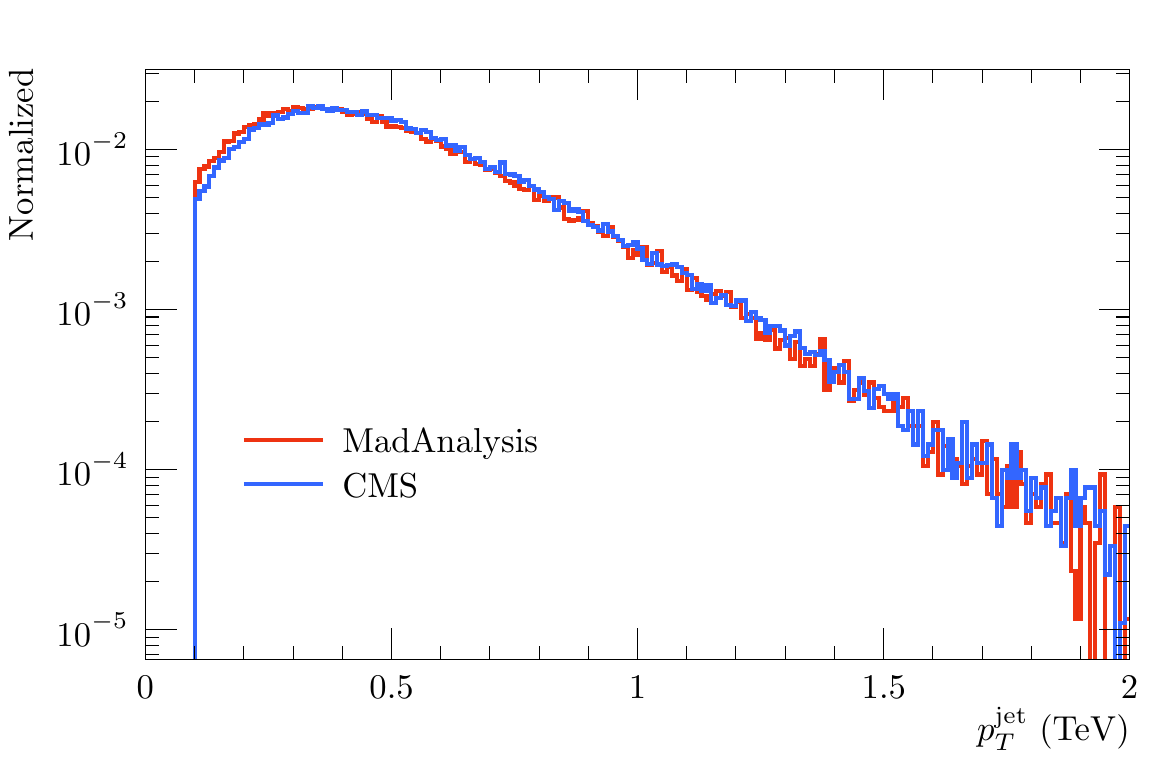}
  \includegraphics[width=0.49\textwidth]{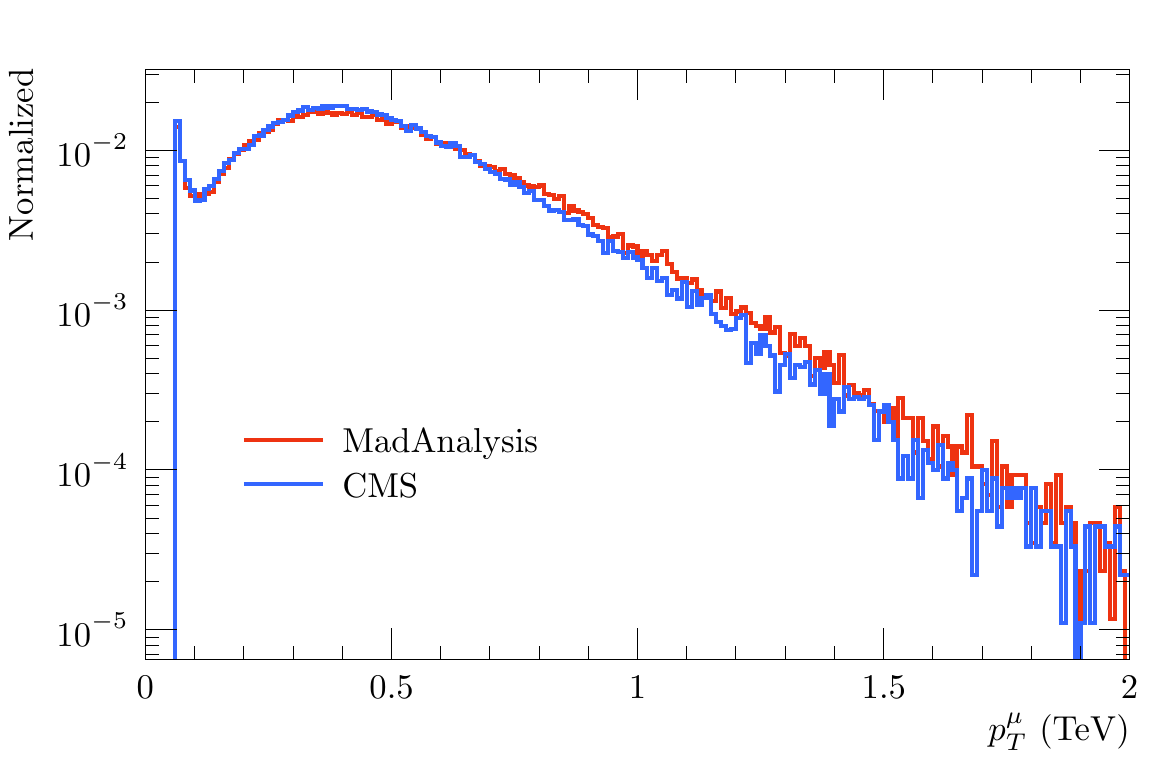}
  \includegraphics[width=0.49\textwidth]{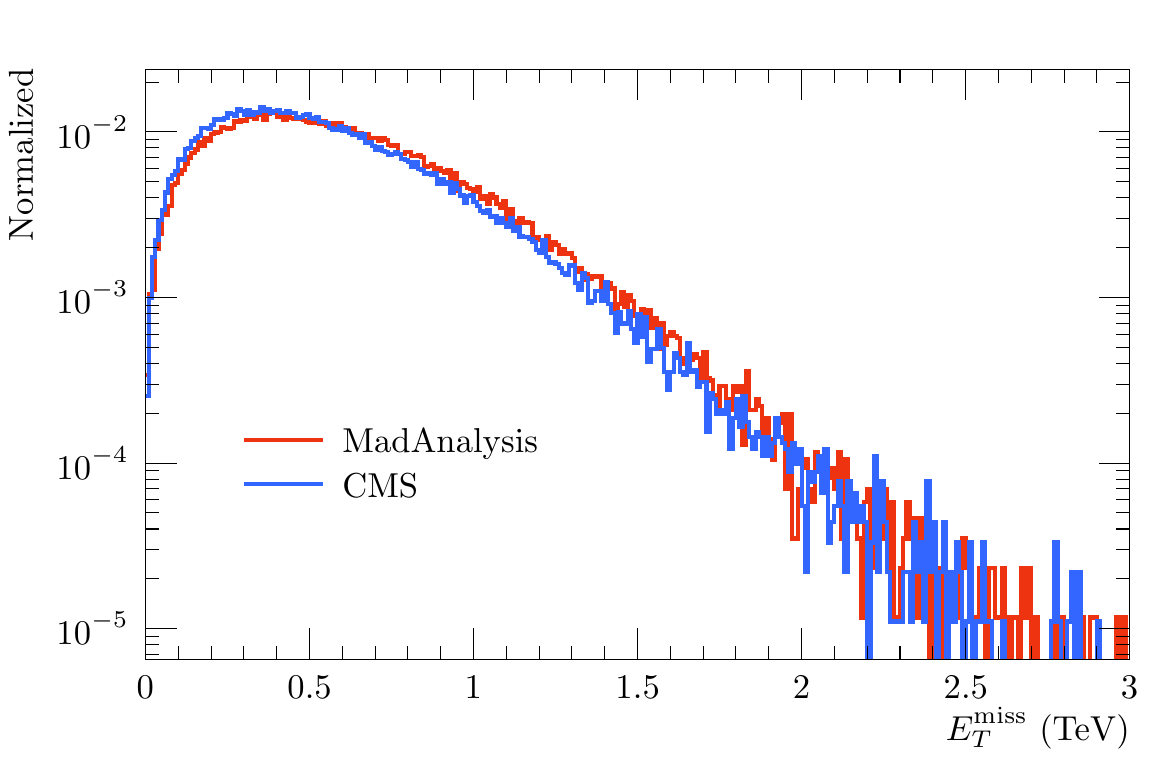}
  \includegraphics[width=0.49\textwidth]{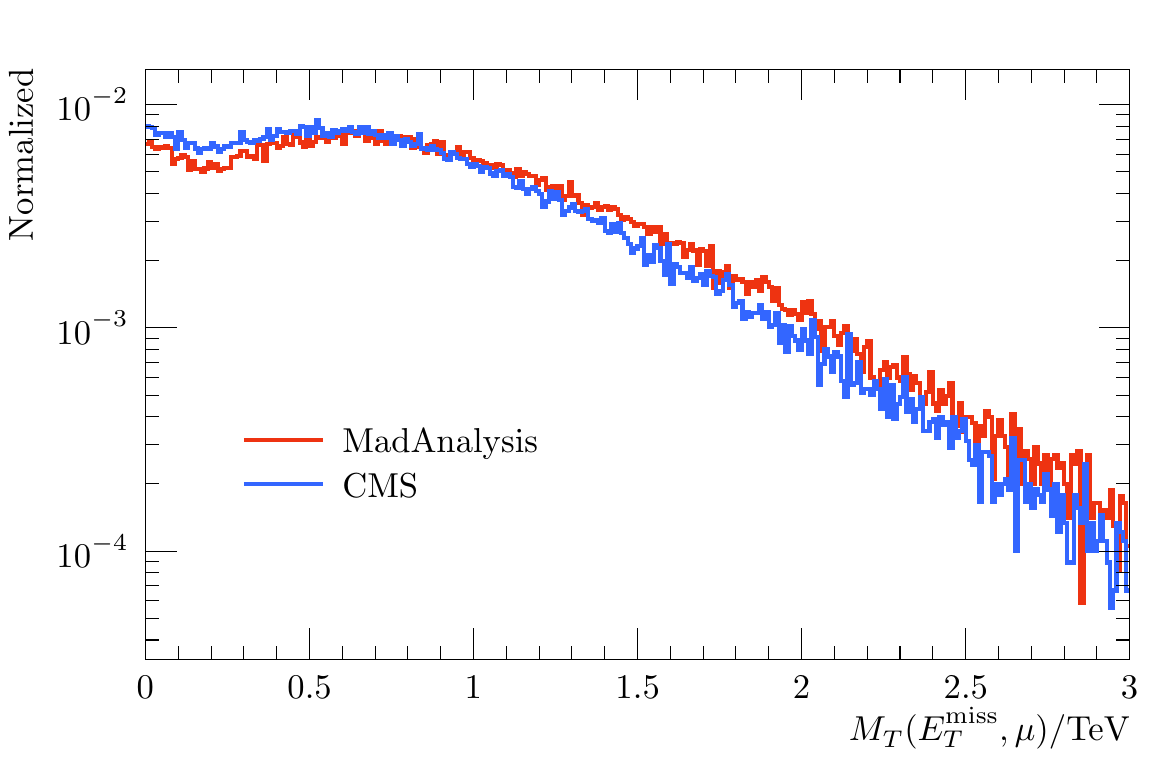}
  \vspace*{8pt}
  \caption{Normalised distributions for some key kinematical quantities used in
    the CMS-EXO-17-015 analysis. We show predictions for the scenario {\tt BP1}
    with $M_{\textrm{LQ}} = 1000~\gev$. The \ma~ predictions are shown in red
    while the CMS ones are given in blue. We consider the hadronic tau multiplicity
    (upper left), the $b$-jet multiplicity (upper right), the transverse momentum of
    the leading jet (centre left), the transverse momentum of the leading muon
    (centre right), the missing transverse energy (lower left) and the transverse
    mass of the leading muon and missing momentum system (lower right).}
\label{fig:comparison}
\end{figure}

Moreover, we confront results at the differential level in
Fig.~\ref{fig:comparison} for different observables relevant for the considered
analysis. An excellent agreement is again found.

\subsection{Conclusions}
\label{sec:conclusion}
In this note, we have made a detailed description of our implementation of
the CMS-EXO-17-015 analysis in the \ma\ framework. This analysis can be used in
particular to constrain models containing scalar or vector leptoquarks that
decay primarily into muons and jets. However, the signal region is not defined by
relying on the leptoquark invariant mass (to be reconstructed from the leading
muon and jet), so that the analysis can in fact be used to probe any model giving
rise to muons, jets and missing transverse energy. For given benchmark scenarios,
we have found an excellent agreement between our predictions with \ma\ and the
official results provided by the CMS collaboration. This validated analysis is available 
on the public \ma~database and can be found from the \ma\ dataverse, at
\url{https://doi.org/10.14428/DVN/ICOXG9}\cite{ICOXG9_2020}, together with relevant
validation material.

\subsection*{Acknowledgments}
The authors would like to thank Abdollah Mohammadi for kindly providing official
CMS cut-flow tables and for his assistance in understanding the event selection
used in the CMS-EXO-17-015 analysis. The work of AJ is sponsored by the
National Research Foundation of Korea, Grant No. NRF-2019R1A2C1009419.

\cleardoublepage
\markboth{Yechan Kang, Jihun Kim, Jin Choi and Soohyun Yun}{Implementation of the CMS-EXO-17-030 analysis}

\section{Implementation of the CMS-EXO-17-030 analysis (pairs of trijet resonasnces)}
  \vspace*{-.1cm}\footnotesize{\hspace{.5cm}By Yechan Kang, Jihun Kim, Jin Choi and Soohyun Yun}
\label{sec:ditrijet}


\subsection{Introduction}
Events associated with a multijet final state at hadron colliders provide a unique window to investigate various beyond standard model (BSM) physics.
Typically, in the Standard Model, pair-produced heavy resonances each decaying into three jets only originate from the production of a pair of hadronically decaying top quarks.
Therefore, if a particle heavier than the top quark exists, and manifests itself as a narrow resonance, then one should be able to see a clean high mass resonance peak in multijet invariant mass distributions.

\begin{figure}[t]
  \centerline{\includegraphics[width=3in]{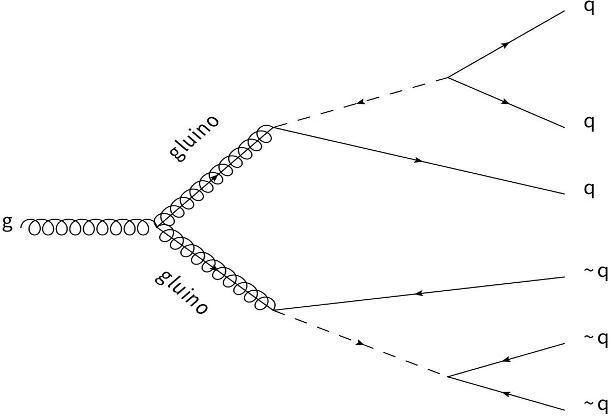}}
  \vspace*{8pt}
  \caption{Feynman diagram representative of pair-produced gluinos decaying into six jets.\protect\label{exo1730_fig1}}
\end{figure}

We present the results of the recast of the CMS-EXO-17-030 three-jet analysis\cite{Sirunyan:2018duw} which targets pair-produced resonances in proton-proton ($pp$) collisions, in a case where each resonance decays into three quarks.
In this search, the RPV SUSY model\cite{Barbier:2004ez} is used as a benchmark, with a varying gluino mass.
This allows for the modeling of high mass resonances pair production, followed by subsequent gluino decays into three jets.
Moreover, this leads to a final state comprising six quarks at the parton level.
In this model, a new quantum number $R$ is defined as \[ R=(-1)^{2S+3B+L}, \] where $S$ is the spin, $B$ is the baryon number, and $L$ is the lepton number.
In this search, we consider a model in which $R$-parity is broken via baryon number violation, so that squarks can decay into two quarks (Fig.~\ref{exo1730_fig1}).
For our recast implementation and its validation, we follow the interpretation of the experimental analysis and the resonance is assumed to be a gluino.

The analysis is divided into four separate regions depending on the mass of the gluino.
It exploits the geometrical event topology to discriminate signal events from background events.
In order to improve the sensitivity to a wide range of resonance masses, the analysis includes signal regions that are each dedicated to a specific resonance mass, the associated topology and kinematics of the final-state jet activity.
This separation is further necessary to manage the estimation of the background properly.
In the low mass regions, the main background comes from top quark decays, whereas it comes from QCD events for the high mass regions.
By defining different signal regions depending on the gluino mass, we can handle the background properly with different strategies.
To perform the validation of our implementation, we select four benchmark gluino mass points representing each signal region, the gluino mass being respectively set to 200 GeV, 500 GeV, 900 GeV, and 1600 GeV.
This enables the direct comparison between the recast and the result of the experimental publication in terms of acceptance and therefore allows us to validate our implementation.

In the rest of this note, we present the recast of the CMS-EXO-17-030 analysis in the {\sc MadAnalysis}~5 framework~\cite{Conte:2012fm,Conte:2014zja,Dumont:2014tja,Conte:2018vmg}, which is now available from the \href{http://madanalysis.irmp.ucl.ac.be/wiki/PublicAnalysisDatabase}{{\sc Madanalysis}~5 Public Analysis Database} and the {\sc MadAnalysis}~5 dataverse \cite{GAZACQ_2020}.

\subsection{Description of the analysis}\label{sec:trijetanalysis}
To identify pair-produced high mass resonances decaying into multiple jets in LHC events, the jet ensemble technique\cite{Aaltonen:2011sg} is applied.
This examines all possible combinatorial triplets that could be formed from a jet collection in each event.
As a concrete example, we consider an event including 6 jets.
First, we collect every possible set of 3 jets into a triplet.
There should be 20 combinations of such triplets, and therefore 10 pairs of triplets in each event.
All such triplet pairs and triplets are candidates for pair-produced gluinos and their decay.
Then, to discriminate the `correct' triplets (which originate from gluino decays) from wrongly combined triplets, and to reject the QCD background as well, we apply cuts on variables that embed the topology expected from the signal events.
The cuts are categorized into three stages and applied step by step: event level, triplet pair level, and triplet level.
The definition of each variable and the motivation to use them are described in section~\ref{sec:eventselection} in detail.

\subsubsection{Object definitions}
Jet candidates are reconstructed using the anti-$k_T$ algorithm\cite{Cacciari:2008gp} with a radius parameter $ R=\sqrt{(\Delta\phi)^2+(\Delta\eta)^2}=0.4 $.
Jets in the detector are required to have a transverse momentum, $p_T$, larger than 20 GeV and an absolute value of the pseudorapidity, $|\eta|$, of at most 2.4.
This analysis neither considers nor vetoes the presence of other objects like hard leptons or photons, so that their precise definition is irrelevant.

\subsubsection{Event selection}\label{sec:eventselection}
Four separate signal regions have been defined to target all possible gluino masses in the range of 200 - 2000 GeV: SR1 (200-400 GeV), SR2 (400-700 GeV), SR3 (700-1200 GeV), and SR4 (1200-2000 GeV).
The requirements in each signal region are described below.

First of all, each event is required to contain at least six reconstructed jets. From the entire set of jets, only the six jets with the highest $p_T$ are considered.
Then four selections based on event-level variables are applied.
For the low mass regions targeting gluino masses below 700 GeV, all jets in the event must have a $p_T$ larger than 30 GeV and the $H_T$ variable, defined as the scalar sum of the $p_T$s of all jets, is imposed to be larger than 650 GeV.
For the high-mass regions dedicated to gluino masses beyond 700 GeV, the $p_T$ of all jets must be larger than 50 GeV and the $H_T$ variable must be greater than 900 GeV.
Jets are arranged in descending order of $p_T$, and the $p_T$ of the sixth jet is required to be larger than 40 GeV, 50 GeV, 125~GeV, or 175 GeV for the SR1, SR2, SR3, and SR4 signal region respectively.

To discriminate the signal from the QCD main background and wrongly combined triplets, Dalitz variables are adopted.
Dalitz variables are effective discriminants for studying three-body decays. They were initially introduced by Dalitz in kaon to three pions decays\cite{Dalitz:1954cq}.
The Dalitz variables for a triplet are defined as
\[ \hat{m}(3,2)^2_{ij}=  \frac{m^2_{ij}}{m^2_{ijk}+m^2_i+m^2_j+m^2_k},\]
where $m_i, m_{ij}$ and $m_{ijk}$ are respectively the invariant mass of the individual jet $j_i$, of the dijet system made of the jets $j_i$ and $j_j$, and of the triplet.
Here, indices refer to the jets in the triplet, where $i,j,k \in \{1,2,3\}.$
These variables have good discriminating power as follows from our signal topology.
In signal events for which a massive particle decays into three quarks, the angular distribution of the jets should be even in the center-of-mass frame.
Therefore we expect the Dalitz variable to be close to 1/3 for each jet pair ($m_{ij}$).

By utilizing the above property of Dalitz variables, a new variable called the mass distance squared of a triplet is defined as
\[ D^2_{[3,2]}=\sum_{i>j}\left(\hat{m}(3,2)_{ij}-\frac{1}{\sqrt{3}}\right)^2.\]
This variable must be close to zero for symmetrically decaying signal triplets but deviates from zero for wrongly combined triplets and QCD backgrounds which may exhibit an asymmetric topology.

A generalized Dalitz variable is introduced as an extension of the original Dalitz variable for a six-jet topology, which should be close to $1/20$ in the case of even angular distributions.
It is defined from the normalized invariant mass of jet triplets,
\[ \hat{m}(6,3)^2_{ijk}= \frac{m^2_{ijk}}{{4m^2_{ijklmn}+6\Sigma_i  m^2_i}}.\]
Here, $m_{ijklmn}$ refers to the invariant mass of the leading six jets, where $i,j,k,l,m,n \in \{1,2,3,4,5,6\}.$

Using the generalized Dalitz variables and the $D^2_{[3,2]}$ value associated with a triplet, the six-jet distance squared of an event is defined as
\[ D^2_{[(6,3)+(3,2)]}=\sum_{i<j<k} \bigg(\sqrt{\hat{m}(6,3)^2_{ijk}+D^2_{[3,2]ijk}}-\frac{1}{\sqrt{20}}\bigg)^2.\]
For signal events, each pair-produced gluino is expected to decay symmetrically, which leads to small values of $D^2_{[3,2]}$.
Furthermore, each generalized Dalitz variable ($m(6,3)^2_{ijk}$) is expected to be close to 1/20.
Therefore, signal events are likely to feature $D^2_{[(6,3)+(3,2)]}$ close to zero.
On the other hand, the events containing triplets originating from QCD multijet production will have an asymmetric angular distribution, and thus have values relatively far from zero.
The official analysis has shown that the distribution of $D^2_{[(6,3)+(3,2)]}$ for QCD multijet events peaks at a farther point than the gluino events, as expected.
The $D^2_{[(6,3)+(3,2)]}$ variable is used for the last selection at the event level and is required to be smaller than 1.25, 1.00, 0.9, or 0.75 for the SR1, SR2, SR3, and SR4 signal region respectively.

Furthermore, the masses of two distinct triplets are expected to be symmetric in the case of the signal, as originating from the decay of the same particle.
Thus the mass asymmetry defined as
\[A_m=\frac{|m_{ijk}-m_{lmn}|}{m_{ijk}+m_{lmn}},\]
where $m_{ijk}$ and $m_{lmn}$ are the masses of the two distinct triplets in a triplet pair, is expected to be closer to zero for the correctly combined triplet pairs in the signal case.
The mass asymmetry of a triplet pair is required to be smaller than 0.25 or 0.175 for the SR1 and SR2, or 0.15 for the SR3 and SR4.

Finally, selections at the triplet-level are applied.
The variable $\Delta$ of a triplet is defined as the sum of the $p_T$ of the jets in the triplet ($|p_T|_{ijk}$), after subtracting the triplet invariant mass ($m_{ijk}$):
\[\Delta=|p_T|_{ijk}-m_{ijk}.\]
In the official analysis, it has been shown that correctly combined triplets have a constant distribution in the mass vs $p_T$ plane, whereas in cases of wrongly combined triplets and QCD backgrounds their $p_T$ and mass are proportional to each other.
Therefore the $\Delta$ observable has good discriminating power between wrongly combined triplets, QCD backgrounds and correctly combined triplets.
This value is required to be larger than 250 GeV, 180 GeV, 20~GeV, or -120 GeV for the SR1, SR2, SR3, and SR4 region respectively.
For the very last selection, the mass distance squared of a triplet $(D^2_{[3,2]})$ is required to be smaller than 0.05, 0.175, 0.2, or 0.25 for each region.

The actual cuts for each variable for the event, triplet pair, and triplet levels are summarized in table~\ref{ditrijet_ta1}.

\begin{table}[t]
  \renewcommand{\arraystretch}{1.2}
  \tbl{Selection criteria}
  {\begin{tabular}{@{}cc|cccc|c|cc@{}}
  \toprule
  &&\multicolumn{4}{c|}{Events}  &Triplet Pairs &\multicolumn{2}{c}{Triplets}\\
 Region&Gluino Mass& Jet $p_T$ & $H_T$ & $p_T(j_6)$ & $D^2_{[(6,3)+(3,2)]}$ & $A_m$ & $\Delta$ & $D^2_{[3,2]}$ \vspace{1px} \\
   \hline  
   1 & 200-400 GeV & $>$30 GeV & $>$650 GeV & $>$40 GeV & $<$1.25 & $<$0.25 & $>$250 GeV & $<$0.05
   \\2 &400-700 GeV & $>$30 GeV & $>$650 GeV & $>$50 GeV & $<$1.00 & $<$0.175 & $>$180 GeV & $<$0.175
   \\3&700-1200 GeV& $>$50 GeV & $>$900 GeV & $>$125 GeV & $<$0.9 & $<$0.15 & $>$20 GeV & $<$0.2
   \\4 &1200-2000 GeV& $>$50 GeV & $>$900 GeV & $>$175 GeV & $<$0.75 & $<$0.15 &$>$-120 GeV & $<$0.25\\
  \botrule
  \end{tabular}\label{ditrijet_ta1} }
\end{table}

\subsection{Validation}

\subsubsection{Event generation}

Simulation of double-trijet resonance events is done by making use of the MadGraph5\_aMC@NLO version 2.7.3 Monte Carlo generator\cite{Alwall:2014hca}, using the  RPVMSSM\_UFO model file\cite{Degrande:2011ua,Fuks:2012im}.
For the parton distribution functions, the LO set of NNPDF3.0\cite{Ball:2014uwa} parton densities with $\alpha_s=0.130$, as implemented in LHAPDF6\cite{Buckley:2014ana}, is used.
To avoid any squark contribution to gluino production, all the masses of squarks are set to be 2.5 TeV, and the masses of gluinos are set to be 200, 500, 900, and 1600 GeV to target the signal regions resulting from the cuts described in section~\ref{sec:eventselection}.
Based on the pair production of gluinos, we used {\sc MadSpin}\cite{Artoisenet:2012st} and {\sc MadWidth}\cite{Alwall:2014bza} without spin correlations to simulate the gluino decays into three jets.
We compared the acceptance resulting from the cuts described in the next section, using signal samples with and without spin correlation, and found that there is negligible difference in the final acceptance.
Here, we thus present the results without any spin correlation.

After the simulation of the hard-scattering process, Pythia8\cite{Sjostrand:2014zea} is used for parton showering and hadronization, followed by Delphes3\cite{deFavereau:2013fsa} for the fast simulation of the CMS detector response.

\subsubsection{Comparison with the official results}
As using combined triplets of jets for the final selection, the analysis suffers from two major backgrounds, irreducible QCD backgrounds and a unique background not originating from a specific physical process: wrongly combined triplets.
Since the invariant mass distribution is similar for QCD backgrounds and wrongly combined triplets \cite{Sirunyan:2018duw}, the CMS collaboration made signal and background fitting templates from those distributions and proceed with signal to background fitting directly to the data to calculate the final signal significance.
Therefore, the number of triplets that pass all cuts is used indirectly for the final result.
To see how many correct triplets survive in each signal region, the signal acceptance has been defined based on the triplet selection described in section~\ref{sec:trijetanalysis}:
\[ {\rm Acc.} = \frac{{\rm Number~of~surviving~triplets}}{{\rm Number~of~generated~events}} . \]
Here, the acceptance is defined as the ratio of the number of triplets and the number of total events, and not the number of events passing the selections and the total number of events.
We hence collect all possible combinations of triplets out of 6 jets and have 20 triplets (or 10 triplet pairs) per single event.
In this analysis, we have cuts at the event level, triplet-pair level, and triplet level.

Since the analysis has a distinctive definition of acceptance based on the number of triplets, one of the major difficulties in using the {\sc MadAnalysis}~5 framework was the implementation of counting the triplets passing the different cuts in each signal region, as there are diverse triplet-level cut thresholds for each region.
In {\sc MadAnalysis}~5, the framework provides cutflows based on the event selection, which makes it hard to count the number of surviving triplets in each signal region.
To overcome this problem, we made four collections of triplets, {\it i.e.}~one for each signal region, and updated each collection with the different cuts for each signal region.
Finally, we multiplied each event weight by the number of triplets (for each region), which makes {\sc MadAnalysis}~5 generating cutflows on triplet level.

The acceptance numbers officially calculated by CMS are of 0.00024, 0.084, 0.17 for SR1, SR3, and SR4.
There is no result provided for SR2. For the purpose of recast, we define the difference as
\[ {\rm Diff.} = \frac{{\rm Acc.(recast)} - {\rm Acc.(CMS~official)}}{{\rm Acc.(CMS~official)}} \]
to compare the recast values with the official results.

Comparing with the official results, the recast showed a large discrepancy.
Acceptances (differences) we calculated are \( 6.25 \times 10^{-4} (140\%), 6.5 \times 10^{-1} (674\%)\) and \( 1.71 (906\%)\) for SR1, SR3 and SR4.
We found out that many wrongly combined triplets not originating from the same gluino still pass the final selection.
Since there is no way to calculate the acceptance of the correctly matched triplets as originally performed through the template fit to the CMS data, we chose an alternative approach, using generator level information to check how many triplets from the same gluino can survive after all cuts.
Therefore, we require that the correct triplets should be matched to their mother gluino as
\begin{itemize}
    \item All jets should be matched to generator level partons within a distance in the transverse plane of \( \Delta R(j, q) < 0.3 \), where $q$ generically stands for \( u, c, d, s \) and the corresponding antiparticles.
    \item Matched partons in a triplet should all be quarks, or all be antiquarks.
    \item All matched (anti)quarks in the triplet should have the same gluino as their mother.
\end{itemize}

Here, we required the jets to be matched to their mother gluino using the truth level information.
For the purpose of generalization, any recasting analysis that wishes to use the truth information should change the Particle Data Group identifier (PID) of the mother particle.
We defined the PID of this mother particle by using the \#DEFINE preprocessor method, so the user can change the value of the EXO\_17\_030\_PID variable to any other value relevant for the signal of their interest.
Therefore, this implementation can be further tested with various other BSM models that allow resonance with three jet decay signature, {\it e.g.}~searches based on composite quark model\cite{Redi:2013eaa} or extra dimensional model\cite{Agashe:2016kfr}.

The final acceptances that we obtain, for the considered benchmark scenarios, are of $2.8 \times 10^{-4}$, $7.3 \times 10^{-2}$, and $1.55 \times 10^{-1}$ for the SR1, SR3 and SR4 regions.
Our predictions show good agreements with the CMS official results, at the level of 8\%,  13\%, and 8.8\% for the SR1, SR3 and SR4 regions.
For the SR2 region, the final acceptance is $1.5 \times 10^{-2}$. This value has no comparison target because the official acceptance for the SR2 region has not been provided by the CMS collaboration. Detailed results are provided in tables~\ref{ditrijet_ta2} and \ref{ditrijet_ta3}.

\begin{table}[t]
  \renewcommand{\arraystretch}{1.3}
  \setlength\tabcolsep{8pt}
\begin{center}
\tbl{Cutflows in the Low-Mass Regions. The initial number of triplets that could be reconstructed from each event is assumed to be 20. All triplets are matched to their mother particle. Since there is no official CMS result for SR2, we did not calculate the difference for that region.}
{
    \begin{tabular}{ @{} p{3cm}   p{1.5cm} p{2cm}   p{1.5cm} p{2cm} @{} }
    \toprule
    & \multicolumn{2}{c }{Signal Region 1} & \multicolumn{2}{c}{Signal Region 2} \\
    Cut & Events & Triplets  & Events & Triplets \\
    \hline
    Initial events      & 400,000 & 8,000,000 & 400,000 & 8,000,000 \\
    Njets$\geq$6        & 231,863 & 4,637,261 & 367,491 & 7,349,821 \\
    preselection        & 148,090 & 2,961,800 & 341,054 & 6,821,079 \\
    HT                  & 38,434  & 768,680   & 329,561 & 6,591,218 \\
    Sixth jet $p_T$       & 29,611  & 592,220   & 242,511 & 4,850,220 \\
    $D^2_{[(6,3)+(3,2)]}$ & 23,296  & 465,920   & 186,731 & 3,734,618 \\
    $A_m$                 & 3,982   & 4,630     & 89,853  & 118,285   \\
    $\Delta$              & 187     & 199       & 5,534   & 6,501     \\
    $D^2_{[3,2]}$         & 108     & 112       & 5,145   & 5,995     \\
    \hline
    Acc.                &         & 0.028\%   &         & 1.50\%     \\
    Acc.(CMS official)  &         & 0.026\%   &         &           \\
    Diff.               &         & 8\%   &         &           \\
    \botrule
    \end{tabular} \label{ditrijet_ta2} }
\end{center}
\end{table}

\begin{table}[t]
  \renewcommand{\arraystretch}{1.3}
  \setlength\tabcolsep{8pt}
\begin{center}
\tbl{Cutflows in the High Mass Region. The initial number of triplets reconstructed from each event is assumed to be 20. All triplets are matched to their mother particle. }
{
    \begin{tabular}{ p{3cm}  p{1.5cm} p{2cm}  p{1.5cm} p{2cm}  }
        \toprule
        & \multicolumn{2}{c}{Signal Region 3} & \multicolumn{2}{c}{Signal Region 4} \\
        Cut & Events & Triplets  & Events & Triplets \\
        \hline
        Initial events      & 400,000 & 8,000,000 & 400,000 & 8,000,000 \\
        Njets$\geq$6        & 388,119 & 7,762,382 & 394,516 & 7,890,321 \\
        preselection        & 340,320 & 6,806,404 & 373,669 & 7,473,380 \\
        HT                  & 339,303 & 6,786,064 & 373,661 & 7,473,221 \\
        Sixth jet $p_T$       & 120,141 & 2,402,821 & 166,877 & 3,337,540 \\
        $D^2_{[(6,3)+(3,2)]}$ & 100,349 & 2,006,981 & 113,436 & 2,268,721 \\
        $A_m$                 & 52,205  & 72,806    & 69,080  & 100,637   \\
        $\Delta$              & 25,465  & 31,320    & 49,767  & 62,731    \\
        $D^2_{[3,2]}$         & 23,948  & 29,025    & 49,309  & 61,959    \\
        \hline
        Acc.                &         & 7.3\%     &         & 15.5\%    \\
        Acc.(CMS official)  &         & 8.4\%     &         & 17.0\%    \\
        Diff.               &         & -13\%   &         & -8.8\%    \\
        \botrule
    \end{tabular}\label{ditrijet_ta3}
} \end{center}
\end{table}

\subsection{Conclusion}
A recast of the CMS-EXO-17-030 double-three-jet analysis has been performed within the {\sc MadAnalysis}~5 framework.
To validate our implementation, we choose four gluino RPV SUSY scenario with masses ranging from 200 to 2000 GeV. The four masses that we selected are 200~GeV, 500~GeV, 900~GeV, and 1600~GeV, and represent each signal region.
In this note, the event selection is described in detail, and corresponding cutflows for each benchmark point are presented.
We exhibit the difficulties that are inherent to the usage of {\sc MadAnalysis}~5 for the CMS-EXO-17-030 recast, as non-event based acceptance calculations are in order. We moreover explain our method to overcome them.
The signal events are simulated under the same condition as for the official CMS result, which corresponds to an integrated luminosity of 35.9 fb$^{-1}$ of collisions at a center-of-mass energy of 13 TeV, but with a CMS detector configuration based on {\sc Delphes}~3.
The validation is performed in terms of the acceptance for each signal region.
The recast and the official results show good agreement, resulting in differences from a minimum of 8\% to a maximum of 13\%.

The code is available online from the {\sc MadAnalysis}~5 dataverse~\cite{GAZACQ_2020}, at \href{https://doi.org/10.14428/DVN/GAZACQ}{https://doi.org/10.14428/DVN/GAZACQ}, on which we also provide cards that were relevant for the validation of this implementation.

\subsection*{Acknowledgments}
We congratulate all our colleagues who have participated in the second {\sc MadAnalysis}~5 workshop on LHC recasting in Korea and thank the organizers and the tutors of the workshop for their sincere support.
In addition, Soohyun Yun is grateful for financial support from Hyundai Motor Chung Mong-Koo Foundation.

\cleardoublepage
\markboth{Eric Conte and Robin Ducrocq}
  {Implementation of the CMS-EXO-19-002 analysis}

\section{Implementation of the CMS-EXO-19-002 analysis (physics beyond the Standard Model with multileptons; 137~fb$^{-1}$)}
  \vspace*{-.1cm}\footnotesize{\hspace{.5cm}By Eric Conte and Robin Ducrocq}
\label{sec:multilep}

%


\subsection{Introduction}
In this document, we present the implementation of the CMS-EXO-19-002 analysis~\cite{Sirunyan:2019bgz} in the {\sc \rm MadAnalysis 5} framework \cite{Conte:2012fm,Conte:2014zja,Dumont:2014tja,Conte:2018vmg}. It consists of a search for events featuring multiple charged leptons, and relies on an integrated luminosity of $137~\textrm{fb}^{-1}$ of LHC proton-proton collisions, with a center-of-mass energy $\sqrt{s}=13~\textrm{TeV}$.

In this analysis, two classes of models are targetted, which leads to the definition of two categories of signal regions. These consist of a type-III seesaw model~\cite{Biggio:2011ja} including three heavy fermions mediator $\Sigma^\pm$ and $\Sigma^0$, and a simple extension of the Standard model, called $t\bar{t}\phi$, with one scalar (or pseudoscalar) $\phi$ that can be produced in association with a top-antitop pair~\cite{Casolino:2015cza,Chang:2017ynj}. The type-III seesaw signal under consideration arises from the production and decay of $(\Sigma^\pm \Sigma^0)$ and $(\Sigma^\pm \Sigma^\mp)$ pairs ($\Sigma^0 \Sigma^0$ being neglected) in a multilepton final state. On the other hand, the $t\bar{t}\phi$ process with a $\phi\rightarrow l^+l^-$ decay induces a signal comprising additional $b$-jets originating from the top decays. The search for such signals is done in three $(3L)$ and four $(4L)$ leptons channels, with extra $b$-jets in the case of the $t\bar{t}\phi$ signal. For the validation of the implementation, we take into account predictions and official results with heavy fermion masses of $m_\Sigma=300~\textrm{GeV}$ and $m_\Sigma=700~\textrm{ GeV}$ for the type-III seesaw benchmark, and masses of $m_\phi=20~\textrm{ GeV}$ ($m_\phi=70~\textrm{GeV}$) for the scalar (pseudoscalar) $t\bar{t}\phi$ model, as the CMS collaboration only provided material for those cases.

In section~\ref{sec2}, we describe the selection and the manner in which the analysis is implemented in \textsc{MadAnalysis~5}. In particular, we present all signal regions defined in the CMS paper. Sections~\ref{val1} and \ref{val2} are devoted to the validation of the implementation of the type-III seesaw and $t\bar{t}\phi$ signal regions respectively. We summarize our main results in section~\ref{sec4}.

\subsection{Description of the analysis}\label{sec2}

\subsubsection{Object definitions}\label{sec:obj}

Muons are required to have a transverse momentum $p_T > 10~\textrm{GeV}$ and a pseudorapidity $|\eta|<2.4$. Requirements on the tracking quality are not implemented, as the package {\sc Delphes}~3 that we use for the fast simulation of the CMS detector~\cite{deFavereau:2013fsa} is not able to reproduce it. To suppress the background, an isolation criterion is applied on the muons. The corresponding procedure relies on a relative isolation variable, Isol$(l)$, defined as the scalar $p_T$ sum of all particle-flow objects in a cone of $\Delta R=0.4$ around the lepton direction and normalized to the lepton $p_T$. This variable, 
\begin{equation}
   {\rm Isol}(l)=\frac{1}{p_T(l)} \sum_{j \neq l}^{\Delta R<0.4}{p_T(j)} \hspace{0.3cm} \textrm{with}\ l\ \textrm{being the lepton and}\ j\ \textrm{any particle flow object},
\end{equation}
must be smaller than $15\%$. The displacement of the muon track with respect to the primary vertex is also constrained,
\begin{equation}
 |d_z|<0.1~\textrm{cm}\ , \qquad |d_{xy}|<0.05~\textrm{cm}.
\end{equation}

Electrons are required to have a $p_T > 10~\textrm{GeV}$, and a pseudorapidity $|\eta|<2.5$ that is consistent with the tracking system acceptance. Requirements on the electron shower shape and track quality are not implemented.  The relative isolation ratio as been chosen to be smaller than $15\%$, and is calculated with a cone of $\Delta R=0.3$ around the electron. The displacement of the electron track with respect to the primary vertex is also constrained:  
    \begin{itemize}
        \item $|d_z|<0.1~\textrm{cm}$ and $|d_{xy}|<0.05~\textrm{cm}$ when the electron is in the eletromagnetic calorimeter (ECAL) barrel acceptance ($|\eta|<1.479$);
        \item $|d_z|<0.2~\textrm{cm}$ and $|d_{xy}|<0.1~\textrm{cm}$ when electron is in the ECAL barrel endcap ($|\eta|>1.479$).
    \end{itemize}
Finally, electrons that are too close to a muon (possibly due to bremsstrahlung from the muon) must be rejected. It is done by searching if there is a muon track in a cone around the electron track with a radius $\Delta R=0.05$.

Jets are defined by using the anti-$k_T$ algorithm~\cite{Cacciari:2008gp} with a distance parameter of 0.4, as provided by the \textsc{FastJet} package~\cite{Cacciari:2005hq,Cacciari:2011ma}. They must have a $p_T>30~\textrm{GeV}$ and a $|\eta|<2.1$. No pile-up simulation has been encapsulated because we assume that pile-up suppression algorithms are good enough to get rid of all related soft contamination. Moreover, all jets which are inside a cone of radius $\Delta R=0.4$ around a selected charged lepton are discarded.

We call '$b$-jets' the reconstructed jets originating from $B$-hadrons. The $b$-tagging performance in this CMS analysis corresponds to the medium working point of the DeepCSV algorithm~\cite{Sirunyan:2017ezt}, with an efficiency of 60--75\% and a misidentification rate of 10\% for $c$-quark jets and 1\% for lighter jets.

In trilepton events, additional constraints are imposed on the charged leptons in order to reduce misidentified-background contributions. If a charged lepton can be matched which a loose $b$-jet (defined by a jet with $p_T$ greater than 10 GeV, $|\eta|\leq 2.5$ and a medium $b$-tag), by using a matching cone of $\Delta R < 0.4$ around the lepton, the lepton is rejected. Besides, an additional selection cut based on a tri-dimensional impact parameter is also applied on the leptons. The lack of information relative to this quantity implies that we have not implemented this cut in the recast analysis.

Finally, the missing transverse momentum, noted $MET$ or $p_T^{miss}$, is taken as the negative vector sum of all particle-flow objects $p_T$.

\subsubsection{Common event selection}\label{sec:common}

The trigger requirements imply an online selection of the events. This two-stage selection requires at least one electron or one muon in the event with a large $p_T$ value. The first step of the offline selection consists of requiring one leading lepton with a threshold a little bit greater than the online selection threshold, and thus encapsulates the online selection. The $p_T$ threshold value used in the offline selection depends on the year of data acquirement. For muons, the threshold is 26 GeV for 2016, 29 GeV for 2017 and goes back to 26 GeV for 2018. For electrons, the threshold is 30 GeV for 2016 and 35 GeV for 2017 and 2018. Considering the integrated luminosity recorded by CMS ($37.80~\textrm{fb}^{-1}$ for 2016, $44.98~\textrm{fb}^{-1}$ for 2017 and $63.67~\textrm{fb}^{-1}$ for 2018), we apply a threshold of 26 GeV for 69\% of the events (randomly chosen according to a flat distribution) and 29 GeV for the remaining events. Similarly, an electron threshold of 35 GeV is fixed for 74\% of the events and 30 GeV for the remaining events. 

We select events with three leptons (electrons or muons) or more. In the case where we have four leptons or more, we only keep the four leading leptons and label those events as ``4 leptons" events.

All events containing a lepton pair where the two leptons are distant by $\Delta R < 0.4$ are rejected. We also remove events that contain a same-flavor lepton pair (independent of the charge) whose invariant mass is below 12 GeV. These two selection cuts allow us to remove low-mass resonances and final-state radiation background contributions.

In the case of a trilepton event, an additional constraint is applied. If the invariant mass of the three leptons is within the $Z$ mass window ($91\pm 15~\textrm{GeV}$), the presence of an opposite-sign same-flavor (OSSF) lepton pair with an invariant mass below 76 GeV yields the rejection of the event. This procedure allows us to remove $Z\rightarrow l^+l^-\gamma$ background contributions where the photon converts into two additional leptons, with one of which being lost.

\subsubsection{Event selection and categorization devoted to the type-III seesaw signal}\label{sec:sel1}

\noindent The events are categorised in 7 signal regions according to:
\begin{itemize}
    \item the number of selected leptons (three or four leptons) in the event,
    \item the number of opposite-sign same-flavor lepton pairs (OSSF multiplicity),
    \item the value of the invariant mass of the OSSF lepton pair relative to the $Z$ mass window ($91\pm15~\textrm{GeV}$). If there are several OSSF pairs, the considered invariant mass is the one which is the closest to the $Z$ nominal mass. We refer the three cases as below-Z, on-Z and above-Z.
\end{itemize}
Table~\ref{tab:region1} collects the definition of the different signal regions.

\begin{table}[t]
  \renewcommand{\arraystretch}{1.2}
  \setlength\tabcolsep{7pt}
\tbl{List of the signal regions dedicated to probing the type-III seesaw model.}
{\begin{tabular}{@{}l c c c l@{}}\toprule
Label & $N_l$ & $N_{OSSF}$ & $M_{OSSF}$ & additional cut \\
\hline
3L below-Z & 3 & 1 & $< 76~\textrm{GeV}$ & -\\
3L on-Z    & 3 & 1 & $\in[76;106]$ GeV & MET $>$ 100 GeV \\
3L above-Z & 3 & 1 & $> 106~\textrm{GeV}$ & - \\
3L OSSF0   & 3 & 0 & - & - \\
4L OSSF0   & $\ge 4$ & 0 & - & -\\
4L OSSF1   & $\ge 4$ & 1 & - & -\\
4L OSSF2   & $\ge 4$ & 2 & - & MET $>$ 100 GeV or no double OSSF on-Z\\
\botrule
\end{tabular}\label{tab:region1}}
\end{table}

Some signal regions involve an extra cut. The region called `\textsl{3L on-Z}' is populated by events that must feature a missing transverse energy MET greater than 100 GeV. Moreover, the region called `\textsl{4L OSSF2}' is populated by events containing either a missing transverse energy MET greater than 100 GeV, or no double OSSF lepton pairs on-Z.

\subsubsection{Event selection and categorisation devoted to the $t\bar{t}\phi$ signal}\label{sec:sel2}

First, events that contain no opposite-sign same-flavor charged lepton pairs are rejected. Then, we denote the invariant mass of the OSSF pair $M_{OSSF}$. If there are several OSSF pairs, then we consider the invariant mass that is the closest to the $Z$ nominal mass. Events that feature an $M_{OSSF}$ in the $Z$ mass window ($91\pm15~\textrm{GeV}$) are rejected. Theses requirements make the signal regions orthogonal to all control regions defined in the considered CMS analysis.

Then, events are categorised into 18 signal regions according to:
\begin{itemize}
    \item the number of selected leptons (three or four leptons),
    \item the number of opposite-sign same-flavor lepton pair (OSSF multiplicity),
    \item the flavor of the leptons involved in the computation of $M_{OSSF}$,
    \item the $b$-jet multiplicity,
    \item the observable $S_T$ defined as the scalar $p_T$ sum of all jets, all charged leptons and the missing transverse momentum.
\end{itemize}
Table~\ref{tab:region2} collects the definition of the different signal regions.

\begin{table}[t]
  \renewcommand{\arraystretch}{1.2}
  \setlength\tabcolsep{7pt}
\tbl{List of the signal regions dedicated to probing the $t\bar t\phi$ model.}
{\begin{tabular}{@{} l c c c c l @{}}\toprule
Label & OSSF flavor & $N_l$ & $N_b$ & $S_T$ \\
\hline
3L($ee$)     0B ST$<$400 & $e$ & 3 & 0 & $<$400 GeV \\
3L($\mu\mu$) 0B ST$<$400 & $\mu$ & 3 & 0 & $<$400 GeV  \\
3L($ee$)     0B 400$<$ST$<$800 & $e$ & 3 & 0 & $\in$[400;800] GeV \\
3L($\mu\mu$) 0B 400$<$ST$<$800 & $\mu$ & 3 & 0 & $\in$[400;800] GeV  \\
3L($ee$)     0B ST$>$800 & $e$ & 3 & 0 & $>$800 GeV\\
3L($\mu\mu$) 0B ST$>$800 & $\mu$ & 3 &  1 & $<$400 GeV \\
3L($ee$)     1B ST$<$400 & $e$ & 3 & 1 & $<$400 GeV  \\
3L($\mu\mu$) 1B ST$<$400 & $\mu$ & 3 & 1 & $<$400 GeV  \\
3L($ee$)     1B 400$<$ST$<$800 & $e$ &  3 & 1 & $\in$[400;800] GeV \\
3L($\mu\mu$) 1B 400$<$ST$<$800 & $\mu$ & 3 & 1 & $\in$[400;800] GeV  \\
3L($ee$)     1B ST$>$800 & $e$ & 3 & 1 & $>$800 GeV \\
3L($\mu\mu$) 1B ST$>$800 & $\mu$ & 3 & 1 & $>$800 GeV \\
4L($ee$)     0B ST$<$400 & $e$ &  $\geq 4$  & 0 & $<$400  \\
4L($\mu\mu$) 0B ST$<$400 & $\mu$ & $\geq 4$  & 0 & $<$400   \\
4L($ee$)     0B ST$>$400 & $e$ &  $\geq 4$  & 0 & $>$400  \\
4L($\mu\mu$) 0B ST$>$400 & $\mu$ & $\geq 4$  & 0 & $>$400   \\
4L($ee$)     1B  & $e$ &  $\geq 4$  & 1 & - \\
4L($\mu\mu$) 1B  & $\mu$ & $\geq 4$  & 1 & -  \\
\botrule\end{tabular}\label{tab:region2}}
\end{table}

\subsection{Validation of the implementation of the type-III seesaw signal regions}\label{val1}

\subsubsection{Event generation}

In the context of the type-III seesaw model, neutrinos are Majorana particles whose mass arises from interactions with new massive fermions organized in an $SU(2)$ triplet comprising heavy Dirac charged leptons ($\Sigma^\pm$) and a heavy Majorana neutral lepton ($\Sigma^0$).

The model has been already implemented in {\sc FeynRules}~\cite{Alloul:2013bka} and is available in the form of a UFO~\cite{Degrande:2011ua,Biggio:2011ja} model. With \textsc{MadGraph5\_aMC@NLO}~\cite{Alwall:2014hca}, we consider the leading-order (LO) production of pairs of new fermions $\Sigma^\pm\Sigma^0$ and $\Sigma^+\Sigma^-$, $\Sigma^0 \Sigma^0$ production being neglected. We assume two different values for the new physics masses of 300 GeV and 700 GeV, and we use the NNPDF3.0~LO~\cite{Ball:2014uwa} parton distribution functions (PDFs) provided by the LHAPDF package~\cite{Buckley:2014ana}.
The cross section is rescaled at NLO+NLL and set to $0.5771 \pm 0.0398$~pb for the 300 GeV case and $0.01395 \pm 0.00150$~pb for the 700 GeV case~\cite{Biggio:2011ja,1764474}.

Each new particle can then decay into a boson $V = h$, $W^\pm$ or $Z$, and a lepton of flavor $l$ through a coupling denoted $V_l$. Following this scheme, a $\Sigma^\pm$ can decay into a $Z+l^\pm$, $h+l^\pm$ or $W^\pm \nu$ system, and a $\Sigma^0$ can decay into a $Z+\nu$, $h+\nu$ and a $W^\pm l^\mp$ system. The branching ratios are identical across all leptons flavors according to the \textsl{flavor-democratic scenario} obtained by taking the coupling $V_e$, $V_\mu$ and $V_\tau$ all equal to $10^{-4}$. Their values have been computed and are given by Table~\ref{tab:decay}. The tau channels are also considered through their leptonic decays. This decay has been implemented at the {\sc MadGraph5\_MC@NLO} level, but all boson decays are handled by {\sc Pythia}~8~\cite{Sjostrand:2014zea}.

{\sc Pythia~8} also handles parton showering, hadronization, and the underlying events (multiple interaction and beam remnant interactions). We choose the CUETP8M1 tune~\cite{Khachatryan:2015pea} and the simulation of the detector response is handled by \textsc{Delphes}~3~\cite{deFavereau:2013fsa}, as driven through the \textsc{MadAnalysis~5} platform.

The list of produced samples and the number of generated events for our validation procedure are given in Table~\ref{tab:samples1}.

\begin{table}[t]
  \renewcommand{\arraystretch}{1.6}
  \setlength\tabcolsep{10pt}
\tbl{Width expression~\cite{Biggio:2011ja} and branching ratio values relative to the new massive fermions $\Sigma$ decay into a boson and a lepton (in the case where all $V_l$ are equal)}  
{\begin{tabular}{@{} l l c c @{}} \toprule
Process & Decay width formula & BR & BR   \\
       &                      & $m_\Sigma = 300$ GeV & $m_\Sigma = 700$ GeV \\
\hline
$\Gamma\left( \Sigma^0 \rightarrow l^\pm W^\pm \right)$ & $\frac{g^2}{32\pi} |V_l|^2 \frac{M_\Sigma^3}{M_W^2} \left( 1 - \frac{M^2_W}{M^2_\Sigma} \right)^2 \left( 1 +2 \frac{M^2_W}{M^2_\Sigma} \right)$ & 71.6 \% & 66.2\% \\
$\Gamma \left( \Sigma^0 \rightarrow \nu_l Z \right)$ & $\frac{g^2}{64\pi c_W^2} |V_l|^2 \frac{M_\Sigma^3}{M_Z^2} \left( 1 - \frac{M^2_Z}{M^2_\Sigma} \right)^2 \left( 1 +2 \frac{M^2_Z}{M^2_\Sigma} \right)$ & 3.6\% & 2.8\% \\
$\Gamma \left( \Sigma^0 \rightarrow \nu_l H \right)$ & $\frac{g^2}{64\pi} |V_l|^2 \frac{M_\Sigma^3}{M_W^2} \left( 1 - \frac{M^2_H}{M^2_\Sigma} \right)^2$ & 24.8\% & 31.0\% \\
\hline
$\Gamma \left( \Sigma^\pm \rightarrow \nu_l W^\pm \right)$ & $\frac{g^2}{32\pi} |V_l|^2 \frac{M_\Sigma^3}{M_W^2} \left( 1 - \frac{M^2_W}{M^2_\Sigma} \right)^2\left( 1 +2 \frac{M^2_W}{M^2_\Sigma} \right)$ & 35.0\% & 31.1\% \\
$\Gamma\left( \Sigma^\pm \rightarrow l^\pm Z \right)$ & $\frac{g^2}{64\pi c_W^2} |V_l|^2 \frac{M_\Sigma^3}{M_Z^2} \left( 1 - \frac{M^2_Z}{M^2_\Sigma} \right)^2 \left( 1 +2 \frac{M^2_Z}{M^2_\Sigma} \right)$ & 52.9\% & 54.3\% \\
$\Gamma\left( \Sigma^\pm \rightarrow l^\pm H \right)$ & $\frac{g^2}{64\pi} |V_l|^2 \frac{M_\Sigma^3}{M_W^2} \left( 1 - \frac{M^2_H}{M^2_\Sigma} \right)^2 $ & 12.1\% & 14.6\% \\
\botrule\end{tabular}\label{tab:decay}}
\end{table}

\begin{table}[t]
  \renewcommand{\arraystretch}{1.2}
  \setlength\tabcolsep{10pt}
\tbl{List of produced signal samples for the validation of the type-III seesaw signal regions.}
{\begin{tabular}{@{} c c  c c@{}} \toprule
$\Sigma$ mass& lepton flavor & Number of produced events \\
\hline
 \multirow{6}{*}{300 GeV} & $\mu\mu$ & 1,000,000 \\
 & $ee$ & 1,000,000 \\
 & $\tau\tau$ & 1,000,000 \\  
 & $e\mu$ & 1,000,000 \\  
 & $\tau\mu$ & 1,000,000 \\
 & $\tau e$ & 1,000,000 \\   
\hline
  \multirow{6}{*}{700 GeV} & $\mu\mu$ & 1,000,000 \\
 & $ee$ & 1,000,000 \\
 & $\tau\tau$ & 1,000,000 \\  
 & $e\mu$ & 1,000,000 \\  
 & $\tau\mu$ & 1,000,000 \\
 & $\tau e$ & 1,000,000 \\ 
\botrule\end{tabular}\label{tab:samples1}}
\end{table}

\subsubsection{Comparison with the official CMS results}
\begin{figure}
    \centering
    \begin{tabular}{cc}
    \includegraphics[trim=0 10 25 50,clip,width=0.45\textwidth]{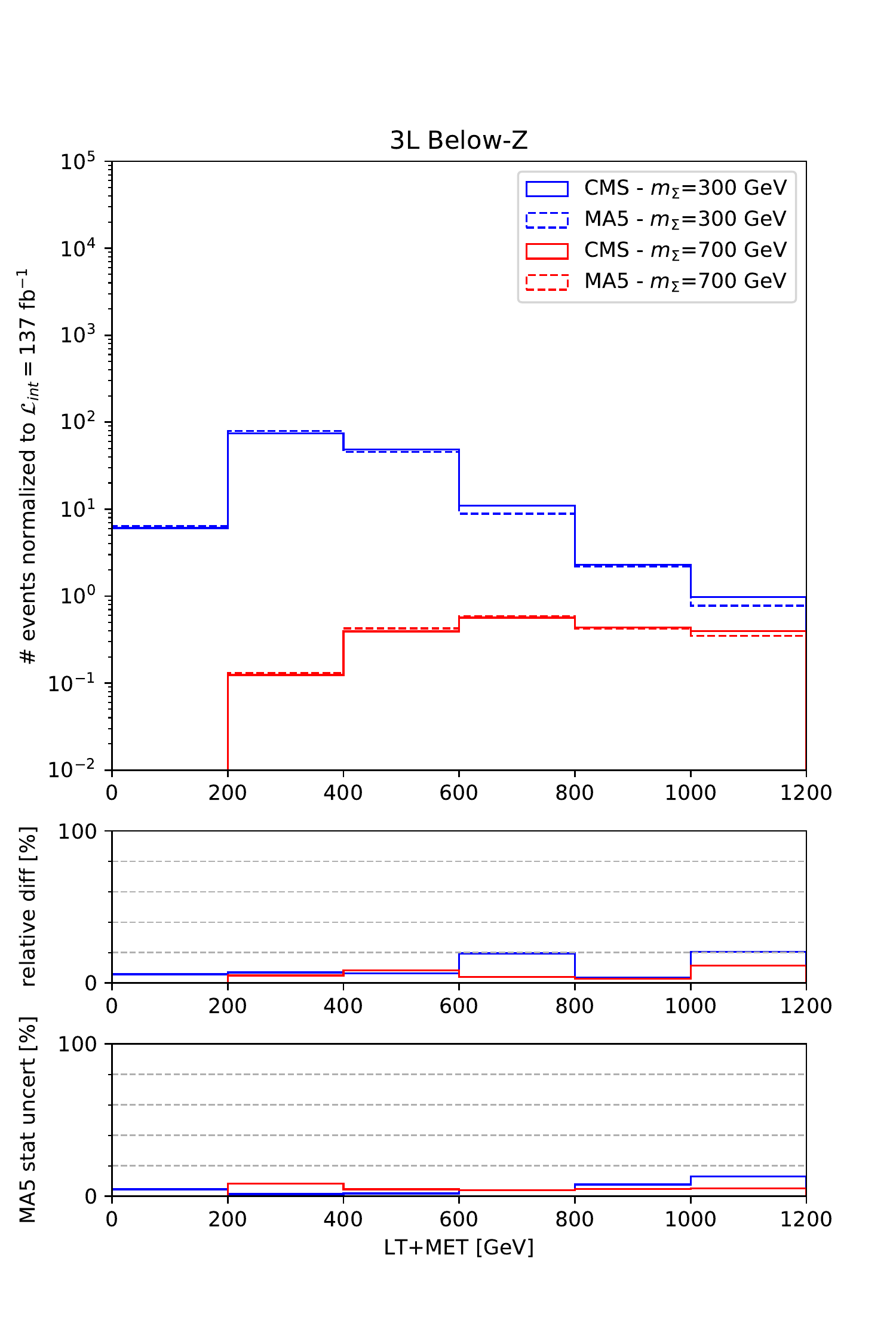} & \includegraphics[trim=0 10 25 50,clip,width=0.45\textwidth]{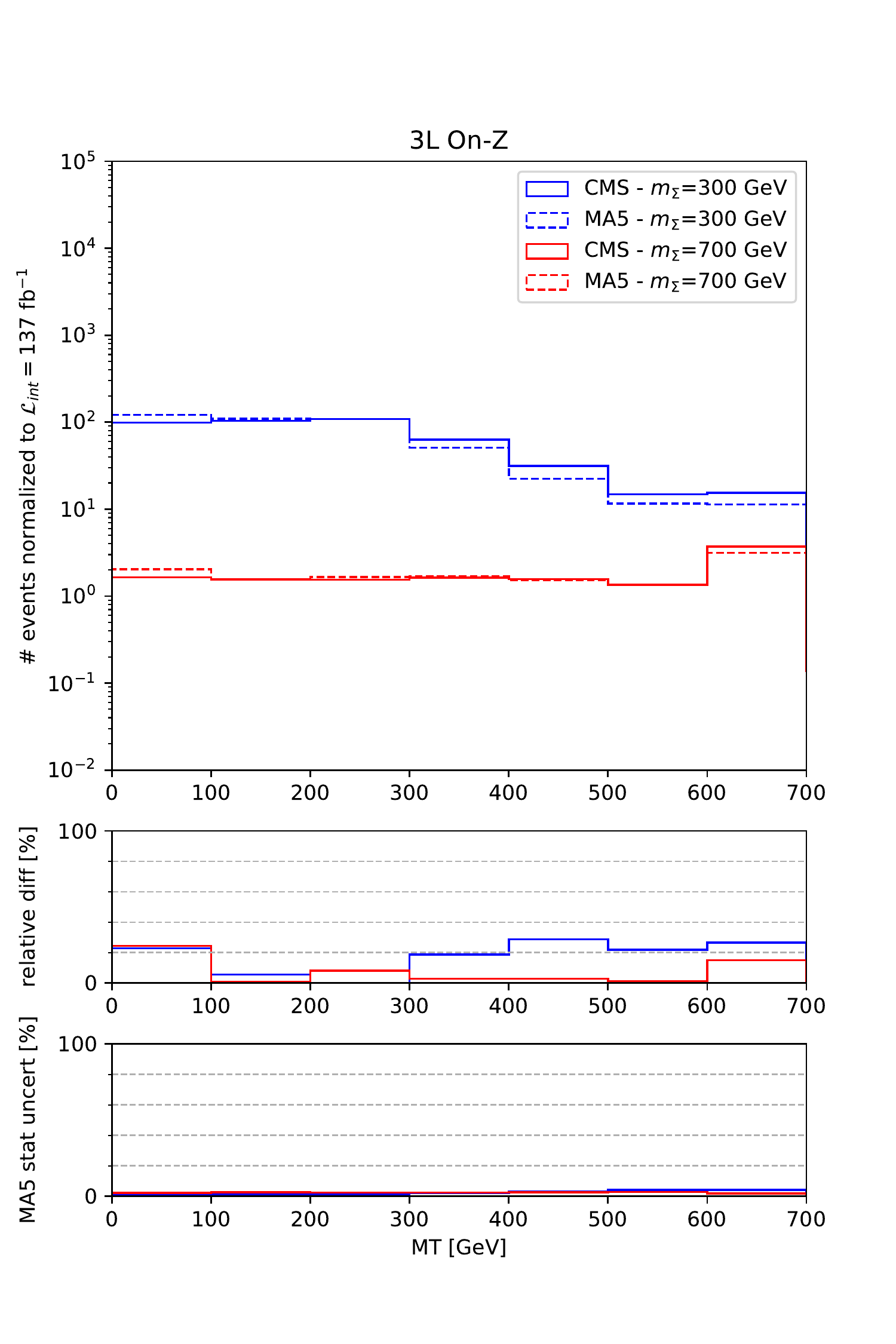} \\
    \includegraphics[trim=0 10 25 50,clip,width=0.45\textwidth]{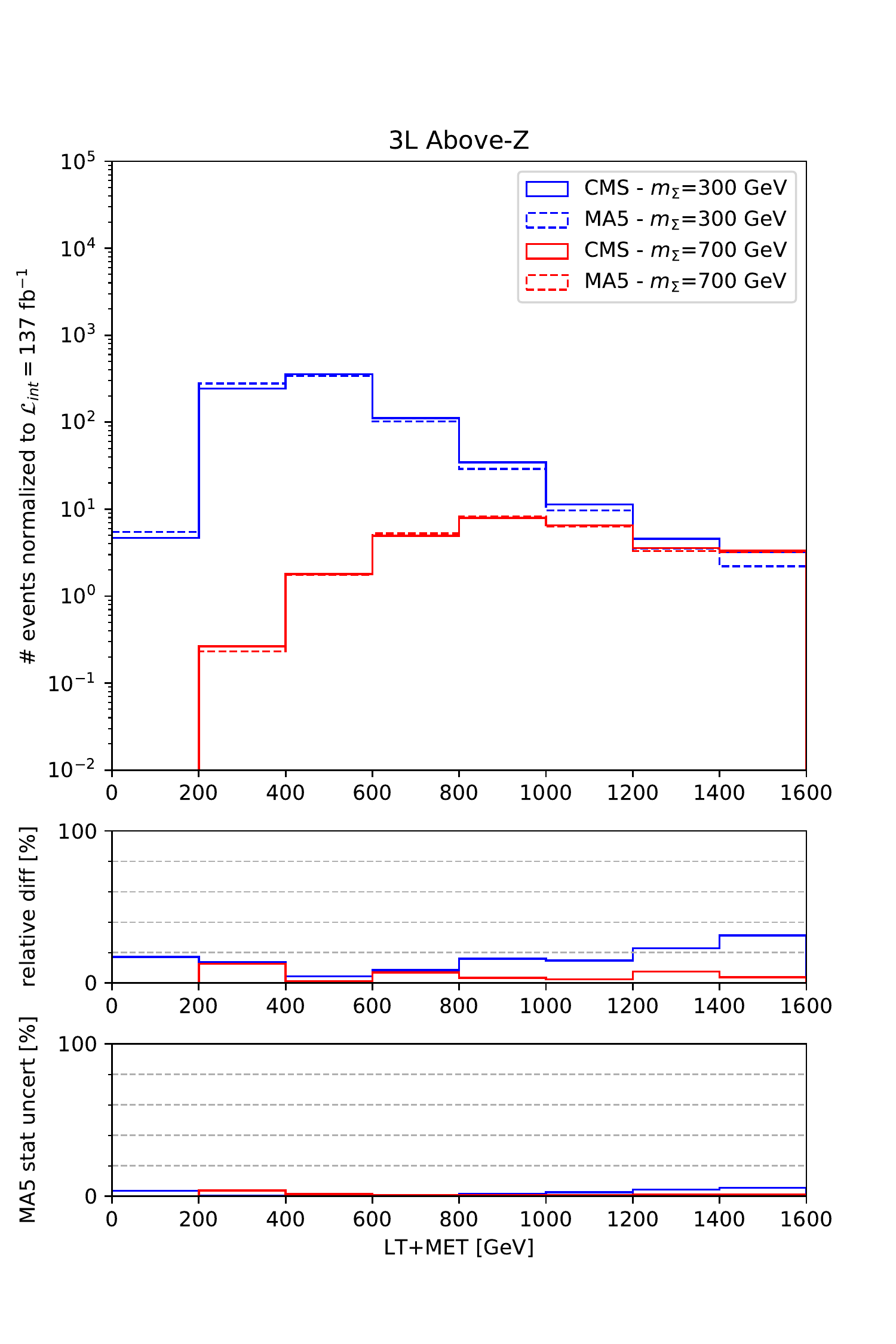} & \includegraphics[trim=0 10 25 50,clip,width=0.45\textwidth]{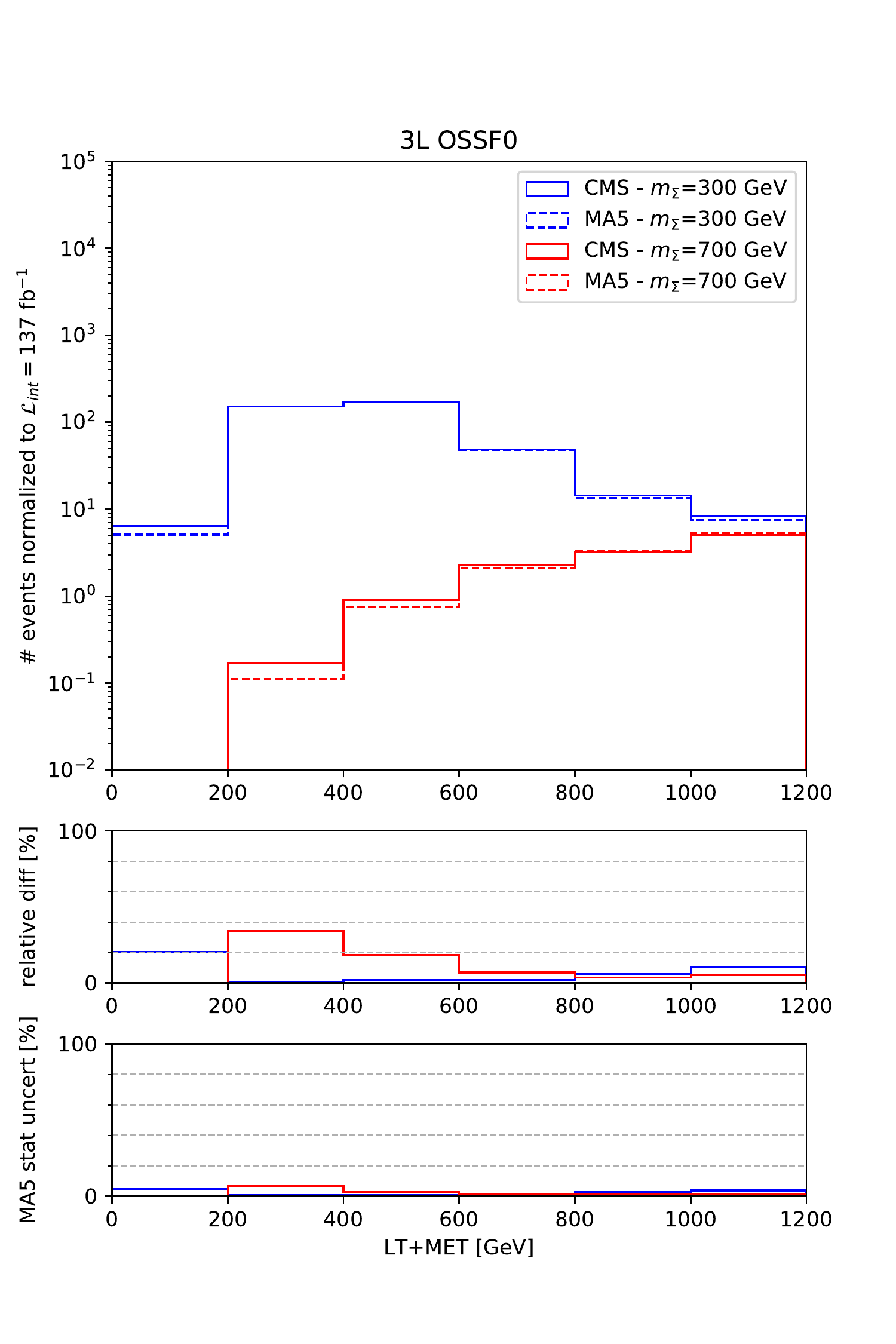} 
    \end{tabular}
    \caption{Comparison between CMS official distributions and the corresponding {\sc MadAnalysis}~5 predictions. In the main panel, we present distributions for the trilepton signal regions dedicated to probing type-III seesaw models. The last bins contain the overflow. In the central insets, we show the bin-to-bin relative difference $\delta({\rm bin})$ in percent between the CMS and {\sc MadAnalysis}~5 values, and in the lower insets, we indicate the statistical uncertainty $\delta_{MC}({\rm bin})$ in percent related to the Monte Carlo samples used for the {\sc MadAnalysis}~5 predictions. The distributions in red and blue correspond respectively to scenarios with a $\Sigma$ mass set to 700 GeV and 300 GeV.}
    \label{fig:plotsSeesaw1}
\end{figure}

\begin{figure}
    \centering
    \begin{tabular}{cc}
    \includegraphics[trim=0 10 25 50,clip,width=0.48\textwidth]{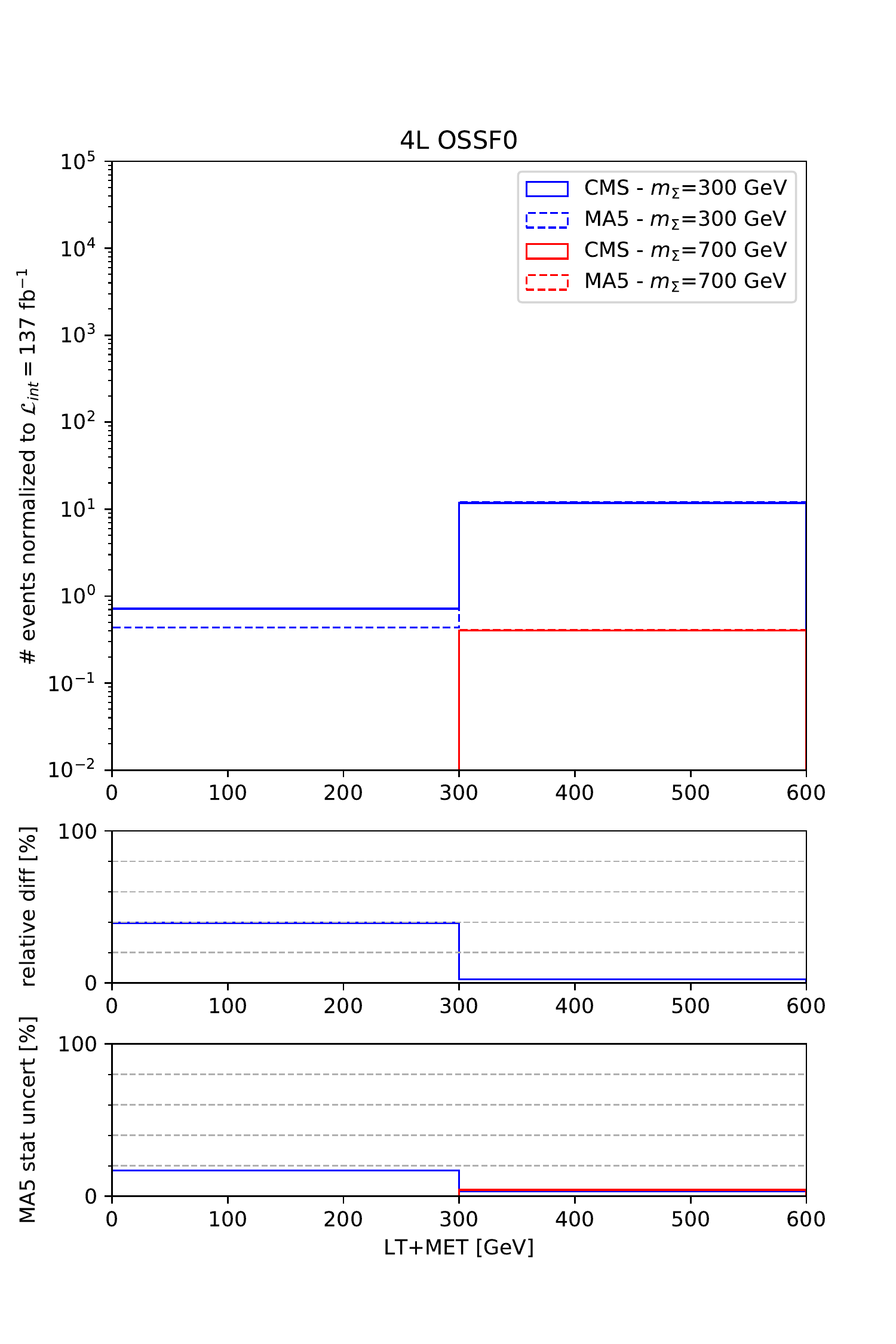} &
    \includegraphics[trim=0 10 25 50,clip,width=0.48\textwidth]{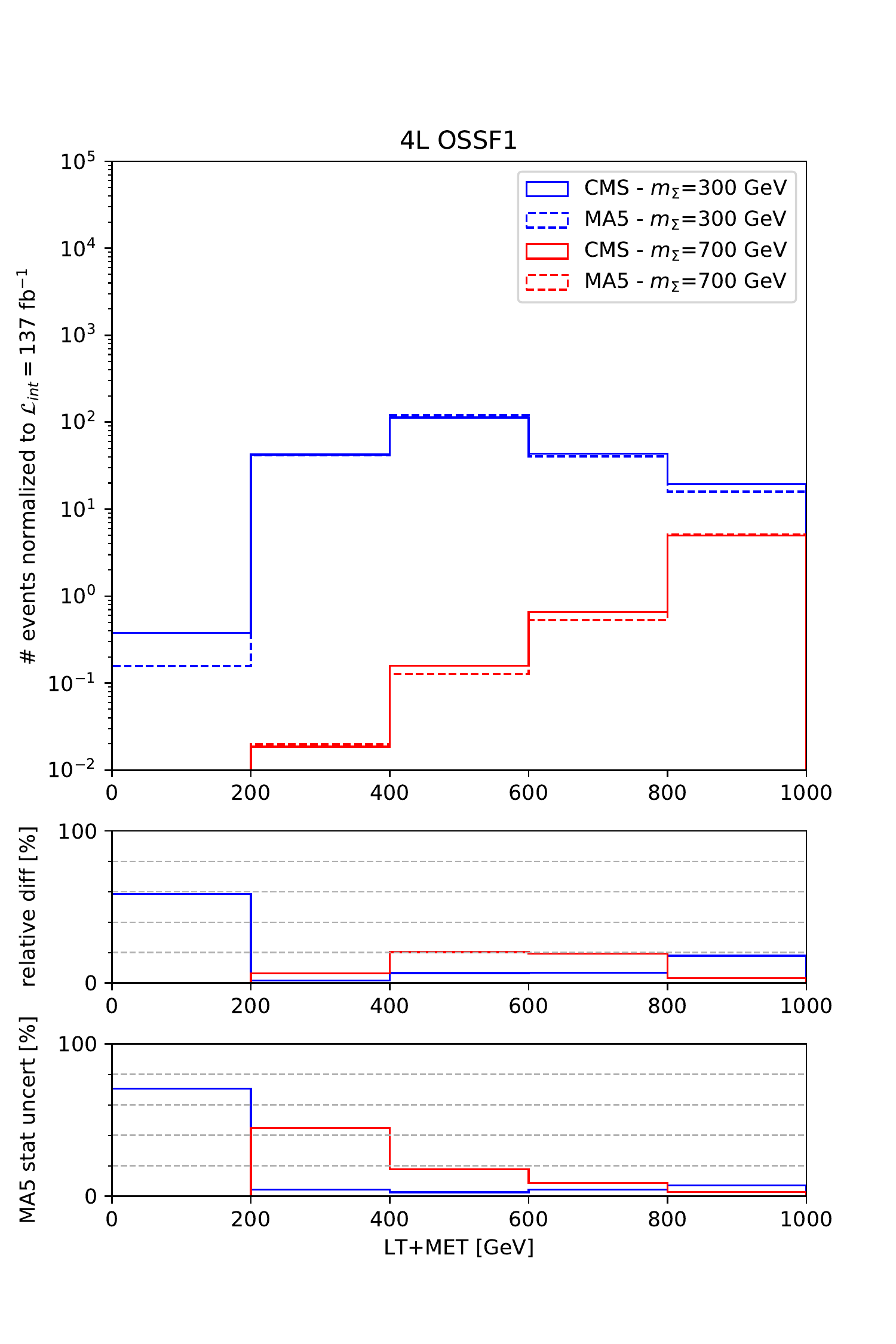}  
    \end{tabular}
    \includegraphics[trim=0 10 25 50,clip,width=0.48\textwidth]{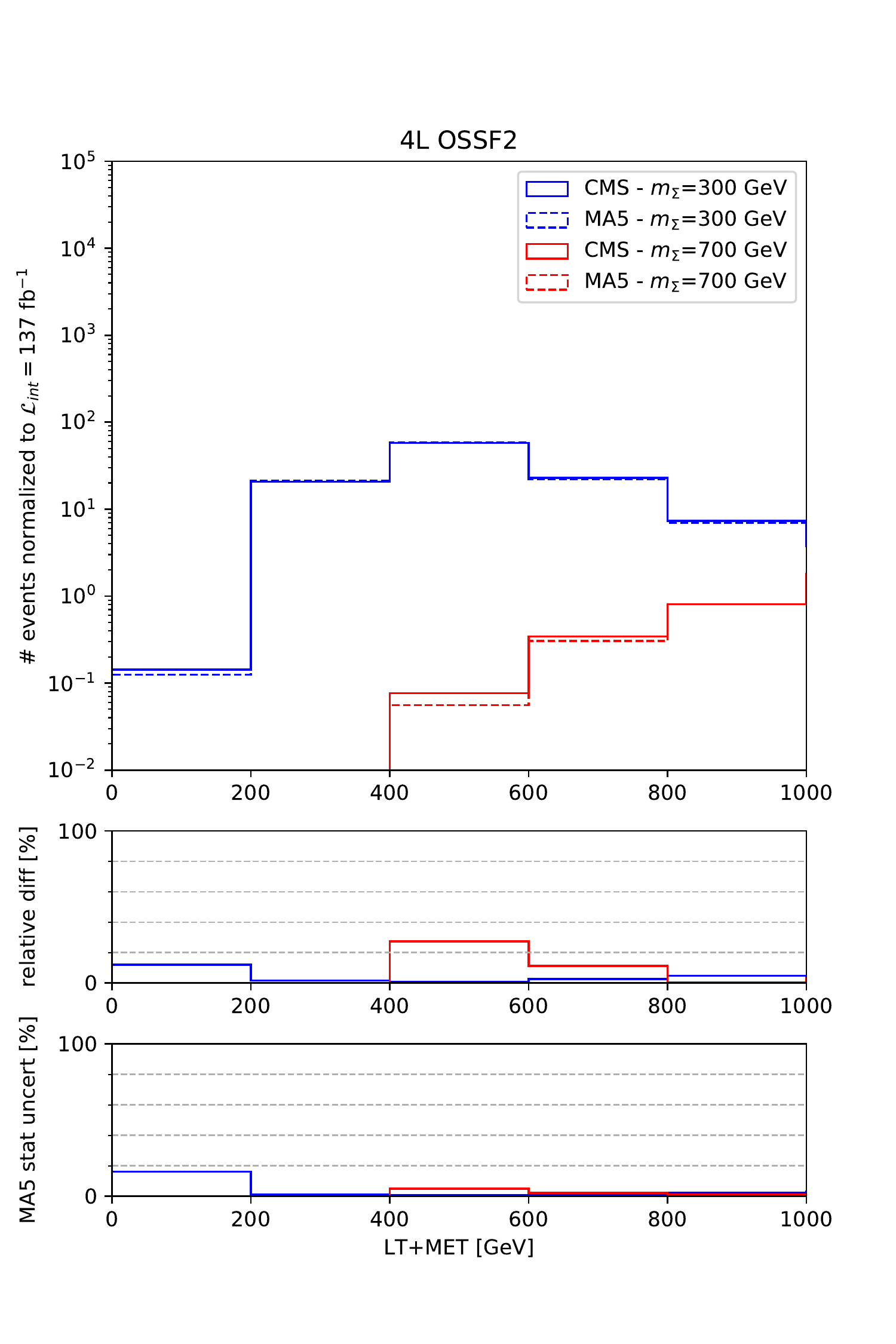}
    \caption{Same as in Figure~\ref{fig:plotsSeesaw1} but for the four-lepton seesaw signal regions.}
    \label{fig:plotsSeesaw2}
\end{figure}

The CMS paper does contain any cutflow-chart for the validation of the recast. This is the reason why the validation will be performed below on the principle of comparison of distributions of key observables at the end of the selection. All data used to build these plots is available from the {\sc HepData} service~\cite{Maguire:2017ypu,1764474} and is used for the validation of the recast analysis. In other words, we will compare the distributions obtained at the end of the selection with the ones presented in Figures~3 and 4 of the CMS analysis note. For 6 out of 7 signal regions, we consider the distribution of the quantity $L_T+p_T^{miss}$, where $L_T$ is defined as the scalar sum of the $p_T$ of all selected charged leptons and the missing transverse momentum. For the remaining \textsl{3L on-Z} signal region, we consider instead the transverse mass of the system made of the missing momentum and the lepton that is not part of any OSSF pair, 
\begin{equation}
  M_T=\sqrt{2p_T^{miss}p_T^l\left(1-\cos(\Delta\phi(\vec{p}_l,\vec{p}_T^{miss}))\right)}.
\end{equation}
\label{sec:indicators} The comparison of the CMS distributions with those obtained with our \textsc{MadAnalysis~5} reinterpretation is presented in Figure~\ref{fig:plotsSeesaw1} and Figure~\ref{fig:plotsSeesaw2}. For interpreting properly the results of the shape comparison, we make use of two indicators.
\begin{itemize}
    \item[$\bullet$] We first rely on the relative difference, on a bin-by-bin basis, between the number of events selected by the CMS analysis ($N_{\rm CMS}$) and the one selected in the recast analysis ($N_{\rm MA5}$). This difference is normalized with respect to the CMS predictions,
\begin{equation}\delta({\rm bin})=\frac{N_{\rm MA5}({\rm bin})-N_{\rm CMS}({\rm bin})}{N_{\rm CMS}({\rm bin})}.\end{equation} 
Such an indicator allows us to quantify the deviations between the CMS results and the recast predictions. We must however keep in mind that a large value in this indicator may not only be explained by the difference in the fast detector simulation or in the analysis implementation in \textsc{MadAnalysis~5}, but also by the statistical uncertainties inherent both to the samples used by CMS for the extraction of the official results (which we have no information on), and the validation samples.
    \item[$\bullet$] As mentionned in the previous item, the CMS official paper does not include information on the statistical uncertainties on the signal events. It is therefore impossible to assess the precision of their predictions. For the recast analysis, the bin-to-bin statistical uncertainties related to the amount of generated signal events at the end of the selection can be evaluated according to a Poisson distribution with variance
\begin{equation}\sigma({\rm bin})= \frac{N({\rm bin})}{\sqrt{N_{MC}({\rm bin})}}, \end{equation}
where $N_{MC}({\rm bin})$ is the number of surviving unweighted events in a specific bin at the end of the selection. We choose to define a relative indicator $\delta_{MC}$ quantifying the statistical uncertainties as
    \begin{equation}\delta_{MC}({\rm bin})= \frac{1}{\sqrt{N_{MC}({\rm bin})}}.\end{equation}
\end{itemize}

In the results shown in Figure~\ref{fig:plotsSeesaw1} and Figure~\ref{fig:plotsSeesaw2}, we can see that the shapes of the distribution are generally quite well reproduced. For all signal regions but the \textsl{4L OSSF1} one, the relative difference is less than 20--30\%. Such an order of magnitude is consistent with the theoretical and statistical uncertainties related to the signal, and the built-in differences in the analysis code and the detector simulation. For the signal region \textsl{4L OSSF1}, a larger difference is observed for the first bin, but it also corresponds to a configuration in which the recast analysis lacks statistics. The indicator $\delta_{MC}$ indeed exhibits a high statistical uncertainty. The differences between CMS and {\sc MadAnalysis}~5 are therefore considered as non-significant, an agreement being found in all the other bins, and we consider the implementation of the type-III seesaw signal regions as validated.

\subsection{Validation of the implementation of the $t\bar{t}\phi$ signal regions}\label{val2}

\subsubsection{Event generation}

To validate our implementation of the $t\bar{t}\phi$ signal regions, we consider a simple model implemented in {\sc FeynRules}. It includes a new light $CP$-even scalar or $CP$-odd pseudoscalar boson, labeled $\phi$, which can is produced at the LHC through its Yukawa coupling $g_t$ to top quarks. The corresponding UFO model~\cite{Chang:2017ynj,cmsmodelbase} has been connected to MG\_aMC@NLO in order to produce events at LO in QCD.

We produce the new boson $\phi$ in association with a top-antitop pair via its coupling $g_t$, and we assume that $\phi$ decays into a pair of charged leptons (electrons or muons) via a Yukawa coupling labeled $g_l$. The cross sections are calculated with the NNPDF3.0 LO set of PDF in the case where the product $g_t \cdot BR(\phi\rightarrow l^+l^-)$ is equal to 0.05, and read $0.02160 \pm 0.00216$~pb for a pseudoscalar boson with a mass of 20 GeV, and to $0.06597 \pm 0.00660$~pb for a scalar boson with a mass of 70 GeV~\cite{1764474}. The associated theory errors are taken as reported by the CMS collaboration as no information is provided on how they have been evaluated. Concerning the (anti-)top quark, the decay into $Wb$ is forced with a branching ratio of 1, and the $W$ decay is handled by {\sc Pythia}~8. Trilepton and four-lepton final state can arise from leptonic $W$-boson decays.

{\sc Pythia}~8 is used in order to handle parton showering, hadronization, and the simulation of the underlying events (multiple interactions and beam remnant interactions). The underlying events tune is chosen to be CP5~\cite{Sirunyan:2019dfx}.

A large statistics of events have been generated for each $\phi$ mass value and for each decay channel, as listed in Table~\ref{tab:samples2}.

\begin{table}[t]
  \renewcommand{\arraystretch}{1.2}
  \setlength\tabcolsep{10pt}
\tbl{List of produced signal samples for the validation of the $t\bar{t}\phi$ signal regions.}
{\begin{tabular}{@{} c c c c @{}}\toprule
$\phi$ scalar/pseudoscalar & $\phi$ mass & $\phi$ decay & Number of produced events \\
\hline
\multirow{2}{*}{pseudoscalar} &  \multirow{2}{*}{20 GeV} & $\phi \rightarrow \mu\mu$ & 2,400,000 \\
                                   &  & $\phi \rightarrow ee$ & 4,400,000 \\ \hline
                                   \multirow{2}{*}{scalar} &  \multirow{2}{*}{70 GeV} & $\phi \rightarrow \mu\mu$ & 2,400,000 \\
                                   &  & $\phi \rightarrow ee$ & 3,200,000 \\
\botrule\end{tabular}\label{tab:samples2}}
\end{table}

\subsubsection{Comparison with CMS results}

\begin{figure}
    \centering
    \begin{tabular}{cc}
    \includegraphics[trim=0 10 25 50,clip,width=0.48\textwidth]{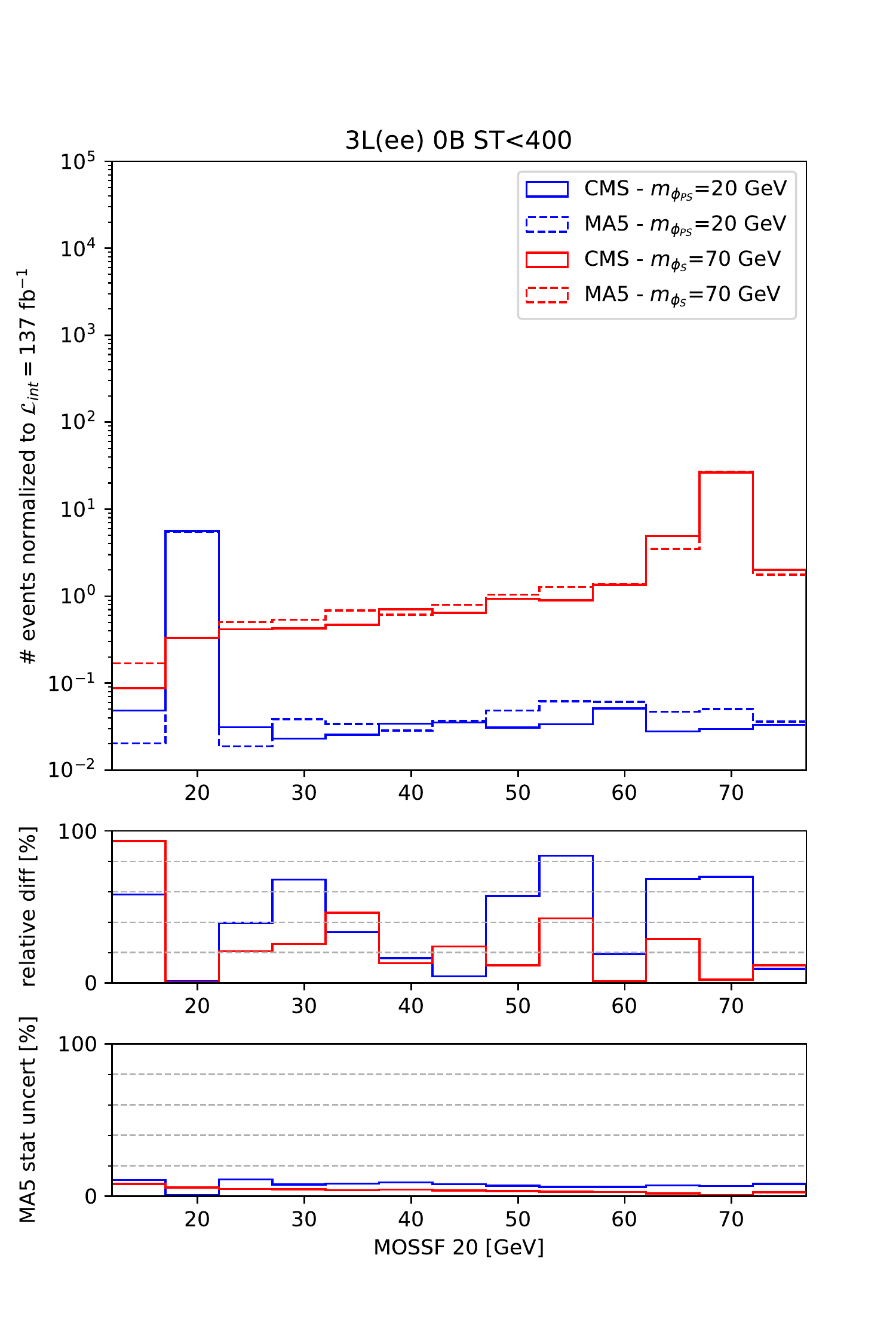} & \includegraphics[trim=0 10 25 50,clip,width=0.48\textwidth]{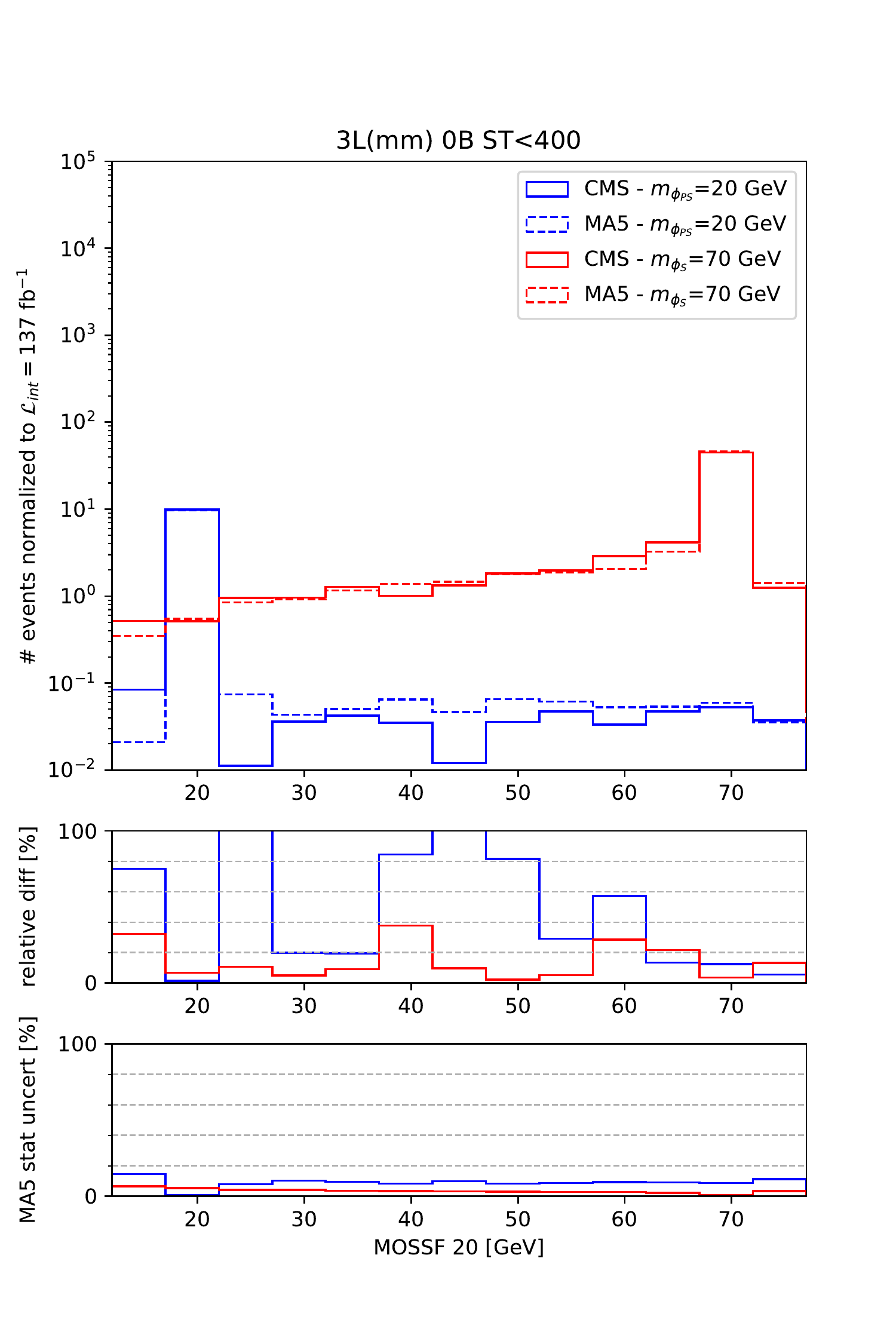} \\
    \includegraphics[trim=0 10 25 50,clip,width=0.48\textwidth]{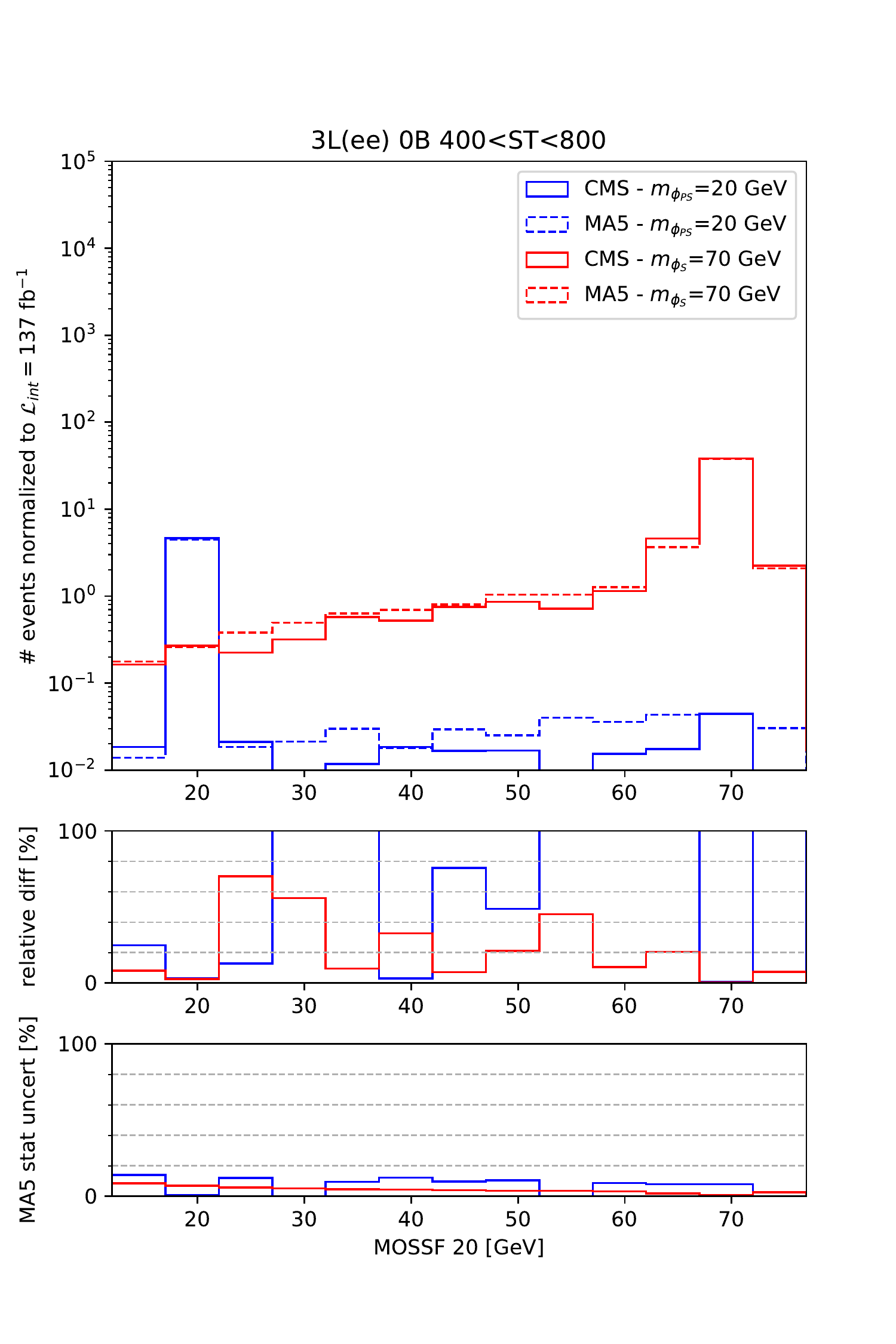} & \includegraphics[trim=0 10 25 50,clip,width=0.48\textwidth]{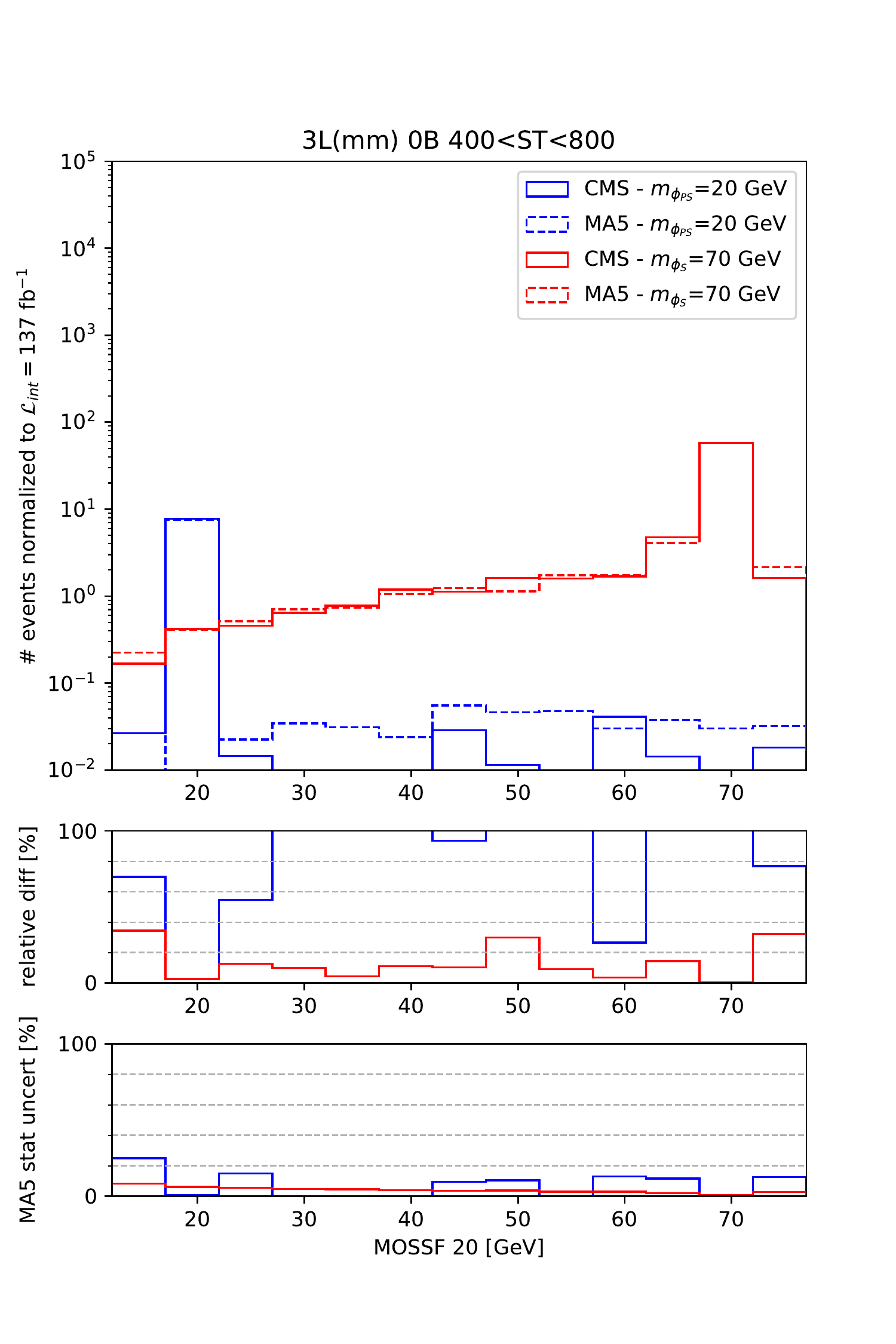}
    \end{tabular}
    \caption{Same as in Figure~\ref{fig:plotsSeesaw1} but for the first four trilepton signal regions dedicated to probing the $t\bar t\phi$ model. The distributions in red and blue correspond respectively to scenarios with a scalar of mass of 70 GeV, and a pseudoscalar of mass of 20 GeV.}
    \label{fig:plotsPhi1}
\end{figure}

\begin{figure}
    \centering
    \begin{tabular}{cc}
    \includegraphics[trim=0 10 25 50,clip,width=0.48\textwidth]{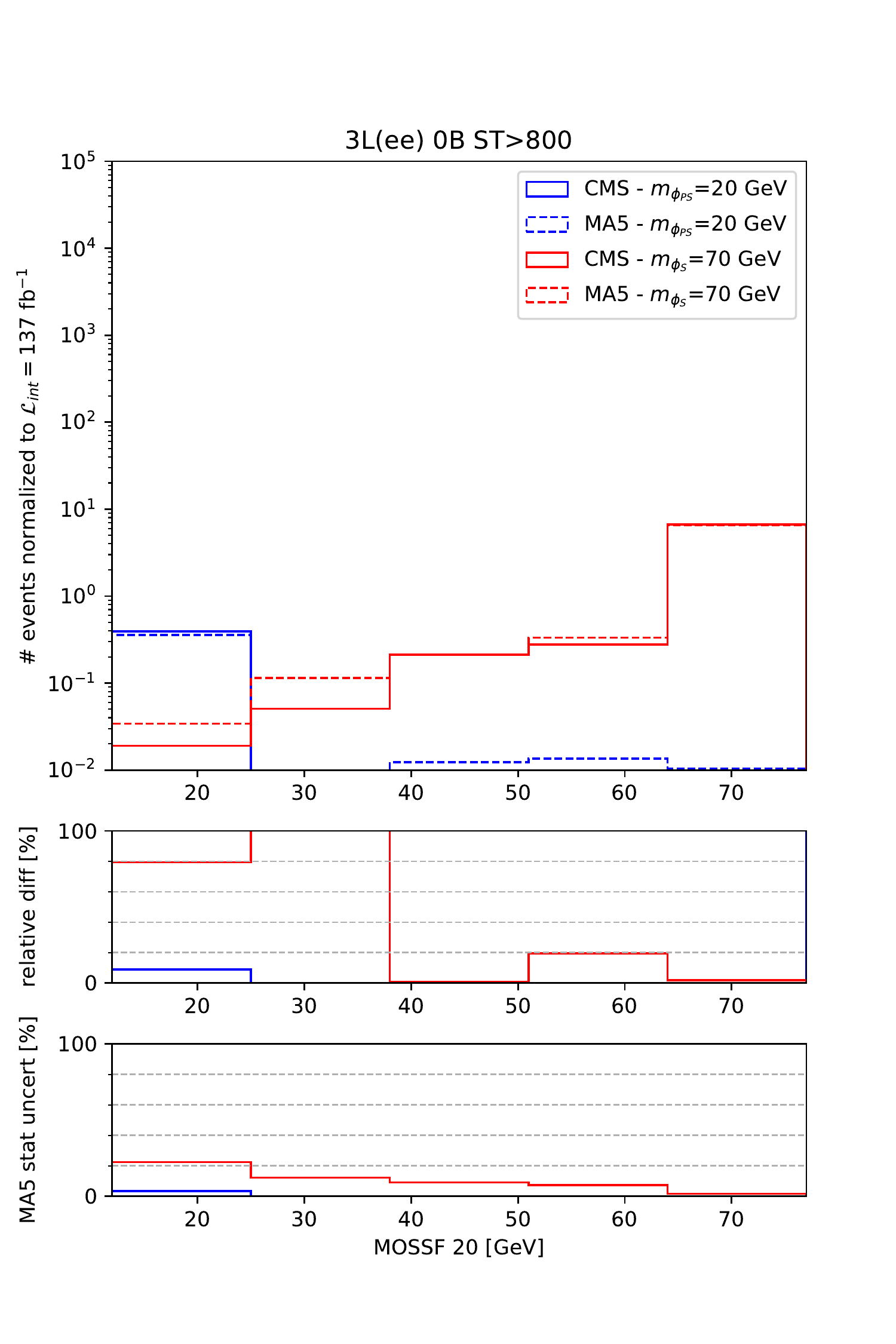} & 
    \includegraphics[trim=0 10 25 50,clip,width=0.48\textwidth]{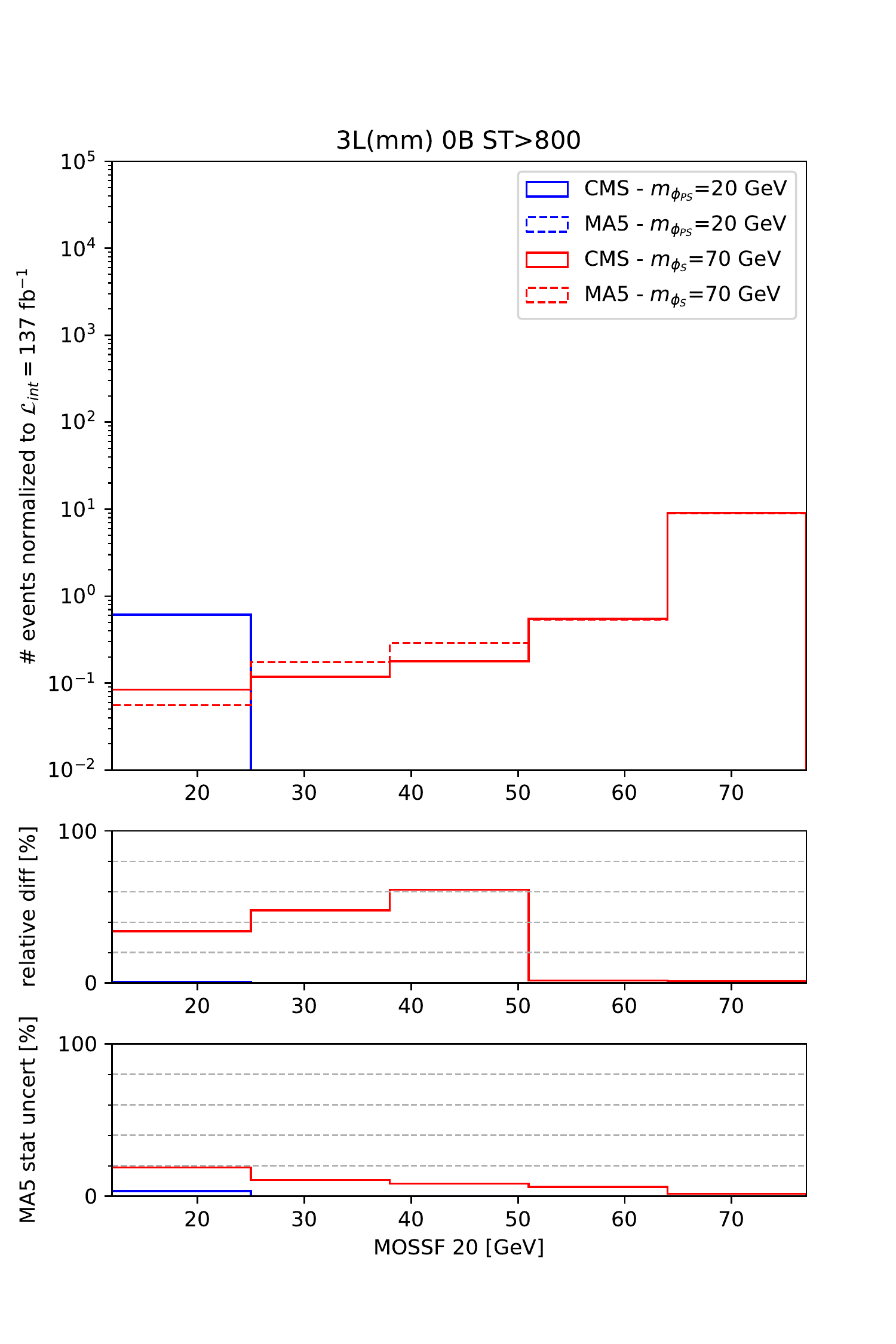} \\
    \includegraphics[trim=0 10 25 50,clip,width=0.48\textwidth]{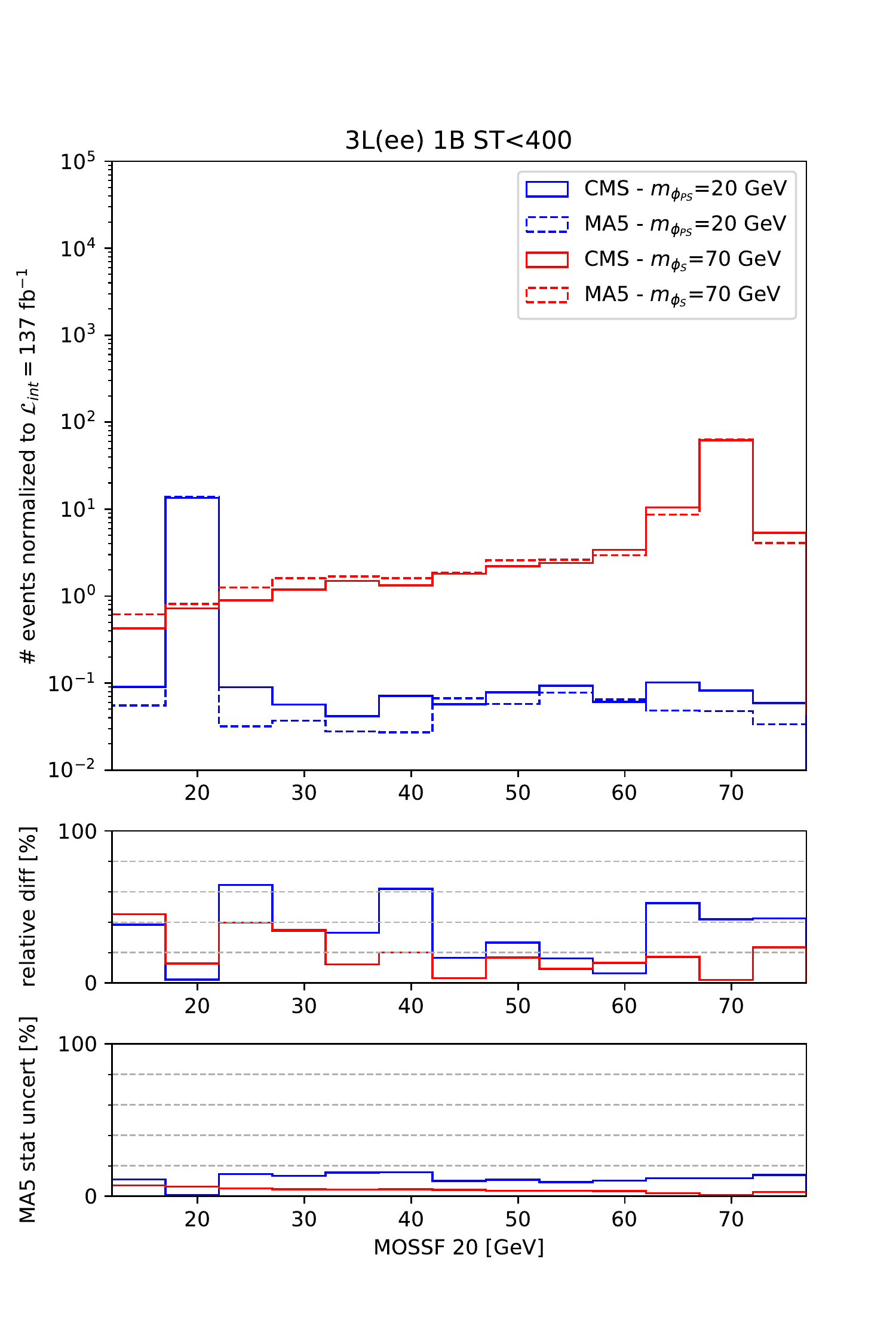} &
    \includegraphics[trim=0 10 25 50,clip,width=0.48\textwidth]{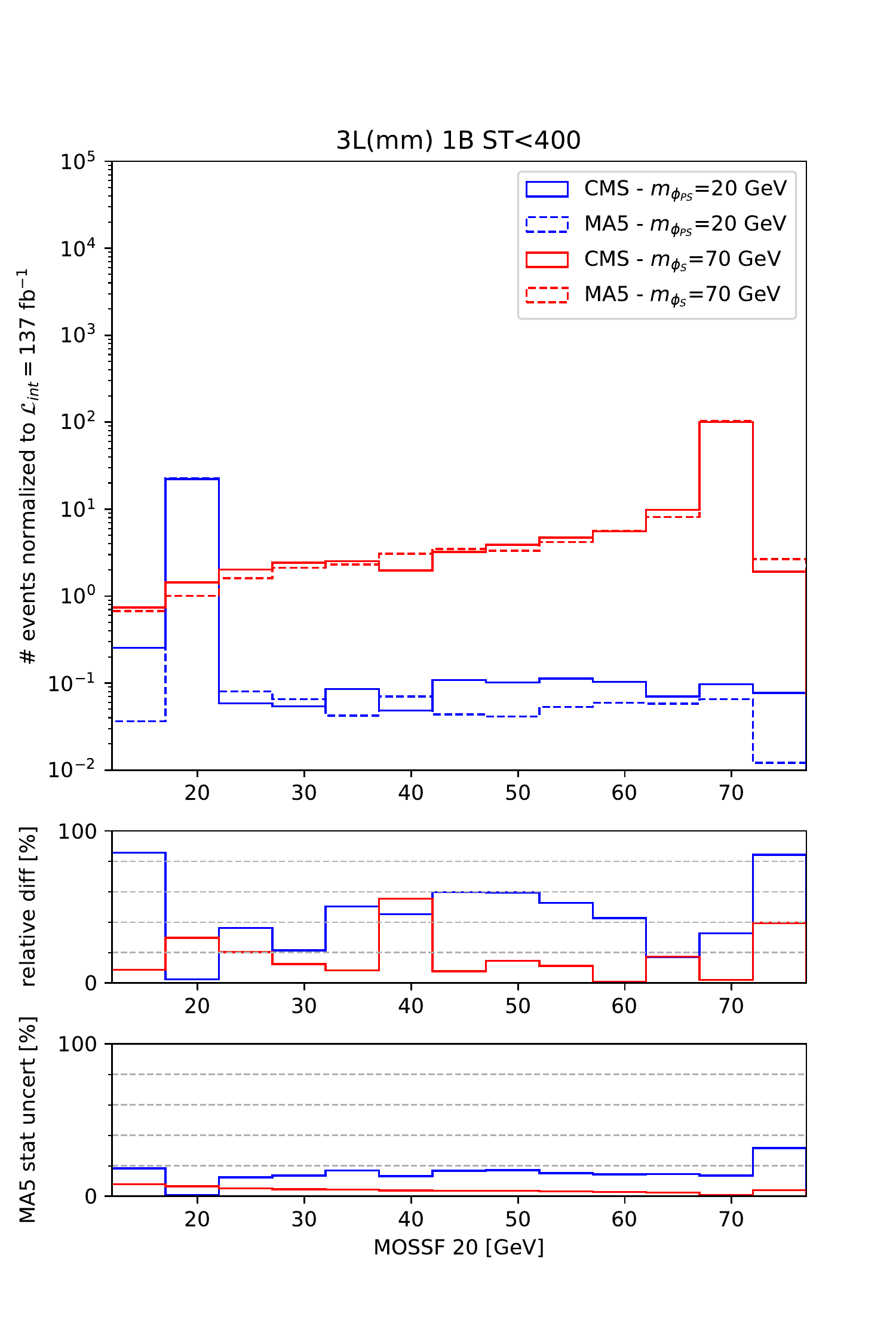} 
    \end{tabular}
    \caption{Same as in Figure~\ref{fig:plotsSeesaw1} but for the next four trilepton signal regions dedicated to probing the $t\bar t\phi$ model. The distributions in red and blue correspond respectively to scenarios with a scalar of mass of 70 GeV, and a pseudoscalar of mass of 20 GeV.}
    \label{fig:plotsPhi2}
\end{figure}

\begin{figure}
    \centering
    \begin{tabular}{cc}
    \includegraphics[trim=0 10 25 50,clip,width=0.48\textwidth]{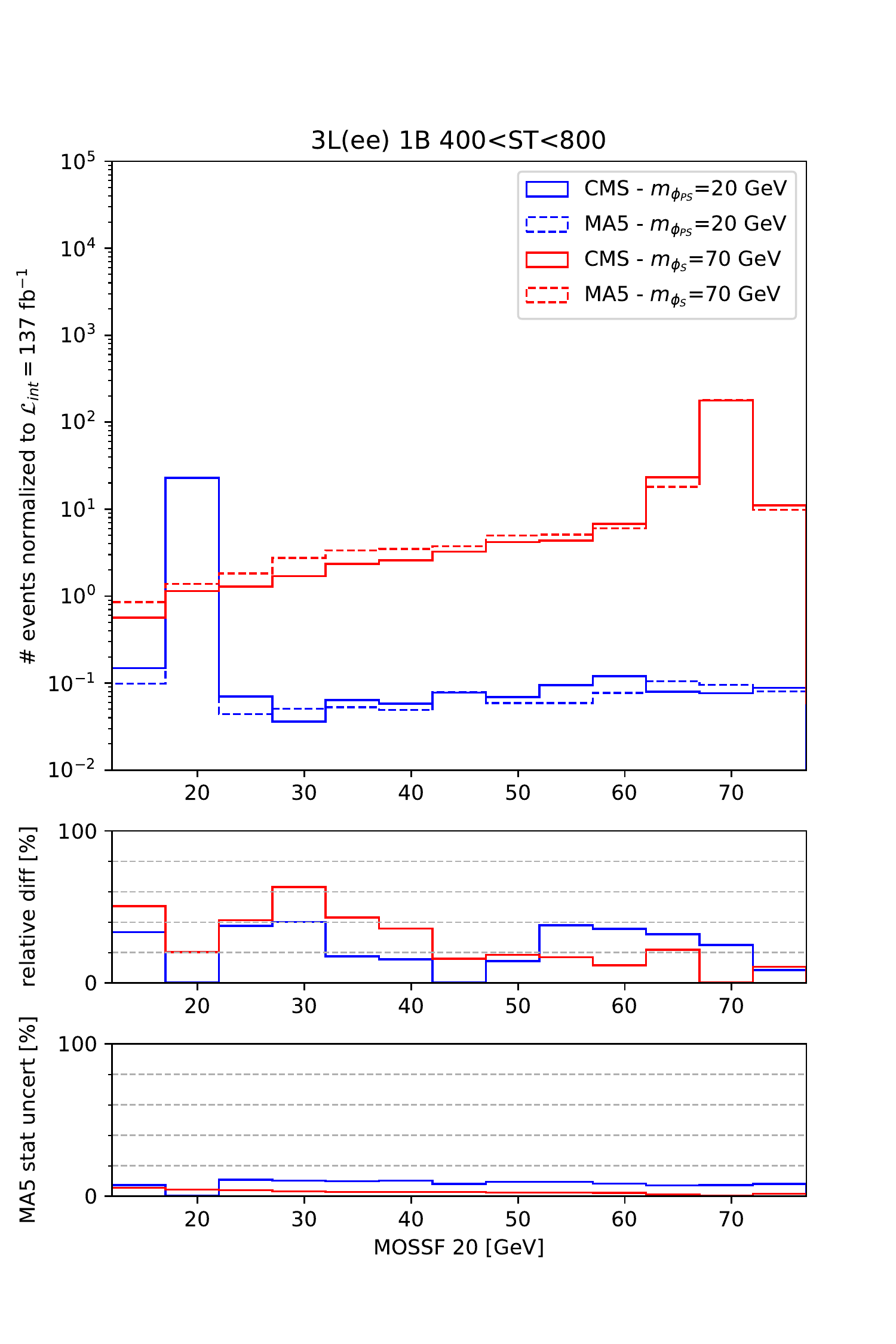} & \includegraphics[trim=0 10 25 50,clip,width=0.48\textwidth]{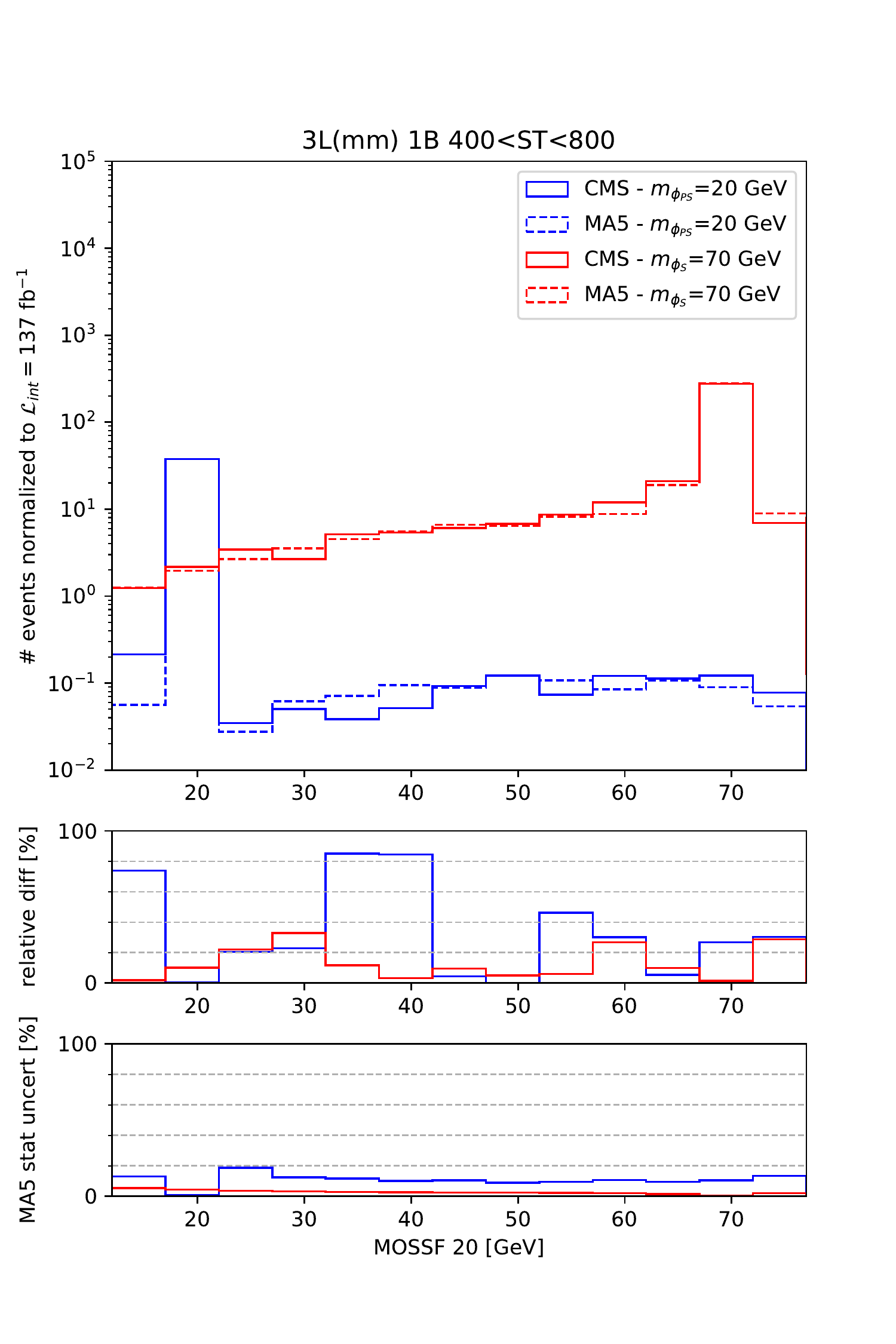} \\
    \includegraphics[trim=0 10 25 50,clip,width=0.48\textwidth]{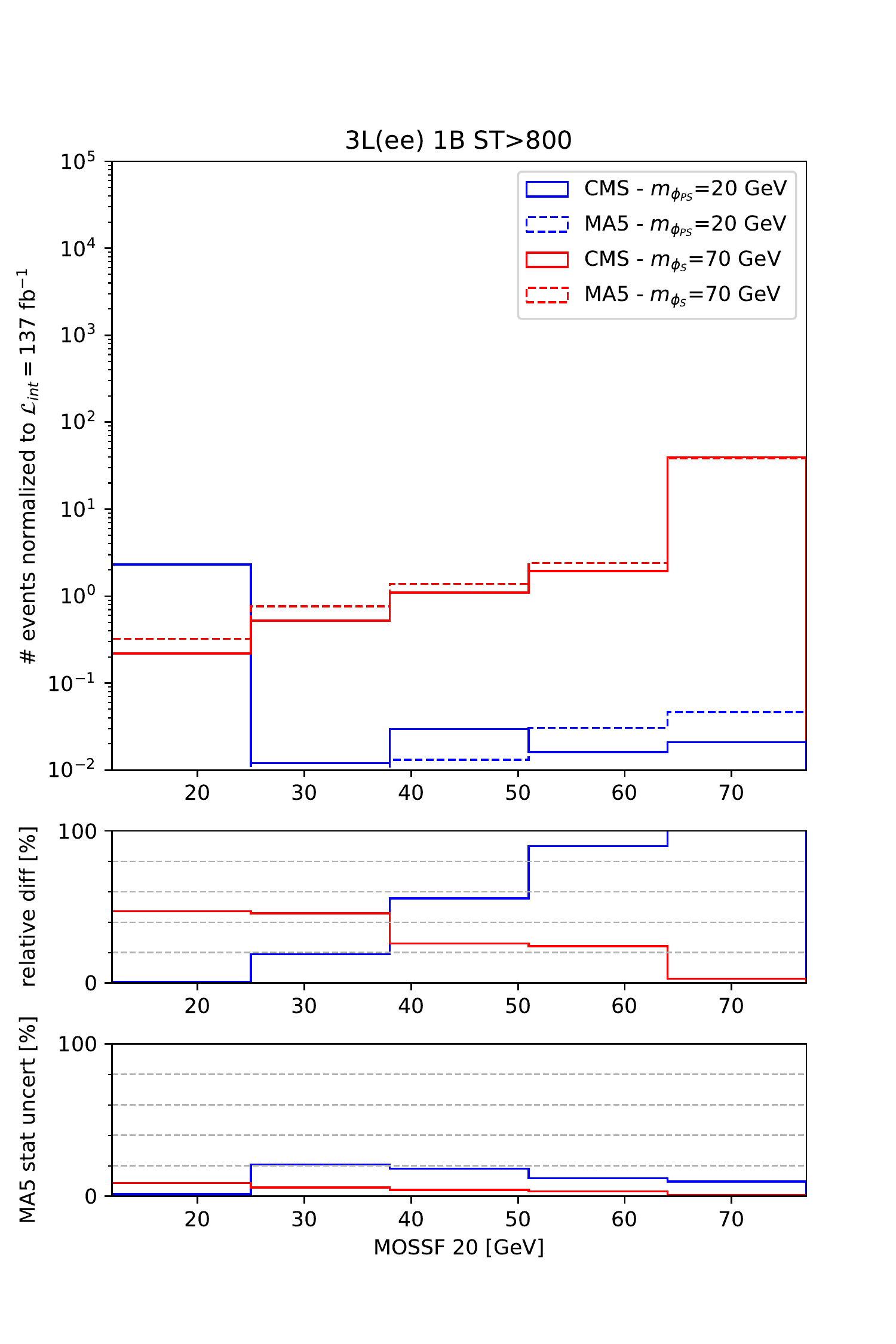} &
    \includegraphics[trim=0 10 25 50,clip,width=0.48\textwidth]{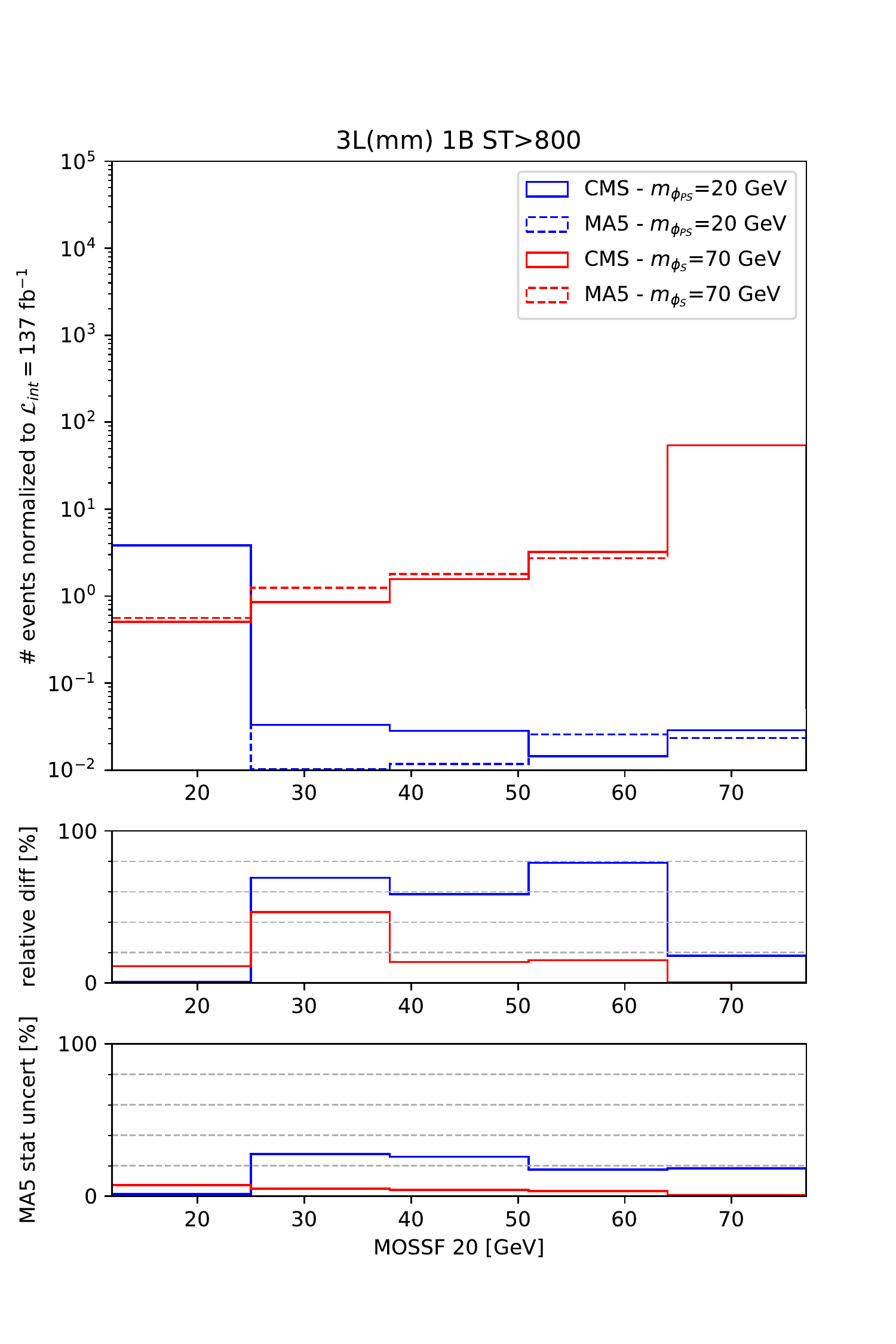} 
    \end{tabular}
    \caption{Same as in Figure~\ref{fig:plotsSeesaw1} but for the last four trilepton signal regions dedicated to probing the $t\bar t\phi$ model. The distributions in red and blue correspond respectively to scenarios with a scalar of mass of 70 GeV, and a pseudoscalar of mass of 20 GeV.}
    \label{fig:plotsPhi3}
\end{figure}

\begin{figure}
    \centering
    \begin{tabular}{cc}
    \includegraphics[trim=0 10 25 50,clip,width=0.48\textwidth]{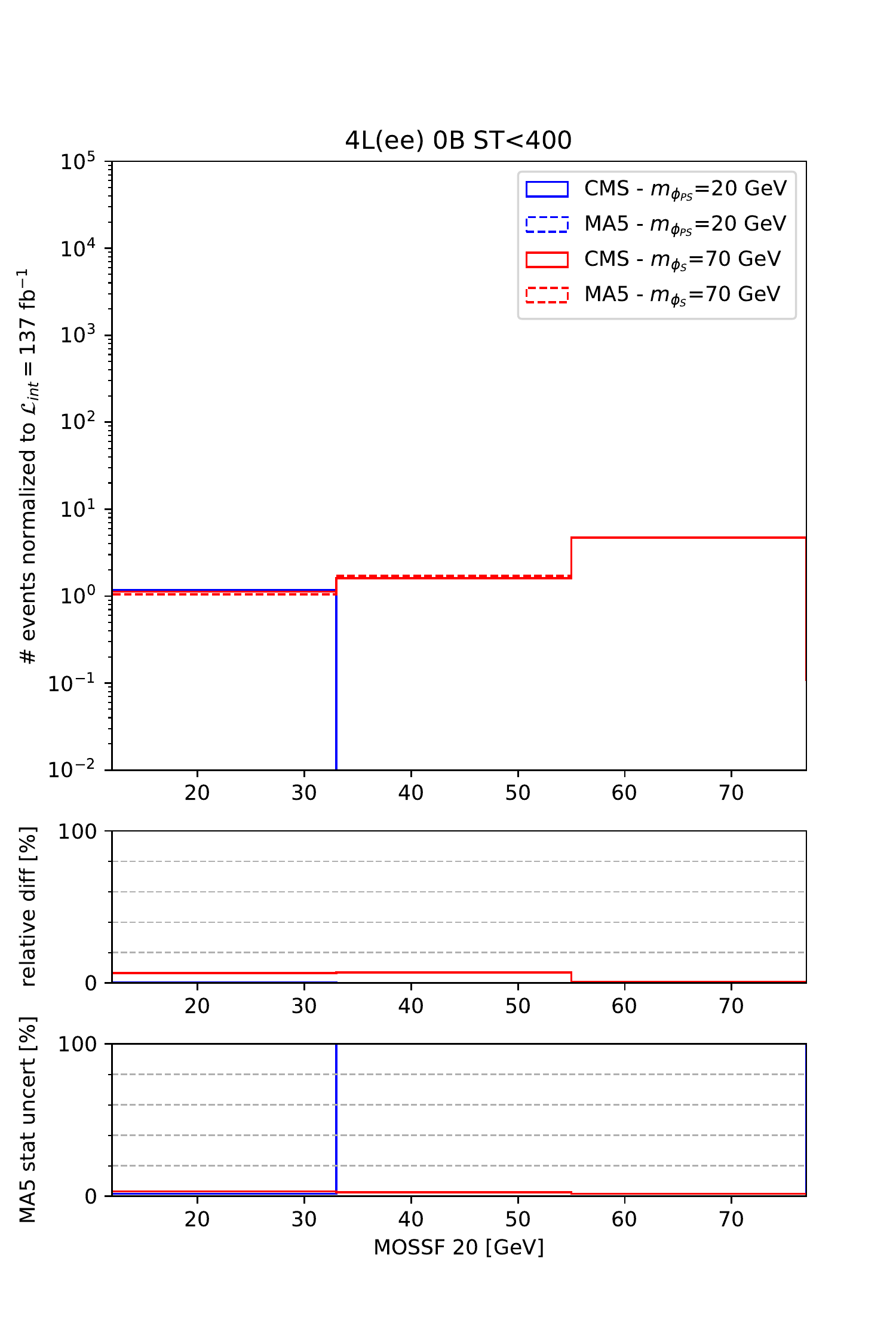} & \includegraphics[trim=0 10 25 50,clip,width=0.48\textwidth]{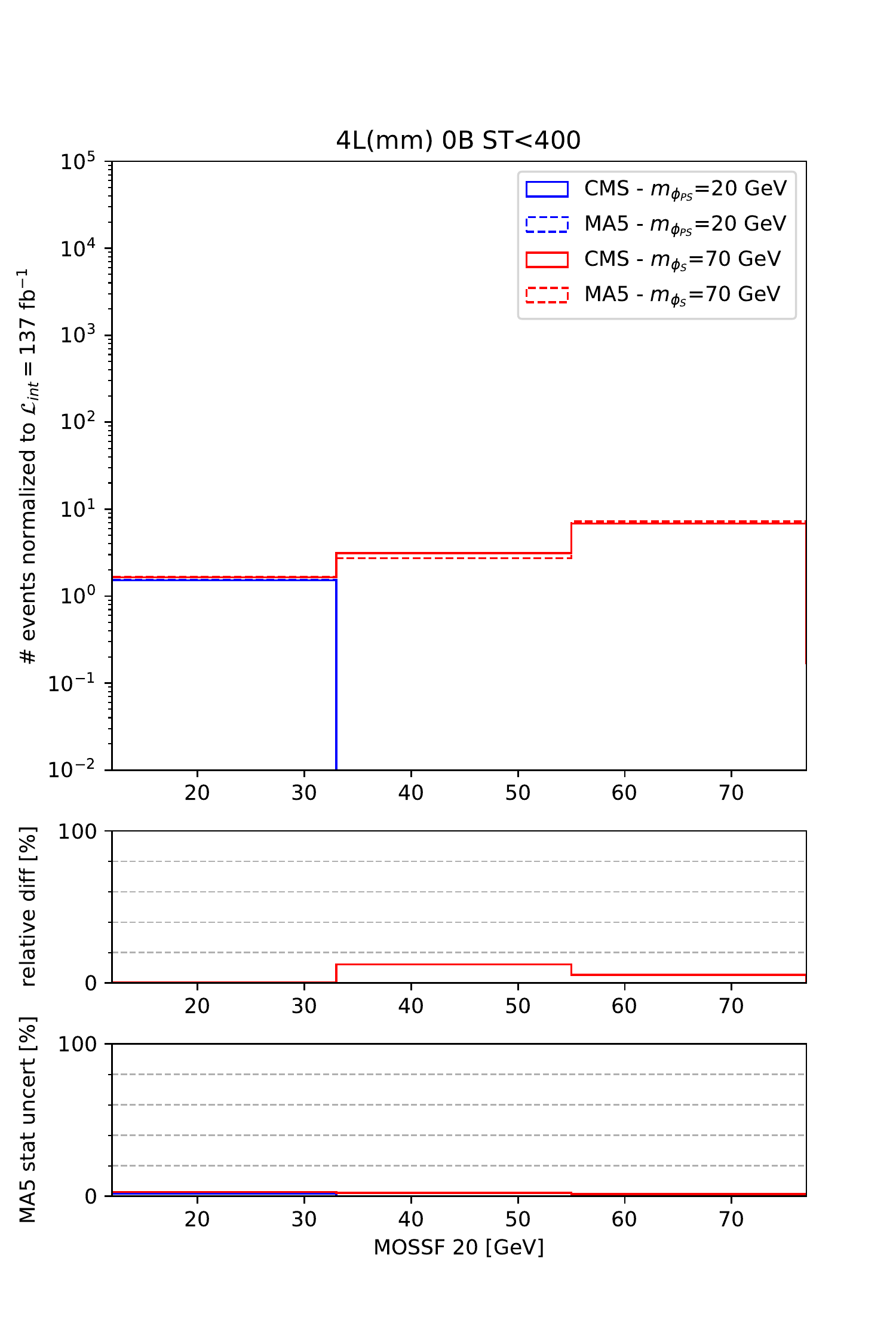} \\
    \includegraphics[trim=0 10 25 50,clip,width=0.48\textwidth]{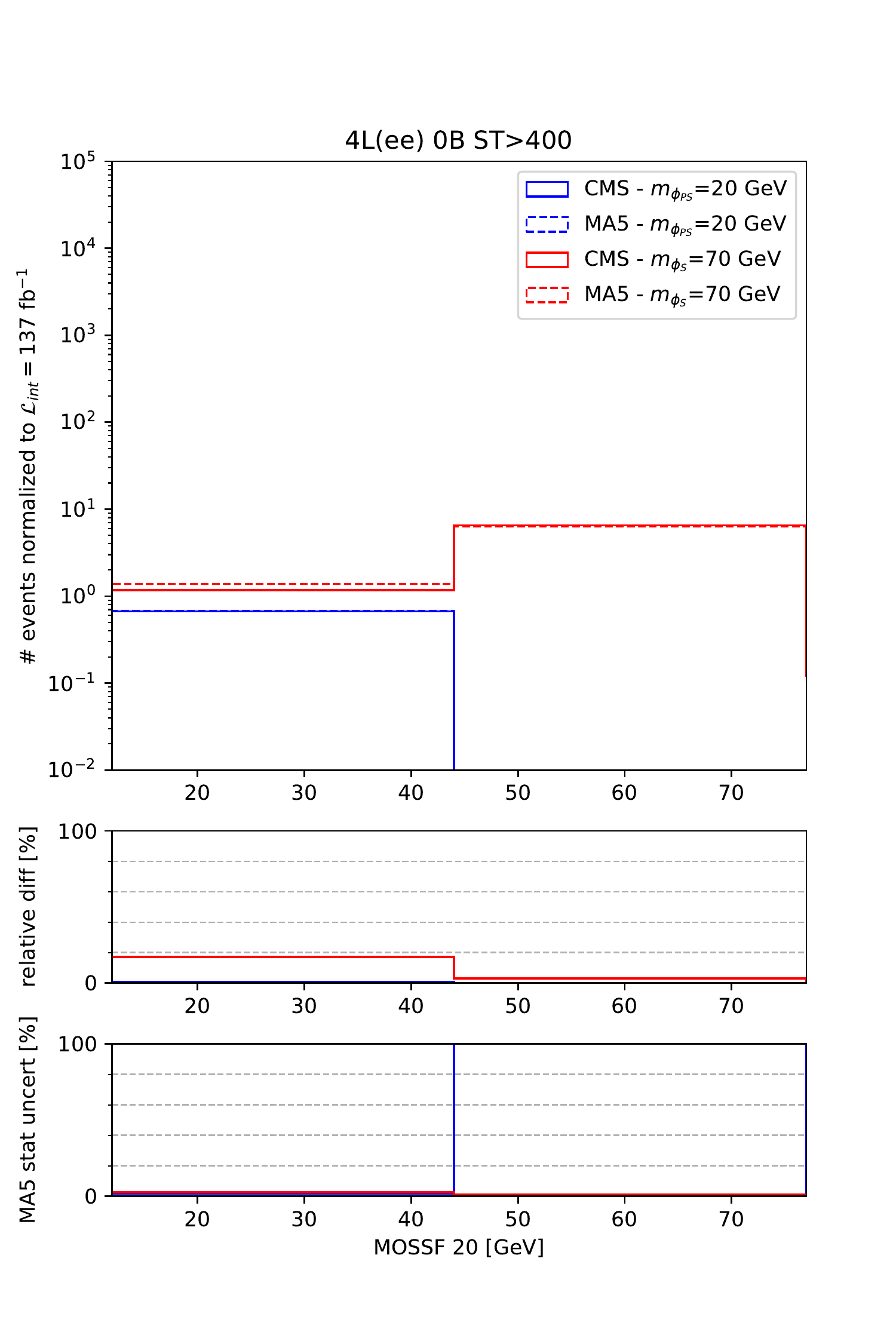} &
    \includegraphics[trim=0 10 25 50,clip,width=0.48\textwidth]{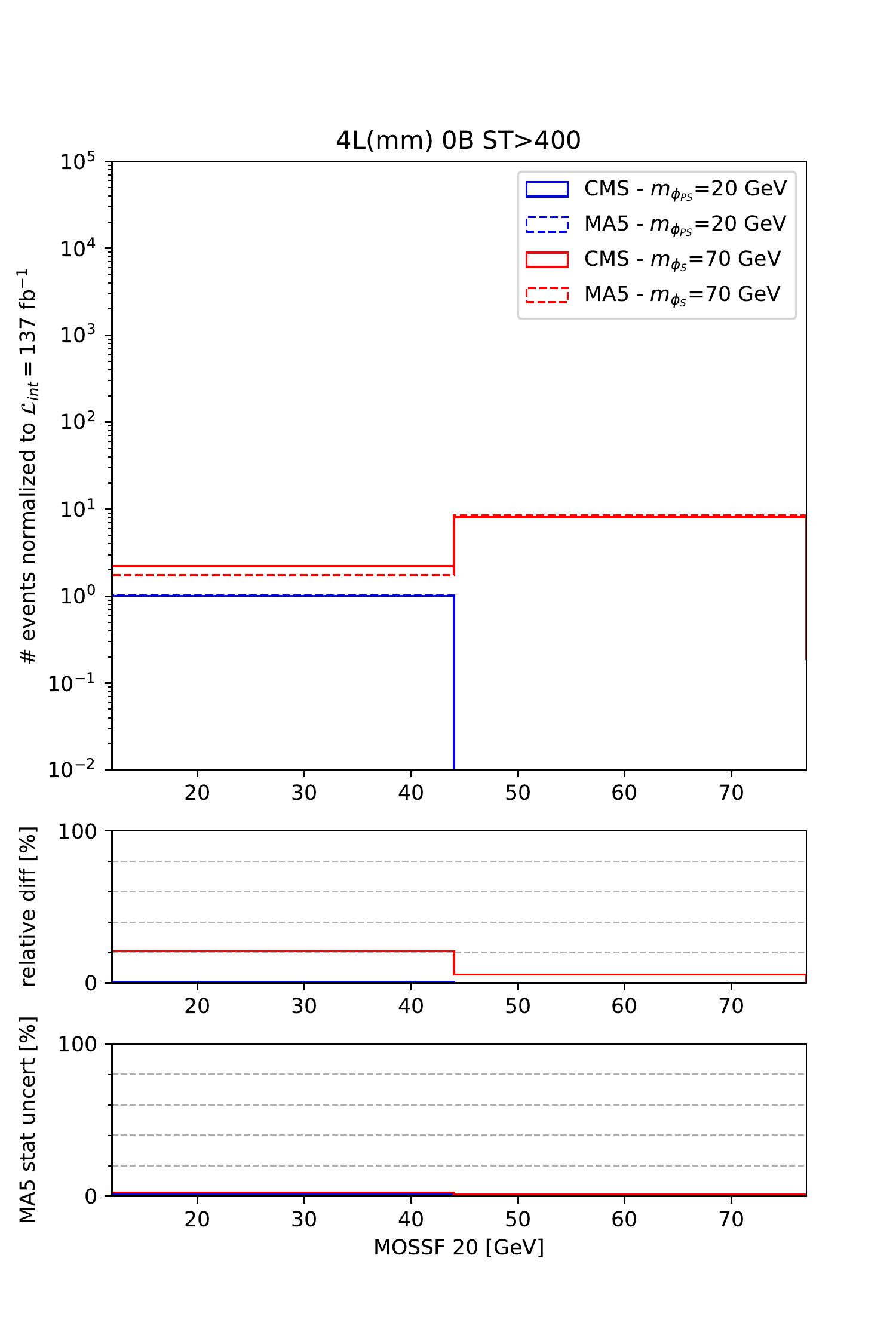} 
    \end{tabular}
    \caption{Same as in Figure~\ref{fig:plotsSeesaw1} but for the first four four-lepton and no $b$-jet signal regions dedicated to probing the $t\bar t\phi$ model. The distributions in red and blue correspond respectively to scenarios with a scalar of mass of 70 GeV, and a pseudoscalar of mass of 20 GeV.}
    \label{fig:plotsPhi4}
\end{figure}

\begin{figure}
    \centering
    \begin{tabular}{cc}
    \includegraphics[trim=0 10 25 50,clip,width=0.48\textwidth]{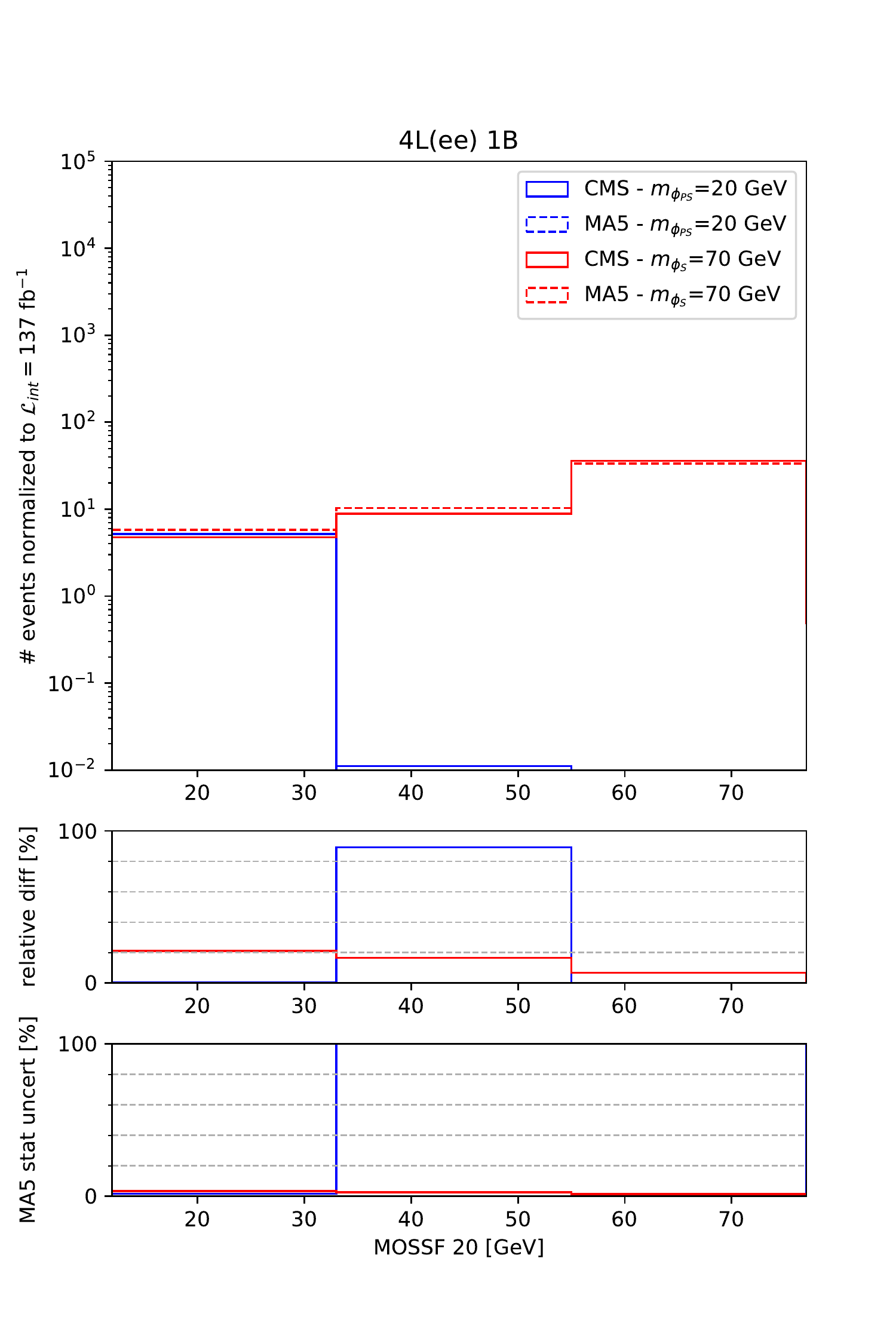} &
    \includegraphics[trim=0 10 25 50,clip,width=0.48\textwidth]{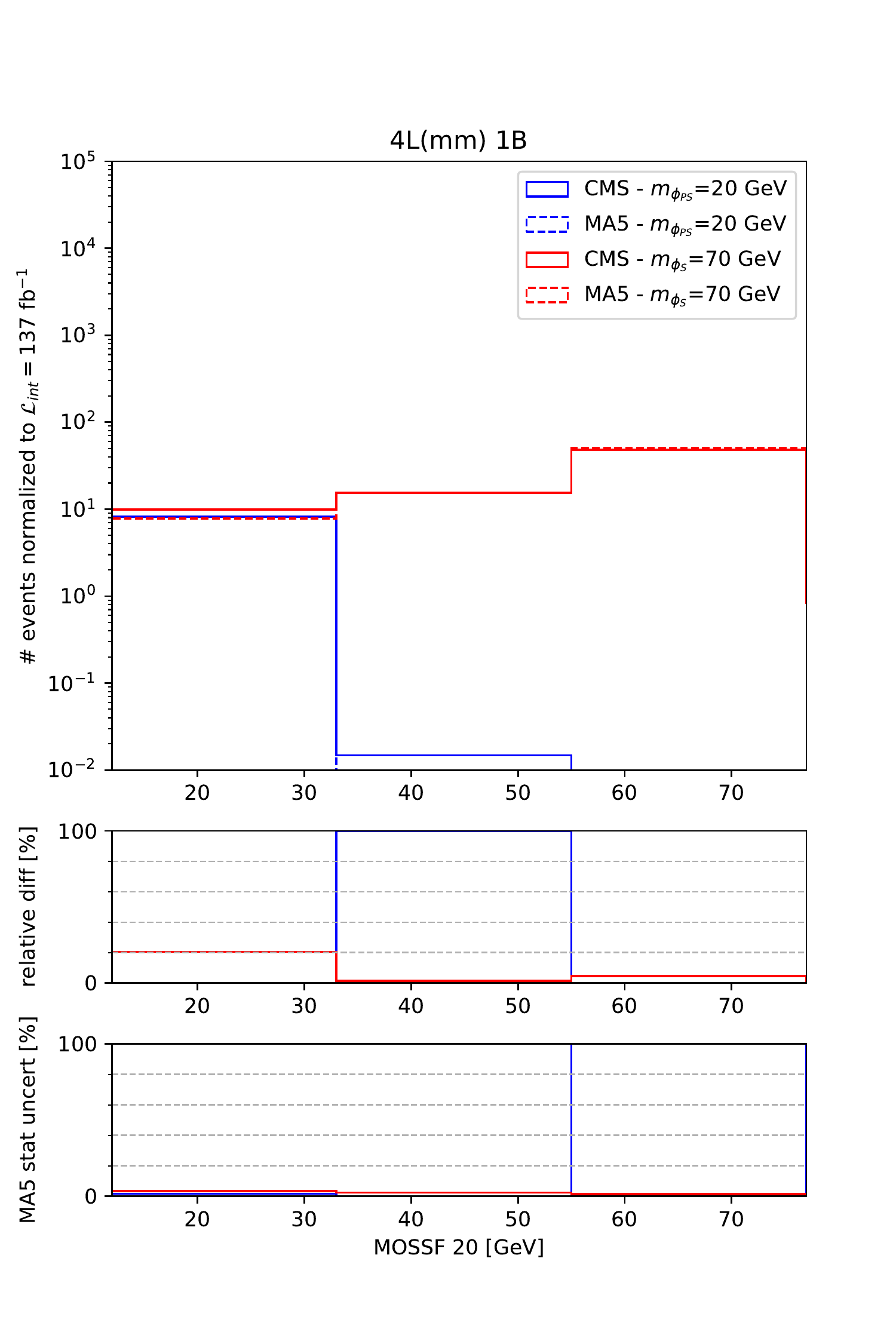} 
    \end{tabular}
    \caption{Same as in Figure~\ref{fig:plotsSeesaw1} but for the last two four-lepton and no $b$-jet signal regions dedicated to probing the $t\bar t\phi$ model. The distributions in red and blue correspond respectively to scenarios with a scalar of mass of 70 GeV, and a pseudoscalar of mass of 20 GeV.}
    \label{fig:plotsPhi5}
\end{figure}

Public results provided by the CMS collaboration only consist of spectra of observables at the end of the selection. We therefore validate our implementation by comparing the distributions obtained with \textsc{MadAnalysis~5} with the ones presented by CMS in Figures~5--10 of the analysis note. For low-mass $\phi$ (it is the case of our validation sample), the represented quantity is chosen to be the single attractor mass $M^{20}_{OSSF}$. The latter is defined as the invariant mass of the opposite-sign same-flavor lepton pair (OSSF) that is the closest to 20 GeV. Comparisons are performed in Figures~\ref{fig:plotsPhi1}, \ref{fig:plotsPhi2}, \ref{fig:plotsPhi3}, \ref{fig:plotsPhi4}  and \ref{fig:plotsPhi5} for all $t\bar t\phi$ signal regions.

At first order, the recast analysis manages to reproduce quite well the distributions presented in the CMS paper. There are however noticeable differences. The two indicators $\delta({\rm bin})$ and $\delta_{MC}({\rm bin})$ defined in Section~\ref{sec:indicators} are once again used for the interpretation of our findings and to quantify the level of agreement.

The statistics used for the validation of the recast analysis seems to be enough because the $\delta_{MC}({\rm bin})$ indicator is less than 10\% for all signal regions. For signal regions in which the relative difference between the CMS and the recast predictions is large, we find first that the issue holds independently of the $\phi$ decay channel. The findings however allow us to interpret this difference as a consequence of a lack of statistics in the events used by the CMS collaboration (on which information is not provided). We can indeed observe that the CMS predictions are plagued with important statistical fluctuations, that are much larger than in the recast analysis. We therefore consider our implementation validated, at least at a level representative of what could be done with the information made public by the CMS collaboration.

\subsection{Conclusions}\label{sec4}
We have presented the implementation of the multileptons search CMS-EXO-19-002 in the \textsc{MadAnalysis 5} framework. This search considers proton-proton collisions at $\sqrt{s}=13~\textrm{TeV}$ and an integrated luminosity of 137~\textrm{fb}$^{-1}$. Samples of signal events relevant for both the type-III seesaw and $t\bar{t}\phi$ signal regions have been generated with \textsc{MadGraph5\_aMC@NLO} at LO, then proccessed by \textsc{Pythia~8} for parton showering, hadronization and mutiple parton interactions, and by \textsc{Delphes}~3 for the detector simulation. We have compared predictions made by
\textsc{MadAnalysis 5} with the official results provided by the CMS collaboration. The only public material for validation consist in key-observable distributions at the end of selection. We have considered various benchmark scenarios in both the electron and muon channel. The shapes of the distributions have been compared and are correctly reproduced for the seesaw signal regions. Discrepancies are found in the case of the $t\bar{t}\phi$ events, in particular in the trilepton channels. These can however be explained mainly by a lack of statistics of the CMS paper.

The {\sc MadAnalysis}~5 C++ code is available, together with the material used for the validation of this implementation, from the MA5 dataverse (\href{https://doi.org/10.14428/DVN/DTYUUE}{https://doi.org/10.14428/DVN/DTYUUE})~\cite{DTYUUE_2021}.

\subsection*{Acknowledgments}
We are very grateful to Yeonsu Ryou Juhee and Song and Kihong Park who have in the first place begun to work on this implementation. We are also indebted to Benjamin Fuks for his help and patience. We sincerely thank the organizers of the second {\sc MadAnalysis}~5 workshop for their warm welcome in Seoul and the success of the event.

\cleardoublepage
\renewcommand{\ma}{{\sc MadAnalysis~5}}

\markboth{Joon-Bin~Lee and Jehyun~Lee}{Implementation of the CMS-HIG-18-011 analysis}

\section{Implementation of the CMS-HIG-18-011 analysis (exotic Higgs decays via two pseudoscalars
with two muons and two $b$-jets; 35.9~fb$^{-1}$)}
  \vspace*{-.1cm}\footnotesize{\hspace{.5cm}By Joon-Bin~Lee and Jehyun~Lee}
\label{sec:nmssm}


\subsection{Introduction}

In this note, we describe the validation of our implementation, in the \ma\
framework~\cite{Conte:2012fm,Conte:2014zja,Dumont:2014tja,Conte:2018vmg}, of the CMS-HIG-18-011 search~\cite{Sirunyan:2018mot}
for exotic decays of the Standard Model Higgs boson into a pair of light
pseudoscalar particles $a_{1}$, where one of the pseudoscalar decays to a pair
of opposite-sign muons and the other one decays into a pair of $b$-quarks.
This analysis focuses on 13 TeV LHC data and an integrated luminosity of 35.9 $\rm fb^{-1}$.  

The considered exotic decay of the Higgs boson is predicted in a variety of models,
including the next-to-minimal supersymmetric extension of the Standard Model
(NMSSM)~\cite{Ellwanger:2009dp}, as well as models with additional scalar doublet and singlet
(2HDM+S)~\cite{Dine:1981rt,Kim:1983dt,Curtin:2013fra}. To validate our implementation, we focus 
on an NMSSM setup in which one decouples most particles, except for the above-mentioned
pseudoscalar states. Such a scenario has been studied in particular in the
CMS-HIG-18-011 analysis that we implemented in this work.

In the rest of this note, we present a brief description of the CMS-HIG-18-011 analysis in
section~\ref{analysisdescription}. Section~\ref{sec:validation} consists in the
core of our work, and contains extensive information about the validation of our
implementation. In particular, the presence of two $b$-jets in the final state
makes this analysis particularly sensitive to the exact details of the $b$-jet
identification algorithm. However, the $b$-jet identification efficiency provided
by the CMS collaboration is not sufficient for a precise enough modeling in
{\sc Delphes}~3. The method that we used to model in an accurate manner the
CMS $b$-tagging algorithm is therefore explained in details in Section~\ref{refinedselection}.
We summarise our work and results in section~\ref{sec:concl}.

\subsection{Description of the analysis} \label{analysisdescription}
The CMS-HIG-18-011 analysis performs a search for the Higgs boson decay chain
$h \rightarrow a_{1}a_{1} \rightarrow \mu^{+}\mu^{-}b\bar{b}$. 
This analysis hence targets a final state containing two opposite-sign muons and two $b$-tagged jets.
In the next subsection, we present the definition of the muon and jet candidates that are used in this analysis, as well as the preselection cuts of the analysis. Then, in Section~\ref{eventselection}, we explain the event selection requirements leading to a good background rejection while preserving as many expected signal events as possible.

\subsubsection{Object definitions and preselection} \label{objectdefinitions}
This analysis requires the presence of at least two final-state muons and two final-state $b$-jets.
Two oppositely charged muons are required to conservatively satisfy an online selection based on the CMS muon triggering system. This enforces that the final state includes two muons with a transverse momentum $p_T >$ 17~GeV (leading muon $\mu_1$ ) and 8~GeV (subleading muon $\mu_2$). Moreover, the geometrical limitations of the CMS muon system leads to the following extra requirements on the muon's $p_T$ and pseudorapidities $\eta$,
\begin{equation}
    p_{T}({\mu_1}) \ > \ 20 \textrm{ GeV },\
    p_{T}({\mu_2}) \ > \ 9 \textrm{ GeV } \ \text{and} \ \ |\eta({\mu_{1,2}})| \ < \ 2.4 .
\end{equation}
Additionally, a particle-flow-based relative isolation is enforced. This requires that the sum of the transverse energy of any detector-level object present in a cone of radius $R = 0.4$ centered on the muon is smaller than 0.15 times the muon $p_{T}^{\mu}$,
\begin{equation}
    I_{rel} = \frac{1}{p_{T}^{\mu}}\sum_{i} (p_{T})_{i} < 0.15.
\end{equation}
The CMS-HIG-18-011 analysis moreover targets a signal scenario in which the narrow width approximation is valid for the new pseudoscalar $a_1$, and its mass is considered to fulfil 20~GeV $\le m_{a_1} \le$ 62.5 GeV.
The invariant mass of the two-muon system is, therefore, restricted to lie within a slightly wider mass range,
\begin{equation}
    19.5 \ \textrm{GeV} \ < m_{\mu\mu} \ < \  63.5 \ \textrm{GeV}.
\end{equation}

Jets are reconstructed by clustering detector-level objects with the anti-$k_{T}$ algorithm~\cite{Cacciari:2008gp} with a distance parameter of 0.4. The transverse momentum $p_{T}$ and pseudorapidity ($\eta$) of the leading jet $j_1$ and subleading jets $j_i$ (with $i\neq 1$) are imposed to satisfy
\begin{equation}
    p_{T}({j_1}) \ > \ 20 \textrm{ GeV},\
   p_T(j_i) >  15 \textrm{ GeV  and } |\eta(j_{1,i})| \ < \ 2.4 .
\end{equation}
Events must contain at least two jets, with the leading two jets having to be well separated from the selected muons in the transverse plane, by a distance $\Delta R > 0.5$. $b$-tagging makes use of the CSVv2 algorithm~\cite{Sirunyan:2017ezt} that relies on secondary vertex information. One of the jets must satisfy tight working point criteria, whereas another one has to satisfy loose working point requirements.
The misidentification rate of light jets as $b$-jets is in average of 10\% (0.1\%) for the tight (loose) working point, and that of $c$-jets as $b$-jets is of 30\% (2\%), for a tagging efficiency of about 80\% (40\%).
If there are more than two $b$-jets in the event, the two with the largest $p_{T}$ are considered as
originating from a pseudoscalar $a_1$ decay.

Finally, the missing transverse momentum vector $\mathbf{p_{T}^{miss}}$ is defined as the opposite of the vector sum of the momentum of all reconstructed physics object candidates, and the missing transverse energy is defined by the norm of this vector,
\begin{equation}
    E_{T}^{miss} = |\mathbf{p_{T}^{miss}}|.
\end{equation}

\subsubsection{Event Selection} \label{eventselection}
Following the preselection described in the previous section, signal events are subjected to additional selection cuts to minimise the Standard Model background contamination. As the transverse momentum of the neutrinos arising from semi-leptonic $B$-hadron decays is small, we require,
\begin{equation}
    E_{T}^{miss} \ < \ 60 \ \textrm{GeV}.
\end{equation}
Next, as both the muons and $b$-jets are the decay products of a pseudoscalar boson $a_{1}$, one requires that the values of the invariant mass of the dimuon system ($m_{\mu\mu}$) and that of the two-$b$-jet system ($m_{bb}$) are close to each other. Therefore, the relative difference between these two invariant masses is evaluated through a quantity $\chi_{bb}$ defined by
\begin{equation} \label{chibb}
    \chi_{bb} \ = \ \frac{(m_{bb}-m_{\mu\mu})}{\sigma_{bb}},
\end{equation}
where $\sigma_{bb}$ is a mass resolution associated with the reconstruction of the two-$b$-jet system.
The invariant mass of the whole system comprising those four objects ($m_{\mu \mu bb}$) should moreover be compatible with the Higgs-boson mass $m_{h}$. 
One subsequently defines the relative difference $\chi_h$,
\begin{equation} \label{chih}
    \chi_{h} \ = \ \frac{(m_{\mu \mu bb} - m_{h})}{\sigma_{h}}, 
\end{equation}
where $\sigma_{h}$ is the mass resolution associated with the reconstruction of the Higgs boson candidate ($\mu\mu bb$).
Events are selected by requiring that the squared sum of these two variables, $\chi^2 = \chi_{bb}^2 + \chi_{h}^2$, is smaller than 5,
\begin{equation} \label{chi}
    \chi^2 \ < \ 5.
\end{equation}

\subsection{Validation}\label{sec:validation}

\subsubsection{Event generation} \label{eventgeneration}
In order to generate events necessary to valide our implementation, we use the NMSSMHET simplified model\cite{Curtin:2013fra}. The latter involves two free parameters, the mass of pseudoscalar $m_{a_{1}}$ and tan$\beta$, that is defined as the ratio of the vacuum expectation values of the two Higgs doublets of the model. tan$\beta$ moreover determines the branching fraction of the $a_{1}$ boson to Standard Model particles.
Following the CMS-HIG-18-011 analysis and the corresponding publicly available validation material, tan$\beta$ is set to 2 and we consider three pseudoscalar mass points,
\begin{equation}
    m_{a_{1}} = 20,\ 40,\ 60 \ \textrm{GeV}.
\end{equation}
As there is no strong dependence of the branching ratio $\mathcal{B}(a_{1} \rightarrow b\bar{b})$ and $\mathcal{B}(a_{1} \rightarrow \mu^{+}\mu^{-})$ on $m_{a_{1}}$~\cite{Curtin:2013fra},
the total signal cross section defined as the product of the Standard Model Higgs boson production cross section ($\sigma_{h}$) and the relevant branching fractions is set to a constant value,
\begin{equation}
   \sigma_{h} \times \mathcal{B}(h \rightarrow a_{1}a_{1} \rightarrow \mu^{+}\mu^{-}b\bar{b}) \approx 8 \ \textrm{\rm fb}.
\end{equation}

\begin{figure}
  \centering
        \includegraphics[width=0.32\textwidth]{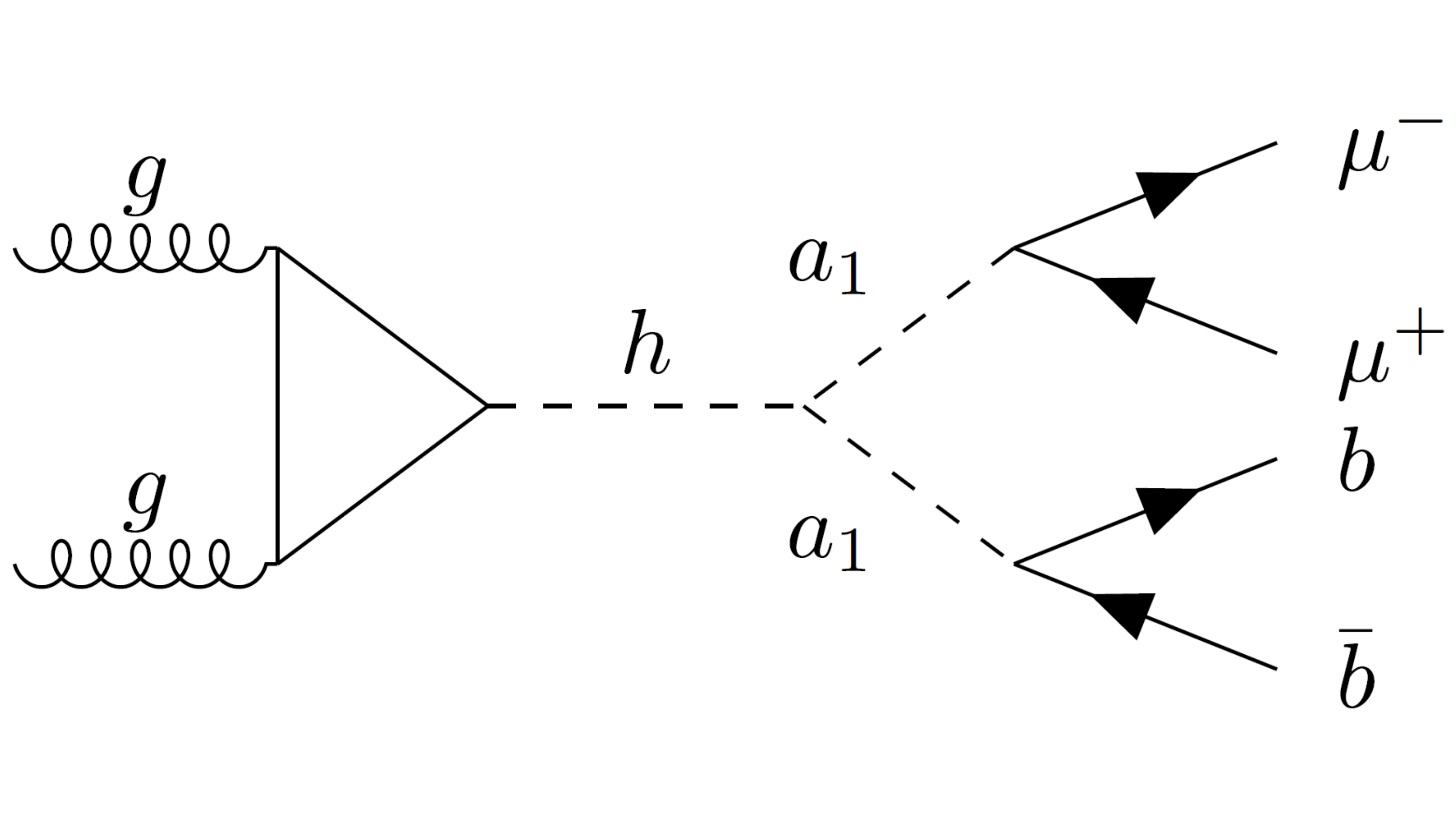}\hspace{2cm}
        \includegraphics[width=0.32\textwidth]{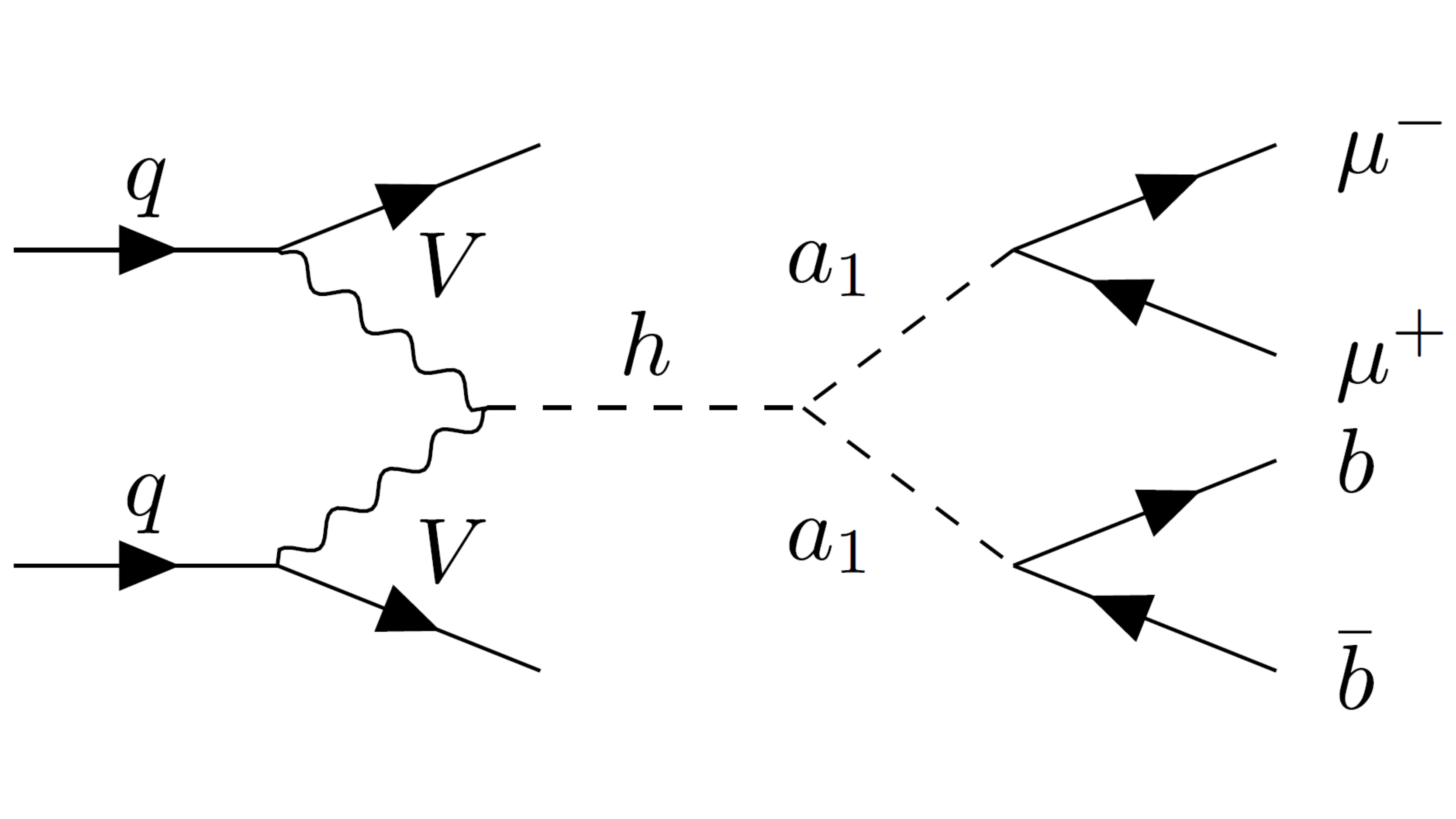}
  \caption{Feynman diagrams illustrating the two hard process considered in the event generation processes relevant for the validation of the implementation in the \ma\ framework of the CMS-HIG-18-011 analysis.  The signal comprises a gluon fusion component (left) and vector-boson fusion component (right).}
  \label{diagrams}
\end{figure}

We generate 1,000,000 events for each test sample. In order to mimic CMS signal event generation (so that we could compare our predictions to public material), we consider Higgs boson production via gluon fusion and vector boson fusion, as illustrated by the two Feynman diagrams shown in Figure~\ref{diagrams}. The total production cross sections resulting from a leading-order (LO) calculation achieved within the {\sc MadGraph5\_aMC@NLO}~\cite{Alwall:2014hca} framework and that are used for our signal normalisation, are, for each of the two subprocesses, 48.58~pb and 3.78~pb respectively.


The generation of the hard process is performed by using the NMSSMHET model implementation~\cite{Curtin:2013fra} in the UFO format~\cite{Degrande:2011ua}, that can be used with the {\sc MadGraph5\_aMC@NLO} framework~\cite{Alwall:2014hca} at LO in QCD.
Our matrix elements are convoluted with the NNPDF3.0 set of parton densities~\cite{Ball:2014uwa}, and the
{\sc Pythia}~8.212 package~\cite{Sjostrand:2014zea} with the CUETP8M1 tune~\cite{Khachatryan:2015pea} is used to model parton showering and hadronisation.

The simulation of the response of the CMS detector is based on the {\sc Delphes}~3 program~\cite{deFavereau:2013fsa}, which internally relies on {\sc Fastjet}~\cite{Cacciari:2011ma} for object reconstruction. We start from the default CMS detector parametrisation and then impose modifications as follows.

First, the mimimum $p_T$ thresholds for muons and jets are reduced to 5 and 10 GeV respectively, in order to cover the full signal region. 

Second, the muon and jet reconstruction efficiencies contained in the Run II CMS card in Delphes version 3.4.2 cannot cover such a small $p_T$ region. Therefore, they are extrapolated from the default ones to conservatively accept all objects used in this analysis.

Finally, the $b$-jet identification efficiencies based on the CSVv2 algorithm~\cite{Sirunyan:2017ezt}, which is used in CMS-HIG-18-011 analysis, have a large dependence on the jet transverse momentum. However, only average efficiency values are provided by the CMS collaboration. To approximatively model the $p_T$ dependence of the combined secondary vertex algorithm used in this work, we have used the efficiency functions associated with the loose working point of the deep combined secondary vertex (DeepCSV) algorithm described in the CMS $b$-jet identification paper~\cite{Sirunyan:2017ezt}. These are then re-weighted via the average tagging efficiencies of the CSVv2 algorithm, as further described in the next section. This re-weighting method has the great advantage of reflecting not only the overall $b$-tagging power of the CSVv2 algorithm, but also the $p_T$ dependence of this general CMS $b$-tagging algorithm.

\subsubsection{Refinements of our event selection} \label{refinedselection}
This work uses the same event selection as described in Section~\ref{analysisdescription}.
However, additional details are necessary to reproduce the CMS-HIG-18-011 results.
This section first describes the re-weighting method that we used to improve the modeling of the $b$-tagging performance in {\sc Delphes}~3, and then explains how to estimate the mass resolutions $\sigma_{bb}$ and $\sigma_h$ that are needed to calculate the $\chi_{bb}$ and $\chi_h$ quantities of Section~\ref{analysisdescription}.

As noted in Section~\ref{eventgeneration}, the HIG-18-011 analysis used the CSVv2 $b$-tagging algorithm.
The average efficiencies and mistagging rates of this algorithm are provided in the CMS $b$-jet identification publication~\cite{Sirunyan:2017ezt}, but there is no information about their $p_T$ dependence.
We have however found out that ignoring this $p_T$ dependence can make a difference of about 20 \% in the final results.

To recover this, we assume that the $p_T$ dependence of the DeepCSV and CSVv2 algorithms is similar.
This assumption is justified as both methods use an almost identical approach based on combined information originating from displaced tracks and secondary vertices.
We hence implement the publicly available loose $b$-tagging efficiency and mis-tagging rates of the DeepCSV algorithm in {\sc Delphes}~3, with their full dependence on the jet's transverse momentum.
In a second step, we re-weight the generated events to account for the different average efficiencies associated with the loose working points of the DeepCSV and CSVv2 algorithms. In practice, we use as a re-weghting factor the squared ratio of the $p_T$-independent efficiencies,
 $(\bar{\epsilon}^{v2}_{L}/\bar{\epsilon}^{D}_{L})^2$, where $\bar{\epsilon}^{v2}_{L}$ and $\bar{\epsilon}^{D}_{L}$ are respectively the CSVv2 and DeepCSV algorithm efficiencies as provided in the CMS $b$-tagging performance publication~\cite{Sirunyan:2017ezt}.

At last, the signal region event selection requires that at least one jet passes the requirement of the tight $b$-jet discriminator. The final event weight is therefore calculated as
\begin{equation}
    w = \Bigg(\frac{\bar{\epsilon}^{v2}_{L}}{\bar{\epsilon}^{D}_{L}}\Bigg)^2 \cdot \Bigg[1 - \bigg(1-\frac{\epsilon^{v2}_T(j_i)}{\epsilon^{v2}_L(j_i)}\bigg) \cdot \bigg(1-\frac{\epsilon^{v2}_T(j_j)}{\epsilon^{v2}_L(j_j)}\bigg)\Bigg].
\end{equation}
In this equation, $\epsilon^{v2}_{T(L)}(j_i)$ represents the efficiency that the $i^{\rm th}$ jet ($j_i$) satisfies the tight (loose) CSVv2 selection criteria, that we estimate again from DeepCSV public information,
\begin{equation}
    \epsilon^{v2}_{T,L}(j_i) = \frac{\bar\epsilon^{v2}_{T,L}}{\bar\epsilon^{D}_{T,L}} \cdot \epsilon^{D}_{T,L}(j_i).
\end{equation}
Here, $\epsilon^{D}_{T,L}(j_i)$ are the DeepCSV tight and loose $p_T$-dependent efficiencies.

\begin{figure}
  \centering
        \includegraphics[width=0.32\textwidth]{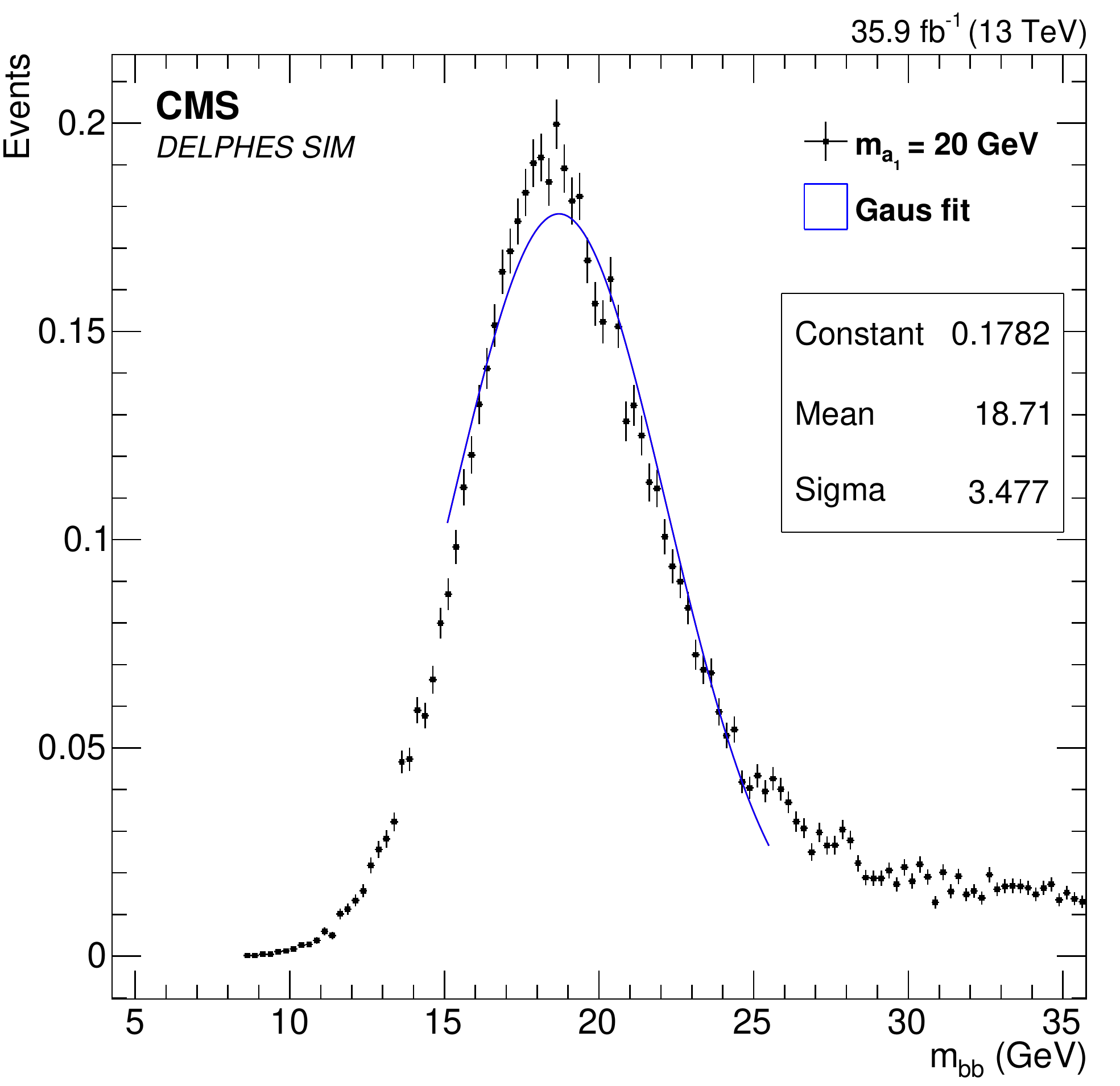}
        \includegraphics[width=0.32\textwidth]{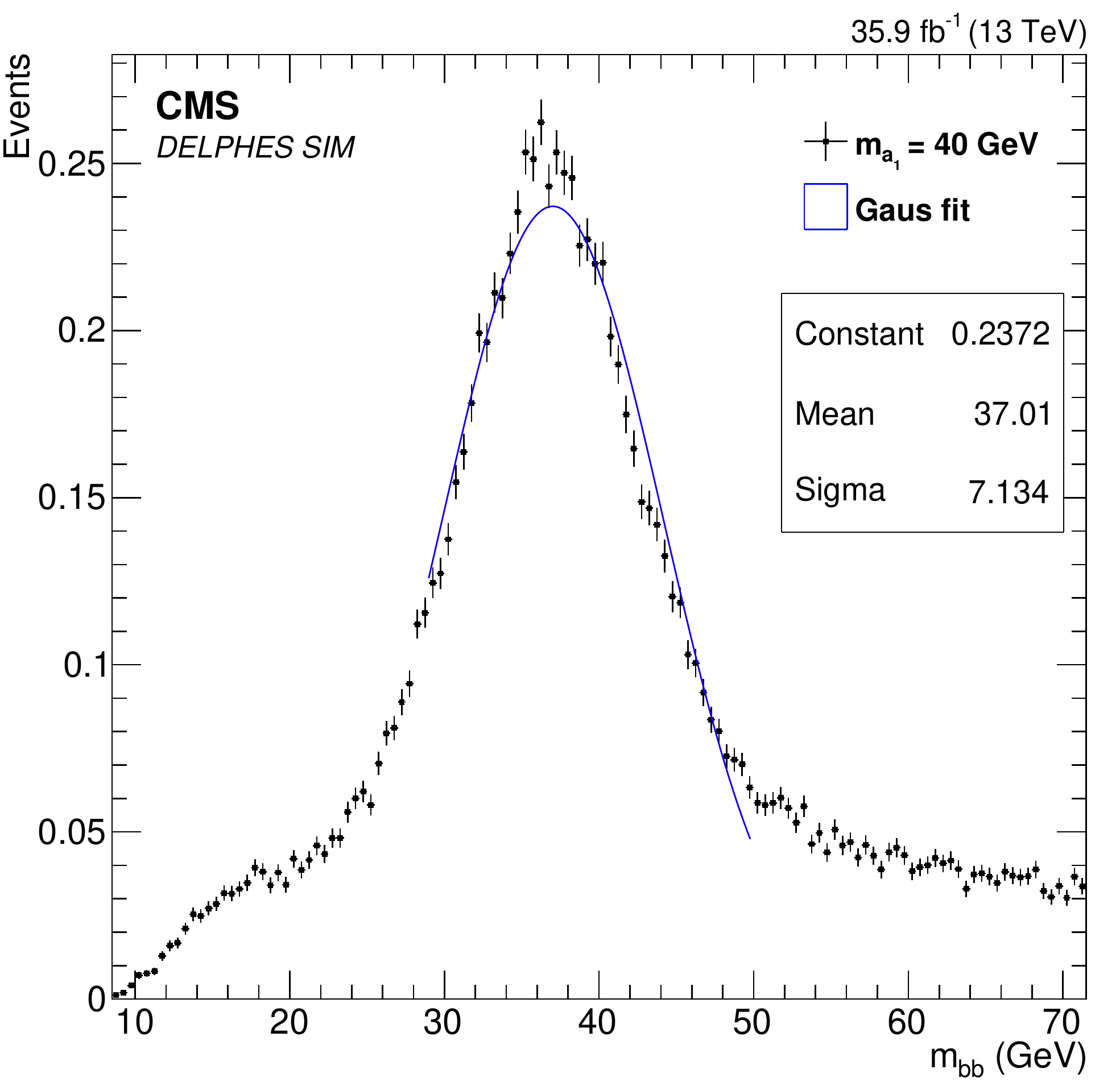}
        \includegraphics[width=0.32\textwidth]{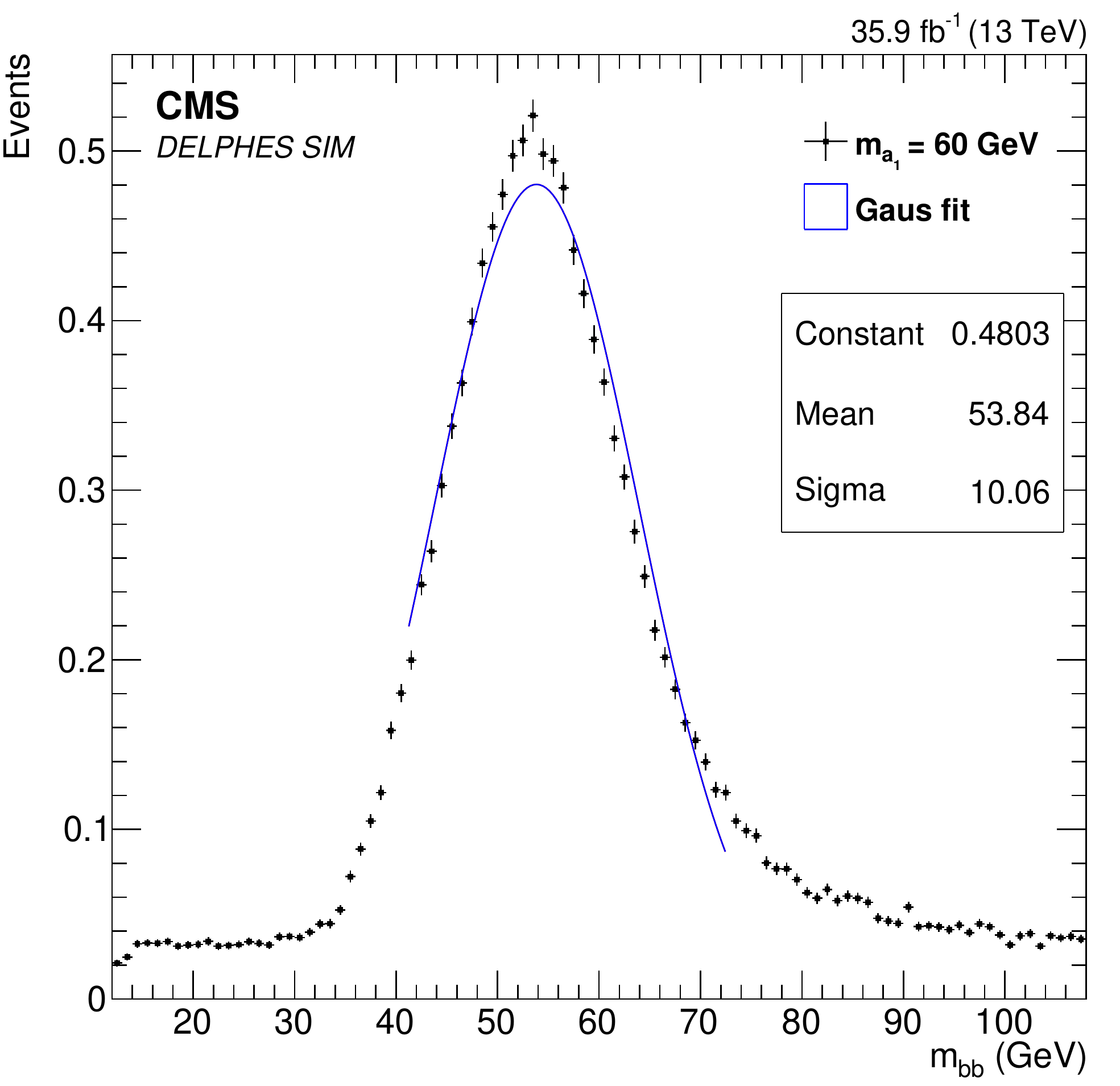}
  \caption{The di-$b$-jet invariant-mass distribution with its Gaussian fitting, for pseudoscalar mass scenarios of $m_{a_{1}}$ = 20 (left), 40 (center), and 60 (right) GeV.\label{bbmassresolution}}\vspace{.3cm}
        \includegraphics[width=0.32\textwidth]{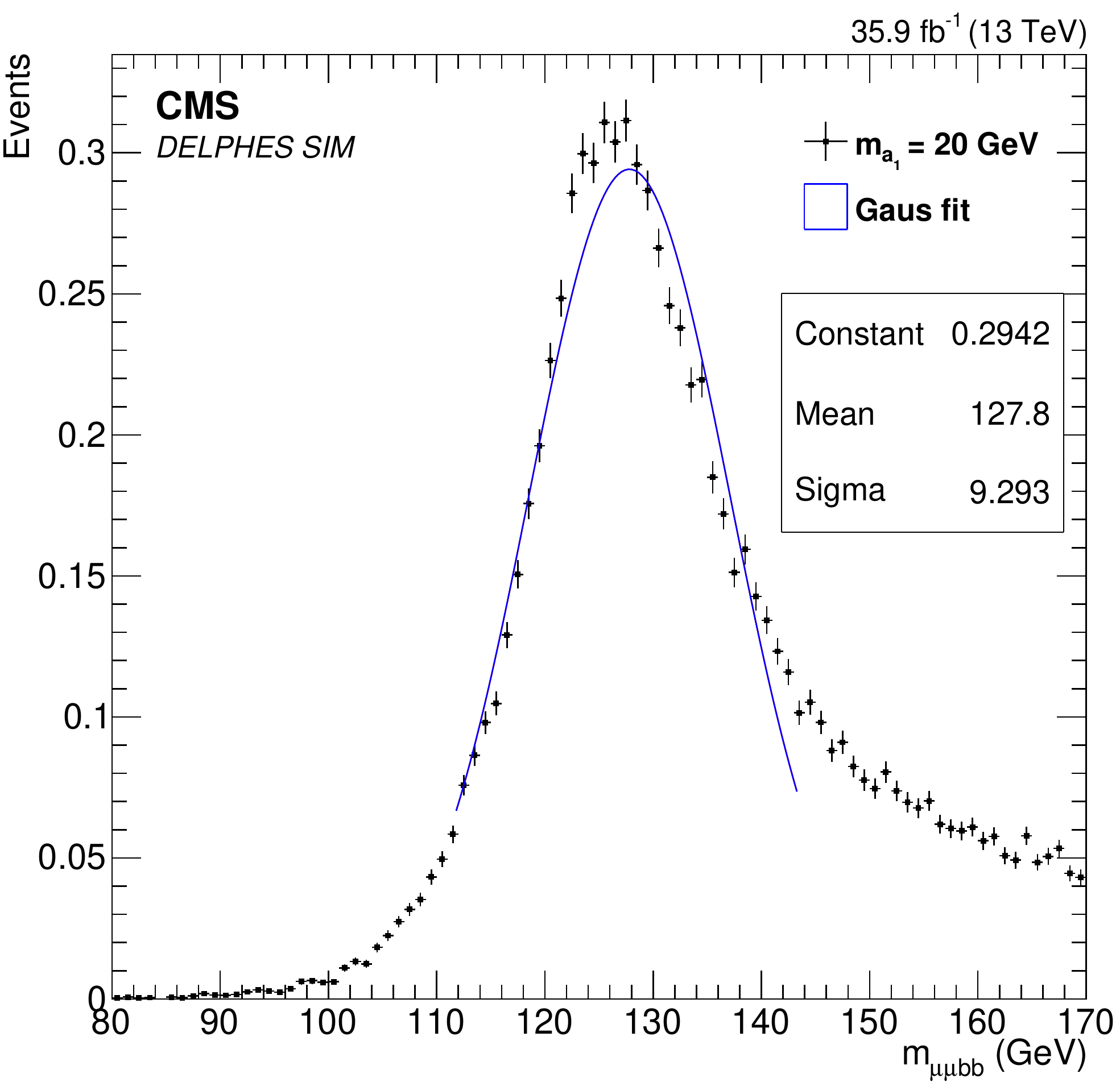}
        \includegraphics[width=0.32\textwidth]{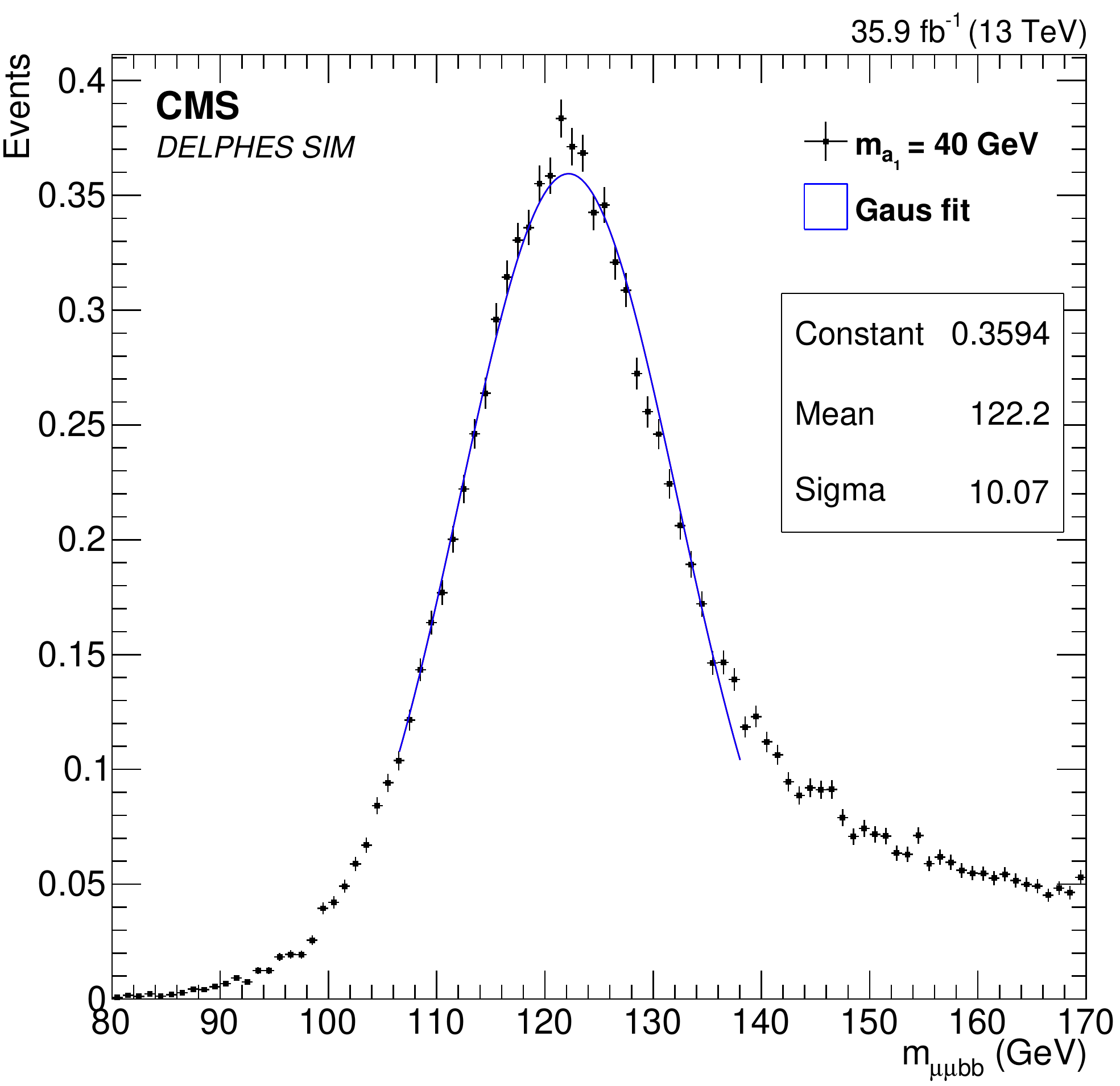}
        \includegraphics[width=0.32\textwidth]{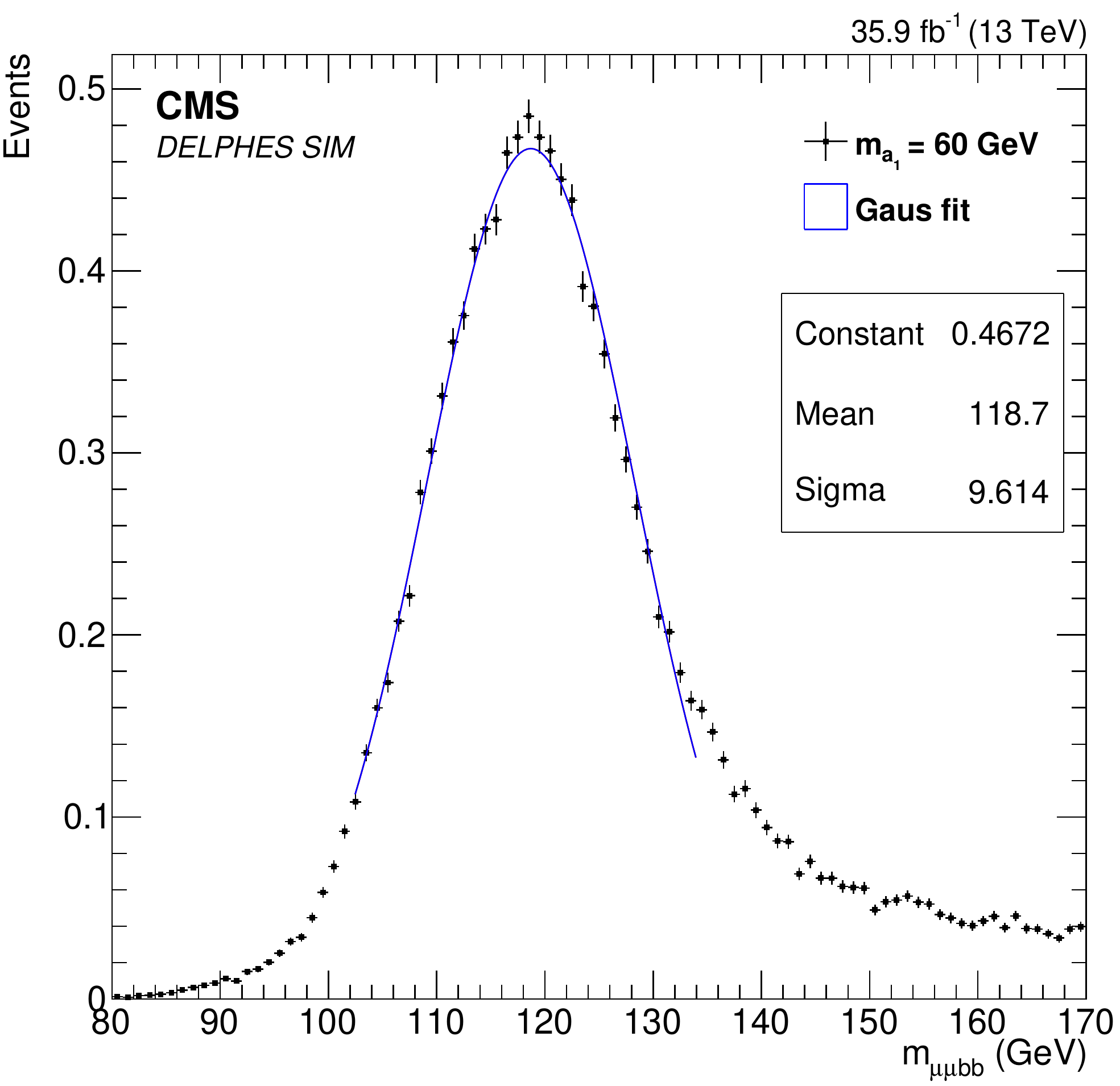}
  \caption{The invariant-mass distribution of the reconstructed Higgs boson candidate, with its Gaussian fitting for pseudoscalar mass scenarios of $m_{a_{1}}$ = 20 (left), 40 (center), and 60 (right) GeV.\label{hmassresolution}}
\end{figure}

After the signal region selection of two jets and two muons, the mass resolutions of the di-$b$-jet ($\sigma_{bb}$) and reconstructed Higgs boson candidate ($\sigma_h$) are estimated by fitting the corresponding invariant-mass distributions with Gaussian functions. However, the CMS analysis note does not provide the exact values of these resolutions as obtained from the fit. We have therefore estimated these values by performing our own fit of the invariant-mass distributions.

As a rough approximation, we estimate the input values of the mass resolutions $\sigma_{bb}^i$ and $\sigma_h^i$ that are used in our Gaussian fitting procedure from the muon and jet $p_T$ resolutions of the CMS detector.
Each muon and jet originating from the Higgs boson decay has an average transverse momentum of about 30~GeV. The momentum resolution of a 30~GeV muon is expected to be of about 1\%~\cite{Chatrchyan:2012xi}, whereas that of a 30~GeV jet is expected to be of about 17\%~\cite{Khachatryan:2016kdb}.
Based on these values, the initial input mass resolution of the di-$b$-jet system is set to $\sigma_{bb}^i = 0.17 m_{a_1}$, and the one of the reconstructed Higgs boson candidate is fixed to $\sigma_h^i=9.3$~GeV.
Our Gaussian fitting is then performed within a fitting range of $m_{a_1 (h)}\pm 1.5 \sigma_{bb (h)}^i$ from the above input values.

Figure~\ref{bbmassresolution} and~\ref{hmassresolution} show the invariant-mass distributions of the di-$b$-jet system and of the reconstructed Higgs boson candidate, together with the corresponding Gaussian fit results. The average mass resolution of the di-$b$-jet system is found to be $0.173 m_{a_1}$, whilst that of the Higgs boson candidate is equal to 9.66 GeV.
As a result, the $\chi_{bb}$, $\chi_h$ and $\chi^2 = \chi_{bb}^2+\chi_{h}^2$ quantities can be calculated by using eqs.~\eqref{chibb} and \eqref{chih}.

\begin{figure}
  \centering
        \includegraphics[width=0.46\textwidth]{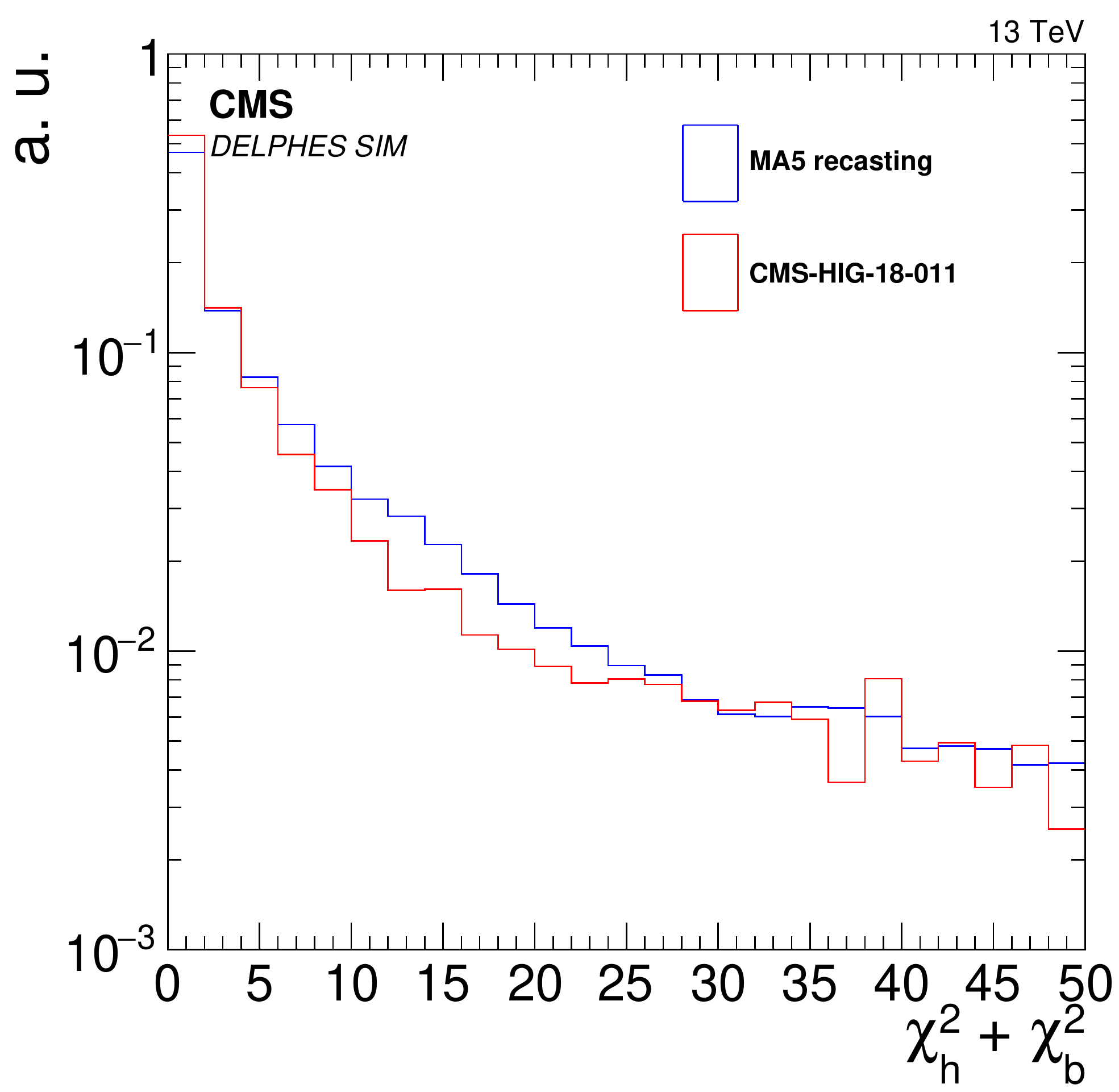}\hspace{.4cm}
        \includegraphics[width=0.46\textwidth]{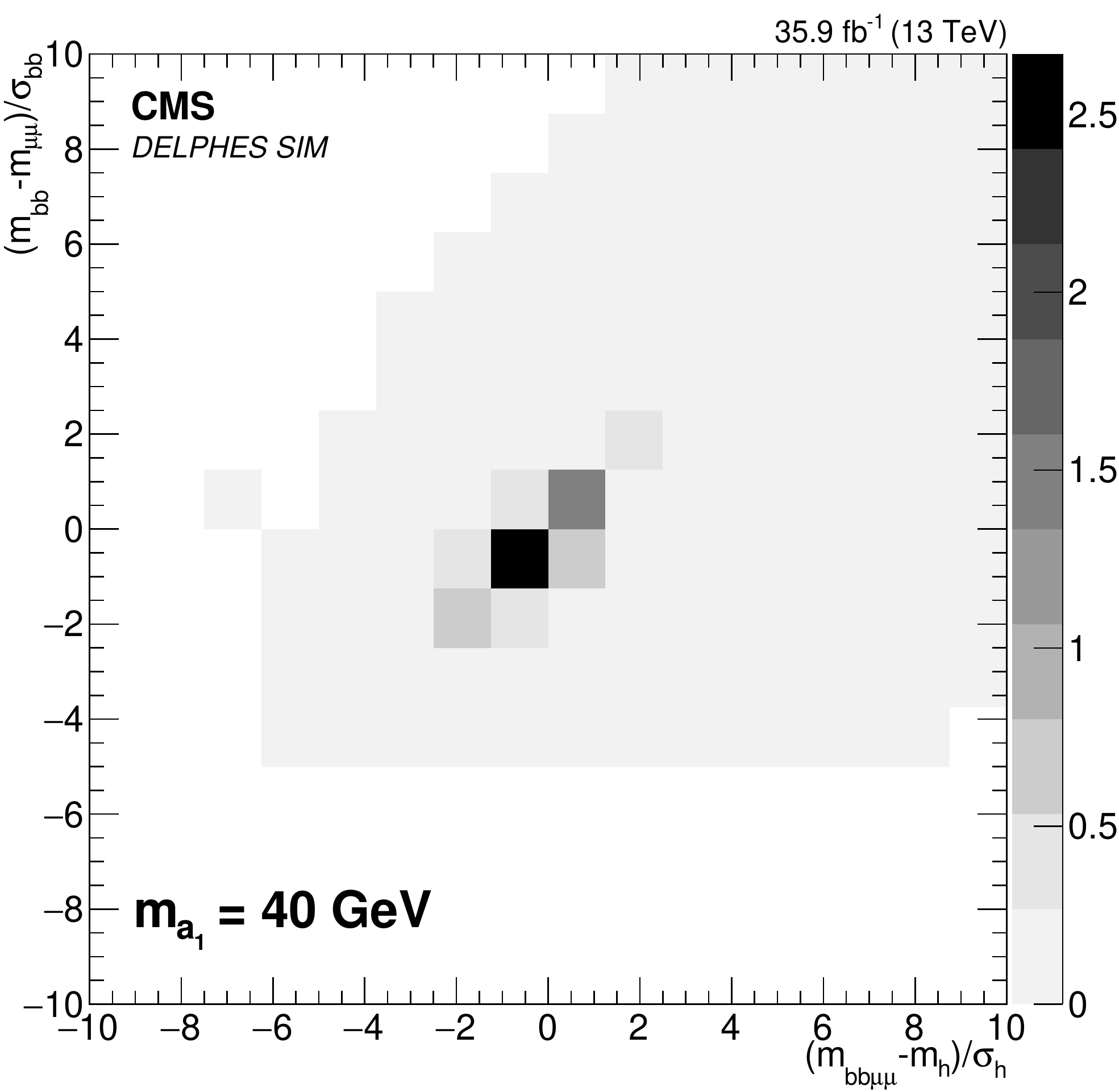}
  \caption{{\bf Left:} The $\chi^2$ distribution associated with the scenario in which $m_{a_{1}}$ = 40~GeV, once all analysis selections except the $\chi^2 <5$ requirement are applied. {\bf Right:}
  Two-dimensional distribution of the $(m_{bb}-m_{\mu\mu})/\sigma_{bb}$ and $(m_{\mu\mu bb}-125~{\rm GeV})/\sigma_{h}$ quantities for the same scenario and after applying all the cuts of the analysis with the exception of the $\chi^2 <$ 5 requirement.}
  \label{chichi}
\end{figure}

In the left panel of Figure~\ref{chichi}, we show the $\chi^2$ distribution that is obtained after applying all selections except the $\chi^2 < 5$ cut. The results are presented for a scenario in which the mass of the pseudoscalar $a_1$ is fixed to 40~GeV.
In the right panel of the figure, we moreover show the two-dimensional distribution of the $\chi_{bb}$ and $\chi_h$ quantities for the same scenario and after applying again all analysis cuts but the last one. This illustrates the quality of our fit and its impact on the signal reconstruction.

\subsubsection{Comparison with official results}
To validate our results, we compare predictions obtained with our \ma\ implementation (and our tuned detector simulation based on {\sc Delphes}~3) to the CMS official results presented in the CMS-HIG-18-011 analysis note for the three considered new physics scenarios.
As a first test, we compare the curve shown in the left panel of Figure~\ref{chichi} to the first figure (Fig.~1) of the CMS note. The shape of the two distributions are similar, the values in the most populated first bins being found to differ by at most a few percent.
We additionally compare the shape of the contours shown in the right panel of Figure~\ref{chichi} with the one exhibited in the second figure (Fig.~2) of the CMS publication. Here, the central bin are even populated equally.  As already mentioned in the previous section, this validates our fitting procedure.

\begin{table}[t]
  \renewcommand{\arraystretch}{1.5}
  \setlength\tabcolsep{9.0pt}
  \tbl{Event yields resulting from the object selection of Section~\ref{objectdefinitions} (the so-called $\mu^+ \mu^- b \bar{b}$ selection) and after the final selection cuts presented in Section~\ref{eventselection}. The results are normalised to an integrated luminosity of 35.9~$\rm fb^{-1}$. We moreover include the total event selection efficiencies ($\epsilon$) for three mass scenarios. We compare predictions obtained with \ma\ and the public CMS results.}
  {\begin{tabular}{@{}ll|lll@{}} \toprule
  & & $\mu^+ \mu^- b \bar{b}$ selection & Final selection & $\epsilon$ (\%) \rule{0pt}{4.2ex}\\ [1.0ex] 
\hline
  \multirow{3}{2.4cm}{$m_{a_{1}}$ = 20 GeV} 
    & CMS-HIG-18-011  & \ $14.0 \pm 0.1$ & \ \ $6.0 \pm 0.1$ & $42.9$ \rule{0pt}{2.5ex}\\ [0.2ex] 
    & MA5 Recasting   & \ $13.2$         & \ \ $5.6$         & $42.5$ \\ [0.2ex] 
    & Difference $\delta$ (\%) & \ $5.7$          & \ \ $6.7$         & $0.4$  \\ [0.3ex]
 \hline
 \multirow{3}{2.4cm}{$m_{a_{1}}$ = 40 GeV} 
     & CMS-HIG-18-011  & \ $14.8 \pm 0.1$ & \ \ $ 7.5 \pm 0.1 $ & $50.7$ \rule{0pt}{2.5ex}\\ [0.2ex] 
     & MA5 Recasting   & \ $15.9$         & \ \ $7.4$           & $46.2$ \\ [0.2ex] 
     & Difference $\delta$ (\%) & \ $7.4$          & \ \ $1.3$           & $4.5$  \\ [0.3ex]
 \hline
 \multirow{3}{2.4cm}{$m_{a_{1}}$ = 60 GeV} 
     & CMS-HIG-18-011  & \ $16.7 \pm 0.1$ & \ \ $10.1 \pm 0.1 $ & $60.5$ \rule{0pt}{2.5ex}\\ [0.2ex] 
     & MA5 Recasting   & \ $16.9$         & \ \ $10.1$          & $60.0$ \\ [0.2ex] 
     & Difference $\delta$ (\%) & \ $1.2$          & \ \ $0.3$           & $0.5$  \\ [0.3ex]
 \botrule\end{tabular}\label{yields} }
\end{table}

In Table~\ref{yields}, we compare CMS public yields with the recasting results predicted with \ma\ after the object definition selection of Section~\ref{objectdefinitions}, {\it i.e.}~before applying the $p_T^{miss} <$ 60 GeV and $\chi^2 <$ 5 requirements, and after the full analysis selection.
The differences between the CMS and \ma\ event yields ($N^{\rm CMS}$ and $N^{\rm MA5}$) is quantified through relative differences,
\begin{equation}
  \delta = \frac{|N^{\rm CMS} - N^{\rm MA5}|}{N^{\rm CMS}}
\end{equation}
Overall results agree at the level of the few percent, where the best agreement is achieved for the final selection with the 60 GeV pseudoscalar mass scenario which only shows a 0.3 \% difference with the CMS-HIG-18-011 results from Ref.~\refcite{Sirunyan:2018mot}.
Event selection efficiencies ($\epsilon$) and the corresponding differences are also computed. The level of agreement between the CMS results and the \ma predictions is again found to lie at the percent level.

Although the event yields exhibit a larger difference after the object definition selection (ranging up to 7.4\% for the $m_{a_1} = 40$~GeV scenario), we consider that such a feature should be expected as resulting from our approximate modeling of the $p_T$ dependence of the $b$-tagging performance. Such an order of magnitude is indeed typical from the differences originating from the use of the DeepCSV and CSVv2 algorithms.
We nevertheless consider this as a minor effect stemming from the lack of public knowledge about the new $b$-tagging algorithms used by CMS. 

Even after adding the impact of our method to estimate the mass resolutions used in the CMS-HIG-18-011 analysis, only a small difference between \ma\ predictions and CMS official results remains. We take it as sufficiently acceptable to guarantee the validation of our recast.
Unfortunately, our validation cannot be performed further because of the lack of available public information.

\subsection{Conclusion}\label{sec:concl}
In this note, we have documented a recast in the \ma\ frsmework of the CMS-HG-18-011 search for light pseudoscalar particles originating from an exotic decay of the Standard Model Higgs boson. This search considers a final state comprising two $b$-jets and a pair of opposite-sign muons, and an integrated luminosity of 35.9 $\rm fb^{-1}$ of data collected at a centre-of-mass energy of 13~TeV.

Our work features two important differences with respect to what CMS has done. First, due to the lack of public knowledge about the transverse momentum dependence on the CSVv2 $b$-tagging algorithm performances, we have modeled our $b$-tagging efficiencies and mistagging rates in {\sc Delphes}~3 by using the dependence of the DeepCSV algorithm performances on the transverse momentum of the jets. We have moreover included an event re-weighting procedure dealing with the differences between the average tagging efficiencies of the two algorithms. Second, we had to implement our own Gaussian fitting procedure to recover the invariant-mass resolutions expected from the signal, in the case of the reconstructed Higgs boson and pseudoscalar boson $a_1$. These are extensively detailed in Section~\ref{refinedselection}.

To validate our implementation of the above search, we generated three signal samples in accordance with the CMS prescriptions. We have found that our approximate treatment of the mass resolutions and the CMS $b$-tagging performance are reasonable enough. These has allowed us to obtained an agreement with the CMS results at the level of a few percent. In contrast, only a poor level of agreement of about 20\% can be reached without implementing our two classes of changes.

Subsequently to the lack of public CMS information for this analysis, we have only validated our code by comparing a few differential distributions and event yields at two stages of the full event selection. Our results exhibit a reliable agreement at the percent level. The implemented code is available online from the \ma\ dataverse~\cite{UOH6BF_2020}, at \href{https://doi.org/10.14428/DVN/UOH6BF}{https://doi.org/10.14428/DVN/UOH6BF}, which also includes the cards and UFO model that have been used in our validation procedure.

\cleardoublepage
\newcommand{\madanalysis}{{\sc MadAnalysis~5}}

\markboth{Jongwon Lim, Chih-Ting Lu, Jae-Hyeon Park and Jiwon Park}{Implementation of the ATLAS-SUSY-2018-04 analysis}

\section{Implementation of the ATLAS-SUSY-2018-04 analysis (stau pairs with two taus and missing transverse energy; 139~fb$^{-1}$)}
  \vspace*{-.1cm}\footnotesize{\hspace{.5cm}By Jongwon Lim, Chih-Ting Lu, Jae-Hyeon Park and Jiwon Park}
\label{sec:stau}


\subsection{Introduction}

\begin{figure}[t]
  \centerline{\includegraphics[width=2.0in]{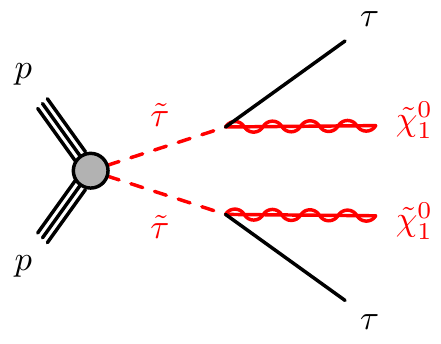}}
  \vspace*{8pt}
  \caption{The Feynman diagram for the process $pp\rightarrow\tilde{\tau}\tilde{\tau}\rightarrow\tilde{\chi}^0_1\tilde{\chi}^0_1\tau\tau$.\protect\label{fig:fig_01}}
\end{figure}

In this note, we describe the validation of the implementation, in the \madanalysis\ framework~\cite{Conte:2018vmg,Dumont:2014tja,Conte:2014zja,Conte:2012fm}, of the ATLAS-SUSY-2018-04 search~\cite{Aad:2019byo} for direct stau production in events featuring two hadronic tau leptons and a large amount of missing transverse energy ($E^{miss}_T$).
This analysis focuses on LHC proton-proton collisions at a center-of-mass energy of 13 TeV, and considers an integrated luminosity of $139 {\rm fb}^{-1}$.
The typical supersymmetric signal which this analysis is dedicated to is illustrated by the representative Feynman diagram shown in Fig.~\ref{fig:fig_01}. 

For the validation of our re-implementation, we have focused on a simplified model in which only a few electroweakly-interacting superpartners are relevant.
The lightest neutralino ($\tilde{\chi}^0_1$) is taken as the lightest supersymmetric particle (LSP).
The stau-left ($\tilde{\tau}_L$) and stau-right ($\tilde{\tau}_R$) sleptons are moreover assumed to be mass degenerate and they do not mix. Therefore the gauge eigenstates ($\tilde{\tau}_L$,$\tilde{\tau}_R$) coincide with the mass eigenstates ($\tilde{\tau}_1$,$\tilde{\tau}_2$) in this theoretical framework.
Furthermore, in order to suppress any other decay modes of the tau sleptons, the masses of all charginos and neutralinos are set to 2.5 TeV except for the $\tilde{\chi}^0_1$ neutralino. 
Hence, the single kinematically allowed decay mode of the staus is 
\begin{equation}
\tilde{\tau}\rightarrow\tilde{\chi}^0_1 \tau 
\end{equation}
Finally, all squarks, that do not contribute at leading-order, are decoupled as well.

This note is organized as follows. In Sec.~\ref{sec:descript}, we present an outline of the analysis under consideration. It in particular includes definitions for the physics objects and event selections that we have implemented for our recasting exercise. Sec.~\ref{sec:val} is dedicated to event generation in the context of the two considered benchmark for the validation or our re-implementation, and includes a comparison with official ATLAS results. In Sec.~\ref{sec:conc}, we summarize our work.

\subsection{Description of the analysis}\label{sec:descript}

This analysis targets a final state containing two hadronic tau leptons with a certain amount of missing transverse energy. 
The kinematics of the di-$\tau +E^{miss}_T$ system is used to reduce the contributions from Standard Model backgrounds. 
In Sec.~\ref{sec:stau_obj}, we first detail how the objects relevant for the analysis are reconstructed and defined. Then, in Sec.~\ref{sec:selection}, we discuss the sequence of event selections that are applied in the aim of unravelling the signal from the background.

\subsubsection{Object definitions}\label{sec:stau_obj}

Jets are reconstructed by means of the anti-$k_t$ algorithm~\cite{Cacciari:2008gp} with a radius parameter set to $R=0.4$. This analysis focuses on jets whose transverse momentum $p^j_T$ and pseudorapidity $\eta^j$ fulfill
\begin{equation}
p^j_T > 20 \textrm{ GeV}\quad \textrm{and}\quad |\eta^j| < 2.8.
\end{equation} 
Moreover, the selected jets that are tagged as originating from the fragmentation of a $b$-quark must satisfy the stronger requirements
\begin{equation}
p^b_T > 20 \textrm{ GeV}\quad \textrm{and}\quad |\eta^b| < 2.5.
\end{equation}
In the considered analysis, a $b$-tagging working point with an average efficiency of $77\%$ is used. This working point corresponds to $c$-jet and light-jet rejection rates of $4.9$ and $110$, respectively.

Electron candidates are required to have a transverse momentum $p^e_T$ and pseudorapidity $\eta^e$ obeying
\begin{equation}
p^e_T > 17 \textrm{ GeV}\quad \textrm{and}\quad |\eta^e| < 2.47.
\end{equation}
Furthermore, all electron candidates are required to have both track and calorimeter isolations. The condition of the track isolation is
\begin{equation}
\sum p_{T,\textrm{tracks}}/p^e_T < 0.15\quad \textrm{with}\quad \Delta R=\min(10\textrm{ GeV}/p^e_T,0.2),
\end{equation}
the condition of the calorimeter isolation is
\begin{equation}
\sum E_{T,\textrm{calorimeter}}/p^e_T < 0.2\quad \textrm{with}\quad \Delta R=0.2,
\end{equation}
and for high transverse momentum electron, we use instead of the two above conditions
\begin{equation}
\sum E_{T,\textrm{calorimeter}} < \max(0.015\times p^e_T,3.5\textrm{ GeV})\ \
  \textrm{with}\ \ \Delta R=0.2\ \ \textrm{if}\ \ p^e_T > 200\textrm{ GeV}.
\end{equation}

Muon candidate definition is similar, although with slightly looser thresholds,
\begin{equation}
p^{\mu}_T > 14 \textrm{ GeV}\quad \textrm{and}\quad |\eta^{\mu}| < 2.7,
\end{equation}
The condition of the track isolation is 
\begin{equation}
\sum p_{T,\textrm{tracks}}/p^{\mu}_T < 0.15\quad \textrm{with}\quad \Delta R=\min(10\textrm{ GeV}/p^{\mu}_T,0.3),
\end{equation}
and the condition of the calorimeter isolation is
\begin{equation}
\sum E_{T,\textrm{tracks}}/p^{\mu}_T < 0.3\quad \textrm{with}\quad \Delta R=0.2.
\end{equation}

In the ATLAS experiment, hadronically decaying tau lepton ($\tau_{had}$) candidates are reconstructed with one or three associated charged pion tracks (prongs).
For 1-prong (3-prong) $\tau$ lepton candidates, the signal efficiencies are $75\%$ and $60\%$ for the \textit{medium} working point respectively.
In the recasting based on \madanalysis\ that we implement in this work, the simulation of the detector response is performed with the {\sc Delphes}~3~\cite{deFavereau:2013fsa} software. We consider a tau-tagging efficiency of $100\%$ with a misidentification probability of $0\%$ at the level of {\sc Delphes}~3, and handle medium and tight tau-tagging efficiencies through event reweighting factors extracted from the official ATLAS cutflow tables. Those factors are evaluated and included at the level of the analysis. Further details are given in Sec.~\ref{sec:selection}.

Baseline tau lepton candidates are required to have 
\begin{equation}
p^{\tau}_T > 50\ (40) \textrm{ GeV}\quad \textrm{and}\quad |\eta^{\tau}| < 2.5
\end{equation}
for the leading (subleading) candidates, and the transition region between the barrel and endcap calorimeters ($ 1.37 < |\eta^{\tau}| < 1.52 $) is excluded.

The object definition ends with some overlap removal conditions. The latter are implemented consistently to the analysis code provided through HEPData\cite{1765529}.
Tau leptons are removed if they are too close to an electron or a muon, with $\Delta R(\tau,e/\mu) < 0.2$. Electrons are then removed if they are too close to a muon, with $\Delta R(e,\mu) < 0.01$. Next, the jet collection is cleaned from those jets lying at an angular distance $\Delta R(j,e/\mu) < 0.2$ of a muon or an electron, and the electrons and muons that are too close to any of the remaining jets are removed if $\Delta R(e/\mu,j) < 0.4$. Finally, jets are removed if they are too close to one of the tau lepton candidates, with $\Delta R(j,\tau) < 0.2$.

\subsubsection{Event selection}\label{sec:selection}
Because {\sc Delphes}~3 utilizes simplified and parameterized approaches to simulate different elements of the detector response, it is hard to emulate some of the properties relevant for the ATLAS-SUSY-2018-04 analysis, and therefore implement certain cut steps precisely. As a consequence, we have modelled several selections through event reweighting. This concerns first the trigger efficiency. Next, several reweighting factors are included to model specific features of the tau-tagging-based selections. This allows us to define the so-called \textit{medium} and \textit{tight} tau lepton cuts in our implementation, from an ideal detector parameterization in {\sc Delphes}~3.

After the object definitions introduced in the previous subsection, events with exactly two baseline tau leptons are selected.
All events are required to pass either an \textit{asymmetric di-$\tau$} trigger for the low stau mass region (SR-lowMass) or a combined \textit{di-$\tau +E^{miss}_T$} ($E^{miss}_T > 150$ GeV) trigger for the high stau mass region (SR-highMass). This is coined \textit{trigger and offline cuts} below.
A trigger efficiency of $80\%$ is applied in our recasting, after that we impose that the transverse momenta of the two leading tau candidates are larger than the offline $p_T$ thresholds given in Table~\ref{tab:trig-eff}.
In order to deal with the different tau candidate kinematic cuts that are applied in the 2015--2017 and 2018 data-taking periods, we randomly tag each event as originating from the 2015--2017 or 2018 data set. In practice, the probability of imposing the 2015--2017 (2018) data set thresholds is calculated from the ratio of the 2015--2017 (2018) integrated luminosity to the total luminosity of $139~{\rm fb}^{-1}$.

Moreover, we assume that the tau leptons which fired the triggers are those selected through the offline cuts. A trigger-level $\tau_{had}$ identification efficiency of 0.9 is correspondingly applied for each reconstructed tau lepton, which mimics the medium tau identification procedure for a tau lepton passing both online and offline requirements~\cite{ATLAS:2017mpa}.
This leads to a total trigger reweighting factor of 64.8\% that includes a global trigger efficiency of 80\% and individual tau reconstruction efficiencies of 90\%.

\begin{table}[t]
  \tbl{Offline $p_T$ thresholds for the leading (subleading) tau lepton candidate, in the case of the \textit{asymmetric di-$\tau$} (second column) and \textit{di-$\tau +E^{miss}_T$} (third column) triggers. This corresponds to a ditau efficiencies of about $80\%$.}
  {\begin{tabular}{@{}c c c@{}} \toprule
  Year & \textit{asymmetric di-$\tau$} & \textit{di-$\tau +E^{miss}_T$} \\
  \colrule
 2015-2017 & 95 (60) GeV & 50 (40) GeV \\
 2018 & 95 (75) GeV & 75 (40) GeV \\ 
  \botrule
  \end{tabular}\label{tab:trig-eff} }
\end{table}

After the handling of the triggers described above, events with exactly two \textit{medium} tau lepton candidates with opposite-sign (OS) electric charges are selected. To treat the efficiency of selecting two offline mediumly tagged OS taus on top of a \textit{di-tau(+$E^{miss}_T$)} trigger selection (as {\sc Delphes} does not simulate charge misidentification), an additional event reweighting factor of 0.7 is enforced.
This number is evaluated from the average ratio of the cut efficiencies provided by the ATLAS collaboration and those predicted by \madanalysis\ when the identification of two medium taus is not included at the cutflow step called \textit{2 medium $\tau$ (OS) and 3rd medium $\tau$ veto} below.

In the next selection steps, a $b$-jet veto is enforced to reject events originating from top quark processes.
Also, events featuring any additional light leptons (muons or electrons) are rejected.
Finally, selection cuts common to both signal regions also include constraints on the reconstructed invariant mass of the two leading tau lepton system, $m(\tau_1,\tau_2)$. The latter is required to be larger than $120$ GeV, in order to remove events exhibiting a pair of tau leptons stemming from low-mass resonances, $Z$ boson, and Higgs boson decays ($Z/H$ veto).

In the SR-lowMass region, a missing energy constraint of 75~GeV $< E^{miss}_T < 150$~GeV is imposed to increase the signal sensitivity. Moreover, the two selected tau leptons are required to be tight tagged.
The selection efficiency $p_{tight}$ associated with two {\it medium} taus passing the \textit{tight} working point requirements is extracted from the official ATLAS cutflow tables. We rely on the ratio of the number of surviving weighted events before applying the tight tau lepton requirement, and after applying it. We use $p_{tight}\simeq 0.70$.

In the SR-highMass region, the tight tagging efficiency is extracted similarly, with the exception that at least one of two tau leptons should pass the tight selection requirements and not both of them). We use here $p_{tight} + 2\sqrt{p_{tight}}(1-\sqrt{p_{tight}})\simeq 0.91$.

The \textit{stransverse mass} $m_{T2}$ variable~\cite{Lester:1999tx,Cheng:2008hk} is defined as
\begin{equation}
m_{T2} =min_{\mathbf{q}_T}
\left[
max(m_{T,\tau_1}(\mathbf{p}_{T,\tau_1},\mathbf{q}_T),m_{T,\tau_2}(\mathbf{p}_{T,\tau_2},\mathbf{p}^{miss}_T -\mathbf{q}_T))
\right],
\end{equation}   
where $\mathbf{p}_{T,\tau_1}$ and $\mathbf{p}_{T,\tau_2}$ are the transverse momenta of the two tau lepton candidates. The transverse momentum vector of one of the invisible particle, $\mathbf{q}_T$, is chosen to minimize the larger of the two transverse mass $m_{T,\tau_1}$ and $m_{T,\tau_2}$. The transverse mass $m_T$ is defined by
\begin{equation}
m_{T}(\mathbf{p}_T,\mathbf{q}_T) = \sqrt{2(p_T q_T -\mathbf{p}_T\cdot\mathbf{q}_T)}.
\end{equation} 
In \madanalysis, the $m_{T2}$ calculation can be done automatically through the function {\tt PHYSICS->Transverse->MT2(vec1,vec2,ETmiss,Minvisible)}. In this expression, {\tt vec1} and {\tt vec2} stand for the two visible momenta, {\tt ETmiss} for the miissing transverse momentum and {\tt Minvisible} for a test mass that should map the expected mass of the invisible state.

A lower bound on the $m_{T2}$ variable of 70~GeV is imposed, in order to reduce the contamination from $t\overline{t}$ and $WW$ events.
Finally, the two tau lepton candidates are required to be well separated in the transverse plane, by $\Delta R(\tau_1,\tau_2) < 3.2$ and $|\Delta\phi (\tau_1,\tau_2)| > 0.8$ to further suppress the contributions of the Standard Model backgrounds.

\subsection{Validation}\label{sec:val}
\begin{table}[t]
\renewcommand{\arraystretch}{1.3}
  \tbl{Cut-flow associated with a simplified model benchmark scenario defined by $ m(\tilde{\tau},\tilde{\chi}^0_1) = (120,1) $~GeV and for the $ pp\to \tilde{\tau}\tilde{\tau} $ production process. We compare ATLAS official results and \madanalysis\ predictions through the expected number of events after each cut and the corresponding efficiencies, and indicate their difference $\delta$.}
  {\begin{tabular}{@{}c c c c c c@{}} \toprule
\multicolumn{6}{c}{ \textbf{$ \tilde{\tau}\tilde{\tau} $ production with $ m(\tilde{\tau},\tilde{\chi}^0_1) = (120,1) $ GeV} }\\
 & ATLAS ($N_{weighted}$) & $\epsilon_i$($\%$) & MA5 ($N_{weighted}$) & $\epsilon_i$($\%$) & diff.($\%$) \\
\hline
Baseline cut                                    &  1686.80 &     - &  1686.80 &     - &      - \\[.4cm]
\multicolumn{5}{c}{ \textbf{SR-low Mass} }\\\hline
Trigger and offline cuts                        &   390.46 & 23.15 &   410.01 & 24.31 &   5.01 \\
2 medium $\tau$ (OS) and 3rd medium $\tau$ veto &   256.01 & 65.57 &   269.37 & 65.70 &   0.20 \\
$b$-jet veto                                    &   250.59 & 97.88 &   263.66 & 97.88 &  -0.00 \\
Light lepton veto                               &   250.12 & 99.81 &   263.66 &   100 &   0.19 \\
$Z/H$-veto                                      &   248.93 & 99.52 &   262.14 & 99.42 &  -0.10 \\
$ 75 < E^{miss}_T < 150 $ GeV                   &    85.70 & 34.43 &    89.90 & 34.30 &  -0.38 \\
2 tight $\tau$                                  &    60.19 & 70.23 &    62.93 & 70.00 &  -0.33 \\
$ |\Delta\phi(\tau,\tau)| > 0.8 $               &    60.14 & 99.92 &    62.75 & 99.72 &  -0.20 \\
$ |\Delta R(\tau,\tau)| < 3.2 $                 &    54.73 & 91.00 &    57.10 & 90.99 &  -0.01 \\
$ m_{T2} > 70 $ GeV                             &     9.78 & 17.87 &    14.65 & 25.66 &  43.58 \\ \hline
All                                             &        - &  0.58 &        - &  0.87 &  49.80 \\[.4cm]
\multicolumn{5}{c}{ \textbf{SR-high Mass} }\\\hline
Trigger and offline cuts                        &   101.23 &  6.00 &    96.35 &  5.71 &  -4.82 \\
2 medium $\tau$ (OS) and 3rd medium $\tau$ veto &    67.04 & 66.23 &    63.23 & 65.62 &  -0.91 \\
$b$-jet veto                                    &    63.98 & 95.44 &    60.37 & 95.47 &   0.04 \\
Light lepton veto                               &    63.87 & 99.83 &    60.36 & 99.99 &   0.16 \\
$Z/H$-veto                                      &    58.33 & 91.33 &    55.70 & 92.28 &   1.04 \\
$ \geq 1 $ tight $\tau$                         &    57.29 & 98.22 &    50.69 & 91.00 &  -7.35 \\
$ |\Delta\phi(\tau,\tau)| > 0.8 $               &    56.71 & 98.99 &    49.99 & 98.63 &  -0.36 \\
$ |\Delta R(\tau,\tau)| < 3.2 $                 &    51.74 & 91.24 &    45.41 & 90.84 &  -0.43 \\
$ m_{T2} > 70 $ GeV                             &     7.18 & 13.88 &     8.24 & 18.14 &  30.75 \\ \hline
All                                             &        - &  0.43 &        - &  0.49 &  14.76 \\[.2cm] \botrule
\end{tabular}
\label{tab:120GeV} }
\end{table}

\subsubsection{Event generation}

In order to validate our analysis, we rely on the MSSM implementation~\cite{Duhr:2011se} available in the {\sc Feynrules}~\cite{Alloul:2013bka} model database and shipped with thew {\sc MadGraph5\_aMC@NLO} event generator~\cite{Alwall:2014hca} as a UFO library~\cite{Degrande:2011ua}.

We consider two benchmark points with masses $m(\tilde{\tau},\tilde{\chi}^0_1)=(120,1)$ GeV and $(280,1)$ GeV to illustrate the validation of our re-implementation, as those correspond two scenarios for which official ATLAS cutflows and differential distributions are provided.
The stau mixing matrix is additionally set to a unity matrix, so that the stau mass-eigenstates correspond to the right-handed and left-handed stau flavor-eigenstates.

We make use of {\sc Madgraph5\_aMC@NLO} version 2.6.7~\cite{Alwall:2014hca} for hard-scattering event generation for each of the two stau eigenstates, in which we convolute leading-order matrix elements with the NNPDF23LO~\cite{Ball:2012cx} set of parton distribution function.
Our signal matrix elements include the potential emission of up to two additional partons, and the different contributions are merged according to the MLM scheme~\cite{Mangano:2006rw,Alwall:2008qv}. We use a merging scale defined through the hard-scattering level parameter of {\sc MadGraph5\_aMC2NLO} {\tt xqcut} $= m_{\tilde{\tau}}/4$.
\begin{table}[t]
\renewcommand{\arraystretch}{1.3}
  \tbl{Same as in Table~\ref{tab:120GeV} but for a scenario with supersymmetric masses $ m(\tilde{\tau},\tilde{\chi}^0_1) = (280,1) $ GeV.}
  {\begin{tabular}{@{}c c c c c c@{}} \toprule
\multicolumn{6}{c}{ \textbf{$ \tilde{\tau}\tilde{\tau} $ production with $ m(\tilde{\tau},\tilde{\chi}^0_1) = (280,1) $ GeV} }\\
 & ATLAS ($N_{weighted}$) & $\epsilon_i$($\%$) & MA5 ($N_{weighted}$) & $\epsilon_i$($\%$) & diff.($\%$) \\
\hline
Baseline cut                                    &   184.36 &     - &   184.36 &     - &      - \\[.4cm]
\multicolumn{5}{c}{ \textbf{SR-low Mass} }\\\hline
Trigger and offline cuts                        &    73.74 & 40.00 &    69.97 & 37.95 &  -5.12 \\
2 medium $\tau$ (OS) and 3rd medium $\tau$ veto &    47.86 & 64.90 &    46.23 & 66.08 &   1.81 \\
$b$-jet veto                                    &    46.63 & 97.43 &    44.94 & 97.20 &  -0.24 \\
Light lepton veto                               &    46.49 & 99.70 &    44.94 & 99.99 &   0.30 \\
$Z/H$-veto                                      &    44.84 & 96.45 &    43.83 & 97.54 &   1.13 \\
$ 75 < E^{miss}_T < 150 $ GeV                   &    17.48 & 38.98 &    16.26 & 37.10 &  -4.83 \\
2 tight $\tau$                                  &    12.04 & 68.88 &    11.38 & 70.00 &   1.63 \\
$ |\Delta\phi(\tau,\tau)| > 0.8 $               &    12.04 &   100 &    11.33 & 99.55 &  -0.45 \\
$ |\Delta R(\tau,\tau)| < 3.2 $                 &    11.08 & 92.03 &    10.35 & 91.32 &  -0.77 \\
$ m_{T2} > 70 $ GeV                             &     6.08 & 54.87 &     5.64 & 54.50 &  -0.68 \\ \hline
All                                             &        - &  3.30 &        - &  3.06 &  -7.24 \\[.4cm]
\multicolumn{5}{c}{ \textbf{SR-high Mass} }\\\hline
Trigger and offline cuts                        &    47.64 & 25.84 &    42.10 & 22.83 & -11.64 \\
2 medium $\tau$ (OS) and 3rd medium $\tau$ veto &    30.72 & 64.48 &    27.80 & 66.03 &   2.40 \\
$b$-jet veto                                    &    29.34 & 95.51 &    26.83 & 96.52 &   1.06 \\
Light lepton veto                               &    29.27 & 99.76 &    26.83 & 99.99 &   0.23 \\
$Z/H$-veto                                      &    24.88 & 85.00 &    24.01 & 89.50 &   5.30 \\
$ \geq 1 $ tight $\tau$                         &    24.21 & 97.31 &    21.85 & 91.00 &  -6.48 \\
$ |\Delta\phi(\tau,\tau)| > 0.8 $               &    23.29 & 96.20 &    21.19 & 96.96 &   0.79 \\
$ |\Delta R(\tau,\tau)| < 3.2 $                 &    21.95 & 94.25 &    19.68 & 92.91 &  -1.42 \\
$ m_{T2} > 70 $ GeV                             &    14.35 & 65.38 &    13.37 & 67.91 &   3.88 \\ \hline
All                                             &        - &  7.78 &        - &  7.25 &  -6.84 \\[.2cm] \botrule
\end{tabular}
\label{tab:280GeV} }
\end{table}

The {\sc Pythia} package version 8.244~\cite{Sjostrand:2014zea} with the so-called $A14$ tune\cite{TheATLAScollaboration:2014rfk} has been used for the simulation of parton showering and hadronization. The simulation of the detector response has been performed by using {\sc Delphes}~3.4.2~\cite{deFavereau:2013fsa}, that relies on {\sc FastJet}~\cite{Cacciari:2011ma} for object reconstruction.

We have tuned the ATLAS detector parameterization in {\sc Delphes}~3 appropriately, according to the needs of the analysis.
For example, loosened isolation criteria are applied so that isolation could be implemented fully at the analysis level.
Moreover, the radius parameter and minimum transverse momentum used for jet reconstruction are reduced to 0.4 and 15~GeV respectively, and we have updated the $b$-tagging and tau-tagging performance.
Finally, the {\tt UniqueObjectFinder} module has been disabled as object overlap removal has been implemented at the level of the analysis.

\subsubsection{Comparison with the official results}

In Tables~\ref{tab:120GeV} and~\ref{tab:280GeV}, we compare predictions obtained with our implementation to the official results provided in the form of auxiliary tables by the ATLAS collaboration, for the two considered benchmark points with masses $m(\tilde{\tau},\tilde{\chi}^0_1)=(120,1)$ and $(280,1)$ GeV respectively.
For each cut, we have calculated the related efficiency
\begin{equation}
\epsilon_i =\frac{n_i}{n_{i-1}}
\end{equation}
where $ n_i $ and $ n_{i-1} $ correspond to the number of events after and before the considered cut respectively.
In our comparison, we have normalized the number of events surviving the baseline cut $n_0$ as in the ATLAS cutflow.
On the other hand, we have also evaluated the differences between the \madanalysis\ ($\epsilon_i ({\rm MA5})$) and ATLAS ($\epsilon_i ({\rm ATLAS})$) cut efficiencies through the quantity
\begin{equation}
\delta_i = \frac{\epsilon_i ({\rm MA5})-\epsilon_i ({\rm ATLAS})}{\epsilon_i ({\rm ATLAS})}.
\end{equation}

\begin{figure}[t]
  \centerline{\includegraphics[width=2.5in]{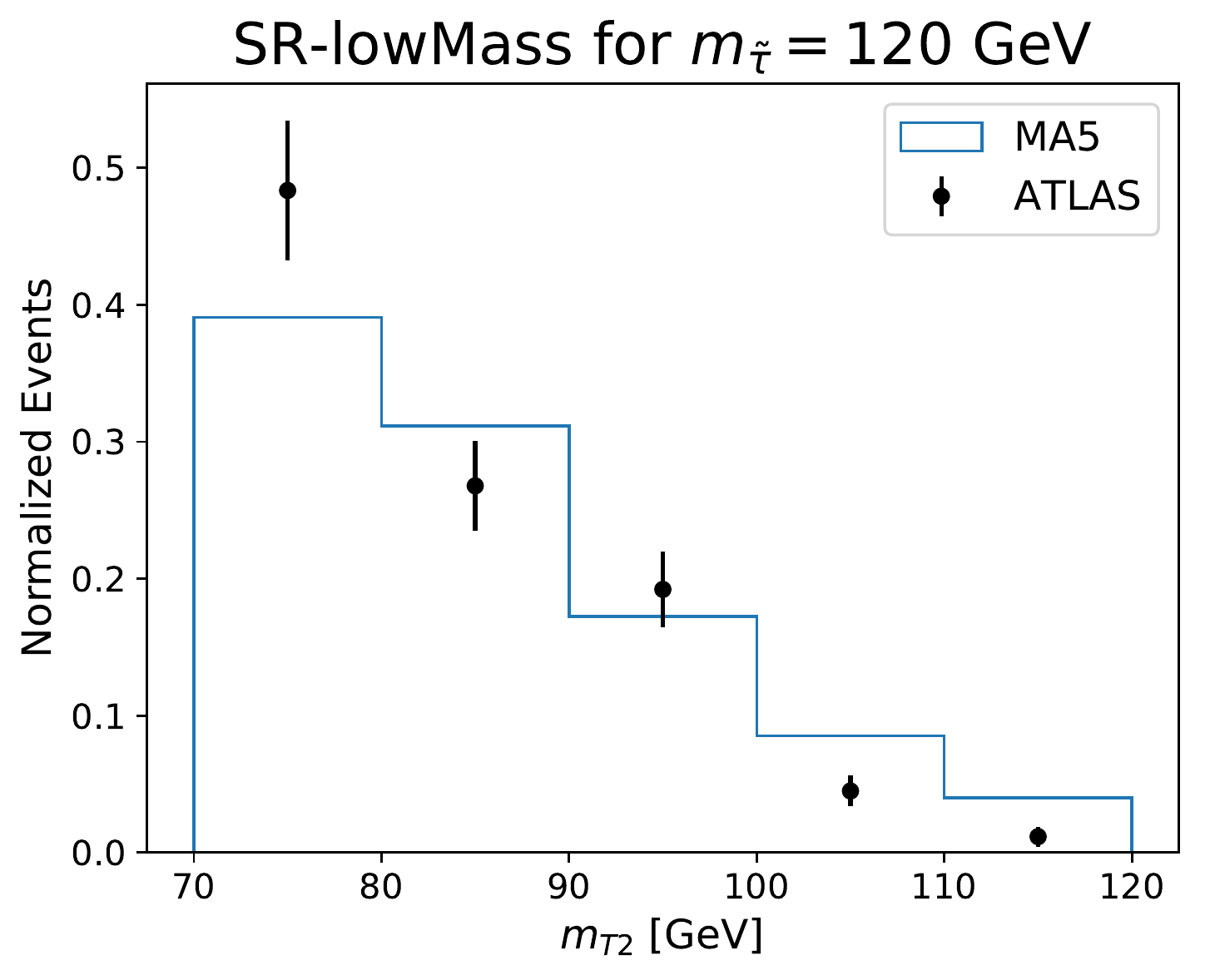}\includegraphics[width=2.5in]{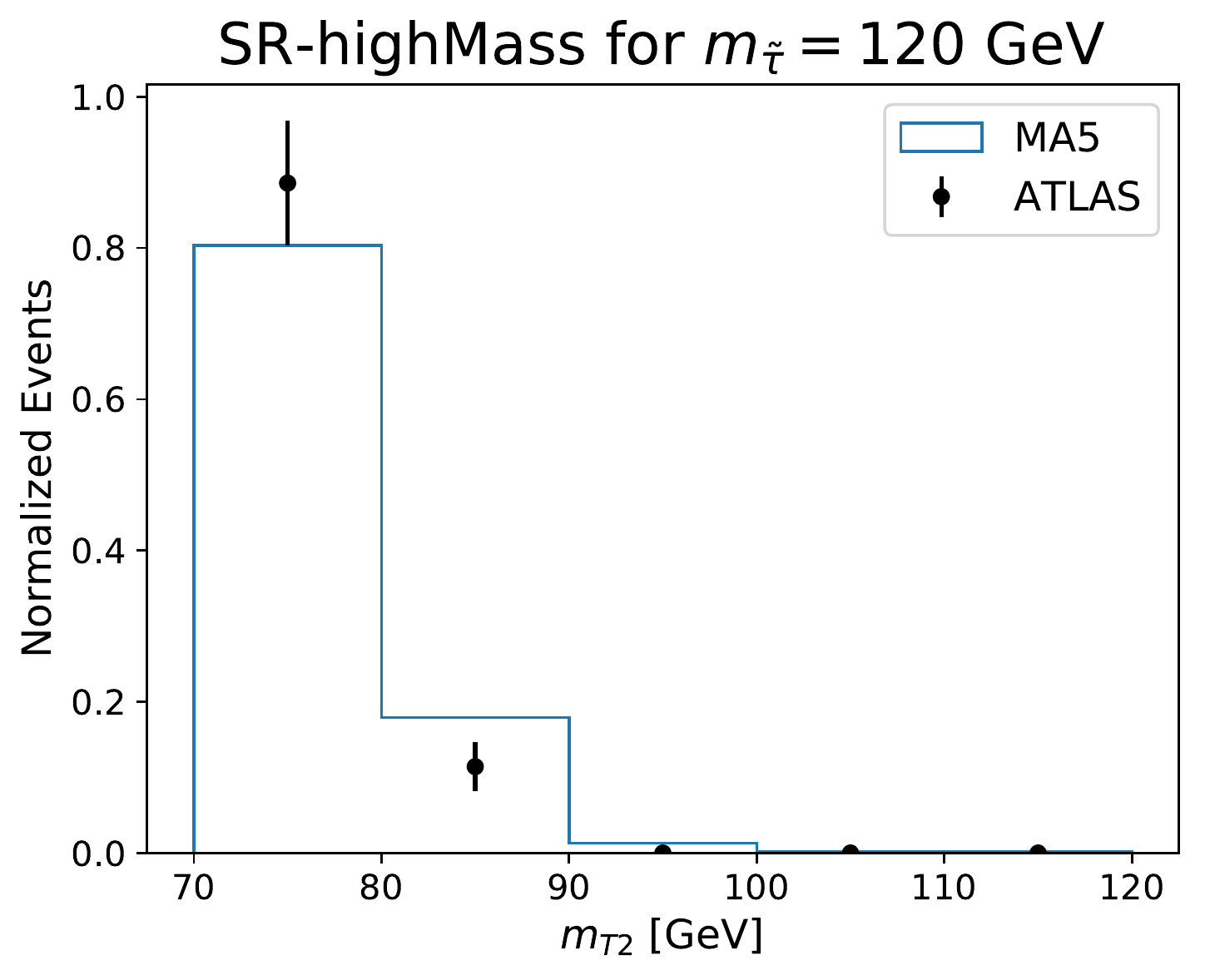}}
  \vspace*{8pt}
  \centerline{\includegraphics[width=2.5in]{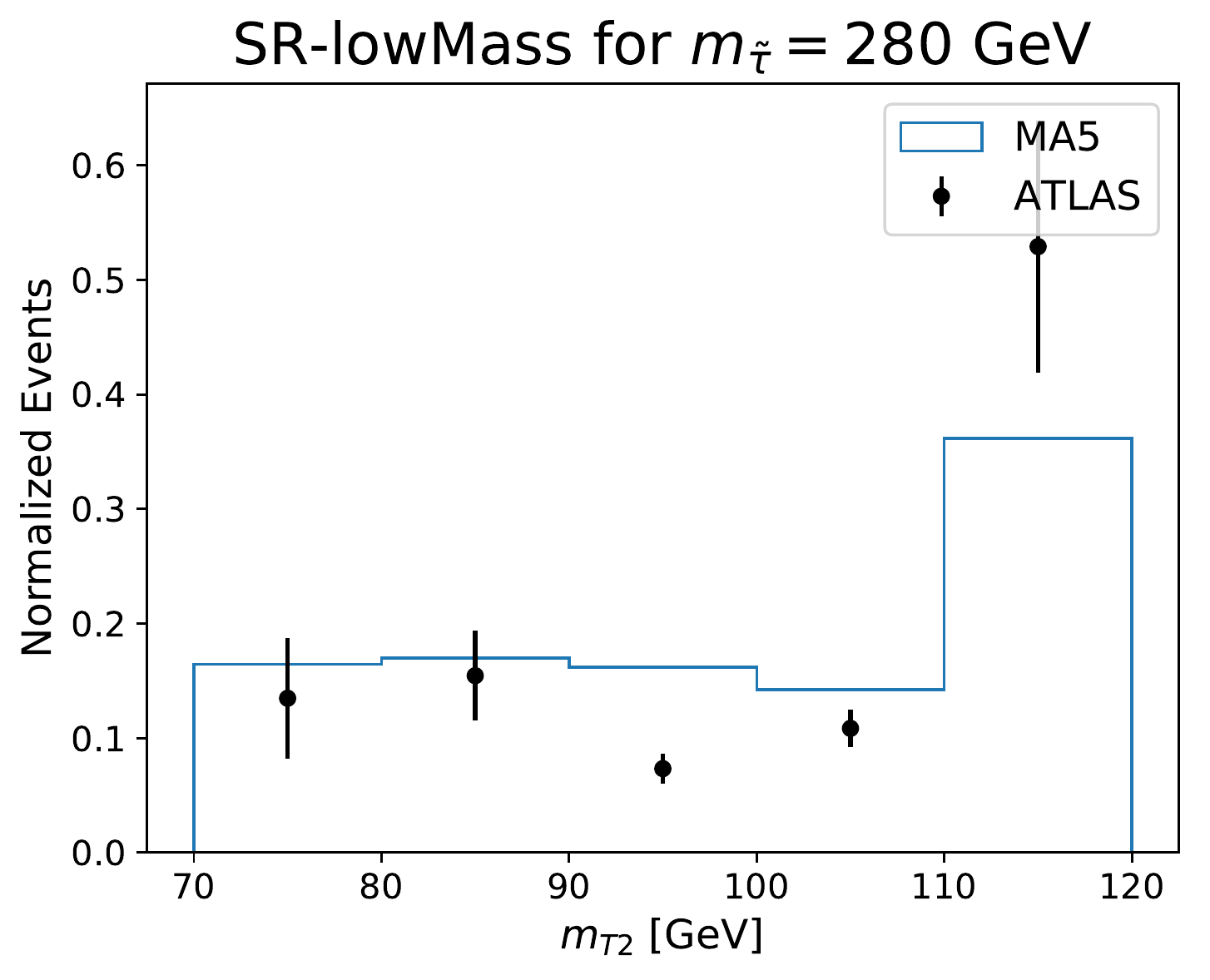}\includegraphics[width=2.5in]{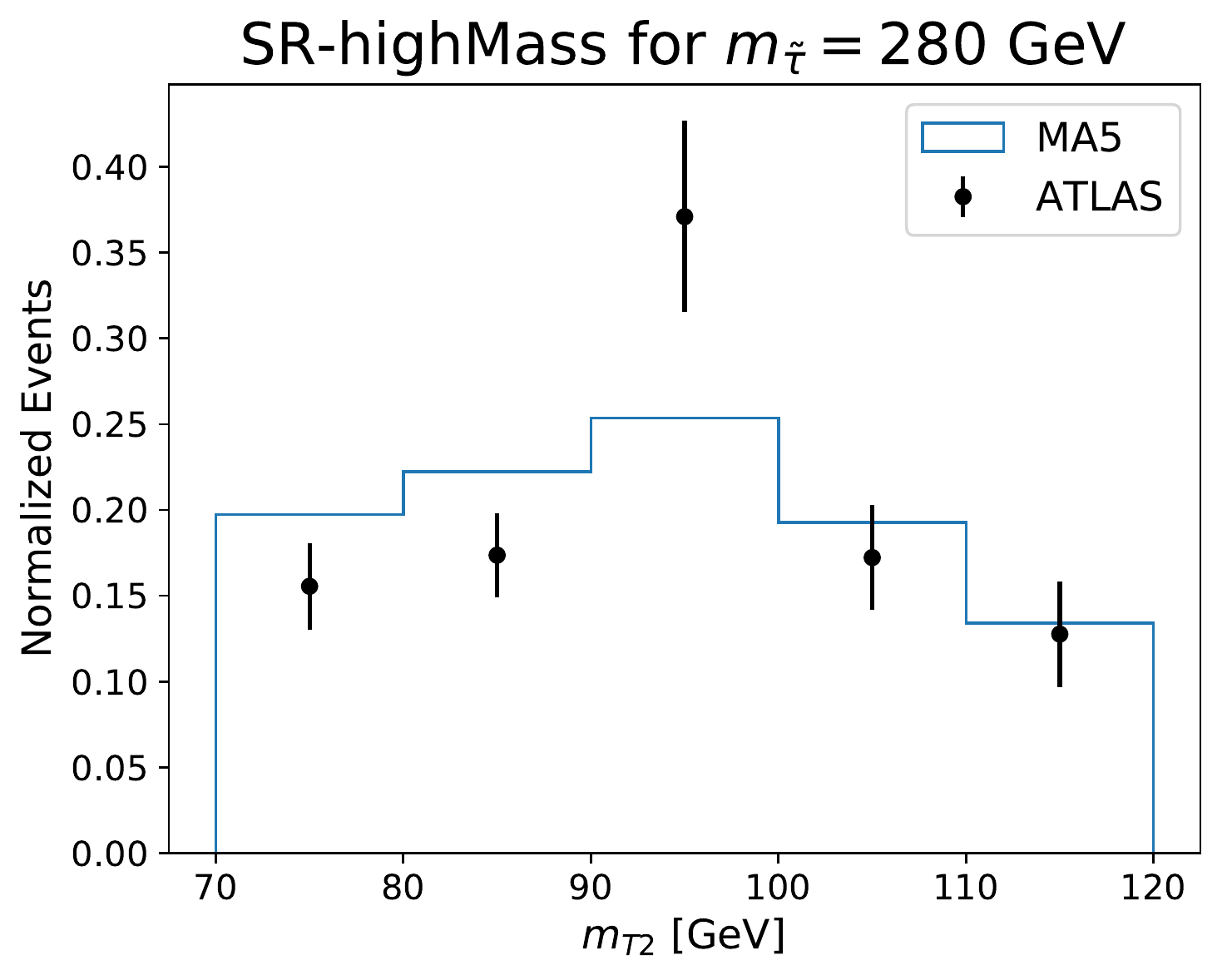}}
  \caption{The $m_{T2}$ distributions after all cuts, for the SR-low Mass (left) and SR-high Mass (right) signal regions, and for the $m(\tilde{\tau},\tilde{\chi}^0_1)=(120,1)$~GeV (top panel) and $m(\tilde{\tau},\tilde{\chi}^0_1)=(280,1)$~GeV (bottom panel) scenarios.\protect\label{fig:mt2}}
\end{figure}

We observe that for both scenarios, a good agreement is obtained at each step of the cutflow, with the exception of the last cut on the $M_{T2}$ variable. For the $m(\tilde{\tau},\tilde{\chi}^0_1)=(120,1) $~GeV scenario, we hence obtain a disagreement of 30\%--40\% for both signal regions. In constrast, for the $m(\tilde{\tau},\tilde{\chi}^0_1)=(280,1)$~GeV scenario does not feature any strong issue at all, the two cutflow agreeing at the level of a few percent.

By lack of additional publicly available experimental information, we have not been able to investigate this issue at a very deep level. We have nevertheless compared $M_{T2}$ distributions as predicted by \madanalysis\ after all cuts, to those released by the ATLAS collaboration\cite{1765529}. We have considered the two signal regions and both scenarios. Our results are shown in Fig.~\ref{fig:mt2}.

We observe that the global shape of the distribution is generally well reproduced, although the curves exhibit large differences that explain our findings at the level of the cutflow tables. However, one must note that the differences concern cases where a not so large number of (unweighted) events survive. Large Monte Carlo uncertainties of 10\%--20\% of percent are thus expected, both for our predictions and the ATLAS results.

\begin{figure}[t]
  \centering
  \includegraphics[width=4.0in]{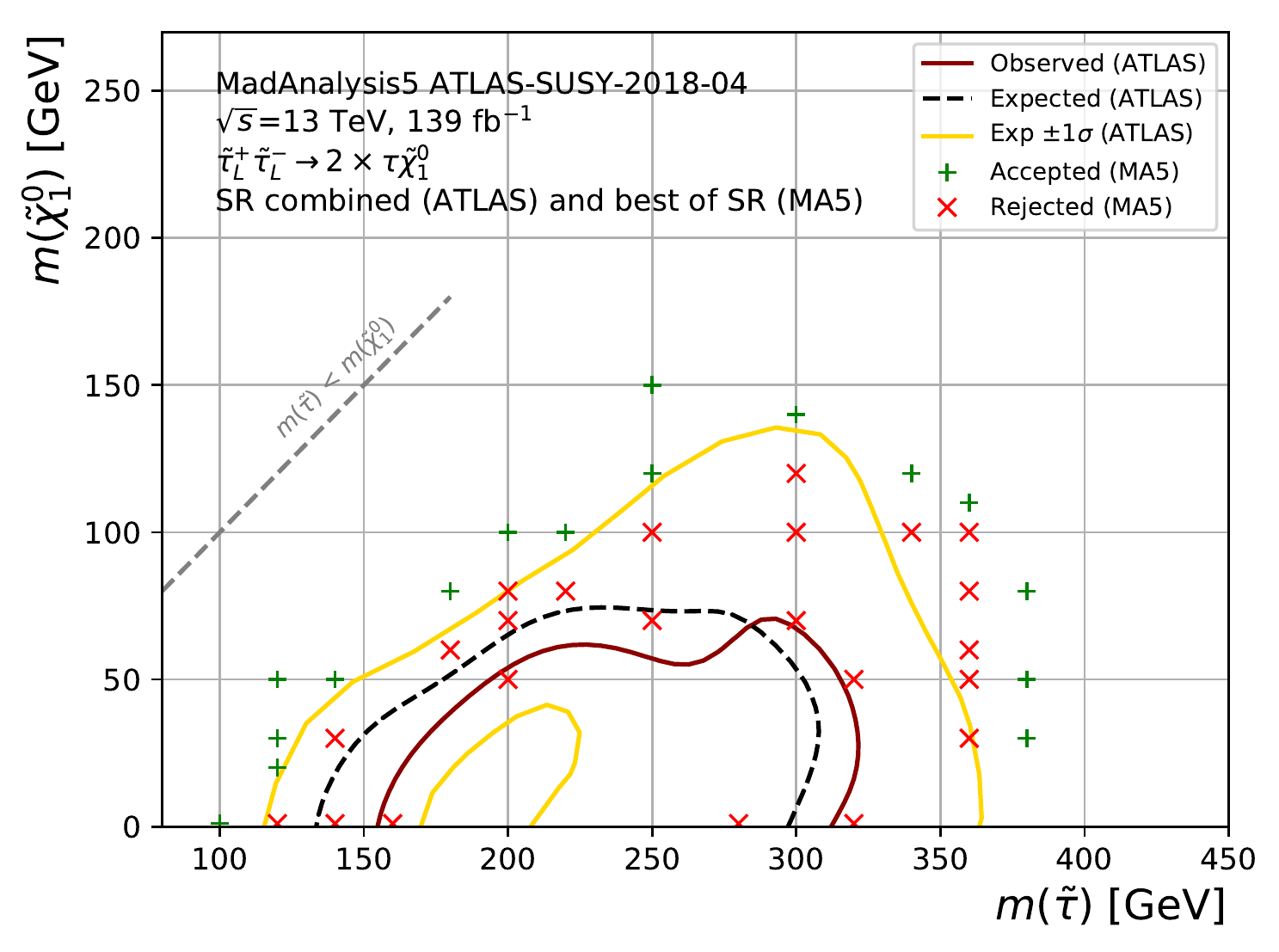}
  \caption{Comparison of \madanalysis\ and official 95\% CL exclusion contours. The point marked with red cross is excluded if one of signal region is rejected at $1-\textrm{CL}_s > 0.95$.\protect\label{fig:limit}}
\end{figure}

To examine the impact of those yield differences on limit setting, we peformed a set of statistical analyses for various points in the stau and neutralino mass parameter space. The yields are normalized to NLO+NLL prediction~\cite{Fuks:2013vua,Fuks:2013lya}, and limits are calculated using the CL$_s$ method~\cite{Read:2002hq}. The mass point is determined to be excluded conservatively if one of the two signal regions leads to a 1-CL$_s$ value greater than 0.95. This contrasts with the interpretations provided in the ATLAS publication~\cite{Aad:2019byo}, that rely on the combination of both signal regions. The results obtained with \madanalysis\ are presented in Fig.~\ref{fig:limit}, along with the official ATLAS results.

The reproduced and official results agree generally well within $1\sigma$. This suggests that the discrepancies related to the  $M_{T2}$ spectrum and the corresponding cut efficiency only affect the limits mildly. This is expected to be improved after combining the SR-lowMass and SR-highMass regions, as shown in the {\sc SModelS} study of Ref.~\cite{Khosa:2020zar} that demonstrated that limits calculated from a single signal region of the considered analysis are overly enthusiastic.

Consequently, we consider our re-implementation as validated.

\subsection{Conclusions}\label{sec:conc}

In this note, we detail our implementation of the ATLAS-SUSY-2018-04 search in the \madanalysis\ framework. Our analysis has been validated in the context of a supersymmetry-inspired simplified benchmark model in which the Standard Model is extended by a neutralino and a stau. Both stau chiralities are considered, as the stau is considered to decay into a tau lepton and a neutralino. Our validation relies on two different benchmark points in the parameter space.

By comparing our predictions for different cutflows for the two benchmarks with the official ones provided by the ATLAS collaboration in Ref.~[\refcite{Aad:2019byo}], we have found an agreement at each step of the analysis, except for the last cut on the stranverse mass variable $m_{T2}$ cut for the light stau scenario. While the shape of the distribution is correctly reproduced, large difference leads to a quite different cut efficiency. Due to the lack of more information, we have however not been able to investigate the issue more precisely.
By further comparing exclusion contours obtained with \madanalysis\ to those provided by the ATLAS collaboration, we however observe that this only impacts the limits at the level of $1\sigma$.
As a consequence, we have considered our re-implementation as validated.

The {\sc MadAnalysis}~5 C++ code is available for download from the \madanalysis\ dataverse
(\href{https://doi.org/10.14428/DVN/UN3NND}{https://doi.org/10.14428/DVN/UN3NND})~\cite{UN3NND_2020}. The material relevant for the validation benchmarks has been obtained from \href{https://www.hepdata.net/record/ins1750597}{HEPData}~\cite{1765529}.

\cleardoublepage
\markboth{Jinheung Kim, Taegyu Lee, Jeongwoo Kim and Ho Jang}{Reimplementation of the ATLAS-SUSY-2018-06 analysis}


\section{Implementation of the ATLAS-SUSY-2018-06 analysis (electroweakinos with Jigsaw variables; 139~fb$^{-1}$)}
  \vspace*{-.1cm}\footnotesize{\hspace{.5cm}By Jinheung Kim, Taegyu Lee, Jeongwoo Kim and Ho Jang}
\label{sec:jigsaw}

\subsection{Introduction}

Supersymmetry (SUSY) \cite{Nilles:1983ge,Haber:1984rc} is an extension of Standard Model (SM) which predicts the existence of spin partners for each SM particle. Some of these partners are the so-called electroweakinos that consist in admixtures of the partners of the neutral and charged gauge and Higgs bosons of the model. In general and as in the analysis considered in this work, the lightest supersymmetric particle (LSP), a candidate of dark matter, is assumed to be the lightest neutralino, $\tilde{\chi}^0_{1}$.

\begin{figure}[t!]
    \centering
    \includegraphics[scale=0.65]{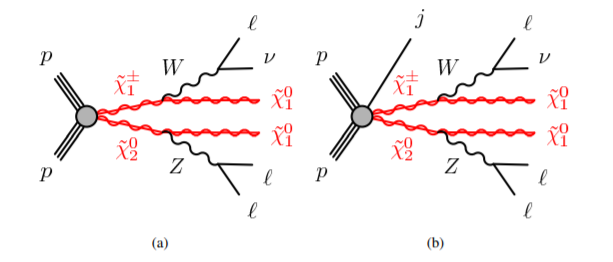}
    \caption{Representative Feynman diagram corresponding to $\tilde{\chi}^\pm_{1}$ $\tilde{\chi}^0_{2}$ production with subsequent decays into two $\tilde{\chi}^0_1$ via leptonically decaying $W$ and $Z$ bosons. The final-state signature is thus comprised of three leptons and a neutrino. Diagrams are shown both (a) without a jet and (b) with a jet originating from initial-state QCD radiation.}
    \label{feynman diagram}
\end{figure}

In this contribution, we present the {\sc MadAnalysis~5} \cite{Conte:2018vmg, Dumont:2014tja, Conte:2014zja, Conte:2012fm} implementation of the ATLAS-SUSY-2018-06 \cite{Aad:2019vvi} search for electroweakinos, together with its validation. The ATLAS analysis is targeting chargino ($\tilde{\chi}^\pm_{1}$) and neutralino ($\tilde{\chi}^0_{2}$) production in the case where the spectrum features mass splittings $\Delta m = m(\tilde{\chi}^\pm_{1}/\tilde{\chi}^0_{2})-m(\tilde{\chi}^0_{1})$ larger than the $Z$ boson mass, and where the chargino and the second neutralino are mass degenerate.
The analysis focuses on a decay chain in which the produced electroweakinos decay into the invisible LSP $\tilde{\chi}^0_{1}$ and either a $W$ or $Z$ gauge boson. The full decay processes are then $\chi^\pm_1 \to \chi^0_1W^\pm \to \chi^0_1l^\pm\nu$ and $\chi^0_2 \to \chi^0_1Z \to \chi^0_1l^+l^-$. Finally, both the $W$ and $Z$ bosons decay leptonically, which leads to a final-state signature comprised of three leptons and the missing transverse momentum originating from two LSPs and a neutrino. This process is illustrated in Figure \ref{feynman diagram}.

The conventional method which uses laboratory-frame when reconstructing the SUSY particles has some ambiguities. If the LSP is produced at colliders, it will leave the detectors without interacting. Thus, its presence can only be inferred from the missing momentum vector. However, this is problematic in that, not only the information of the properties of the final state is lost, but also the information of the intermediate particles is lost. For the SUSY particles with multiple decays, it can be difficult to match the decay products, which are indistinguishable, and reconstruct the originally produced particles without this information. This results in ambiguities when reconstructing the potentially produced electroweakinos $\tilde{\chi}^\pm_{1},\,\tilde{\chi}^0_{2}$.

The recursive jigsaw reconstruction (RJR) technique\cite{Jackson:2016mfb,Jackson:2017gcy} has been proposed to resolve these ambiguities by analyzing each event starting from the laboratory-frame particles and boosting back to the rest frames of the pair-produced parent sparticles (PP frame). Using this technique, the ATLAS collaboration found excesses of three lepton events in two regions in 36.1 fb$^{-1}$ of data collected between 2015 and 2016~\cite{Aaboud:2018sua}. In this last analysis, one region, named SR-LOW, has led to a local significance of 2.1$\sigma$ and targeted low-mass ($\tilde{\chi}^\pm_{1}$ $\tilde{\chi}^0_{2}$) production. Another region, named SR-ISR (initial state radiation), yielded a local significance of 3.0$\sigma$ and targeted $\tilde{\chi}^\pm_{1}$ $\tilde{\chi}^0_{2}$ production associated with an initial-state radiation (ISR). Thus, further analysis was made in ATLAS-SUSY-2018-06 with higher luminosity of 139~fb$^{-1}$.

In the ATLAS-SUSY-2018-06 analysis, a new approach was made to find the intersection between the conventional and the RJR approach. This new technique emulates the variables used in the RJR approach with laboratory frame variables and using minimal assumptions about the mass of the invisible system.
This technique provides a simple set of variables that are easily reproducible. When defining the object and region, the emulated recursive jigsaw reconstruction (eRJR) variables are kept as close as possible with \cite{Aaboud:2018sua}. The eRJR technique was validated by reproducing the excesses of three-lepton events using the same 36.1 fb$^{-1}$ of $pp$ collision data. In the ATLAS-SUSY-2018-06 analysis, this technique is applied to an integrated luminosity of 139 fb$^{-1}$ of $pp$ collision data collected between 2015 and 2018. In this higher luminosity upgrade of the work of \cite{Aaboud:2018sua}, the number of events and the number of expected background events in the SR-LOW region are $51$ and $46\pm5$, whereas in the SR-ISR region, the number of events and the number of expected background events are $30$ and $23.4\pm2.1$.

\subsection{Description of the analysis}
The analysis is targeting chargino ($\tilde{\chi}^\pm_{1}$) and neutralino ($\tilde{\chi}^0_{2}$) production, where they are assumed to respectively decay with a 100\% branching ratio into $W$ and $Z$ bosons. Thus, the analysis requires three leptons in the final state. After defining the signal objects, eRJR variables are computed for each event passing some preselection. Those different eRJR variables are then used to define two different classes of signal regions, SR-LOW and SR-ISR, as further detailed in section~\ref{sec:2.2}.

\subsubsection{Object definitions}
 Electron candidates are required to have a transverse momentum $p_{\rm T}$ and pseudorapidity $\eta$ satisfying
 \begin{equation}
     p_{\rm T}\,>\,10\;\mbox{GeV}\qquad\mbox{and}\qquad |\eta|\,<\,2.47.
 \end{equation}
To isolate electrons from any additional activity, requirements are imposed using energy clusters in the electromagnetic calorimeter and restrict the activity in a cone of radius $\Delta R = 0.2$ around the electron.
Moreover, the sum of transverse energy of the calorimeter energy clusters and the sum of the $p_{\rm T}$ of all tracks within the cone is constrained to be below 6\% of the electron $p_{\rm T}$.

Muon candidates are reconstructed with the following requirement,
 \begin{equation}
     p_{\rm T}\,>\,10\;\mbox{GeV}\qquad\mbox{and}\qquad |\eta|\,<\,2.4.
 \end{equation}
Muons must also be isolated from any additional activity and their isolation is defined similarly as for the electrons. For muons with $p_{\rm T}\,<\,33$\;\mbox{GeV}, the isolation cone radius is $\Delta R\,=\,0.3$. In the case of muons with $p_{\rm T}$ larger than 33 GeV, $\Delta R$ decreases linearly as a function of $p_{\rm T}$, and terminates to $\Delta R\,=\,0.2$ at $p_{\rm T}\,=\,50$\;\mbox{GeV}. For muons with $p_{\rm T}$ larger than 50 GeV, the isolation cone is maintained as $\Delta R = 0.2$. The sum of transverse energy of the calorimeter energy clusters within the cone should be below 15\% of the muon $p_{\rm T}$. Furthermore, the sum of the $p_{\rm T}$ of all tracks within the cone is constrained to be below 4\% of the muon $p_{\rm T}$.

Jets are reconstructed using the anti-$k_t$ algorithm\cite{Cacciari:2008gp} with reconstruction radius of $R=0.4$. Jets whose $p_{\rm T}$ is larger than 20 $\mbox{GeV}$ and $|\eta| < 2.4$ are considered as signal jets.

For the unique identification of leptons and jets, an overlap removal procedure is implemented. Electrons are removed if they are within $\Delta R\,=\,0.2$ of a muon. Jets are discarded if they are within $\Delta R\,=\,0.2$ of a lepton. Finally, leptons with $p_{\rm T} \leq 25\;\mbox{GeV}$ are removed if their angular distance from jet is $\Delta R < 0.4$. For leptons with $p_{\rm T} \geq 25\;\mbox{GeV}$, this angular distance ($\Delta R$) decreases as a linear function of $p_{\rm T}$ to $\Delta R < 0.2$ as $p_{\rm T}$ increases from $p_{\rm T}\;=\;25\;\mbox{GeV}$ to $p_{\rm T}\;=\;50\;\mbox{GeV}$.

\subsubsection{Event selection}\label{sec:2.2}
Event selection is performed using different eRJR variables. Five different eRJR variables are used to define the SR-ISR region. The first of them consists in the missing transverse energy $E^{\rm {miss}}_{\rm T}$ that is defined as the magnitude of the missing transverse momentum. The second variable is the magnitude of the transverse momentum of jets $p^{\rm {jets}}_{\rm T}$ that is defined as the magnitude of the vector sum of the signal jets' transverse momenta. This variable is calculated as
\begin{equation}
    p^{\rm {jets}}_{\rm T} = p\left(\sum^{N}_{i=1}{ {\bf p}^{\rm jets}_i}\right)_{\rm T},
\end{equation}
where $N$ is the number of reconstructed signal jets. We defined $p({\bf X})_{\rm T}$ as the magnitude of the transverse momentum of the vector ${\bf X}$ in the parentheses. The third variable $\Delta\phi(E^{\rm miss}_{\rm T},\rm {jets})$ is the azimuthal angle between the missing transverse momentum vector and the vector sum of the signal jets' momenta. The fourth variable is $R(E^{\rm miss}_{\rm T},\rm {jets})$, the ratio of the missing transverse momentum to the total transverse momenta of jets, is calculated as
\begin{equation}
    R(E^{\rm miss}_{\rm T},{\rm {jets}})\;=\;|\bf p^{\rm miss}_{\rm T}\cdot\bf p^{\rm jets}_{\rm T}|/(p^{\rm jets}_{\rm T}\cdot p^{\rm jets}_{\rm T}),
\end{equation}
where ${\bf p^{\rm miss}_{\rm T}}$ is the missing transverse momentum and ${\bf p^{\rm jets}_{\rm T}}$ is the vector sum of transverse momenta of jets.
The last variable that is used to define the SR-ISR is $p^{\rm {soft}}_{\rm T}$ which is defined as the $p_T$ in the vector sum of the four-momenta of the signal jets, leptons and the missing transverse momentum,
\begin{equation}
    p_{\rm T}^{\rm soft} = p({\bf p^{\rm leptons} + p^{\rm jets} + p^{\rm miss}_{\rm T}})_{\rm T},
\end{equation}
where ${\bf p^{\rm leptons}}$ is the vector sum of the four-momenta of the leptons.

There are also three different eRJR variables that are used to define the SR-LOW region. The first variable is $p^{\rm {soft}}_{\rm T}$ which has the same name as the one introduced in the SR-ISR region. However, it is defined differently in the SR-LOW region due to a jet veto. $p^{\rm {soft}}_{\rm T}$ is defined as the $p_{\rm T}$ of the vector sum of the four-momenta of the signal leptons and the missing transverse momentum,
\begin{equation}
    p_{\rm T}^{\rm soft} = p({\bf p^{\rm leptons} + p^{\rm miss}_{\rm T}})_{\rm T}.
\end{equation}
The second variable is $m^{\rm {3l}}_{\rm eff}$ and it is defined as the scalar sum of the $p_{\rm T}$ of the signal leptons and the missing transverse energy,
\begin{equation}
    m^{\rm {3l}}_{\rm eff} = \left(\sum_{i}^{3} p^{\rm lepton}_{i, \rm T}\right) + E^{\rm miss}_{\rm T},
\end{equation}
where $p^{\rm lepton}_{i, \rm T}$ is the transverse momentum of each lepton
The last variable that is used in the SR-LOW region is $H^{\rm boost}$ that is defined as the scalar sum of the momentum of the signal leptons and the missing-momentum vector after applying a boost to the rest frame of the pair-produced parent sparticles (PP frame),
\begin{equation}
    H^{\rm boost} = \left(\sum p^{\rm PP, lepton}_{i, \rm T}\right) + E^{\rm PP,  miss}_{\rm T},
\end{equation}
where $p^{\rm PP, lepton}_{i, \rm T}$ and $ E^{\rm PP,  miss}_{\rm T}$ is the transverse momentum of each lepton and the missing transverse momentum in the PP frame.
In order to calculate $H^{\rm boost}$, two more variables should be determined. Firstly, the longitudinal component of the missing-momentum vector, $p^{\rm miss}_{\rm ||}$, is calculated as
\begin{equation}
    p^{\rm miss}_{\rm ||} = p_{\rm V,\rm ||}\frac{|\bf{p^{\rm miss}_{\rm T}}|}{\sqrt{({p_{\rm V,\rm T}})^2+(m_{\rm V})^2}},
\end{equation}
where $p_{\rm V,\rm ||}$ is the $z$-component of the vector sum of the four-momenta of the three signal leptons, $p_{\rm V,\rm T}$ is the $p_{\rm T}$ of the vector sum of the four-momenta of three leptons, and $m_{\rm V}$ is the mass of the three-lepton system. Secondly, the boost of the system is calculated as 
\begin{equation}
    \bf\beta = \frac{\bf{p}}{\mbox{E}} = \frac{\bf{p^{\rm V}} + \bf{p^{\rm miss}}}{\mbox{E}^{\rm V} + |\bf p^{\rm miss}|},
\end{equation}
where $\bf p^{\rm V}$ is the vector sum of the three-momenta of three leptons, which is calculated in the laboratory frame.

\paragraph{Preselection\\}

Two different signal regions are defined with different constraints. However, there are some common constraints for both regions. Events are required to feature three signal leptons with at least one pair of leptons with the same-flavor and a different electric charge (SFOS). In addition, the invariant mass $m_{\rm ll}$ of this lepton pair is required to be within $[75 , 105]\;\mbox{GeV}$. If there exists more than one SFOS pair, the pair is chosen so that its invariant mass is the closest to the $Z$ boson mass. Moreover, the invariant mass of the three-lepton system should be larger than $105\;\mbox{GeV}$. In order to reduce the contribution from the top backgrounds, a $b$-jet veto is applied. Finally, to reduce the contribution from the $W$ and $Z$ boson backgrounds, $m_{\rm T}$ should be larger than 100 GeV, where $m_{\rm T}$ is the transverse momentum of the system that is made of the unpaired third lepton and $E^{\rm miss}_{\rm T}$. It is calculated as
\begin{equation}
    m_{\rm T}\,=\,\sqrt{2p_{\rm T}E_{\rm T}^{\rm miss}(1\,-\,\cos{(\Delta \phi)})},
\end{equation}
where $\Delta \phi$ is the azimuthal separation between the unpaired third lepton and the missing transverse momentum.

\paragraph{The SR-LOW low-mass region\\}

In the low-mass region, SR-LOW, there should be no signal jet. To reduce the background from fake or non-prompt leptons, the transverse momenta of three leptons should satisfy
\begin{equation}
    p_{\rm T}^{\rm l_1}\,>\,60\;\mbox{GeV}\,\quad\mbox{,}\quad\,p_{\rm T}^{\rm l_2}\,>\,40\;\mbox{GeV}\,\quad\mbox{and}\quad\,p_{\rm T}^{\rm l_3}\,>\,30\;\mbox{GeV}.
\end{equation}
Moreover, to reduce the contribution from the $W$ and $Z$ boson backgrounds, $H^{\rm boost}$ should be larger than 250 GeV. To further reduce the $WZ$ contribution, $p_{\rm T}^{\rm soft}$ and $m^{\rm 3l}_{\rm eff}$ should satisfy
\begin{equation}
    \frac{p_{\rm T}^{\rm soft}}{p_{\rm T}^{\rm soft}\,+\,m^{\rm 3l}_{\rm eff}}\,<\,0.05\,\quad\mbox{and}\quad\,\frac{m^{\rm 3l}_{\rm eff}}{H^{\rm boost}} > 0.9.
\end{equation}

\paragraph{The SR-ISR initial-state radiation region\\}
In the SR-ISR region, there should be at least one signal jet as this region has been designed to target the production of an electroweakino pair in association with a hard initial-state radiation jet. $p_{\rm T}^{\rm soft}\,<\,25\;\mbox{GeV}$ together with jet multiplicity smaller than four ($N_{\rm jets}<4$) is imposed to reduce the contribution from the WZ backgrounds. The transverse momenta of three leptons should satisfy
\begin{equation}
    p_{\rm T}^{\rm l_1}\,>\,25\;\mbox{GeV}\,\quad\mbox{,}\quad\,p_{\rm T}^{\rm l_2}\,>\,25\;\mbox{GeV}\,\quad\mbox{and}\quad\,p_{\rm T}^{\rm l_3}\,>\,20\;\mbox{GeV}.
\end{equation}
In addition, $E^{\rm miss}_{\rm T} \geq 80 \;\mbox{GeV}$ reduces the contamination from the $Z$+jets events. Moreover, the azimuthal separation between the missing transverse momentum and the vector sum of the momenta of the signal jets, $\Delta\phi(E_{\rm T}^{\rm miss}, {\rm jets})$, should be larger than 2.
To reduce the contribution from the $WZ$ backgrounds, the ratio of the transverse momentum to the total transverse momenta of jets, $R(E_{\rm T}^{\rm miss}, {\rm jets})$, should be within (0.55 , 1.0). Background contamination is further reduced by the requirement of $p^{\rm jets}_{\rm T} > 100 \;\mbox{GeV}$.

\subsection{Validation}
\subsubsection{Event generation}

Our benchmark points are defined by a spectrum featuring $M(\tilde{\chi}^{\pm}_1/\tilde{\chi}^0_2) = 200 \;\mbox{GeV}$, $M(\tilde{\chi}^0_1) = 100\;\mbox{GeV}$ and the masses of the other SUSY particles in the model are set to 5 TeV. In order to validate our analysis, we generated signal samples relevant for a few benchmark scenarios and compared the predictions with the cutflows included in the original paper. Signal samples were generated via MG5\_aMC v2.6.7\cite{Alwall:2014hca} at leading order(LO) with the LO sets of NNPDF2.3 parton densities \cite{Ball:2014uwa}. We used the MSSM-SLHA2\cite{Duhr:2011se} model file shipped with MG5\_aMC.
We generated events by combining samples associated with the $p\ p \to \chi^\pm_1\chi^0_2$, $\chi^\pm_1\chi^0_2+j$ and $\chi^\pm_1\chi^0_2+ 2j$ processes, the $\chi^\pm_1\chi^0_2$ decay to $\chi^0_1W^\pm\chi^0_1Z$ being enforced to occur with a branching ratio of 1.
The decay process were done when simulating parton showering. Parton showering was simulated by {\sc Pythia} 8.244\cite{Sjostrand:2014zea}. For the merging of the multi-partonic matrix elements, we used the MLM technique with the parameters {\tt Xqcut} = 50 GeV in MG5\_aMC and {\tt qCut} = 75 GeV in {\sc Pythia}~8. {\sc Delphes}~3.4.2\cite{deFavereau:2013fsa}  was used to emulate the ATLAS detector. The isolation and unique identification modules in {\sc Delphes}~3 were not used and those were done at the analysis level. For jet clustering, 
we used {\sc FastJet} \cite{Cacciari:2011ma} and its implementation of the anti-$k_t$ algorithm with a radius parameter R = 0.4. The b-tagging efficiency input in {\sc Delphes} was provided as a function of $p_{\rm T}$ extracted from the data collected from 2015 to 2017\cite{Aad:2019aic}.

\subsubsection{Comparison with the official results}
\begin{table}[t]
  \setlength\tabcolsep{10pt}
  \renewcommand{\arraystretch}{1.25}
  \tbl{Comparison of the cutflow predicted by MA5 with the results provided by the ATLAS collaboration for the SR-LOW region.}
  {\begin{tabular}{p{4.0cm}|p{2.5cm}p{2.3cm}p{2.3cm}}\toprule
    Cuts  & ATLAS(Official)&MA5 & difference ($\Delta$)\\
    \hline
    3 leptons \& SFOS&-&-&-\\
    $b$-jet veto&0.963&0.992&3.0\%\\
    $m_{\rm 3l}\,>\,105\;\mbox{GeV}$&0.970&0.959&1.1\%\\
    $\mbox{lepton } p_{\rm T}\,>\,60,40,30\;\mbox{GeV}$&0.352&0.301&14.5\%\\
    $m_{\rm ll}\,\in\,[75,105]\;\mbox{GeV}$&0.982&0.985&0.3\%\\
    jet veto&0.485&0.564&16.3\%\\
    $H^{\rm boost}\,>\,250\;\mbox{GeV}$&0.712&0.724&1.7\%\\
    ${p^{\rm soft}_{\rm T}}/{(p^{\rm soft}_{\rm T} + m^{\rm 3l}_{\rm eff})}\,<\,0.05$&0.712&0.590&17.1\%\\
    ${m^{\rm 3l}_{\rm eff}}/{H^{\rm boost}}\,>\,0.9$&0.651&0.595&8.6\%\\
    $m_{\rm T}\,>\,100$&0.392&0.356&9.2\%\\
    \multicolumn{4}{c}{}
    \botrule
    \end{tabular}\label{SR-LOW} }
    \end{table}

\begin{table}[t]
  \setlength\tabcolsep{10pt}
  \renewcommand{\arraystretch}{1.25}
    \tbl{Comparison of the cutflow predicted by MA5 with the results provided by the ATLAS collaboration for the SR-ISR region. }
    {\begin{tabular}{p{4cm}p{2.5cm}p{2.3cm}p{2.3cm}}
    \toprule
    Cuts & ATLAS(Official)&MA5 & difference ($\Delta$)\\
    \hline
    3 leptons \& SFOS&-&-&-\\
    $b$-jet veto&0.963&0.992&3.0\%\\
    $m_{\rm 3l}\,>\,105\;\mbox{GeV}$&0.970&0.959&1.1\%\\
    $\mbox{lepton} p_{\rm T}\,>\,25,25,20\;\mbox{GeV}$&0.800&0.749&6.4\%\\
    $m_{\rm ll}\,\in\,[75,105]\;\mbox{GeV}$&0.977&0.976&0.1\%\\
    $N_{\rm jets}\,\in\,[1,3]$&0.467&0.408&12.6\%\\
    $|\Delta \phi(E_{\rm T}^{\rm miss}, {\rm jets})|\,>\,2.0$&0.672&0.671&0.1\%\\
    $R(E_{\rm T}^{\rm miss},{\rm jets})\,\in\,[0.55,1.0]$&0.331&0.355&7.3\%\\
    $p_{\rm T}^{\rm jets}\,>\,100\;\mbox{GeV}$&0.551&0.509&7.6\%\\
    $E_{\rm T}^{\rm miss}\,>\,80\;\mbox{GeV}$&0.956&0.965&0.9\%\\
    $m_{\rm T}\,>\,100\;\mbox{GeV}$&0.425&0.505&18.8\%\\
    $p_{\rm T}^{\rm soft}\,<\,25\;\mbox{GeV}$&0.764&0.696&8.9\%\\
    \multicolumn{4}{c}{}
    \botrule
    \end{tabular}\label{SR-ISR} }
    \end{table}

Table \ref{SR-LOW} and Table \ref{SR-ISR} compares {\sc MadAnalysis~5} (MA5) cutflow predictions with the ATLAS official results for two signal regions. The relative difference ($\Delta$) between the results from MA5 and the ATLAS ones is computed as,
\begin{equation}
    \left|\Delta\right| = \left|{1 - \frac{\epsilon^{{\rm MA5}}_i}{\epsilon^{{\rm ATLAS}}_i}}\right|\times100.
\end{equation}
The index $i$ corresponds to the cut number, and $\epsilon_i^{\rm MA5}$ and $\epsilon_i^{\rm ATLAS}$ refers to the corresponding efficiencies,
\begin{equation}
    \epsilon_i = \frac{({\rm Cut})_i}{({\rm Cut})_{i-1}},
\end{equation}
where $({\rm Cut})_i$ is the number of events remaining after applying the $i^{th}$ cut. For both signal regions, we observed an agreement of order of up to 10\% at every step of each cutflow. In the case of the SR-LOW region, the largest observed discrepancy is related to $p^{\rm soft}_{\rm T}$, which was 17.1\%. For the case of the SR-ISR region, the largest discrepancy is related to $m_{\rm T}$, which was 18.8\%.
For both regions, large discrepancies are likely to start from the cutflow that strongly rely on jets. That is, the large discrepancies tend to start, in the case of the SR-LOW region, at the jet veto level and in the case of the SR-ISR region, at the cut on $N_{\rm jets}$. Those discrepancies are thought to be related to the lack of exact information about jet energy scale and jet efficiency that is needed when reconstructing jets in the {\sc Delphes}~3 level. We could therefore not further investigate the reasons for the differences.

\begin{figure}[h!]
    \centering
    \includegraphics[width = \textwidth]{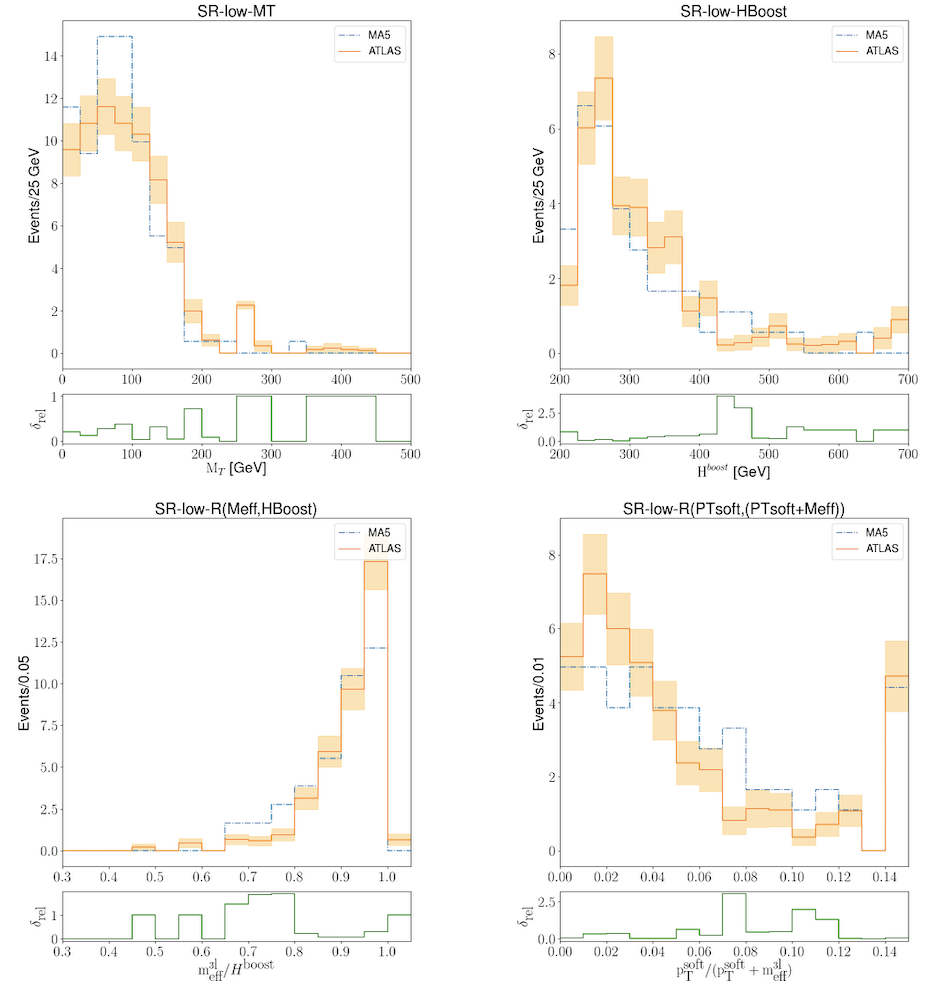}
    \caption{Comparison of several differential distributions as predicted by {\sc MadAnalysis~5} with the official results provided by the ATLAS collaboration, for the SR-LOW region. The official results provided by the ATLAS collaboration are shown with their uncertainties. The SR-LOW event selections are applied for each distribution except for the variable shown.}
    \label{sr-low}
\end{figure}

\begin{figure}[h!]
    \centering
    \includegraphics[width = \textwidth]{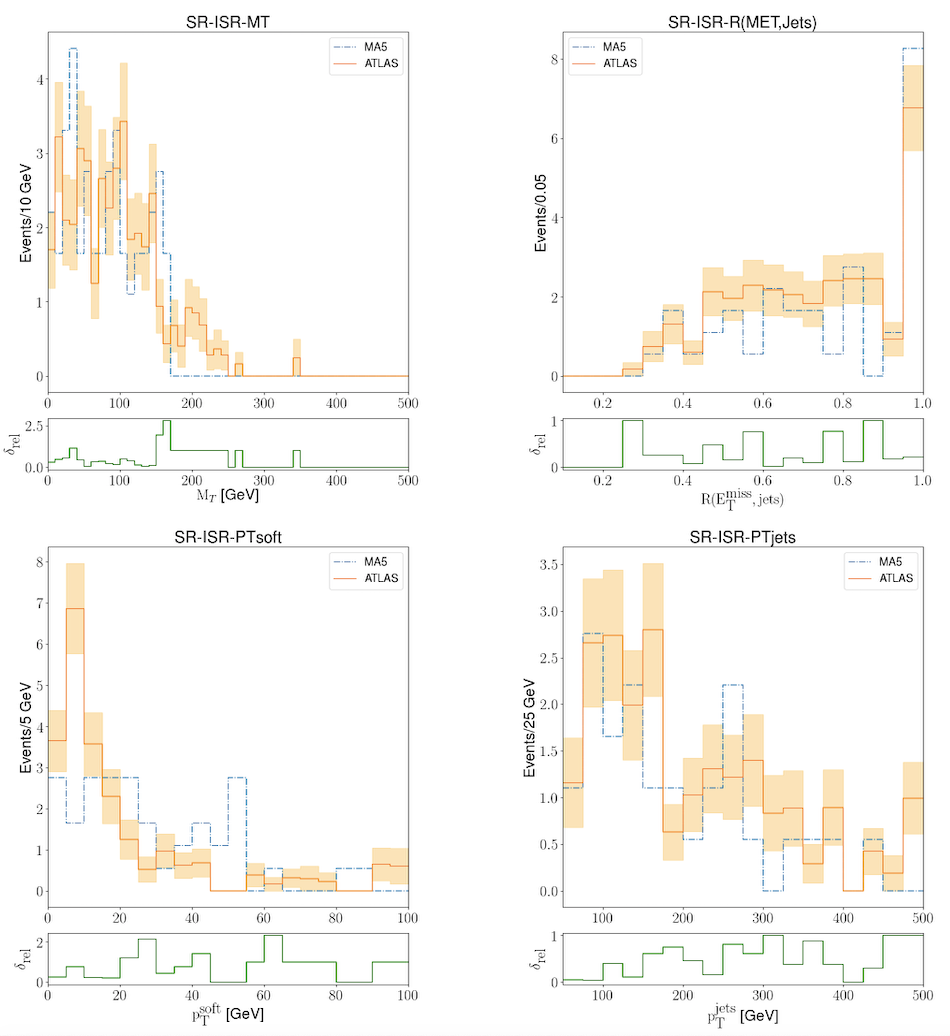}
    \caption{Comparison of several differential distributions as predicted by {\sc MadAnalysis~5} with the official results provided by the ATLAS collaboration, for the SR-ISR region. The official results provided by the ATLAS collaboration are shown with their uncertainties. The SR-ISR event selections are applied for each distribution except for the variable shown.}
    \label{sr-isr}
\end{figure}

Figure \ref{sr-low} and Figure \ref{sr-isr} compare the distribution of variables that is predicted by {\sc MadAnalysis~5} with official data \cite{1771533} that is provided by the ATLAS collaboration for the two regions (SR-LOW and SR-ISR). The dotted lines represents the ATLAS official results and the solid lines are the {\sc MadAnalysis~5} predictions. To be specific, Figure \ref{sr-low} compares the distribution of variables in the SR-LOW region: $m_{\rm T}$, $H^{\rm boost}$, $m^{\rm 3l}_{\rm eff}/H^{\rm boost}$ and $p^{\rm soft}_{\rm T}/(p^{\rm soft}_{\rm T} + m^{\rm 3l}_{\rm eff})$. The entire SR-LOW event selection is applied for each distribution with the exception of the variable shown. Similarly, Figure \ref{sr-isr} compares distributions in the SR-ISR region: $m_{\rm T}$, $R(E_{\rm T}^{\rm miss},{\rm jets})$, $p^{\rm soft}_{\rm T}$ and $p^{\rm jets}_{\rm T}$. The entire SR-ISR event selection is applied for each distribution of variables with the exception of the shown variable. The remaining events after applying the event selections were quite small, which caused large differences at some point. However, the shape of the histograms matched fairly well with the data provided by the ATLAS collaboration, especially for the SR-LOW region. In the case of the SR-ISR region, large discrepancies were again observed for the variables that strongly rely on jets. The limitation of information about jet reconstruction is thought to be an issue, which is consistent with the results of the cutflow. However, considering the large uncertainties of the data that is provided by the ATLAS collaboration and the small number of events remaining after the cutflow, our MA5 data could be said to fit with the data provided by the ATLAS collaboration fairly well.

The figures also include the relative difference ($\delta_{\rm rel}$) between the MA5 and the ATLAS predictions, that is calculated by using
\begin{equation}
    \delta_{\rm rel}\;=\;\left|\frac{N^{\rm ATLAS}-N^{\rm MA5}}{N^{\rm ATLAS}}\right|.
\end{equation}
For the points where $N^{\rm ATLAS}\;=\;0$, we let $\delta_{\rm rel}\;=\;0$ for convenience. $N^{\rm ATLAS}$ and $N^{\rm MA5}$ refers to the number of events in the ATLAS and the MA5 sequentially. \\

\subsection{Conclusions}
The {\sc MadAnalysis~5} implementation of the ATLAS-SUSY-2018-06 analysis, a search for the production of electroweakinos with the three-lepton final state has been presented. The ATLAS collaboration found excesses of three-lepton events in two signal regions, the low-mass region SR-LOW, and the initial-state radiation region SR-ISR. Validation was done by comparing the efficiency of various selection cuts that has been provided by the ATLAS collaboration with the result predicted by the MA5 framework. We found that the difference between the two was quite acceptable with differences at the level of at most 20\%. Thus, we concluded that the analysis reimplementation is validated. The material that has been used for the validation of this implementation is available, together with the {\sc MadAnalysis}~5 C++ code, at the MA5 dataverse (\href{https://doi.org/10.14428/DVN/LYQMUJ}{https://doi.org/10.14428/DVN/LYQMUJ})~\cite{LYQMUJ_2020}.

\subsection*{Acknowledgments}
We thank all the chairpersons for their help. Particularly, special thanks to Benjamin Fuks, Jack Araz for fruitful discussion, as well as Eric Conte, Dipan Sengupta, and Si Hyun Jeon for their help.

\cleardoublepage
\renewcommand{\madanalysis}{{\sc MadAnalysis~5}}
\newcommand{\sfsim}{{\sc simplified--Fast Simulation}}
\newcommand{\sfs}{{\sc SFS}}
\newcommand{\delphes}{{\sc Delphes~3}}
\newcommand{\madgraph}{{\sc MG5\_aMC}}
\newcommand{\pythia}{{\sc Pythia}}
\newcommand{\fastjet}{{\sc FastJet}}

\markboth{Jack Y. Araz and Benjamin Fuks}
  {Implementation of the ATLAS-SUSY-2018-31 analysis}

\section{Implementation of the ATLAS-SUSY-2018-31 analysis (sbottoms with multi-bottoms
  and missing transverse energy; 139~fb$^{-1}$)}
  \vspace*{-.1cm}\footnotesize{\hspace{.5cm}By Jack~Y.~Araz and Benjamin~Fuks}
\label{sec:sbottom}


\subsection{Introduction}
The popularity of supersymmetry (SUSY) mostly arises as it provides, by
extending the Poincar\'e algebra and by linking the fermionic
and bosonic content of the theory, an elegant solution to the hierarchy problem
inherent to the Standard Model (SM). In the Minimal Supersymmetric Standard Model
(MSSM)~\cite{Nilles:1983ge,Haber:1984rc}, each of the SM degree of freedom is
associated with a supersymmetric partner. After the breaking of the electroweak
symmetry, the partners of the gauge and Higgs fields mix into four neutralino
($\tilde{\chi}^0_{1,2,3,4}$) and two chargino ($\tilde{\chi}^\pm_{1,2}$) mass
eigenstates, the lightest neutralino being often taken as a viable candidate for
dark matter. In order for the MSSM to consist of a solution for the hierarchy
problem, the supersymmetric partners of the top and bottom quarks are in general
required to be quite light, so that their quadratic contributions to
the quantum corrections to the Higgs boson mass stay under
control~\cite{Witten:1981nf,Barbieri:1987fn}.
They have thus the possibility to be copiously pair-produced at the LHC.

The ATLAS-SUSY-2018-31 analysis~\cite{Aad:2019pfy} has been designed to
investigate the possibility of such light sbottoms, and probes multi-bottom final
states additionally featuring a large amount of missing transverse energy. The
signature under consideration could arise from sbottom pair production followed
by $\tilde{b}\to\tilde{\chi}^{0}_2 b$ decays, where the second neutralino
further decays into an SM Higgs boson and a lightest neutralino,
\begin{equation}
  p p \to \tilde b \tilde b^* \to (\tilde{\chi}^{0}_2 b)  (\tilde{\chi}^{0}_2 \bar b)
  \to (\tilde{\chi}^{0}_1 h b) (\tilde{\chi}^{0}_1 h \bar b)\ .
\label{eq:proc}\end{equation}
A representative Feynman diagram for the above process is shown in
figure~\ref{fig:09diagrams}. Such a decay pattern is predicted to be
enhanced in MSSM
scenarios in which the lightest state $\tilde{\chi}^{0}_1$ is bino-like and the 
heavier state $\tilde{\chi}^{0}_2$ is wino-like, the $\tilde b\to
\tilde{\chi}^{0}_1 b$ and $\tilde b\to \tilde{\chi}^-_1 t$ decays being in
this case suppressed.
The kinematics of the final-state objects largely
depend on the mass spectrum of the various involved particles. Whilst a rather
split spectrum gives rise to high-$p_T$ $b$-jets, a compressed spectrum leads,
on the other hand, to relatively soft $b$-jets.

The ATLAS-SUSY-2018-31 analysis has been divided into three main signal
regions SRA, SRB and SRC. Region SRA is dedicated to final states including at least
four hard $b$-jets arising both from Higgs-boson and sbottom decays. Region SRB
aims to track relatively softer $b$-jets which are originating from sbottom
decays, together with a harder leading jet dawned from initial-state radiation.
Finally, region SRC targets a topology with softer $b$-jets that are all
well separated from the missing energy, which gives rise to a sizeable
missing energy significance. This analysis has been found to constrain sbottom
masses ranging up to 1.5 TeV in the corresponding simplified models.

In the rest of this note, we present the recast of the ATLAS-SUSY-2018-31
analysis of Ref.~\cite{Aad:2019pfy} in the \madanalysis/\sfs\
framework~\cite{Araz:2020lnp} that relies on smearing and efficiency functions
for the simulation of the detector response. The code of our implementation is
available from the \madanalysis\ dataverse~\cite{IHALED_2020}, and
the Monte Carlo cards relevant for the validation have been obtained from
\href{https://www.hepdata.net/record/ins1748602}{HEPData}\cite{1748602}.

\subsection{Description of the analysis}
\begin{figure}[t]
\centering \includegraphics[scale=.3]{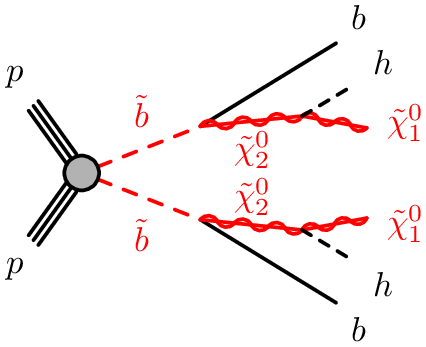}
\caption{Representative Feynman diagram for sbottom pair production, followed by sbottom cascade
  decays through the second neutralino. The figure has been retrieved
from ref.~\cite{Aad:2019pfy}.}\label{fig:09diagrams}
\end{figure}

The considered analysis focuses on a signature made of multiple $b$-jets and
missing transverse energy, which could originate from sbottom pair-production
and cascade decays. As mentioned above, the results are interpreted in three
classes of simplified models, two of which being relevant for compressed mass
spectra and the third one being representative of split spectra. The
topology in question is illustrated by figure~\ref{fig:09diagrams} and by
eq.~\eqref{eq:proc}.

\subsubsection{Object definitions}
Jets are obtained by clustering all final-state objects of a given event, with
the exception of muons (and invisible particles as the latter are present in
typical Monte Carlo event records). Electrons and photons being
included, an overlap removal procedure is in order. This is detailed
below. Jet clustering relies on the anti-$k_T$ algorithm with a
radius parameter set to $R=0.4$~\cite{Cacciari:2008gp}, that is used within the
\fastjet\ package version~3.3.3~\cite{Cacciari:2011ma}.

Two types of jets are
considered in this analysis. First, baseline jets are enforced to have a
transverse momentum $p_T>20$~GeV, and a pseudorapidity satisfying $|\eta|<4.8$.
Signal jets are obtained from this collection, after the object removal
procedure described below.

Hadronic taus are extracted from the full jet collection, analysis-level tau
candidates having a transverse momentum $p_T>2.5$~GeV and a pseudorapidity
fulfiling $ |\eta|<2.5$. The tau-tagging performance are taken from
ref.~\cite{ATLAS:2017mpa}, the tagging efficiency being in average of 60\% for
a mistagging rate of light jets as hadronic taus of 1\%.

The initial sets of leptons are those electrons and muons that are
reconstructed from final-state objects after imposing loose electron
identification requirements~\cite{Aad:2019tso}, and medium muon identification
requirements~\cite{Aad:2016jkr}. Baseline electrons (muons) are then defined
from these collections after enforcing that their transverse momentum obeys
$p_T>4.5\ (4)$~GeV, and their pseudorapidity $|\eta|<2.47\ (2.5) $.

In order to clean the jet and lepton collections, first jets are discarded if a
baseline electron is found within a distance, in the transverse plane, of
$\Delta R \leq 0.2$. Furthermore, baseline electrons and muons are removed if
they are found within a distance $\Delta R \leq 0.4$ of a jet.

Signal light jets and $b$-jets are chosen among the set of cleaned jets. Signal
jets are enforced to have a transverse momentum $p_T>30$~GeV and a
pseudorapidity $|\eta|<2.8$. Any $b$-tagged jet within this set, with
$|\eta|<2.5$, is considered as an element of the signal $b$-jet collection.
The analysis considers a $b$-tagging working point involving an efficiency of
77\%, for corresponding misidentification rates of 20\% for $c$-jets and 0.9\%
for lighter-flavour jets~\cite{Aad:2019aic}. In our simulations, we enforce the
$b$-tagging algorithm to be based on the presence of a true $B$-hadron in a cone
of radius $\Delta R=0.3$ around the jet.

Finally, the $H_T$ variable is defined as the scalar sum of the $p_T$ of all
jets belonging to the signal jet collection, and the missing momentum vector is
defined as the negative vector sum of the $p_T$ of all visible objects.

\subsubsection{Event selection}
The ATLAS-SUSY-2018-31 analysis includes three main non-orthogonal
signal regions denoted SRA, SRB and SRC.
Each signal region is dedicated to a specific configuration of the
final-state multi-bottom system. Region SRA has been designed to probe quite hard
$b$-jets, typical of a split sbottom-neutralino mass spectrum. Region SRB
focuses on softer $b$-jets as arising from a more compressed mass spectrum,
when they are produced in association with a hard initial-state radiation.
Finally, region SRC is also dedicated to softer $b$-jets, but this time when their
properties include a large missing energy significance.

All regions include a lepton veto, so that events featuring any baseline
electron or muon are rejected. Moreover, one requires that events populating the
regions SRA, SRB and SRC feature a number of jets $N_j\geq 6$, 5 and 4, respectively.
While the selection for regions SRA and SRB asks for the presence of a number of
$b$-jets $N_b\geq 4$, imposes a tau veto and requires a missing transverse energy
$\slashed{E}_T$ greater than
350~GeV, the one for region SRC only asks for at least three $b$-jets with a missing
transverse energy greater than 250~GeV. The preselection finally ends with the
requirement that the four leading jets are separated in azimuth from the
missing-momentum vector by at least 0.4, in all regions,
min$[\Delta\phi(j_i,\slashed{E}_T)] >0.4$ with $i=1, 2, 3, 4$.

After this preselection, the region SRC is split into several bins of
object-based missing transverse energy significance, which we define by
$\slashed{E}_T/\sqrt{H_T}$.

\begin{table}[t]
 \renewcommand{\arraystretch}{1.6}
 \setlength\tabcolsep{10pt}
 \tbl{Schematic representation of the cut-flows associated with the three
   classes of signal regions of the ATLAS-SUSY-2018-31 analysis.}
  {\begin{tabular}{@{}c c c@{}} \toprule
     SRA & SRB & SRC\\\hline
     Lepton veto & Lepton veto & Lepton veto\\
     $N_j\geq 6$ & $N_j\geq 5$ & $N_j\geq 4$ \\
     $N_b \geq 4$ & $N_b \geq 4$ & $N_b\geq 3$ \\
     $\slashed{E}_T > 350$~GeV &$\slashed{E}_T > 350$~GeV & $\slashed{E}_T> 250$ GeV\\
     min$[\Delta\phi(j_{1-4},\slashed{E}_T)] >0.4$ & min$[\Delta\phi(j_{1-4},\slashed{E}_T)] >0.4$
       &min$[\Delta\phi(j_{1-4},\slashed{E}_T)] >0.4$\\
     tau veto & tau veto & $\slashed{E}_T/\sqrt{H_T}$ bins\\
     $p_T(b_1) > 200$ GeV & $\bar m_h^{\rm reco}\in[75,175]$ GeV & \\
     $\Delta R_{max}(b,b)>2.5$  & Leading jet non-$b$-tagged  & \\
     $\Delta R_{max-min}(b,b)<2.5$ & $p_T(j_1) > 350$ GeV  & \\
     $m_h^{\rm reco}>80$ GeV  & $\Delta\phi(j_{1},\slashed{E}_T)>2.8$ & \\
      $ m_{\rm eff}  $ bins &$m_{\rm eff} > 1$ TeV &  \\
\botrule\end{tabular}\label{tab09:cut-flow}}
\end{table}

Before being further subdivided into various regions, region SRA selects
events featuring a leading $b$-jet with a $p_T$ of at least
200~GeV, and for which the maximal angular separation between any pair of two
$b$-jets obeys $\Delta R_{max}(b,b)>2.5$. This defines the pair of $b$-jets
originating from the bottom squark decays. Considering the collection made of
all the remaining $b$-jets, one restricts the minimal angular separation in the
transverse plane between any pair made of those $b$-jets to satisfy
$\Delta R_{max-min}(b,b)<2.5$. This tags the $b$-jets that are considered to
originate from a Higgs-boson decay, the corresponding invariant mass
$m_h^{\rm reco}$ being enforced to be larger than 80~GeV. The signal region
is finally divided into several bins in the effective mass
$m_{\rm eff} = \slashed{E}_T+H_T$.

In contrast to the two other sets of regions, region SRB is not further
subdivided. It however includes extra cuts. First, the $b$-jets are organised to
form di-jet systems compatible with the decay of the SM Higgs boson. The first
Higgs candidate is defined by the pair of $b$-jets featuring the largest
separation $\Delta R$ in the transverse plane. Next, the second Higgs candidate
is similarly defined from the set of remaining $b$-jets. The average
invariant mass of those two Higgs candidates $\bar m_h^{\rm reco}$ is then
imposed to
lie in the $[75, 175]$~GeV range. The leading jet is moreover required not to be
$b$-tagged, and to be consistent with a very hard initial-state radiation. Its
$p_T$ is hence constrained to be greater than 350~GeV, and this jet has to lie
at an azimuthal distance $\Delta\phi(j_{1},\slashed{E}_T)>2.8$ from the missing
momentum vector.
Finally, an effective mass of at least 1 TeV is required.

A schematic representation of the definition of all signal regions is shown in
table~\ref{tab09:cut-flow}.

\subsection{Validation}

\subsubsection{Event generation}\label{sec:event_gen}
In order to validate our implementation, we consider three scenarios, each of
them featuring a spectrum with a different mass splitting and being thus relevant
for the validation of a specific class of signal regions. For the signal regions
of type SRA, a largely split benchmark point has been used, with masses
$m(\tilde{b},\tilde\chi^0_2,\tilde\chi^0_1)=(1100,330,200)$~GeV. A more
compressed spectrum has been choosen for the validation of the implementation of
the single signal region SRB, with
$m(\tilde{b},\tilde\chi^0_2,\tilde\chi^0_1)=(700,680,550)$ GeV. Finally, for the
class of SRC signal regions, we use a spectrum defined by
$m(\tilde{b},\tilde\chi^0_2,\tilde\chi^0_1)=(1200,1150,60)$ GeV. All SLHA mass
spectrum files have been taken from information publicly available from
\href{https://www.hepdata.net/record/ins1748602}{HEPData records} that are
dedicated to this analysis and that have been
provided by the ATLAS collaboration\cite{1748602}.

For our validation, we generate leading-order (LO) event samples for all these
benchmark scenarios with \madgraph\ version 2.7.3~\cite{Alwall:2014bza},
convoluting LO matrix elements with the LO set of NNPDF 2.3 parton
distribution functions~\cite{Ball:2013hta} as driven by the
LHAPDF~6 library~\cite{Buckley:2014ana}. Following the
Multi-Leg Merging (MLM)
prescription~\cite{Mangano:2006rw,Alwall:2008qv}, our samples combine matrix
elements describing sbottom pair
production in association with up to two extra partons, the merging scale being
set to $Q_{\rm match}=m({\tilde b})/4$. Particle decays, parton showering and
hadronisation are dealt by means of \pythia\ version
8.2~\cite{Sjostrand:2014zea}, and the simulation of the response of the ATLAS
detector has been achieved with the \sfs\ module of
\madanalysis~\cite{Araz:2020lnp}. Our recast can then be used with
\madanalysis\ version 1.9.4\footnote{The implementation relies on jet
energy scale corrections, which have been implemented in \madanalysis\ from
version 1.9.4.} (or more recent). All analysis files can be obtained from the
\madanalysis\ dataverse~\cite{IHALED_2020}.
As in the ATLAS-SUSY-2018-31 publication, we normalise our Monte
Carlo samples to cross sections evaluated at the next-to-leading-order in
perturbative QCD after matching with threshold resummation at the
next-to-leading logarithmic accuracy~\cite{Fuks:2012qx,Fuks:2013vua}. The
employed cross section values rely on the PDF4LHC15\_mc parton distribution
functions~\cite{Butterworth:2015oua}.

\subsubsection{Comparison with the official results}

In this section, we compare our predictions with the official ATLAS results.
Although the different signal regions of the considered analysis
overlap, the ATLAS collaboration provides validation material for single
regions. This allows us to handle the validation procedure region by region.
The quality of our implementation is quantified via the parameter $\delta$
defined by
\begin{eqnarray}
\delta_i = \frac{|\varepsilon^{\rm ATLAS}_i - \varepsilon^{\rm MA5}_i |}{\varepsilon^{\rm ATLAS}_i }\ .
\label{eq:delta}\end{eqnarray}
In this expression, $\varepsilon_i=N_i/N_{i-1}$ represents the relative
selection efficiency of the $i^{\rm th}$ cut, with $N_i$ and $N_{i-1}$ being the
number of events surviving this cut and the previous one respectively. This
parameter is required to satisfy $\delta\lesssim
20\%$ for each cut. Such a level of agreement, that is somewhat
arbitrary, is known to only midly impact any limit on a new physics state,
due to the steeply falling nature of the cross section with the new physics
masses~\cite{Dumont:2014tja}. Moreover, a difference of this order
is nevertheless expected from the different detector modeling the simulation
chain used in our recast and in the non-public ATLAS software.

Moreover, it is also important to include a measure of uncertainties pertained to
the Monte Carlo (MC) event generation process. To quantify this, we
define $\Delta_{MC}$ as 
\begin{eqnarray}
\Delta_{\rm MC} = N_i\sqrt{\frac{1-\epsilon_i}{N^{\rm MC}_i}}\ , 
\label{eq:delMC}\end{eqnarray}
where $N^{\rm MC}_i$ is defined as the number of MC events surviving the
$i^{\rm th}$ cut, and $\epsilon_i$ stands for the cumulative selection
efficiency of the $i^{\rm th}$ cut. Here, we emphasise that $N_i$ refers to the
number of events surviving the $i^{\rm th}$ cut, after including the signal
production cross section and the luminosity under consideration.
We aim to constrain $\Delta_{\rm MC} < 10\%$,
as this is comparable with the largest MC uncertainty associated with the
published ATLAS results.

\begin{table}[t]
 \renewcommand{\arraystretch}{1.5}
 \setlength\tabcolsep{12pt}
  \tbl{Cut-flow associated with the SRA class of signal regions. The common SRA
    selection is shown in the first panel of the table, whilst the other panels
    are dedicated to various $m_{\rm eff}$ bins. For each cut, we present the
    level of deviation between our predictions and the official ATLAS results
    $\delta$ defined by eq.~\eqref{eq:delta}. We also report for each
    $m_{\rm eff}$ bin the corresponding MC uncertainty $\Delta_{\rm MC}$ of
    eq,~\eqref{eq:delMC}. We consider a benchmark scenario defined by the mass
    spectrum $m(\tilde{b},\tilde\chi^0_2,\tilde\chi^0_1)=(1100,330,200)$~GeV.
    The ATLAS results correspond to 9,265 MC events prior to any cut,
    whereas our predictions rely on 165,806 events.}
	{\begin{tabular}{@{}lccccc@{}} \toprule
			& \multicolumn{2}{c}{ATLAS} & \multicolumn{3}{c}{\madanalysis--\sfs}\\
			& Events & $\varepsilon$ & Events & $\varepsilon$ & $\delta$ [\%]\\ \hline
			Geneator-level                          & 319.7 & -  & 319.7 & - & - \\
			Initial                                 & 319.7 & -  & 319.7 & - & - \\
			Lepton veto                             & 230.5 & 0.721 & 216.7 & 0.678 & 6.0 \\
			$N_{j} \geq 6$                          & 192.3 & 0.834 & 188.9 & 0.871 & 4.5 \\
			$N_{b} \geq 4$                          & 87.9 & 0.457 & 88.7 & 0.470 & 2.7 \\
			$\slashed{E}_T > 350$ GeV             & 45.1 & 0.513 & 49.7 & 0.560 & 9.1 \\
			min$[\Delta\phi(j_{1-4},\slashed{E}_T)]>0.4$ & 20.9 & 0.463 & 22.8 & 0.459 & 0.9 \\
			Tau veto                           & 19.3 & 0.923 & 21.7 & 0.953 & 3.2 \\
			$p_T(b_1) > 200$ GeV                 & 18.2 & 0.943 & 20.7 & 0.950 & 0.8 \\
			$\Delta R_{max}(b,b)>2.5$               & 17.6 & 0.967 & 20.1 & 0.975 & 0.9 \\
			$\Delta R_{max-min}(b,b)<2.5$           & 15.0 & 0.852 & 19.1 & 0.950 & 11.5 \\
			$m_h^{\rm reco}>80$ GeV                  & 13.7 & 0.913 & 16.1 & 0.839 & 8.2 \\\hline
			$m_{\rm eff} > 1$ TeV                     & 13.7 & 1.000 & 16.1 & 1.000 & 0.0 \\
			$\Delta_{\rm MC}/N_{\rm yield} $ & \multicolumn{2}{c}{$ 5.1\% $} & \multicolumn{3}{c}{$ 1.2\% $} \\\hline
			$m_{\rm eff} \in [1,1.5]$ TeV             & 0.4 & 0.029 & 0.5 & 0.030 & 2.4 \\ 
			$\Delta_{\rm MC}/N_{\rm yield} $ & \multicolumn{2}{c}{$ 28.9\% $} & \multicolumn{3}{c}{$ 6.3\% $} \\\hline
			$m_{\rm eff} \in [1.5,2]$ TeV             & 6.4 & 0.467 & 5.5 & 0.344 & 26.3 \\
			$\Delta_{\rm MC}/N_{\rm yield} $ & \multicolumn{2}{c}{$ 7.8\% $} & \multicolumn{3}{c}{$ 1.8\% $} \\\hline
			$m_{\rm eff} > 2$ TeV                     & 7.0 & 0.511 & 10.0 & 0.626 & 22.5 \\
			$\Delta_{\rm MC}/N_{\rm yield} $ & \multicolumn{2}{c}{$ 7.1\% $} & \multicolumn{3}{c}{$ 1.0\% $} \\
			\botrule\end{tabular}}\label{tab:sra}
\end{table}

\begin{figure}[t]
  \centering
  \includegraphics[scale=0.35]{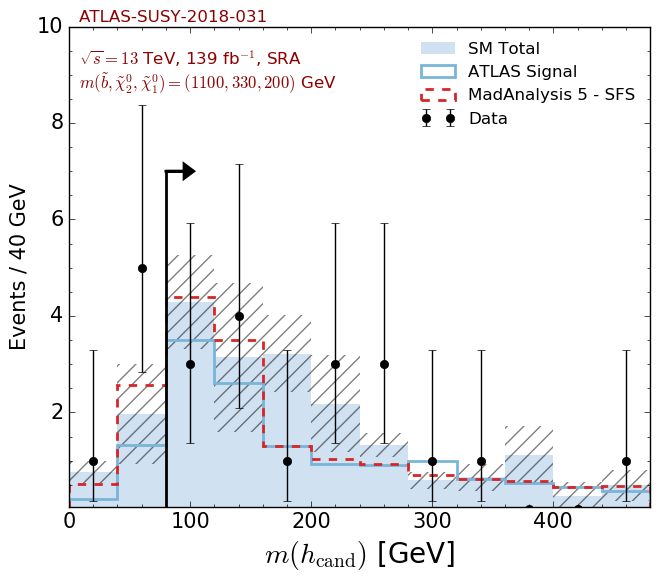}\quad
  \includegraphics[scale=0.35]{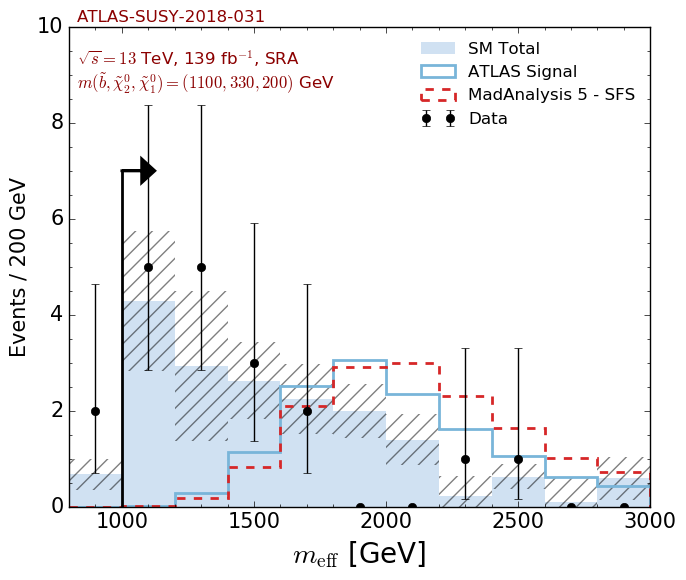}
  \caption{Histograms representative of the SRA cut-flow. We consider the
    distribution in the invariant mass of the reconstructed Higgs boson
    candidate (left) and in the effective mass (right), for a spectrum defined
    by $m(\tilde{b},\tilde\chi^0_2,\tilde\chi^0_1)=(1100,330,200)$ GeV. The
    dashed red line refers to the \madanalysis\ predictions, and the blue solid
    line to the ATLAS official results. As a reference, we show through solid
    blue bars matched with hatched areas the expected SM background and the
    related uncertainties, as well as the results emerging from data
    (black dots).\label{fig:sra}}
\end{figure}

Our validation results are presented in a twofold way. First, we consider
cut-flow tables for all scenarios and the signal region to which they are
dedicated. Second, we present two histograms for each signal region, in
which we compare differential distributions that are critical for each region.
In our comparison, we use official results and digitised histogram information
obtained from
\href{https://www.hepdata.net/record/ins1748602}{HEPData}\cite{1748602}.
All cut-flow tables comprise two main columns, one for the ATLAS predictions and
one for the \madanalysis\ ones. We provide the event counts surviving each cut,
together with the relative cut efficiency ($\varepsilon$) and the $\delta$
difference between ATLAS and \madanalysis\ predictions. We also indicate the
MC uncertainties after all requirements (for each bin in the case of the
SRA and SRC regions). All tables have been prepared with the
\texttt{ma5\_expert} package~\cite{ma5_expert}.

In table \ref{tab:sra} and figure~\ref{fig:sra}, we consider the benchmark
scenario that is defined by the mass spectrum
$m(\tilde{b},\tilde\chi^0_2,\tilde\chi^0_1)=(1100,330,200)$~GeV and that probes
the SRA class of regions. Investigating the cut-flow chart presented in the
table, we observe a generally good agreement between our predictions and the
official ATLAS results. The largest variations arise in the third and fourth
bins in the effective mass $m_{\rm eff}$, with $\delta$ deviations reaching 26\%
and 23\% respectively, for an ATLAS MC uncertainty $\Delta_{\rm MC}$ of about
7\%--8\%. We additionally observe a large MC uncertainty associated with the
ATLAS predictions for the second $m_{\rm eff}$ bin, with $\Delta_{\rm MC}=29\%$,
that is much larger than the difference between the ATLAS and
\madanalysis\ predictions. This large MC uncertainty is also reflected through
shifts in the $m_{\rm eff}$ distributions presented in figure~\ref{fig:sra}
(right panel). In the left panel of the figure, we moreover show the invariant
mass of the reconstructed Higgs boson candidate for which the ATLAS and
\madanalysis\ numbers agree to a good level.
In general, we thus observe a good agreement between the ATLAS and \madanalysis\
predictions, both at the cut-flow and differential distribution levels, after
accounting for the sometimes quite large Monte Carlo uncertainties associated
with the public ATLAS results.

\begin{table}[t]
 \renewcommand{\arraystretch}{1.5}
 \setlength\tabcolsep{12pt}
  \tbl{Same as in table~\ref{tab:sra} but for the signal region SRB and a
    spectrum defined by
    $m(\tilde{b},\tilde\chi^0_2,\tilde\chi^0_1)=(700,680,550)$~GeV.
    The ATLAS results correspond to 3,527 MC events prior to any cut,
    whereas our predictions rely on 149,019 events.}
  {\begin{tabular}{@{}l cc ccc@{}} \toprule
			& \multicolumn{2}{c}{ATLAS} & \multicolumn{3}{c}{\madanalysis--\sfs} \\
			& Events & $\varepsilon$ & Events & $\varepsilon$ & $\delta$ [\%]\\ \hline
			Initial                                 & 2278.6 & -  & 2278.6 & - & - \\
			Lepton veto                             & 1482.6 & 0.651 & 1638.4 & 0.719 & 10.5 \\
			$N_{j} \geq 5$                          & 943.8 & 0.637 & 907.5 & 0.554 & 13.0 \\
			$N_{b} \geq 4$                          & 130.2 & 0.138 & 145.0 & 0.160 & 15.9 \\
			$\slashed{E}_T > 350$ GeV             & 24.1 & 0.185 & 25.1 & 0.173 & 6.5 \\
			min$[\Delta\phi(j_{1-4},\slashed{E}_T)]>0.4$ & 12.8 & 0.531 & 12.8 & 0.510 & 3.9 \\
			Tau veto                           & 12.8 & 1.000 & 12.6 & 0.982 & 1.8 \\
			$\bar m_h^{\rm reco} \in[75,175]$ GeV& 8.5 & 0.664 & 7.0 & 0.559 & 15.8 \\
			Leading jet non-$b$-tagged                & 8.5 & 1.000 & 5.6 & 0.802 & 19.8 \\
			$p_T(j_1) > 350$ GeV                 & 7.4 & 0.871 & 4.1 & 0.724 & 16.9 \\
			$|\Delta\phi(j_{1},\slashed{E}_T)|>2.8$ & 7.4 & 1.000 & 3.2 & 0.787 & 21.3 \\
			$m_{\rm eff} > 1$ TeV                     & 7.4 & 1.000 & 3.2 & 0.990 & 1.0 \\ \hline
			$\Delta_{\rm MC}/N_{\rm yield} $ & \multicolumn{2}{c}{$ 28.8\% $} & \multicolumn{3}{c}{$ 6.9\% $} \\
			\botrule\end{tabular}\label{tab:srb}}
\end{table}

\begin{figure}[t]
  \centering
  \includegraphics[scale=0.35]{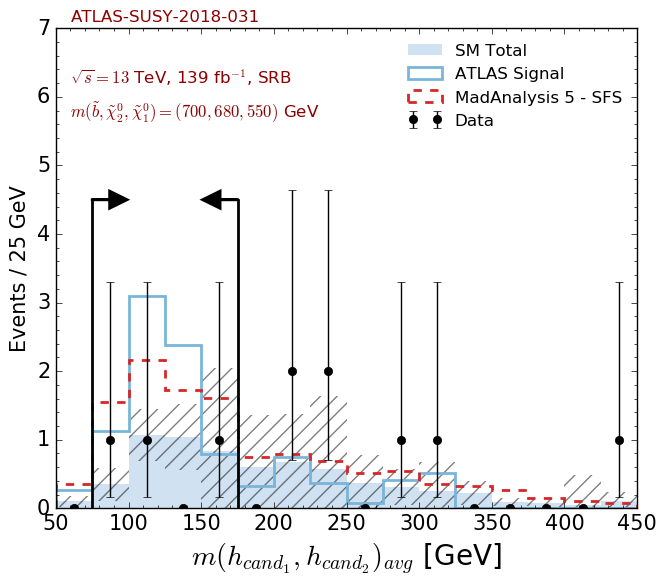}\quad
  \includegraphics[scale=0.35]{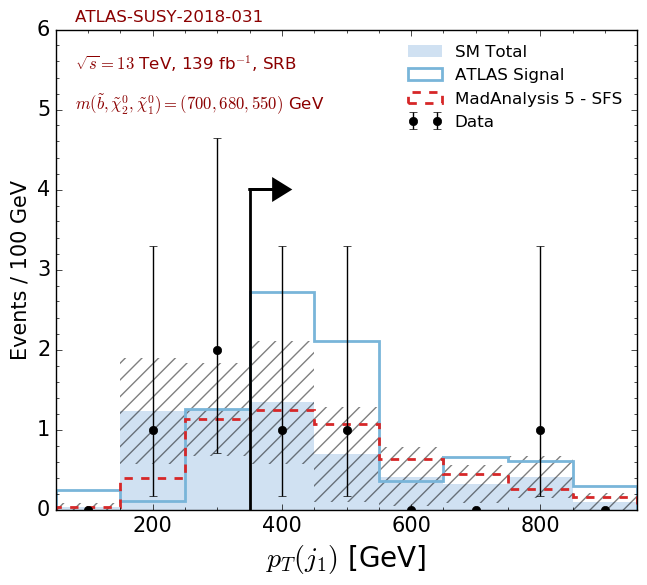}
  \caption{Histograms representative of the SRB cut-flow. We consider the
    distribution in the average invariant mass of the Higgs boson candidates
    (left) and in the $p_T$ of the leading jet (right), for a spectrum defined
     by $m(\tilde{b},\tilde\chi^0_2,\tilde\chi^0_1)=(700,680,550)$~GeV. The
    dashed red line refers to the \madanalysis\ predictions, and the blue solid
    line to the ATLAS official results. As a reference, we show through solid
    blue bars matched with hatched areas the expected SM background and the
    related uncertainties, as well as the results emerging from data
    (black dots).\label{fig:srb}}
\end{figure}

\begin{figure}[t]
	\centering
	\includegraphics[scale=0.35]{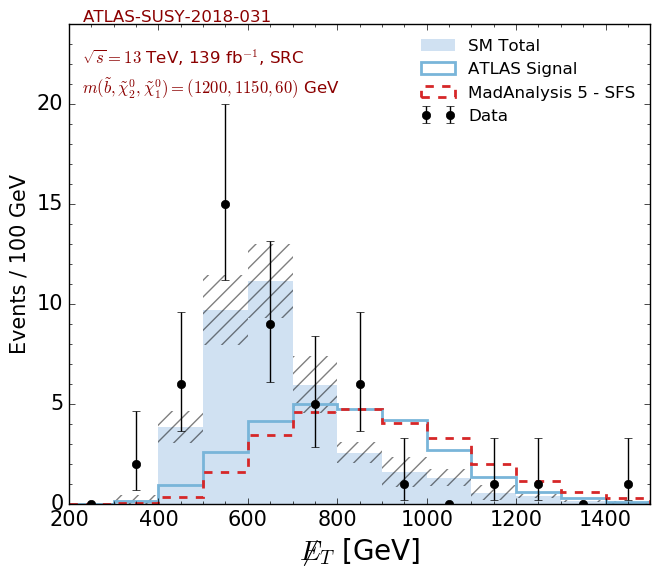}\quad
	\includegraphics[scale=0.35]{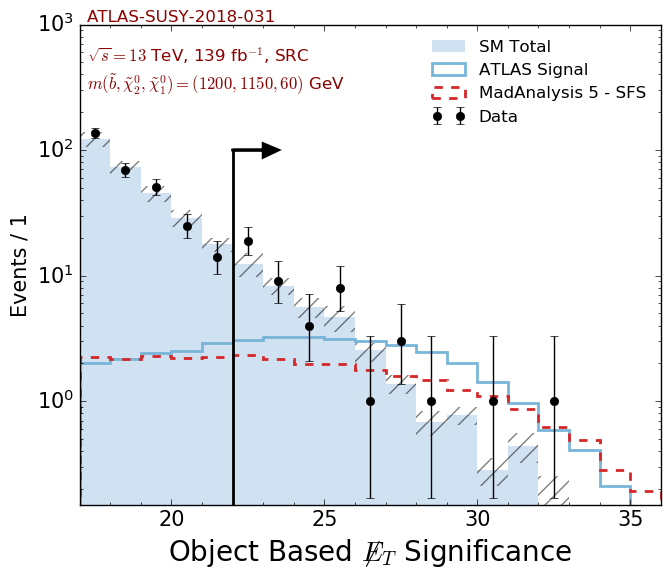}
  \caption{Histograms representative of the SRC cut-flow. We consider the
    distribution in the missing transverse energy (left) and in its
    significance (right), for a spectrum defined
     by $m(\tilde{b},\tilde\chi^0_2,\tilde\chi^0_1)=(1200,1150,60)$~GeV. The
    dashed red line refers to the \madanalysis\ predictions, and the blue solid
    line to the ATLAS official results. As a reference, we show through solid
    blue bars matched with hatched areas the expected SM background and the
    related uncertainties, as well as the results emerging from data
    (black dots).\label{fig:src}}
\end{figure}

In table~\ref{tab:srb} and figure~\ref{fig:srb}, we present results that are
relevant for the validation of our implementation of the SRB region, and that
have been computed in the context of the benchmark scenario defined by the mass
spectrum $m(\tilde{b},\tilde\chi^0_2,\tilde\chi^0_1)=(700,680,550)$~GeV. In
terms of the various cut efficiencies shown in table~\ref{tab:srb}, we observe
deviations $\delta$ between our \madanalysis\ predictions and the ATLAS official
results that reach up to 21\%. However, the ATLAS reference numbers are coming
with a quite large MC uncertainty $\Delta_{\rm MC}$ of 28\%. The situation is
further emphasised in figure~\ref{fig:srb}, in which we present in particular
the transverse momentum distribution of the leading jet (right panel). The
\madanalysis\ (red dashed lines) and ATLAS predictions indeed quite differ in
the low $p_T$ regime that is relevant for the cut-flow. Focusing on the
second distribution shown in the figure, one may be tempted to naively conclude
that the shape of the $\bar m_h^{\rm reco}$ observable is on the contrary pretty
well reproduced (left panel). The ATLAS curve however features
fluctutations and quite differs from the \madanalysis\ results in the region
that is relevant for the
cut-flow. After accounting for the uncertainties on the ATLAS numbers,
we cannot therefore draw any strong conclusion about the validation of
our implementation. As the related \madanalysis\ code is similar to that
relevant for the SRA
and SRC regions for which good agreement is found (see above and below), we
nevertheless consider our implementation as validated. The lack of publc
information prevents us from investigating this problem more deeply.

In table~\ref{tab:src} and figure~\ref{fig:src}, we finally turn to the
validation of the implementation of the last class of signal regions, the SRC
regions, for which we consider a benchmark scenario featuring
$m(\tilde{b},\tilde\chi^0_2,\tilde\chi^0_1)=(1200,1150,60)$~GeV. Both our
predictions and the official ATLAS numbers are here numerically accurate,
$\Delta_{\rm MC}$ being small. Our predictions are found to agree quite well
with the ATLAS predictions, both for the cut-flow tables, the $\slashed{E}_T$
spectrum and the missing transverse energy significance spectrum.

\begin{table}[t]
 \renewcommand{\arraystretch}{1.5}
 \setlength\tabcolsep{12pt}
  \tbl{Same as in table~\ref{tab:sra} but for the signal region SRC and a
  spectrum defined by
  $m(\tilde{b},\tilde\chi^0_2,\tilde\chi^0_1)=(1200,1150,60)$~GeV.
    The ATLAS results correspond to 9,668 MC events prior to any cut,
    whereas our predictions rely on  64,305 events.}
	{\begin{tabular}{@{}lccccc@{}} \toprule
			& \multicolumn{2}{c}{ATLAS} & \multicolumn{3}{c}{\madanalysis--\sfs} \\
			& Events & $\varepsilon$ & Events & $\varepsilon$ & $\delta$ [\%]\\ \hline
			Initial                                 & 180.3 & -  & 180.3 & - & - \\
			Lepton veto                           & 127.5 & 0.707 & 129.6 & 0.719 & 1.7 \\
			$N_{j} \geq 4$                          & 117.1 & 0.918 & 120.8 & 0.932 & 1.5 \\
			$N_{b} \geq 3$                          & 67.9 & 0.580 & 61.9 & 0.513 & 11.6 \\
			$\slashed{E}_T > 250$ GeV             & 61.5 & 0.906 & 56.3 & 0.910 & 0.5 \\
			min$[\Delta\phi(j_{1-4},\slashed{E}_T)]>0.4$ & 50.4 & 0.820 & 45.6 & 0.810 & 1.2 \\\hline
			$\slashed{E}_T/\sqrt{H_T}>22\ \sqrt{\rm GeV}$& 26.7 & 0.530 & 26.3 & 0.577 & 9.0 \\ 
			$\Delta_{\rm MC}/N_{\rm yield} $ & \multicolumn{2}{c}{$ 2.2\% $} & \multicolumn{3}{c}{$ 1.1\% $} \\\hline
			$\slashed{E}_T/\sqrt{H_T}\in [22,24]\ \sqrt{\rm GeV}$& 6.7 & 0.133 & 5.8 & 0.126 & 4.8 \\
			$\Delta_{\rm MC}/N_{\rm yield} $ & \multicolumn{2}{c}{$ 4.5\% $} & \multicolumn{3}{c}{$ 1.7\% $} \\\hline
			$\slashed{E}_T/\sqrt{H_T}\in [24,26]\ \sqrt{\rm GeV}$& 6.4 & 0.127 & 5.8 & 0.126 & 0.7 \\
			$\Delta_{\rm MC}/N_{\rm yield} $ & \multicolumn{2}{c}{$ 4.7\% $} & \multicolumn{3}{c}{$ 1.7\% $} \\\hline
			$\slashed{E}_T/\sqrt{H_T}>\in [26,28]\ \sqrt{\rm GeV}$& 5.5 & 0.109 & 5.1 & 0.112 & 2.9 \\
			$\Delta_{\rm MC}/N_{\rm yield} $ & \multicolumn{2}{c}{$ 5.5\% $} & \multicolumn{3}{c}{$ 1.9\% $} \\\hline
			$\slashed{E}_T/\sqrt{H_T} > 28\ \sqrt{\rm GeV}$& 8.2 & 0.163 & 8.1 & 0.178 & 9.2 \\
			$\Delta_{\rm MC}/N_{\rm yield} $ & \multicolumn{2}{c}{$ 4.9\% $} & \multicolumn{3}{c}{$ 1.2\% $} \\
			\botrule\end{tabular}\label{tab:src}}
\end{table}

\subsection{Conclusions}
In this note, we presented our efforts on the implementation of the
ATLAS-SUSY-2018-31 analysis in the \madanalysis\ framework, using the \sfs\
detector simulation based on smearing and efficiency functions that is shipped with
\madanalysis. We have validated our work in the context of three simplified
models dedicated to the production and decay of supersymmetric partners of the
bottom quark. The validation has been achieved by comparing predictions obtained
with our implementation to official results from the ATLAS collaboration. A
reasonable agreement has been achieved for each signal region, the deviations
remaining in general under a level of 20\%--30\%. The most considerable discrepancies can
be traced to MC uncertainties inherent to the official ATLAS results, hindering hence
our capacity to properly validate the implementation of one of the analysis
signal regions. The good agreement obtained for all other regions, relying on the
same piece of code, nevertheless makes us considering this analysis as validated.

The {\sc MadAnalysis}~5 C++ code is available for download from the \madanalysis\ dataverse
(\href{https://doi.org/10.14428/DVN/IHALED}{https://doi.org/10.14428/DVN/IHALED})~\cite{IHALED_2020}. The material relevant for the validation benchmarks has been obtained from \href{https://www.hepdata.net/record/ins1748602}{HEPData}~\cite{1748602}.

\subsection*{Acknowledgments}
The authors are grateful to Laura Jeanty and Federico Meloni for their help
with understanding the ATLAS analysis considered in this work. JYA has received
funding from the European Union’s Horizon 2020 research and innovation programme as part of the Marie Sklodowska-Curie Innovative Training Network MCnetITN3 (grant agreement no. 722104).

\cleardoublepage
\renewcommand{\madanalysis}{{\sc MadAnalysis~5}}
\renewcommand{\sfsim}{{\sc simplified--Fast Simulation}}
\renewcommand{\sfs}{{\sc sFS}}
\renewcommand{\delphes}{{\sc Delphes~3}}
\newcommand{\herwig}{{\sc Herwig~7}}
\renewcommand{\madgraph}{{\sc MG5\_aMC}}
\renewcommand{\pythia}{{\sc Pythia~8}}
\newcommand{\rivet}{{\sc Rivet}}
\newcommand{\rooot}{{\sc Root}}
\renewcommand{\fastjet}{{\sc FastJet}}

\markboth{Jack Y. Araz and Benjamin Fuks}{Implementation of the ATLAS-SUSY-2018-32 analysis}

\section{Implementation of the ATLAS-SUSY-2018-32 analysis (sleptons and eletroweakinos with two leptons
  and missing transverse energy; 139~fb$^{-1}$)}
  \vspace*{-.1cm}\footnotesize{\hspace{.5cm}By Jack~Y.~Araz and Benjamin~Fuks}
\label{sec:sleptons}

%


\subsection{Introduction}
Supersymmetry (SUSY) is one of the most popular extensions of the Standard Model
(SM). By naturally extending the Poincar\'e algebra and linking the fermionic
and bosonic degrees of freedom of the theory, SUSY could provide, among others,
a solution for the hierarchy problem of the SM and an explanation for the
problematics of dark matter. The so-called Minimal Supersymmetric Standard Model
(MSSM)~\cite{Nilles:1983ge,Haber:1984rc} consists in the direct
supersymmetrisation of the SM, giving thus rise to one SUSY partner to each SM
degrees of freedom. In this framework, the slepton mass eigenstates are the
superpartners of the SM leptons, and the electroweakino mass eigenstates
consist of admixtures of the partners of the SM gauge and Higgs fields. More
precisely, electroweakinos comprise the electrically-charged charginos
($\tilde{\chi}^\pm_i,\ i\in 1,2 $) and the
electrically-neutral neutralinos ($ \tilde{\chi}^0_i,\ i\in 1, 2, 3, 4 $). The
lightest neutralino is often considered to be the lightest supersymmetric
particle (LSP), hence a potentially viable candidate for dark matter.

The ATLAS-SUSY-2018-32 analysis~\cite{Aad:2019vnb} has been structured to look
for new physics signals that might appear due to charginos and sleptons. It
searches for dilepton and missing energy final states which can emerge from the
production of either a pair of lightest charginos, or of a pair of sleptons. Two
separate benchmarks have been constructed for this purpose in the chargino case.
First, a pure wino-like ($ \tilde{W}$) chargino is considered to decay into a
bino-like ($\tilde{B} $) LSP and a lepton via an intermediate $W$-boson. Second,
the same wino-like chargino is assumed to cascade decay to the same final state,
but this time through a slepton exchange. Additionally, a third benchmark
scenario is dedicated to the pair production of sleptons that decay each into a
lepton and an LSP. Those three cases are illustrated with the Feynman diagrams
of Fig.~\ref{fig:diagrams}.

Those three scenarios have been analysed to constrain the corresponding
simplified models. The experimental results have shown that for a massless LSP,
chargino masses up to 420 GeV are excluded at $ 95\% $ confidence level
(CL) in the first class of scenarios (chargino pair production and decay via a
charged weak gauge boson). A more severe bound has been set on the second class
of scenarios, when charginos decay through an intermediate slepton. The masses
are in this case constrained to be larger than 1 TeV at $95\%$ CL. Finally,
slepton mass bounds, as derived in the context of the third class of scenarios,
are of 700 GeV.

In the rest of this note, we present the recast of ATLAS-SUSY-2018-32 analysis
of Ref.~[\refcite{Aad:2019vnb}] in the \madanalysis\ framework, that is now
available from the \href{http://madanalysis.irmp.ucl.ac.be/wiki/PublicAnalysisDatabase}{\madanalysis\ Public Analysis Database} and \href{https://doi.org/10.14428/DVN/EA4S4D}{the \madanalysis\ dataverse}~\cite{EA4S4D_2020}.

\subsection{Description of the analysis}
\begin{figure}[t]
\centering
  \includegraphics[width=0.32\textwidth]{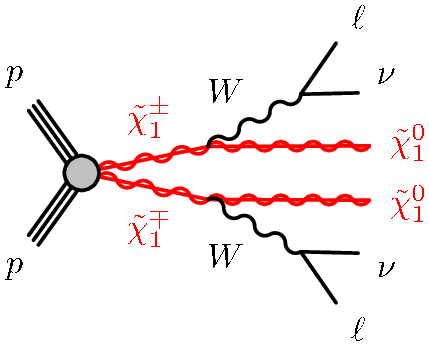}
  \includegraphics[width=0.32\textwidth]{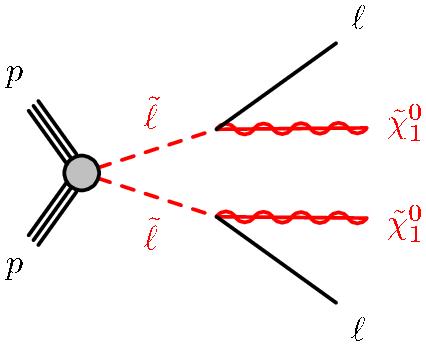}
  \includegraphics[width=0.32\textwidth]{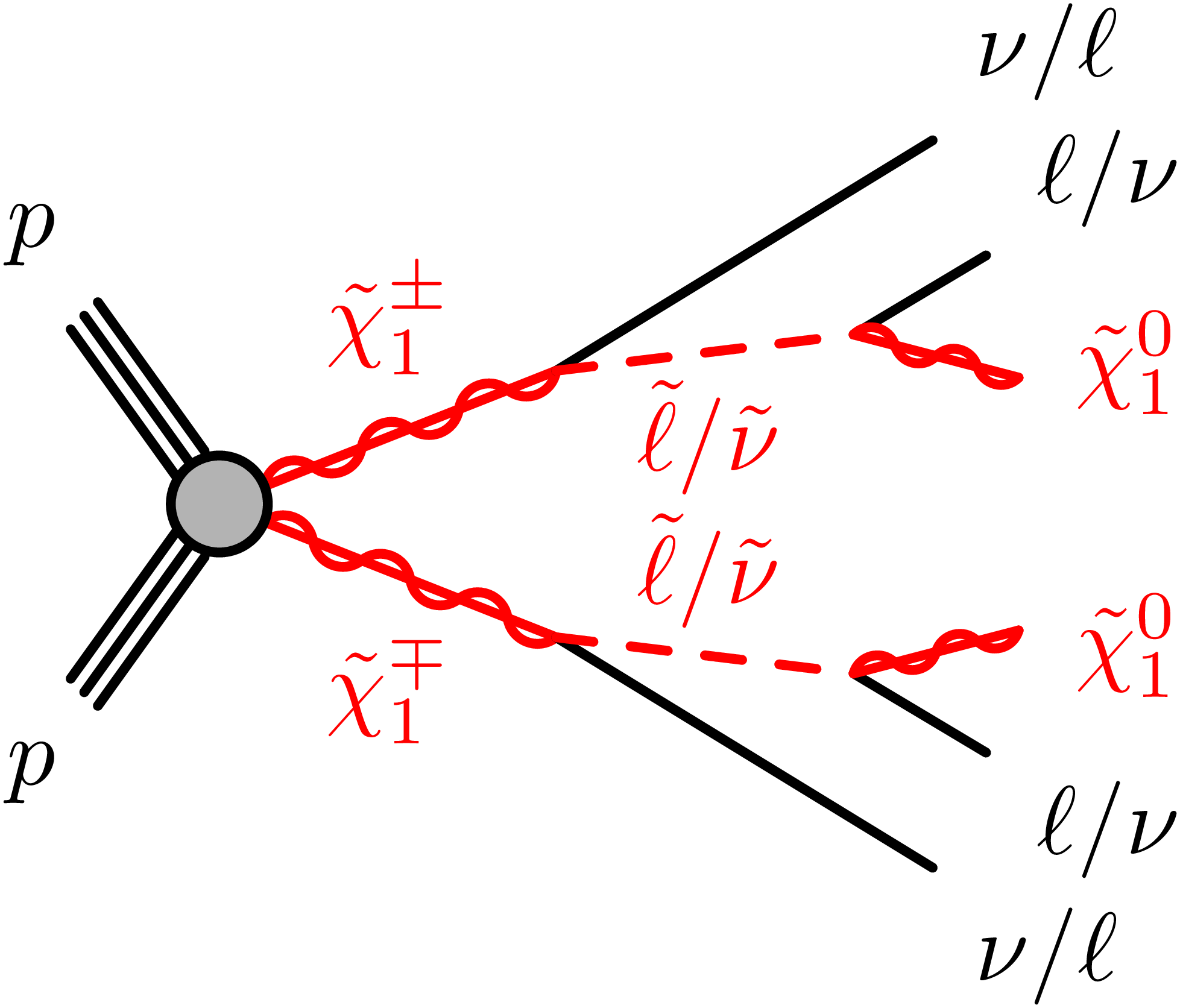}
  \caption{Representative Feynman diagrams for the scenarios used for the
    validation of our re-implementation of the ATLAS-SUSY-2018-32 analysis. We
    consider chargino production followed by decays through electroweak bosons
    (left), slepton production followed by direct decays into leptons and
    missing energy (centre) and chargino production followed by decays mediated
    by an intermediate slepton or sneutrino (right). The diagrams have been
    taken from Ref.~\cite{Aad:2019vnb}.\label{fig:diagrams}}
\end{figure}

This analysis targets a signature made of two lepton and missing transverse
energy, as could arise from the production and decay of a pair lightest
charginos ($\tilde{\chi}^\pm_1 $) or slepton ($ \tilde{l}_i $). As mentioned in
the previous section, the results are interpreted in three classes of simplified
models depicted in the diagrams of Fig~\ref{fig:diagrams}. The first two of
these extend the SM by a chargino and a neutralino LSP,
the difference between them lying at the level of the chargino decay. In the
first setup, charginos decay into a single lepton and missing energy via an
intermediate $W$-boson, whereas in the second setup, they decay via an
intermediate slepton. In the last class of simplified models under
consideration, the SM is supplemented by a charged slepton and a neutralino LSP,
the slepton being taken directly decaying into the LSP and a lepton. The
validation of our re-implementation is achieved in these three cases.

\subsubsection{Object definitions}
Jets are obtained by the clustering all participants to the hadronic activity in
the event, electrons and photons according to the anti-$k_T$
algorithm~\cite{Cacciari:2008gp}, as embedded in the \fastjet\ package
version~3.3.3~\cite{Cacciari:2011ma}, with a radius parameter set to $R=0.4$.
Jet candidates are then extracted by requiring the reconstructed objects to have
a minimum transverse momentum $p_T>20$~GeV and a pseudorapidity satistying
$ |\eta|<2.4 $.

Electron and muons are required to satisfy strong isolation conditions to be 
considered as a signal leptons.

A signal electron is required to have a minimum $p_T$ of 10~GeV and to be within
$ |\eta|<2.47 $. These electrons are then required to be isolated from the
calorimetric activity and any other charged track. This is achieved in practice
by constraining the sum of the $p_T$ of all tracks lying in a cone of radius
$\Delta R$ around the electron to be smaller than 15\% of the electron
transverse momentum, the cone radius being defined by $\Delta R = \min(10/p_T,
0.2)$. Moreover, the calorimetric activity, $I^e_{rel}$, in a cone of radius
$\Delta R = 0.2$ around the electron is restricted to be smaller than 20\% of
the electron $p_T$. For very hard electrons with $p_T > 200$~GeV, a special
isolation treatment is, however, implemented. In this case, one solely imposes
the signal electron to be calorimetrically isolated, requiring $I^e_{rel} <
{\rm max}(0.015 p_T, 3.5) $.

Signal muons are defined similarly. Their transverse momentum is imposed to
fulfil $ p_T > 10 $~GeV and their pseudorapidity $ |\eta| < 2.7 $. Track-based
isolation implies that the sum of the transverse momentum of all tracks lying in
a cone of radius $ \Delta R = {\rm min}(10/p_T, 0.3) $ around the muon is
smaller than 15\% of the muon transverse momentum, whilst calorimetric isolation
enforces  $I^\mu_{rel} < 0.3 p_T$, for a cone of radius $\Delta R =0.2 $ centred
on the muon.

Finally, all jets at a distance in the transverse plane of $\Delta R \leq 0.2$
(0.4) of an electron (muon) are removed, and all electrons are required to be
separated from any muon by least $\Delta R = 0.2$. The collection of $b$-jets
is extracted from the collection of cleaned signal jets.

\subsubsection{Event selection}
The considered ATLAS-SUSY-2018-32 analysis includes four sets of signal regions
differing by the properties of the dilepton system and the jet activity in the
event. Two categories of signal regions feature a pair of leptons of different
flavours (DF). Regions of the first class impose a veto to the presence of any
final-state signal jet, whereas regions of the second sub-category allow for the
presence of one jet in the final state. Similarly, two classes of regions are
designed to probe final states featuring a pair of leptons of the same flavour
(SF), these two sub-categories differing by requiring either zero or one
final-state jet.

After dedicated pre-selection requirements, all signal regions are further
divided into different bins in the $ m_{T2} $ observable defined
by~\cite{Lester:1999tx,Cheng:2008hk}
\begin{eqnarray}
m_{T2}(\mathbf{p}_{T,1},\mathbf{p}_{T,2},\mathbf{p}^{\rm miss}_T) = \min\limits_{\mathbf{q}_{T,1}+\mathbf{q}_{T,2}=\mathbf{p}^{\rm miss}_T} \left\{ \max\left[ m_T(\mathbf{p}_{T,1},\mathbf{q}_{T,1}) ,m_T(\mathbf{p}_{T,2},\mathbf{q}_{T,2}) \right] \right\}\ . \nonumber
\end{eqnarray}
Here $ \mathbf{p}_{T,1} $ and $\mathbf{p}_{T,2} $ are transverse momentum
vectors of the two leptons and $ \mathbf{q}_{T,1} $ and $\mathbf{q}_{T,2} $ are chosen to be a decomposition of the missing momentum vector, $ \mathbf{q}_{T,1} +\mathbf{q}_{T,2} =\mathbf{p}^{\rm miss}_T$. A minimisation is performed over all
possible decompositions of the missing momentum vector. For each decomposition,
we calculate the transverse mass of the system constitued by the first (second)
lepton and the $\mathbf{q}_{T,1}$ ($\mathbf{q}_{T,2} $) vector. The $m_{T2}$
value is then taken as the minimum of the maximum of the two transverse masses
associated with a given $\mathbf{p}^{\rm miss}_T$ decomposition.

In order to take trigger efficiencies into account, events are reweighed by 85\%
before any selection requirement. All signal regions then imposes that events
feature two opposite-sign leptons with a minimum transverse momentum of 25~GeV,
and whose invariant ,ass is larger than 25~GeV. A $b$-jet veto is further
enforced. At this stage, the analysis is split into four categories, as
mentioned above (DF/SF lepton pair, with or without one jet).

The DF signal regions additionally asks that the invariant mass of the lepton
pair satisfies $ m_{l_1,l_2} > 100 $~GeV, whereas the SF ones increase this
threshold to $m_{l_1,l_2} > 121.2 $~GeV. Finally, all signal regions require the
presence of at least 100~GeV of missing transverse energy, and the missing
energy significance, defined by $ \slashed{E}_T/\sqrt{H_T} $, has to be larger
than 10~$\sqrt{\rm GeV}$. In this expression, the denominator involves the $H_T$
variable that consists of the scalar sum of the transverse momenta of all signal
jets. A schematic representation of all signal regions is shown in
Table~\ref{tab:cut-flow}.

\begin{table}[t]
  \renewcommand{\arraystretch}{1.5}
  \tbl{Schematic representation of the ATLAS-SUSY-2018-32 signal region definitions.}
  {\begin{tabular}{@{}cc|cc@{}} \toprule\vspace*{-0.8cm}\\
    \multicolumn{2}{c|}{Different Flavour (DF)} & \multicolumn{2}{c}{Same Flavour (SF)}\\
    \hline
			\multicolumn{4}{c}{OS dilep. with $p^{l_1,l_2}_T>25$ [GeV]}\\
			\multicolumn{4}{c}{$m_{l_1,l_2}>25$ [GeV]}\\
			\multicolumn{4}{c}{b veto}\\\hline
			\multicolumn{1}{c|}{DF dilep. \& $N_j=0$}  & \multicolumn{1}{c|}{DF dilep. \& $N_j=1$}  & \multicolumn{1}{c|}{SF dilep. \& $N_j=0$}  & \multicolumn{1}{c}{SF dilep. \& $N_j=1$} \\\hline
			\multicolumn{2}{c|}{$m_{l_1,l_2} > 100$ [GeV]} & 	\multicolumn{2}{c}{$m_{l_1,l_2} > 121.2$ [GeV]} \\\hline
			\multicolumn{4}{c}{$\slashed{E}_T > 110$ [GeV] }\\
			\multicolumn{4}{c}{$\slashed{E}_T$ Sig. $> 10$ [$\sqrt{\rm GeV}$] }\\		
					\multicolumn{4}{c}{$m_{T2}$ bins}\\
			\botrule\end{tabular}\label{tab:cut-flow}}
\end{table}

\subsection{Validation}

\subsubsection{Event generation}\label{sec:event_gen}
For the validation of the re-implementation of the ATLAS-SUSY-2018-32 analysis,
we study the three different scenarios defined above. For the production of a
pair of charginos that both decay via a $W$-boson, we choose a mass spectrum
such that $m(\tilde\chi^{\pm}_1,\tilde\chi^0_1)=(300,50)$~GeV, all other SUSY
states being decoupled. Similarly, for the scenario focusing on slepton pair
production, we choose $m(\tilde{l}^{\pm},\tilde\chi^0_1)=(400, 200)$~GeV and
decouple the rest of the spectrum. Finally, for the scenario where the two
pair-produced charginos decay via an intermediate slepton, the mass spectrum is
chosen to be $m(\tilde{\chi}^\pm_1,\tilde{l}^{\pm},\tilde\chi^0_1)=
(600,300,1)$~GeV, with again all other superpartners being decoupled. All SLHA
spectrum files can be found in dedicated \href{https://www.hepdata.net/record/ins1750597}{HEPData} records provided by the ATLAS collaboration~\cite{1750597}.

For our validation, we generate various leading-order event samples with
\madgraph\ version 2.7.3~\cite{Alwall:2014bza}. Following the MLM
prescription~\cite{Mangano:2006rw}, we merge samples featuring up to two extra
jets at the matrix-element level, the merging scale being set to one quarter of
the mass of the pair-produced SUSY particle. All events are showered and
hadronised by means of \pythia~\cite{Sjostrand:2014zea}, and the simulation of
the ATLAS detector is performed with the \delphes\
package~\cite{deFavereau:2013fsa}. Through our simulation we used the
leading-order set of NNPDF 2.3 parton distribution functions, as provided by
{\sc LhaPDF}~\cite{Ball:2013hta,Buckley:2014ana}. Our re-implementation can then
be used to investigate the ATLAS-SUSY-2018-32 sensitivity to the simulated
signals, through
\madanalysis\ version 1.8 (or more recent)~\cite{Conte:2018vmg}. All analysis
files can be obtained from the \madanalysis\ Public Analysis
Database~\cite{Dumont:2014tja}.

\subsubsection{Comparison with the official results}
In this section we compare our preductions for all the benchmarks described in
section~\ref{sec:event_gen} with the corresponding official ATLAS results. To
estimate the quality of our re-implementation, we define a variable $\delta$ to
quantify the difference between the relative cut efficiencies as obtained from
the ATLAS and \madanalysis\ results,
\begin{eqnarray}
\delta_i = \frac{|\varepsilon^{\rm ATLAS}_i - \varepsilon^{\rm MA5}_i |}{\varepsilon^{\rm ATLAS}_i }\ . \nonumber
\end{eqnarray}
Here $ \varepsilon_i $ represents the relative cut efficiency which is defined
as $ \varepsilon_i = N_i/N_{i-1} $, $ N_i $ being number of surviving events
after the $i^{\rm th}$ cut. The analysis will be considered as validated
provided that all $\delta_i$ values are found to satisfy $\delta\lesssim20\%$.
In the present recast exercise, it should be noted that the lack of public
information related to how the ATLAS collaboration has prepared its Monte Carlo
production introduces a certain bias and makes the comparison complicated.

Finally, in order to evaluated the statistical power associated with our event
generation procedure, we quantify the Monte Carlo uncertainty through a
$\Delta_{\rm MC}$ quantity defined by
\begin{eqnarray}
\Delta_{\rm MC} = \frac{N_n}{\sqrt{N^{\rm MC}_n}}\ , \nonumber
\end{eqnarray}
where $ N^{\rm MC}_n$ is defined by the number of unweighted Monte Carlo events
surviving the last cut. In our validation, we aim to remain a $10\%$ Monte Carlo
uncertainty, that is found to always be smaller than the magnitude of the
deviation between the \madanalysis\ predsictions and the ATLAS results after the
last cut.

Our results include a comparison between \madanalysis\ predictions and ATLAS
official results for all four considered classes of signal regions, that
respectively target the production of a SF lepton pair with 0 jet, the
production of a SF lepton pair with 1 jet, the production of a DF lepton pair
with 0-jet and the production of a DF lepton pair with 1 jet. As only ATLAS
predictions for the $ m_{T2} \in [100,\infty[ $~GeV bin are available in
\href{https://www.hepdata.net/record/ins1750597}{HEPData}, we accordingly
restrict the discussion to this sole inclusive bin.

Tables~\ref{tab:sf_cha} and \ref{tab:df_cha} include cut-flow results for the
benchmark simplified model featuring chargino production and decay via a
$W$-boson. As can be seen in the lower panel of
table~\ref{tab:df_cha}, the largest variation from the ATLAS results has been
observed to be $ 20.3\% $ for the $ \slashed{E}_T $ significance requirement.
This disagreement stems from potentially genuine differences between the
implementation of this cut in our re-implementation and in the non-public ATLAS
code. However, our definition still gives a reasonably acceptable deviation from
the ATLAS results, especially after accounting for all other cuts.

We then present results for slepton production (third class of benchmarks) in
table~\ref{tab:sf_slep}, and for chargino
pair-production followed by chargino decays via an intermediate slepton (second
class of benchmarks) in tables~\ref{tab:sf_chaslep} and \ref{tab:df_chaslep}.

Tables~\ref{tab:sf_cha}--\ref{tab:df_chaslep} comprise two main columns, one for the ATLAS results and one for
the \madanalysis\ ones. These columns are further divided, so that they include
the number of events surviving each cut, the relative cut efficiencies and the
$\delta_i$ quantities for each cut. All tables have been prepared with the
\texttt{ma5\_expert} package~\cite{ma5_expert}.

\begin{table}[t]
\renewcommand{\arraystretch}{1.3}
  \tbl{Cut-flow associated with the signal region dedicated to the production of
   a SF lepton pair without any jet (upper) or with a single jet (lower), for a
   benchmark scenario of the first category (production of a pair of charginos
   that decay each via a $W$-boson), for a spectrum defined by
   $m(\tilde\chi^{\pm}_1,\tilde\chi^0_1)=(300,50)$~GeV.}
  {\begin{tabular}{@{}l|cc|ccc@{}} \toprule
  & \multicolumn{2}{c|}{ATLAS} & \multicolumn{3}{c}{{\sc MadAnalysis 5}}\\
		& Events & $\varepsilon$ & Events & $\varepsilon$ & $\delta$ [\%]\\ \hline
		Initial                                 & 26432.0 & -  & 26432.0 & - & - \\
		OS dilep. with $p^{l_1,l_2}_T>25$ [GeV] & 565.0 & 0.021 & 570.1 & 0.022 & 0.9 \\
		$m_{l_1,l_2}>25$ [GeV]                  & 559.0 & 0.989 & 564.0 & 0.989 & 0.0 \\
		b veto                                  & 526.0 & 0.941 & 557.7 & 0.989 & 5.1 \\
		SF dilep. \& $N_j=0$                    & 138.7 & 0.264 & 134.0 & 0.240 & 8.9 \\
		$m_{l_1,l_2} > 121.2$ [GeV]             & 92.4 & 0.666 & 81.9 & 0.612 & 8.2 \\
		$\slashed{E}_T > 110$ [GeV]             & 47.1 & 0.510 & 42.4 & 0.518 & 1.5 \\
		$\slashed{E}_T$ Sig. $> 10$ [$\sqrt{\rm GeV}$]& 42.9 & 0.911 & 42.4 & 1.000 & 9.8 \\
		$m_{T2} \in [100, \infty[$ [GeV]        & 25.4 & 0.592 & 21.3 & 0.501 & 15.3 \\\hline
		$\Delta_{\rm MC}/N_{\rm yield} $ & \multicolumn{2}{c|}{$ 4.3\% $} & \multicolumn{3}{c}{$ 7.5\% $} \\
  \botrule
  \multicolumn{6}{c}{}\\
\toprule
  & \multicolumn{2}{c|}{ATLAS} & \multicolumn{3}{c}{{\sc MadAnalysis 5}}\\
		& Events & $\varepsilon$ & Events & $\varepsilon$ & $\delta$ [\%]\\ \hline
		Initial                                 & 26432.0 & -  & 26432.0 & - & - \\
		OS dilep. with $p^{l_1,l_2}_T>25$ [GeV] & 565.0 & 0.021 & 570.1 & 0.022 & 0.9 \\
		$m_{l_1,l_2}>25$ [GeV]                  & 559.0 & 0.989 & 564.0 & 0.989 & 0.0 \\
		b veto                                  & 526.0 & 0.941 & 557.7 & 0.989 & 5.1 \\
		SF dilep. \& $N_j=1$                    & 88.8 & 0.169 & 87.7 & 0.157 & 6.9 \\
		$m_{l_1,l_2} > 121.2$ [GeV]             & 58.9 & 0.663 & 57.5 & 0.656 & 1.1 \\
		$\slashed{E}_T > 110$ [GeV]             & 32.6 & 0.553 & 31.8 & 0.552 & 0.2 \\
		$\slashed{E}_T$ Sig. $> 10$ [$\sqrt{\rm GeV}$]& 26.9 & 0.825 & 30.5 & 0.961 & 16.4 \\
		$m_{T2} \in [100, \infty[$ [GeV]        & 14.0 & 0.520 & 13.9 & 0.455 & 12.6 \\ \hline
			$\Delta_{\rm MC}/N_{\rm yield} $ & \multicolumn{2}{c|}{$ 5.7\% $} & \multicolumn{3}{c}{$ 9.4\% $} \\
  \botrule\\ \end{tabular}\label{tab:sf_cha}}
\end{table}

\begin{table}[t]
 \renewcommand{\arraystretch}{1.3}
   \tbl{Same as in table~\ref{tab:sf_cha} but for a pair of DF leptons produced
   without any jet (upper) and with one jet (lower).}
 {\begin{tabular}{@{}l|cc|ccc@{}} \toprule
   & \multicolumn{2}{c|}{ATLAS} & \multicolumn{3}{c}{{\sc MadAnalysis 5}} \\
   & Events & $\varepsilon$ & Events & $\varepsilon$ & $\delta$ [\%]\\ \hline
	Initial                                 & 26432.0 & -  & 26432.0 & - & - \\
	OS dilep. with $p^{l_1,l_2}_T>25$ [GeV] & 565.0 & 0.021 & 570.1 & 0.022 & 0.9 \\
	$m_{l_1,l_2}>25$ [GeV]                  & 559.0 & 0.989 & 564.0 & 0.989 & 0.0 \\
	b veto                                  & 526.0 & 0.941 & 557.7 & 0.989 & 5.1 \\
	DF dilep. \& $N_j=0$                    & 122.7 & 0.233 & 137.0 & 0.246 & 5.3 \\
	$m_{l_1,l_2} > 100$ [GeV]               & 94.2 & 0.768 & 103.7 & 0.757 & 1.4 \\
	$\slashed{E}_T > 110$ [GeV]             & 46.5 & 0.494 & 52.2 & 0.503 & 1.9 \\
	$\slashed{E}_T$ Sig. $> 10$ [$\sqrt{\rm GeV}$]& 42.2 & 0.908 & 52.2 & 1.000 & 10.2 \\
	$m_{T2} \in [100, \infty[$ [GeV]        & 26.4 & 0.626 & 30.1 & 0.578 & 7.6 \\ \hline
			$\Delta_{\rm MC}/N_{\rm yield} $ & \multicolumn{2}{c|}{$ 4.2\% $} & \multicolumn{3}{c}{$ 6.3\% $} \\
  \botrule
  \multicolumn{6}{c}{}\\
  \toprule
  & \multicolumn{2}{c|}{ATLAS} & \multicolumn{3}{c}{{\sc MadAnalysis 5}}\\
   & Events & $\varepsilon$ & Events & $\varepsilon$ & $\delta$ [\%]\\ \hline
		Initial                                 & 26432.0 & -  & 26432.0 & - & - \\
		OS dilep. with $p^{l_1,l_2}_T>25$ [GeV] & 565.0 & 0.021 & 570.1 & 0.022 & 0.9 \\
		$m_{l_1,l_2}>25$ [GeV]                  & 559.0 & 0.989 & 564.0 & 0.989 & 0.0 \\
		b veto                                  & 526.0 & 0.941 & 557.7 & 0.989 & 5.1 \\
		DF dilep. \& $N_j=1$                    & 81.9 & 0.156 & 88.2 & 0.158 & 1.5 \\
		$m_{l_1,l_2} > 100$ [GeV]               & 62.3 & 0.761 & 65.0 & 0.738 & 3.0 \\
		$\slashed{E}_T > 110$ [GeV]             & 33.8 & 0.543 & 35.4 & 0.544 & 0.3 \\
		$\slashed{E}_T$ Sig. $> 10$ [$\sqrt{\rm GeV}$]& 27.2 & 0.805 & 34.3 & 0.968 & 20.3 \\
		$m_{T2} \in [100, \infty[$ [GeV]        & 15.3 & 0.562 & 15.9 & 0.464 & 17.6 \\ \hline
			$\Delta_{\rm MC}/N_{\rm yield} $ & \multicolumn{2}{c|}{$ 5.2\% $} & \multicolumn{3}{c}{$ 8.8\% $} \\
  \botrule\\\end{tabular}\label{tab:df_cha}}
\end{table}

\begin{table}[t]
 \renewcommand{\arraystretch}{1.3}
  \tbl{Same is in table~\ref{tab:sf_cha} but for a benchmark scenario of
   the third category (production of a pair of sleptons decaying each into a
   lepton and the LSP), for a spectrum defined by
   $m(\tilde{l}^{\pm},\tilde\chi^0_1)=(400,200)$ GeV.}
	{\begin{tabular}{@{}l|cc|ccc@{}} \toprule
		& \multicolumn{2}{c|}{ATLAS} & \multicolumn{3}{c}{{\sc MadAnalysis 5}} \\
				& Events & $\varepsilon$ & Events & $\varepsilon$ & $\delta$ [\%]\\ \hline
      Initial                                 & 503.0 & -  & 503.0 & - & -   \\
OS dilep. with $p^{l_1,l_2}_T>25$ [GeV] & 316.0 & 0.628 & 322.2 & 0.641 & 2.0   \\
$m_{l_1,l_2}>25$ [GeV]                  & 315.0 & 0.997 & 322.1 & 1.000 & 0.3   \\
b veto                                  & 298.0 & 0.946 & 316.6 & 0.983 & 3.9   \\
SF dilep. \& $N_j=0$                    & 136.0 & 0.456 & 141.7 & 0.448 & 1.9  \\
$m_{l_1,l_2} > 121.2$ [GeV]             & 123.5 & 0.908 & 129.8 & 0.916 & 0.8   \\
$\slashed{E}_T > 110$ [GeV]             & 97.5 & 0.789 & 100.1 & 0.771 & 2.3   \\
$\slashed{E}_T$ Sig. $> 10$ [$\sqrt{\rm GeV}$]& 88.5 & 0.908 & 100.1 & 1.000 & 10.2   \\
$m_{T2} \in [100, \infty[$ [GeV]        & 75.1 & 0.849 & 81.1 & 0.811 & 4.5   \\ \hline
		$\Delta_{\rm MC}/N_{\rm yield} $ & \multicolumn{2}{c|}{$ 2.7\% $} & \multicolumn{3}{c}{$ 0.8\% $} \\
  \botrule
  \multicolumn{6}{c}{}\\
  \toprule
  & \multicolumn{2}{c|}{ATLAS} & \multicolumn{3}{c}{{\sc MadAnalysis 5}}\\
					& Events & $\varepsilon$ & Events & $\varepsilon$ & $\delta$ [\%]\\ \hline
      Initial                                 & 503.0 & -  & 503.0 & - & -   \\
OS dilep. with $p^{l_1,l_2}_T>25$ [GeV] & 316.0 & 0.628 & 322.2 & 0.641 & 2.0   \\
$m_{l_1,l_2}>25$ [GeV]                  & 315.0 & 0.997 & 322.1 & 1.000 & 0.3   \\
b veto                                  & 298.0 & 0.946 & 316.6 & 0.983 & 3.9   \\
SF dilep. \& $N_j=1$                    & 99.2 & 0.333 & 102.7 & 0.324 & 2.5   \\
$m_{l_1,l_2} > 121.2$ [GeV]             & 90.3 & 0.910 & 94.1 & 0.916 & 0.6   \\
$\slashed{E}_T > 110$ [GeV]             & 71.8 & 0.795 & 74.3 & 0.790 & 0.6   \\
$\slashed{E}_T$ Sig. $> 10$ [$\sqrt{\rm GeV}$]& 61.3 & 0.854 & 72.9 & 0.981 & 14.9   \\
$m_{T2} \in [100, \infty[$ [GeV]        & 51.1 & 0.834 & 55.7 & 0.764 & 8.4   \\ \hline
			$\Delta_{\rm MC}/N_{\rm yield} $ & \multicolumn{2}{c|}{$ 3.7\% $} & \multicolumn{3}{c}{$ 0.7\% $} \\
\botrule\\\end{tabular}\label{tab:sf_slep}}
\end{table}

\begin{table}[t]
 \renewcommand{\arraystretch}{1.3}
   \tbl{Same is in table~\ref{tab:sf_cha} but for a benchmark scenario of
   the third category (production of a pair of charginos decaying each into a
   lepton and the LSP via an intermediate slepton), for a spectrum defined by
   $m(\tilde{\chi}^\pm_1,\tilde{l}^{\pm},\tilde\chi^0_1)=(600,300,1)$ GeV.}
	{\begin{tabular}{@{}l|cc|ccc@{}} \toprule
			& \multicolumn{2}{c|}{ATLAS} & \multicolumn{3}{c}{{\sc MadAnalysis 5}} \\
					& Events & $\varepsilon$ & Events & $\varepsilon$ & $\delta$ [\%]\\ \hline
      Initial                                 & 1320.0 & -  & 1320.0 & - & - \\
	OS dilep. with $p^{l_1,l_2}_T>25$ [GeV] & 430.0 & 0.326 & 450.7 & 0.341 & 4.8 \\
	$m_{l_1,l_2}>25$ [GeV]                  & 429.0 & 0.998 & 449.9 & 0.998 & 0.1 \\
	b veto                                  & 401.0 & 0.935 & 441.5 & 0.981 & 5.0 \\
	SF dilep. \& $N_j=0$                    & 89.8 & 0.224 & 93.4 & 0.212 & 5.5 \\
	$m_{l_1,l_2} > 121.2$ [GeV]             & 82.2 & 0.915 & 84.8 & 0.907 & 0.9 \\
	$\slashed{E}_T > 110$ [GeV]             & 68.7 & 0.836 & 72.6 & 0.857 & 2.6 \\
	$\slashed{E}_T$ Sig. $> 10$ [$\sqrt{\rm GeV}$]& 63.5 & 0.924 & 72.6 & 1.000 & 8.2 \\
	$m_{T2} \in [100, \infty[$ [GeV]        & 56.0 & 0.882 & 59.1 & 0.814 & 7.7 \\ \hline
			$\Delta_{\rm MC}/N_{\rm yield} $ & \multicolumn{2}{c|}{$ 4.6\% $} & \multicolumn{3}{c}{$ 1.2\% $} \\
  \botrule
  \multicolumn{6}{c}{}\\
  \toprule
  & \multicolumn{2}{c|}{ATLAS} & \multicolumn{3}{c}{{\sc MadAnalysis 5}}\\
					& Events & $\varepsilon$ & Events & $\varepsilon$ & $\delta$ [\%]\\ \hline
      Initial                                 & 1320.0 & -  & 1320.0 & - & - \\
OS dilep. with $p^{l_1,l_2}_T>25$ [GeV] & 430.0 & 0.326 & 450.7 & 0.341 & 4.8 \\
$m_{l_1,l_2}>25$ [GeV]                  & 429.0 & 0.998 & 449.9 & 0.998 & 0.1 \\
b veto                                  & 401.0 & 0.935 & 441.5 & 0.981 & 5.0 \\
SF dilep. \& $N_j=1$                    & 74.0 & 0.185 & 73.5 & 0.166 & 9.8 \\
$m_{l_1,l_2} > 121.2$ [GeV]             & 65.5 & 0.885 & 66.8 & 0.909 & 2.7 \\
$\slashed{E}_T > 110$ [GeV]             & 55.9 & 0.853 & 57.6 & 0.863 & 1.1 \\
$\slashed{E}_T$ Sig. $> 10$ [$\sqrt{\rm GeV}$]& 49.7 & 0.889 & 56.7 & 0.984 & 10.6 \\
$m_{T2} \in [100, \infty[$ [GeV]        & 41.7 & 0.839 & 44.4 & 0.783 & 6.6 \\ \hline
			$\Delta_{\rm MC}/N_{\rm yield} $ & \multicolumn{2}{c|}{$ 5.3\% $} & \multicolumn{3}{c}{$ 1.4\% $} \\
  \botrule\\\end{tabular}\label{tab:sf_chaslep}}
\end{table}

\begin{table}[t]
 \renewcommand{\arraystretch}{1.3}
   \tbl{Same as in table~\ref{tab:sf_chaslep} but for a pair of DF leptons
   produced without any jet (upper) and with one jet (lower).}
	{\begin{tabular}{@{}l|cc|ccc@{}} \toprule
			& \multicolumn{2}{c|}{ATLAS} & \multicolumn{3}{c}{{\sc MadAnalysis 5}} \\
			& Events & $\varepsilon$ & Events & $\varepsilon$ & $\delta$ [\%]\\ \hline
      Initial                                 & 1320.0 & -  & 1320.0 & - & - \\
OS dilep. with $p^{l_1,l_2}_T>25$ [GeV] & 430.0 & 0.326 & 450.7 & 0.341 & 4.8 \\
$m_{l_1,l_2}>25$ [GeV]                  & 429.0 & 0.998 & 449.9 & 0.998 & 0.1 \\
b veto                                  & 401.0 & 0.935 & 441.5 & 0.981 & 5.0 \\
DF dilep. \& $N_j=0$                    & 82.8 & 0.206 & 92.8 & 0.210 & 1.8 \\
$m_{l_1,l_2} > 100$ [GeV]               & 77.8 & 0.940 & 86.8 & 0.935 & 0.5 \\
$\slashed{E}_T > 110$ [GeV]             & 66.8 & 0.859 & 73.8 & 0.850 & 1.0 \\
$\slashed{E}_T$ Sig. $> 10$ [$\sqrt{\rm GeV}$]& 62.9 & 0.942 & 73.8 & 1.000 & 6.2 \\
$m_{T2} \in [100, \infty[$ [GeV]        & 53.8 & 0.855 & 60.3 & 0.817 & 4.5 \\ \hline
			$\Delta_{\rm MC}/N_{\rm yield} $ & \multicolumn{2}{c|}{$ 4.8\% $} & \multicolumn{3}{c}{$ 1.2\% $} \\
  \botrule
  \multicolumn{6}{c}{}\\
  \toprule
  & \multicolumn{2}{c|}{ATLAS} & \multicolumn{3}{c}{{\sc MadAnalysis 5}}\\
			& Events & $\varepsilon$ & Events & $\varepsilon$ & $\delta$ [\%]\\ \hline
      Initial                                 & 1320.0 & -  & 1320.0 & - & - \\
OS dilep. with $p^{l_1,l_2}_T>25$ [GeV] & 430.0 & 0.326 & 450.7 & 0.341 & 4.8 \\
$m_{l_1,l_2}>25$ [GeV]                  & 429.0 & 0.998 & 449.9 & 0.998 & 0.1 \\
b veto                                  & 401.0 & 0.935 & 441.5 & 0.981 & 5.0 \\
DF dilep. \& $N_j=1$                    & 66.3 & 0.165 & 72.7 & 0.165 & 0.4 \\
$m_{l_1,l_2} > 100$ [GeV]               & 61.3 & 0.925 & 68.4 & 0.941 & 1.8 \\
$\slashed{E}_T > 110$ [GeV]             & 53.4 & 0.871 & 59.0 & 0.863 & 1.0 \\
$\slashed{E}_T$ Sig. $> 10$ [$\sqrt{\rm GeV}$]& 48.6 & 0.910 & 58.0 & 0.984 & 8.1 \\
$m_{T2} \in [100, \infty[$ [GeV]        & 40.7 & 0.837 & 45.7 & 0.788 & 5.9 \\\hline
			$\Delta_{\rm MC}/N_{\rm yield} $ & \multicolumn{2}{c|}{$ 5.4\% $} & \multicolumn{3}{c}{$ 1.3\% $} \\
	\botrule\\\end{tabular}\label{tab:df_chaslep}}
\end{table}

\subsection{Conclusions}
In this validation note, we presented our efforts on re-implementing the
ATLAS-SUSY-2018-32 analysis in the \madanalysis\ framework. We have validated
our work in the context of three simplified models dedicated to the production
of electroweakinos and sleptons. The validation has been achieved by comparing
predictions obtained with our code to official results from the ATLAS
collaboration. We have obtained a good agreement at each step of the analysis
and for each of the four considered signal regions, the deviations being usually
smaller than 10\%. The largest discrepancies can be traced to difficulties in
modelling the missing energy significance, which however yields a small impact
when including the entire selection. The re-implementation is therefore
considered as validated.

The {\sc MadAnalysis}~5 C++ code is available for download from the \madanalysis\ dataverse
(\href{https://doi.org/10.14428/DVN/EA4S4D}{https://doi.org/10.14428/DVN/EA4S4D})~\cite{EA4S4D_2020}. The material relevant for the validation benchmarks has been obtained from \href{https://www.hepdata.net/record/ins1750597}{HEPData}~\cite{1750597}.

\subsection*{Acknowledgments}
The authors are grateful to Laura Jeanty and Federico Meloni for their help
with understanding the ATLAS analysis considered in this work. JYA has received
funding from the European Union’s Horizon 2020 research and innovation programme as part of the Marie Sklodowska-Curie Innovative Training Network MCnetITN3 (grant agreement no. 722104).

\cleardoublepage
\def\twomat[#1,#2][#3,#4]{\left( \begin{array}{cc} #1 & #2 \\ #3 & #4 \end{array} \right)}
\def\thv[#1,#2,#3]{\left( \begin{array}{c} #1 \\ #2 \\ #3 \end{array} \right)}
\def\twv[#1,#2]{\left( \begin{array}{c} #1 \\ #2 \end{array} \right)}
\def\lagr{\mathcal{L}}
\def\nn{\nonumber}
\def\ov{\overline}
\newcommand{\SARAH}{{\sc SARAH}\xspace}
\newcommand{\SPheno}{{\sc SPheno}\xspace}
\def\DR{\ensuremath{\overline{\mathrm{DR}}}\xspace}
\def\MS{\ensuremath{\overline{\mathrm{MS}}}\xspace}
\def\MSUSY{\ensuremath{M_{\rm SUSY}}\xspace}
\def\blog{\overline{\log}}
\def\llog{\overline{\log}}
\def\GeV{\ensuremath{\mathrm{GeV}}}
\def\TeV{\ensuremath{\mathrm{TeV}}}
\newcommand{\MG}{{\sc MadGraph5}\_a{\sc MC@NLO}\xspace}
\def\MA{{\sc MadAnalysis\ 5}\xspace}
\def\HepData{{\sc HepData}\xspace}
\def\Pythia{{\sc Pythia8}\xspace}
\def\Delphes{{\sc Delphes}\xspace}
\def\ptl{\ensuremath{p_T^\ell}}
\def\met{\ensuremath{E_T^{\rm miss}}}
\def\mbb{\ensuremath{m_{\rm bb}}}
\def\mt{\ensuremath{m_{\rm T}}}
\def\mct{\ensuremath{m_{\rm CT}}}
\def\mpt{\ensuremath{p_T^{\rm miss}}}
\def\mctbb{\ensuremath{m_{\rm CT}}}
\def\pt{\ensuremath{p_T}}
\def\fb{\mathrm{fb}}
\def\mysizecommand{\scriptsize}

\markboth{Mark D. Goodsell}{Implementation of the ATLAS-SUSY-2019-08 analysis in the \MA framework}

\section{Implementation of the ATLAS-SUSY-2019-08 analysis (electroweakinos with a Higgs decay into a $b\bar b$ pair, one lepton
  and missing transverse energy; 139~fb$^{-1}$)}
  \vspace*{-.1cm}\footnotesize{\hspace{.5cm}By Mark D. Goodsell}
\label{sec:ewkinos_bbl}


\subsection{Introduction}
\label{introduction}
 
\begin{figure}[t]\centering
  \includegraphics[width=0.4\textwidth]{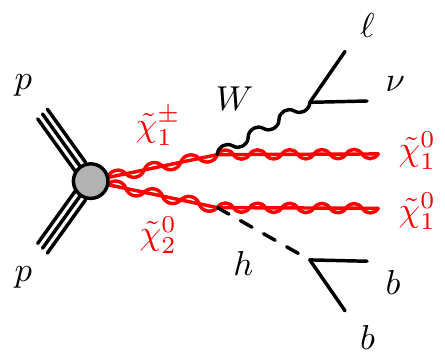} 
  \caption{The signal scenario targeted by \cite{Aad:2019vvf}, taken from that paper. Note that the simulated signal events also include up to two hard jets not shown here.}\label{FIG:SusyProcess}
\end{figure}

This note describes the recasting of the study ATLAS-SUSY-2019-08 \cite{Aad:2019vvf} in \MA~\cite{Conte:2012fm,Conte:2018vmg}, and available in the Public Analysis Database \cite{Dumont:2014tja,DVN/BUN2UX_2020}. This analysis targets electroweakinos produced in the combination of a chargino and a heavy neutralino, where the neutralino decays by emitting an on-shell Higgs, and the chargino decays by emitting a $W$ boson. The Higgs is identified by looking at $b$-jets with an invariant mass in the window $[100,140]$ GeV, while the $W$ boson is identified through leptonic decays. The typical production diagram targeted via the search is shown in figure \ref{FIG:SusyProcess}. The analysis uses $139\, \mathrm{fb}^{-1},$ and is well adapted to search for a relatively flavour-pure wino that can decay to a bino (winos being the fermionic superpartner of $W$-bosons, binos being the partners of the hypercharge, and if they are flavour-pure there is little mixing between the states) with a moderate-to-large mass splitting between the two, since a wino has a large production cross-section, and would occur as a roughly degenerate chargino/neutralino pair.

This search should be particularly effective when other supersymmetric particles (such as sleptons and additional Higgs fields) are heavy; there are other, specifically adapted searches for those cases. However, given constraints on heavy Higgs sectors and colourful particles, this analysis is rather model independent and difficult to evade in a minimal model. The assumption of chargino decay via a $W$ boson is indeed rather a good one, it should proceed typically with branching ratio close to unity, provided: (a) that there is no charged Higgs or slepton channel available, (b) the decay is kinematically allowed, and (c) the chargino is relatively pure wino (with small higgsino component). If we relax assumption (a), then the cascade decay is preferred; if we relax (b) then three-body and loop decays are preferred; if we relax assumption (c) then the decay channel via a $Z$ boson would also have a significant branching ratio. 

The ATLAS collaboration made available substantial additional data via \HepData \cite{MyHepData} at \url{https://www.hepdata.net/record/ins1755298},
in particular including detailed cutflows and tables for the exclusion curves, and full likelihoods, which are relevant for this note. For simplified model analysis they also provided efficiency maps. 

\subsection{Preselection and event cleaning}

This analysis has a number of preselection cuts on the events; I shall first summarise them as presented in the ATLAS paper and in the provided pseudocode; in subsection \ref{sec:mypreselection} I will describe how these are implemented in the recasting code.

\subsubsection{Selections defined in the ATLAS paper}

Jets are reconstructed from using the anti-$k_t$ algorithm with a radius parameter $R=0.4$~\cite{Cacciari:2008gp}, and this is done internally in \Delphes~3 \cite{deFavereau:2013fsa} using {\sc FastJet}~\cite{Cacciari:2011ma}.
Initial `soft' jets are selected in the region $|\eta|<4.5$ and have $\pt>20$~\GeV; initial `soft' leptons are defined according to the baseline kinematic and isolation criteria listed in \ref{APP:isolation} (where the criteria for signal lepton isolation are also given). To suppress jets from pile-up interactions, the jets with $|\eta|<2.8$ and $\pt<120$~\GeV\ are required to satisfy the `medium' working point of the jet vertex tagger (JVT),  a tagging algorithm that identifies jets originating from the Primary Vertex (PV) using track information.

Next, an overlap removal procedure is applied to electrons, muons and jets. First, for overlapping electrons, the electron with the lower $p_T$ is rejected; and any electron overlapping with a muon is rejected (the criterion for overlap is interpreted in the provided pseudocode and therefore in the recasting code as having $\Delta R < 0.01$). Next, electrons and muons within a cone of size $\Delta R = \min(0.4,0.04+10$ \GeV$/\pt)$ around a jet are removed, and jets  are rejected if they lie within  $\Delta R = 0.2$ of a muon. The remaining objects constitute the \emph{baseline} leptons and jets.  

From the baseline objects, signal jets are required to be in the region $|\eta|<2.8$ and have $\pt>30$~\GeV, and of these, $b$-tagged signal jets are reconstructed with $|\eta|<2.5 $\footnote{The first version of the analysis paper incorrectly gave $\pt>20$~\GeV\ for the $b$-jets.}.

\subsubsection{Implementation of preselection}
\label{sec:mypreselection}

This recast relies on detector simulation through \Delphes \cite{deFavereau:2013fsa} with a specially modified card. There are several issues with the standard ATLAS card, uncovered when comparing with the experimental cutflows:
\begin{itemize}
\item The isolation options are too simple (only a fixed $\Delta R$ is possible).
\item Too few signal leptons and jets are reconstructed with the given efficiencies. In particular, the kinematic selections on leptons start at $6$ GeV, whereas the standard reconstruction efficiencies are zero below $10$ GeV.
\item It is not possible to distinguish between ``background'' and ``signal'' leptons in terms of whether they should be clustered with jets. In ATLAS, leptons identified as coming from hadronic decays (so usually clustered into a jet and/or having a displaced vertex associated with e.g. charged pion decays) are designated ``background'' and not considered as part of the baseline leptons. In \Delphes, if we use the isolation routines, the ``unique object identifier'' will decide whether a lepton is part of a jet depending on whether it is isolated -- but isolation criteria prove to be inadequate for this job for this analysis, removing too few leptons. 
\item The $b$-tag algorithm used (MV2c10) has a quoted efficiency $77\%$ independent of $p_T$; it is not clear how this corresponds to the \Delphes $b$-tagging, but certainly the ``standard'' efficiency is much worse than this. Unfortunately, it also appears that even setting a flat $77\%$ efficiency of $b$-tagging also results in too few $b$-jets.
\item There is no default implementation of the jet vertex tagging algorithm in \Delphes. This complicates the situation regarding pile-up: if we include pile-up events in \Delphes, then we will have the wrong number of jets unless we also implement a jet vertex tagger in the analysis.
\end{itemize}

To solve these issues, I modified the \Delphes card and implemented several features directly in the analysis:
\begin{itemize}
\item Electron, muon and photon reconstruction and tracking efficiencies were modified to reflect the improved performance of ATLAS, e.g. given in \cite{Aad:2016jkr,Aad:2019tso}.
\item $\Delta R$ for jet reconstruction was set to $0.4$ as used in the analysis.
\item The Hadron calorimeter (HCAL) minimum energy and energy significance are halved; this way more jets are found and the $m_T$ distribution better matches the cutflows.
\item For $b$-tagging, a flat $77\%$ efficiency is taken for $p_T > 300$ GeV to match the MC2c10 result. For smaller values, $85\%$ is taken. This was done after investigating $b$-tagging performance for $t \ov{t}$ processes and comparing to truth jets; I found that for $p_T \lesssim 300$ GeV the $b$-tagging efficiency in \Delphes was poor. This certainly warrants further investigation.
\item Isolation is deactivated in the \Delphes card and implemented directly in the analysis. This also means that we must identify leptons/photons uniquely in the analysis, through a function labelled {\tt RemoveFakeJets}, very similar to the inbuilt \MA function {\tt JetCleaning}. 
\item To emulate the JVT and effect of pileup, efficiencies are implemented \emph{in the analysis} for jets with $|\eta| <2.8$ and for $p_T < 120$ GeV with $|\eta| < 2.5$. Removed jets have their momentum added to missing $p_T$. In addition, jets missing the baseline criteria and having $|\eta| > 2.5$ add their momentum to missing $p_T$, because they cannot contribute to the ``soft term'' defined in the analysis.
\item To remove ``background'' leptons, since we cannot access jet constituents in \MA (and so determine whether a lepton is clustered with a jet) I use as a proxy the absolute displacement of the lepton creation vertex, in addition to the isolation criteria defined in the analysis and given in the appendix. I define any electron or muon created more than $0.1$ mm from the primary vertex as ``background'' and removed. This is similar to the algorithm used by the experiments which look for the characteristic ``kink'' \cite{Aad:2016jkr} but with a (presumably) unrealistically small cutoff; this is unfortunately the best that can be done in the current framework. It therefore misses a few prompt decays from \emph{neutral} pions (which have only a very small branching ratio to electrons so this is not a problem) but is potentially dangerous in models with non-prompt chargino decays so this analysis should be used with caution in such cases. 
\end{itemize}

Once these have been applied, I then apply the cuts described in the analysis and recalled in the next section.

\subsection{Signal regions and data}
\label{SEC:CUTS}

\begin{table}[]
\begin{center}
\begin{tabular} {l | c c c }
\hline 
&  \textbf{SR-LM} & \textbf{SR-MM} & \textbf{SR-HM} \\
\hline
$N_{\mathrm{lepton}}$ & \multicolumn{3}{c}{$=$ 1}\\
$\ptl$ [\GeV] & \multicolumn{3}{c}{ $>7(6)$ for $e$($\mu$)} \\
$N_\mathrm{jet}$ & \multicolumn{3}{c}{$=$ 2 or 3}\\
$N_{b\textrm{-jet}}$ &\multicolumn{3}{c}{$=$ 2} \\
\met $[\GeV]$ & \multicolumn{3}{c}{$>240$}\\
\mbb  $[\GeV]$ & \multicolumn{3}{c}{$\in [100,140]$}\\
$m(\ell,b_1)$ $[\GeV]$ & -- & -- & $>120$ \\
\hline
\mt $[\GeV]~\mathrm{(excl.)}$&   $\in [100,160]$ & $\in [160,240]$ & $>240$ \\

\mct $[\GeV]~\mathrm{(excl.)}$ &\multicolumn{3}{c}{ $ \{ \in [180,230]$,\,$\in [230,280]$, $>280  \}$}\\
 
\hline
\mt $[\GeV]~\mathrm{(disc.)}$&   $>100$ & $>160$ & $>240$ \\
\mct $[\GeV]~\mathrm{(disc.)}$ & \multicolumn{3}{c}{ $>180$}\\
\hline
\hline 
\end{tabular}
\caption{Overview of the selection criteria for the signal regions. Each of the three `excl.' SRs is binned in three \mct\ regions.}
\label{tab:SignalRegionDef}
\end{center}
\end{table}
The signal regions are summarised in table \ref{tab:SignalRegionDef}. There are therefore 12 signal regions, which are denoted in the recasting code as {\tt XXdisc, XXlowCT, XXmedCT, XXhighCT} for {\tt XX $\in \{$ LM,MM,HM $\}$} respectively corresponding to the \textbf{SR-LM}, \textbf{SR-MM}, \textbf{SR-HM} in the table. They are split into two categories: ``disc.'' (for ``discovery'') and ``excl.'' (for ``exclusion'') which are not independent (as discussed below). In the cuts, several quantities are defined:

\begin{itemize}
\item The invariant mass of the two $b$-jets, \mbb. This targets the main decay channel of the SM Higgs boson. 
\item  $m(\ell, b_1)$, which is the invariant mass of the lepton and the leading $b$-jet.
\item The transverse mass $m_{\rm T}$ is given in the analysis paper as:
  $$ m_{\mathrm{T}} = \sqrt{2 p_{\mathrm{T}}^{\ell} \met (1-\cos[\Delta\phi({\boldsymbol p}_{\mathrm{T}}^{\ell},\mpt)])}, $$
  where $\Delta\phi({\boldsymbol p}_{\mathrm{T}}^{\ell},\mpt)$ is the azimuthal angle between ${\boldsymbol p}_{\mathrm{T}}^{\ell}$ and \mpt. This is not the same as the definition in \cite{Tovey:2008ui, Polesello:2009rn} cited by the analysis, only applying when the lepton is massless. Since the pseudocode provided with the analysis uses a predefined hidden function for the transverse mass, I choose to use the full expression in the analysis even if the effect is irrelevant.  
\item The contransverse mass of two $b$-jets, \mctbb, is defined as:
$$
\mctbb =  \sqrt{2 \pt^{b_1} \pt^{b_2}  \left ( 1+\cos\Delta\phi_{bb} \right )}, \notag
\label{eq:mctbb}
$$
where $\pt^{b_1}$ and $\pt^{b_2}$ are the transverse momenta of the two leading $b$-jets and $\Delta\phi_{bb}$ is the azimuthal angle between them. Again this differs from the cited definition in \cite{Tovey:2008ui, Polesello:2009rn}, being equal only when the $b$-jets are massless. Once again I implemented the function including masses. 
\end{itemize}

The analysis also provides sample cutflows (to which I compare results in section \ref{SEC:CUTFLOWS}) which introduce additional cuts. Most of these are self-explanatory, and I implement them in the same order in the recasting; however, the first cut is simply labelled ``$N_{\mathrm{jets},25} \ge 2 $'' which I interpret as being two jets with $p_T \ge 25$ GeV. An alternative interpretation would be $|\eta| < 2.5$; since the analysis requires two or three signal jets with $p_T > 30$ and two $b$-jets with $|\eta| < 2.5$ these choices make no difference to the final efficiency, and, since there was some difficulty matching the initial number of jets, I take the more permissive choice.

\begin{table}[]
\newcolumntype{C}{@{}>{{}}c<{{}}@{}}
\begin{center}
\setlength{\tabcolsep}{0.0pc}
{\small
\begin{tabular*}{\textwidth}{@{\extracolsep{\fill}}l*4{C}}
\noalign{\smallskip}\hline\noalign{\smallskip}
{\textbf{ SR-LM}}           & LMdisc          & LMlowCT         & LMmedCT        & LMhighCT    \\[-0.05cm]
\noalign{\smallskip}\hline\noalign{\smallskip}
Observed           & $66^*$              & $16$              & $11$              & $7$                    \\
\noalign{\smallskip}\hline\noalign{\smallskip}
Expected          & $47^* \pm 6^*$          & $8.8 \pm 2.8$          & $11.3 \pm 3.1~~$          & $7.3 \pm 1.5$              \\
\noalign{\smallskip}\hline\noalign{\smallskip}
{\textbf{ SR-MM}}           &  MMdisc          & MMlowCT         & MMmedCT        & MMhighCT   \\[-0.05cm]
\noalign{\smallskip}\hline\noalign{\smallskip}
Observed           & $32^*$              & $4$              & $7$              & $2$                    \\
\noalign{\smallskip}\hline\noalign{\smallskip}
Expected          & $20.5^* \pm 4^*$          & $4.6 \pm 1.7$          & $2.6 \pm 1.3$          & $1.4 \pm 0.6$              \\
\noalign{\smallskip}\hline\noalign{\smallskip}
{\textbf{ SR-HM}}           &  HMdisc          & HMlowCT         & HMmedCT        & HMhighCT    \\[-0.05cm]
\noalign{\smallskip}\hline\noalign{\smallskip}
Observed           & $14$              & $6$              & $5$              & $3$                    \\
\noalign{\smallskip}\hline\noalign{\smallskip}
Expected          & $8.1 \pm 2.7$          & $4.1 \pm 1.9$          & $2.9 \pm 1.3$          & $1.1 \pm 0.5$              \\
\noalign{\smallskip}\hline\noalign{\smallskip} 
\end{tabular*}
}
\end{center}
\caption{Expected background and observed events for each signal region, taken from Table 5 of \cite{Aad:2019vvf} for the exclusion regions and HMdisc; for LMdisc and MMdisc the data were scraped from Figure 4 of that reference (and are hence labelled with an asterisk), since tabulated data were not provided.}
\label{TAB:expresults}
\end{table}

The observed and expected background events for each signal region are reproduced in table \ref{TAB:expresults}. The ``disc.'' (for ``discovery'') regions are supposed to be for ``discovery and model-independent limits'' but are not independent of the other regions. The HMdisc region is the sum of all HM bins, and LMdisc includes all of the MMdisc bins as a subset. However, the LMdisc and MMdisc regions cannot be obtained from the exclusion regions, due to the $m(\ell, b_1)$ cut on the HM bins which does not apply to them. ATLAS only use the exclusion bins for setting limits in their exclusion plot; moreover, the discovery regions have excesses, and since the data is not precisely available I do not include it in the ``info'' card for the analysis so that it will not interfere with the setting of limits. However, if the user wants to use these regions, I also provide a card {\tt atlas\_susy\_2019\_08\_with\_disc\_regions.info} which includes them.

\subsection{Generation of signal events}

The signal events simulated in \cite{Aad:2019vvf} assume a simplified model with wino-like $\tilde{\chi}_2^0/\tilde{\chi}^+_1$ which are degenerate and decay to a bino-like lightest supersymmetric particle (LSP) $\tilde{\chi}_1^0$. The branching ratios of the decays $\tilde{\chi}_2^0 \rightarrow \tilde{\chi}_1^0 + h, \tilde{\chi}_1^+ \rightarrow \tilde{\chi}_1^0 + W^+_\mu$ are taken to be $100\%$, which, as described above, may not be far from realistic, although the scenario as a whole would be disfavoured as having an unrealistic relic density of dark matter. On the other hand, in the signal events, the decay $h\rightarrow bb$ and $W^+ \rightarrow \ell \nu$ are specifically selected; in the SM these rates are $58.3\%$ and $10.86\% $ to $\mu^+ \nu_\mu$, $10.71\%$ to $e^+ \nu_e$, so if we  naively simulated a general hard process and shower with the full decay table, then we would only be targeting about $12\%$ of the points before any other cuts are applied.

To reproduce the signal events from  \cite{Aad:2019vvf}, I used the standard MSSM UFO  \cite{Degrande:2011ua} file for the MSSM \cite{Christensen:2009jx,Duhr:2011se} included with \MG {\tt v2.8} \cite{Alwall:2014hca}  and spectrum files provided as auxiliary material by the analysis. The hard process is simulated in \MG {\tt v2.8} and showering is performed in \Pythia \cite{Sjostrand:2014zea}, with detector response simulated in \Delphes \cite{deFavereau:2013fsa} using a card modified as described above. The analysis uses the A14 Pythia tune \cite{TheATLAScollaboration:2014rfk}, so I include those changes in the \Pythia card (summarised in appendix \ref{APP:PYTHIA}) in addition to the choices:
\begin{lstlisting}
24:offIfAny=1 2 3 4 5 6 15 16
25:oneChannel=1 0.5876728 0 -5 5
\end{lstlisting}
These select the $W$ decays to electrons/muons, and Higgs decays to $b$-quarks, while allowing \Pythia to use its inbuilt routines for the phase-space of the decays, rather than using a flat phase-space as would be the case for SLHA decay blocks. Note that this is not the only way the filters could be used in \Pythia, however in \MG the commands are read and then reordered alphabetically (and, in fact, earlier versions would not recognise these sorts of commands) so some care is needed to make sure that only one command per particle is passed!

The simulated signal events in \cite{Aad:2019vvf} involve up to two hard jets, which are then merged with the CKKL algorithm \cite{Lonnblad:2012ix} with a merging scale of one quarter the mass of the $\tilde{\chi}_1^\pm/\tilde{\chi}_2^0$. In the analysis below, I take the default MLM merging algorithm \cite{Mangano:2006rw,Alwall:2008qv} used in \MG, but there the paremeter {\tt xqcut} is used to set the merging scale:
\begin{align}
{\tt qcut} = \frac{3}{2}{\tt xqcut} \longrightarrow {\tt xqcut} = m_{\tilde{\chi}_1^{\pm}}/6.
\end{align}
To match the cutflows provided, I simulated $150$k events at leading order in \MG, which after merging and passing to \Pythia give between $100{\rm k}$ and $120{\rm k}$ merged events depending on the point; for a comparison of the exclusion plot I use $100$k events per point. Both the ATLAS analysis and this recasting use the {\tt NNPDF2.3LO } parton distribution functions (pdfs).

ATLAS use NLO-NLL cross-sections, and so to match the final number of events I interpolate the cross-sections from \cite{Debove:2010kf,Fuks:2012qx,Fuks:2013vua,Fiaschi:2018hgm} tabulated at

\url{https://twiki.cern.ch/twiki/bin/view/LHCPhysics/SUSYCrossSections13TeVn2x1wino}

For other models, the user should use the leading-order \emph{merged} cross-sections unless an improved calculation is available. As an example of the impact of the NLO/NLL corrections, the cross-sections for the example cutflow points are:
\begin{align}
  \begin{array}{|c|c|c|} \hline
    (m (\tilde{\chi}_1^\pm, \tilde{\chi}_1^0) [\mathrm{GeV}] & \sigma^{LO} (\fb) & \sigma^{NLO-NLL} (\fb) \\ \hline
    (300, 75) &  278 &387 \pm 26\\
    (500,0) & 31.8 & 46.4 \pm 4.2\\
    (750,100) & 4.5&  6.7 \pm 0.8\\ \hline
\end{array}
\end{align}
The corrections are therefore consistently around $40\%$ to $50\%$.

\subsection{Cutflows}
\label{SEC:CUTFLOWS}

To validate the recasting, I present here the cutflows compared to all of the examples given in the \HepData repository~\cite{MyHepData}. The cutflows are weighted to match the final number of events predicted, and helpfully include uncertainties.

To compare the cutflows from \cite{Aad:2019vvf} with the recasting presented here, I define the net efficiency of each cut by
\begin{align}
  \epsilon_i^{\rm MA} \equiv&  \frac{\mathrm{sum\ of\ weights\ of\ events\ surviving\ cut\ }i}{\mathrm{sum\ of\ weights\ of\ merged\ events}}, \nn\\
\epsilon_i^{\rm ATLAS} \equiv& \frac{\mathrm{number\ of\ simulated\ events\ surviving\ cut\ }i}{\mathrm{initial\ number\ of\ weighted\ events\ (after\ cleaning)}}.
\end{align}
The analysis also provide uncertainties for their data, which I translate into uncertainties on the efficiency, while for the implementation here I can only calculate Monte-Carlo errors given by
\begin{align}
\sigma (\epsilon_i^{\rm MA}) =& \sqrt{\frac{\epsilon_i^{\rm MA}(1-\epsilon_i^{\rm MA})}{N}},
\end{align}
where $N$ is the initial number of merged events before cuts. In tables \ref{TAB:LM}, \ref{TAB:MM}, \ref{TAB:HM} I give the cutflow comparisons for all available signal regions and list in the final column the percentage error of each cut compared to those provided by ATLAS, defined as
\begin{align}
\delta_i \equiv \frac{\epsilon_i^{\rm MA} - \epsilon_i^{\rm ATLAS}}{\epsilon_i^{\rm ATLAS}} \times 100.
\end{align}
I find very good agreement (to within one standard deviation of the ATLAS result) for each cutflow, with the possible exception of the medium CT bins for the LM and MM points, where the results agree within two standard deviations. Indeed, the points with the poorest agreement also have the largest experimental uncertainties. 

For each point, I also compare the final number of events passing all cuts. This is given as
\begin{align}
  \mathrm{Number\ of\ events\ (MA)} =& 139\, \mathrm{fb}^{-1} \times \sigma (p p \rightarrow \tilde{\chi}^{\pm}_1 + \tilde{\chi}_2^0 + n j, n \le 2)\nn\\
  &\times  \epsilon_{\rm final}^{\rm MA} \times 0.583 \times 0.2157,
\end{align}
where $\epsilon_{\rm final}^{\rm MA}$ refers to the efficiency of the final cut, $\sigma$ is the cross-section for the hard process (obtained from \cite{Debove:2010kf,Fuks:2012qx,Fuks:2013vua,Fiaschi:2018hgm} as described above), and the final two factors account for the SM ratio of $H \rightarrow bb$ and $W \rightarrow \ell \nu$. This number, along with the ATLAS value, is given alongside the cutflows in tables \ref{TAB:LM}-\ref{TAB:HM}, with the Monte-Carlo uncertainty (from ``only'' simulating 150k events) and the cross-section uncertainty given separately.

\begin{table}[]\centering
  \begin{tabular}{|c|c|c|c|} \hline\hline
\multicolumn{4}{|c|}{ LM preselection cuts}\\ \hline \hline 
 Cut & $\epsilon_i^{\rm ATLAS}$ & $\epsilon_i^{\rm MA}$ & $\delta_i$  \\\hline \hline
$N_{\mathrm{jets,25}} \geq 2$           	 &$0.8116 \pm 0.0000$	& $0.7502 \pm 0.0014$&$-7.6$\% \\ \hline
1 signal lepton                         	 &$0.7053 \pm 0.0000$	& $0.6205 \pm 0.0016$&$-12.0$\% \\ \hline
Second baseline lepton veto             	 &$0.6868 \pm 0.0000$	& $0.6205 \pm 0.0016$&$-9.7$\% \\ \hline
$m_{\mathrm{T}} > $ 50 GeV              	 &$0.5601 \pm 0.0000$	& $0.4928 \pm 0.0016$&$-12.0$\% \\ \hline
$E_\mathrm{T}^\mathrm{miss} > $ 180 GeV     	 &$0.1639 \pm 0.0000$	& $0.1341 \pm 0.0011$&$-18.2$\% \\ \hline
$N_{\mathrm{jets}} \leq 3$              	 &$0.1399 \pm 0.0033$	& $0.1135 \pm 0.0010$&$-18.9$\% \\ \hline
$N_{\mathrm{b-jets}}=2$                 	 &$0.0575 \pm 0.0022$	& $0.0520 \pm 0.0007$&$-9.5$\% \\ \hline
$m_{\mathrm{bb}}$ $>$ 50 GeV            	 &$0.0575 \pm 0.0022$	& $0.0509 \pm 0.0007$&$-11.4$\% \\ \hline
$E_\mathrm{T}^\mathrm{miss} > $ 240 GeV     	 &$0.0228 \pm 0.0013$	& $0.0213 \pm 0.0005$&$-6.5$\% \\ \hline
  $m_{\mathrm{bb}}$ $\in$ [100,140] GeV   	 &$0.0175 \pm 0.0012$	& $0.0156 \pm 0.0004$&$-10.6$\% \\ \hline \hline
\multicolumn{4}{|c|}{ Region LMdisc}\\ \hline \hline
$m_{\mathrm{T}}$ $>$ 100 GeV            	 &$0.0123 \pm 0.0010$	& $0.0115 \pm 0.0003$&$-6.6$\% \\ \hline
$m_{\mathrm{CT}}$ $>$ 180 GeV           	 &$0.0097 \pm 0.0008$	& $0.0084 \pm 0.0003$&$-12.8$\% \\ \hline
 \hline Number of events (ATLAS): & \multicolumn{3}{|c|}{$58.0 \pm 5.0$} \\ \hline
    Number of events (MA): & \multicolumn{3}{|c|}{$57.0$  $ \pm 2.0$ (stat) $\pm 3.9$ (xsec)} \\  \hline \hline
\multicolumn{4}{|c|}{ Region LMlow }\\ \hline \hline
$m_{\mathrm{T}}$ $\in$ [100,160] GeV    	 &$0.0050 \pm 0.0007$	& $0.0045 \pm 0.0002$&$-10.5$\% \\ \hline
$m_{\mathrm{CT}}$ $\in$ [180,230] GeV   	 &$0.0010 \pm 0.0003$	& $0.0007 \pm 0.0001$&$-30.4$\% \\ \hline
 \hline Number of events (ATLAS): & \multicolumn{3}{|c|}{$6.2 \pm 1.7$} \\ \hline
 Number of events (MA): & \multicolumn{3}{|c|}{$4.9$  $ \pm 0.6$ (stat) $\pm 0.3$ (xsec)} \\  \hline \hline     
    \multicolumn{4}{|c|}{ Region LMmed }\\ \hline \hline
$m_{\mathrm{T}}$ $\in$ [100,160] GeV    	 &$0.0050 \pm 0.0007$	& $0.0045 \pm 0.0002$&$-10.5$\% \\ \hline
$m_{\mathrm{CT}}$ $\in$ [230,280] GeV   	 &$0.0017 \pm 0.0004$	& $0.0011 \pm 0.0001$&$-39.1$\% \\ \hline
 \hline Number of events (ATLAS): & \multicolumn{3}{|c|}{$10.5 \pm 2.2$} \\ \hline
 Number of events (MA): & \multicolumn{3}{|c|}{$7.2$  $ \pm 0.7$ (stat) $\pm 0.5$ (xsec)} \\  \hline \hline     
    \multicolumn{4}{|c|}{ Region LMhigh }\\ \hline \hline
$m_{\mathrm{T}}$ $\in$ [100,160] GeV    	 &$0.0050 \pm 0.0007$	& $0.0045 \pm 0.0002$&$-10.5$\% \\ \hline
$m_{\mathrm{CT}}$ $>$ 280 GeV           	 &$0.0018 \pm 0.0004$	& $0.0020 \pm 0.0001$&$13.8$\% \\ \hline
 \hline Number of events (ATLAS): & \multicolumn{3}{|c|}{$10.6 \pm 2.3$} \\ \hline
 Number of events (MA): & \multicolumn{3}{|c|}{$13.6$  $ \pm 1.0$ (stat) $\pm 0.9$ (xsec)} \\  \hline \hline     
\end{tabular}
\caption{Cutflow comparison for Low Mass signal regions.}\label{TAB:LM}
\end{table}

\begin{table}[]\centering
  \begin{tabular}{|c|c|c|c|} \hline\hline
      \multicolumn{4}{|c|}{ MM preselection cuts}\\ \hline \hline 
  Cut & $\epsilon_i^{\rm ATLAS}$ & $\epsilon_i^{\rm MA}$ & $\delta_i$  \\\hline \hline
$N_{\mathrm{jets,25}} \geq 2$           	 &$0.8666 \pm 0.0000$	& $0.7859 \pm 0.0012$&$-9.3$\% \\ \hline
1 signal lepton                         	 &$0.7598 \pm 0.0000$	& $0.6784 \pm 0.0014$&$-10.7$\% \\ \hline
Second baseline lepton veto             	 &$0.7370 \pm 0.0000$	& $0.6784 \pm 0.0014$&$-7.9$\% \\ \hline
$m_{\mathrm{T}} > $ 50 GeV              	 &$0.6531 \pm 0.0000$	& $0.5951 \pm 0.0015$&$-8.9$\% \\ \hline
$E_\mathrm{T}^\mathrm{miss} > $ 180 GeV     	 &$0.4574 \pm 0.0000$	& $0.3943 \pm 0.0015$&$-13.8$\% \\ \hline
$N_{\mathrm{jets}} \leq 3$              	 &$0.3837 \pm 0.0089$	& $0.3465 \pm 0.0014$&$-9.7$\% \\ \hline
$N_{\mathrm{b-jets}}=2$                 	 &$0.1601 \pm 0.0051$	& $0.1778 \pm 0.0012$&$11.1$\% \\ \hline
$m_{\mathrm{bb}}$ $>$ 50 GeV            	 &$0.1588 \pm 0.0051$	& $0.1751 \pm 0.0011$&$10.3$\% \\ \hline
$E_\mathrm{T}^\mathrm{miss} > $ 240 GeV     	 &$0.1131 \pm 0.0051$	& $0.1236 \pm 0.0010$&$9.3$\% \\ \hline
    $m_{\mathrm{bb}}$ $\in$ [100,140] GeV   	 &$0.0865 \pm 0.0042$	& $0.0896 \pm 0.0009$&$3.5$\% \\ \hline
\multicolumn{4}{|c|}{Region MMdisc}\\ \hline \hline
$m_{\mathrm{T}}$ $>$ 160 GeV            	 &$0.0665 \pm 0.0037$	& $0.0691 \pm 0.0008$&$4.0$\% \\ \hline
$m_{\mathrm{CT}}$ $>$ 180 GeV           	 &$0.0485 \pm 0.0032$	& $0.0494 \pm 0.0007$&$1.8$\% \\ \hline\hline
Number of events (ATLAS): & \multicolumn{3}{|c|}{$38.2 \pm 2.5$} \\ \hline
Number of events (MA): & \multicolumn{3}{|c|}{$40.0$  $ \pm 0.5$ (stat) $\pm 3.6$ (xsec)} \\  \hline \hline
\multicolumn{4}{|c|}{Region MMlow}\\ \hline \hline
$m_{\mathrm{T}}$ $\in$ [160,240] GeV    	 &$0.0161 \pm 0.0018$	& $0.0143 \pm 0.0004$&$-11.6$\% \\ \hline
$m_{\mathrm{CT}}$ $\in$ [180,230] GeV   	 &$0.0033 \pm 0.0008$	& $0.0025 \pm 0.0001$&$-25.5$\% \\ \hline\hline
Number of events (ATLAS): & \multicolumn{3}{|c|}{$2.6 \pm 0.6$} \\ \hline
Number of events (MA): & \multicolumn{3}{|c|}{$2.0$  $ \pm 0.1$ (stat) $\pm 0.2$ (xsec)} \\  \hline \hline 
\multicolumn{4}{|c|}{Region MMmed}\\ \hline \hline
$m_{\mathrm{T}}$ $\in$ [160,240] GeV    	 &$0.0161 \pm 0.0018$	& $0.0143 \pm 0.0004$&$-11.6$\% \\ \hline
$m_{\mathrm{CT}}$ $\in$ [230,280] GeV   	 &$0.0043 \pm 0.0009$	& $0.0029 \pm 0.0002$&$-32.1$\% \\ \hline
 \hline Number of events (ATLAS): & \multicolumn{3}{|c|}{$3.4 \pm 0.7$} \\ \hline
 Number of events (MA): & \multicolumn{3}{|c|}{$2.4$  $ \pm 0.1$ (stat) $\pm 0.2$ (xsec)} \\  \hline \hline 
\multicolumn{4}{|c|}{Region MMhigh}\\ \hline \hline
$m_{\mathrm{T}}$ $\in$ [160,240] GeV    	 &$0.0161 \pm 0.0018$	& $0.0143 \pm 0.0004$&$-11.6$\% \\ \hline
$m_{\mathrm{CT}}$ $>$ 280 GeV           	 &$0.0069 \pm 0.0011$	& $0.0075 \pm 0.0003$&$9.0$\% \\ \hline
 \hline Number of events (ATLAS): & \multicolumn{3}{|c|}{$5.4 \pm 0.9$} \\ \hline
 Number of events (MA): & \multicolumn{3}{|c|}{$6.1$  $ \pm 0.2$ (stat) $\pm 0.5$ (xsec)} \\  \hline \hline     
\end{tabular}
\caption{Cutflow comparison for Medium Mass signal regions.}\label{TAB:MM}
\end{table}

\begin{table}[]\centering
  \begin{tabular}{|c|c|c|c|} \hline\hline
 \multicolumn{4}{|c|}{ HM preselection cuts}\\ \hline \hline    
 Cut & $\epsilon_i^{\rm ATLAS}$ & $\epsilon_i^{\rm MA}$ & $\delta_i$  \\\hline \hline
$N_{\mathrm{jets,25}} \geq 2$           	 &$0.8833 \pm 0.0000$	& $0.7975 \pm 0.0012$&$-9.7$\% \\ \hline
1 signal lepton                         	 &$0.7917 \pm 0.0000$	& $0.7003 \pm 0.0013$&$-11.5$\% \\ \hline
Second baseline lepton veto             	 &$0.7667 \pm 0.0000$	& $0.7003 \pm 0.0013$&$-8.7$\% \\ \hline
$m_{\mathrm{T}} > $ 50 GeV              	 &$0.7083 \pm 0.0000$	& $0.6413 \pm 0.0014$&$-9.5$\% \\ \hline
$E_\mathrm{T}^\mathrm{miss} > $ 180 GeV     	 &$0.6083 \pm 0.0000$	& $0.5301 \pm 0.0014$&$-12.9$\% \\ \hline
$N_{\mathrm{jets}} \leq 3$              	 &$0.5092 \pm 0.0117$	& $0.4680 \pm 0.0014$&$-8.1$\% \\ \hline
$N_{\mathrm{b-jets}}=2$                 	 &$0.2258 \pm 0.0075$	& $0.2459 \pm 0.0012$&$8.9$\% \\ \hline
$m_{\mathrm{bb}}$ $>$ 50 GeV            	 &$0.2250 \pm 0.0075$	& $0.2432 \pm 0.0012$&$8.1$\% \\ \hline
$E_\mathrm{T}^\mathrm{miss} > $ 240 GeV     	 &$0.1917 \pm 0.0075$	& $0.2072 \pm 0.0012$&$8.1$\% \\ \hline
$m_{\mathrm{bb}}$ $\in$ [100,140] GeV   	 &$0.1450 \pm 0.0067$	& $0.1502 \pm 0.0010$&$3.6$\% \\ \hline
  $m_{\mathrm{\ell,b_1}}$ $>$ 120 GeV     	 &$0.1350 \pm 0.0058$	& $0.1383 \pm 0.0010$&$2.5$\% \\ \hline
$m_{\mathrm{T}}$ $>$ 240 GeV            	 &$0.0967 \pm 0.0050$	& $0.1022 \pm 0.0009$&$5.7$\% \\ \hline
\multicolumn{4}{|c|}{ Region HMdisc}\\ \hline \hline    
$m_{\mathrm{CT}}$ $>$ 180 GeV           	 &$0.0842 \pm 0.0050$	& $0.0868 \pm 0.0008$&$3.2$\% \\ \hline
 \hline Number of events (ATLAS): & \multicolumn{3}{|c|}{$10.1 \pm 0.6$} \\ \hline
  Number of events (MA): & \multicolumn{3}{|c|}{$10.2$  $ \pm 0.1$ (stat) $\pm 1.2$ (xsec)} \\  \hline \hline
  \multicolumn{4}{|c|}{ Region HMlow}\\ \hline \hline
$m_{\mathrm{CT}}$ $\in$ [180,230] GeV   	 &$0.0158 \pm 0.0021$	& $0.0149 \pm 0.0003$&$-5.6$\% \\ \hline
 \hline Number of events (ATLAS): & \multicolumn{3}{|c|}{$1.9 \pm 0.2$} \\ \hline
 Number of events (MA): & \multicolumn{3}{|c|}{$1.7$  $ \pm 0.0$ (stat) $\pm 0.2$ (xsec)} \\  \hline \hline   
  \multicolumn{4}{|c|}{ Region HMmed}\\ \hline \hline
$m_{\mathrm{CT}}$ $\in$ [230,280] GeV   	 &$0.0182 \pm 0.0022$	& $0.0175 \pm 0.0004$&$-4.3$\% \\ \hline
 \hline Number of events (ATLAS): & \multicolumn{3}{|c|}{$2.2 \pm 0.3$} \\ \hline
 Number of events (MA): & \multicolumn{3}{|c|}{$2.0$  $ \pm 0.0$ (stat) $\pm 0.2$ (xsec)} \\  \hline \hline  
  \multicolumn{4}{|c|}{ Region HMhigh}\\ \hline \hline
$m_{\mathrm{CT}}$ $>$ 280 GeV           	 &$0.0500 \pm 0.0042$	& $0.0545 \pm 0.0007$&$9.0$\% \\ \hline
 \hline Number of events (ATLAS): & \multicolumn{3}{|c|}{$6.0 \pm 0.5$} \\ \hline
 Number of events (MA): & \multicolumn{3}{|c|}{$6.4$  $ \pm 0.1$ (stat) $\pm 0.7$ (xsec)} \\  \hline \hline  
\end{tabular}
\caption{Cutflow comparison for High Mass signal regions.}\label{TAB:HM}
\end{table}

\newpage

\subsection{Comparison of exclusion plot}
\label{SEC:PLOT}

\begin{figure}[]\centering
  \includegraphics[width=0.8\textwidth]{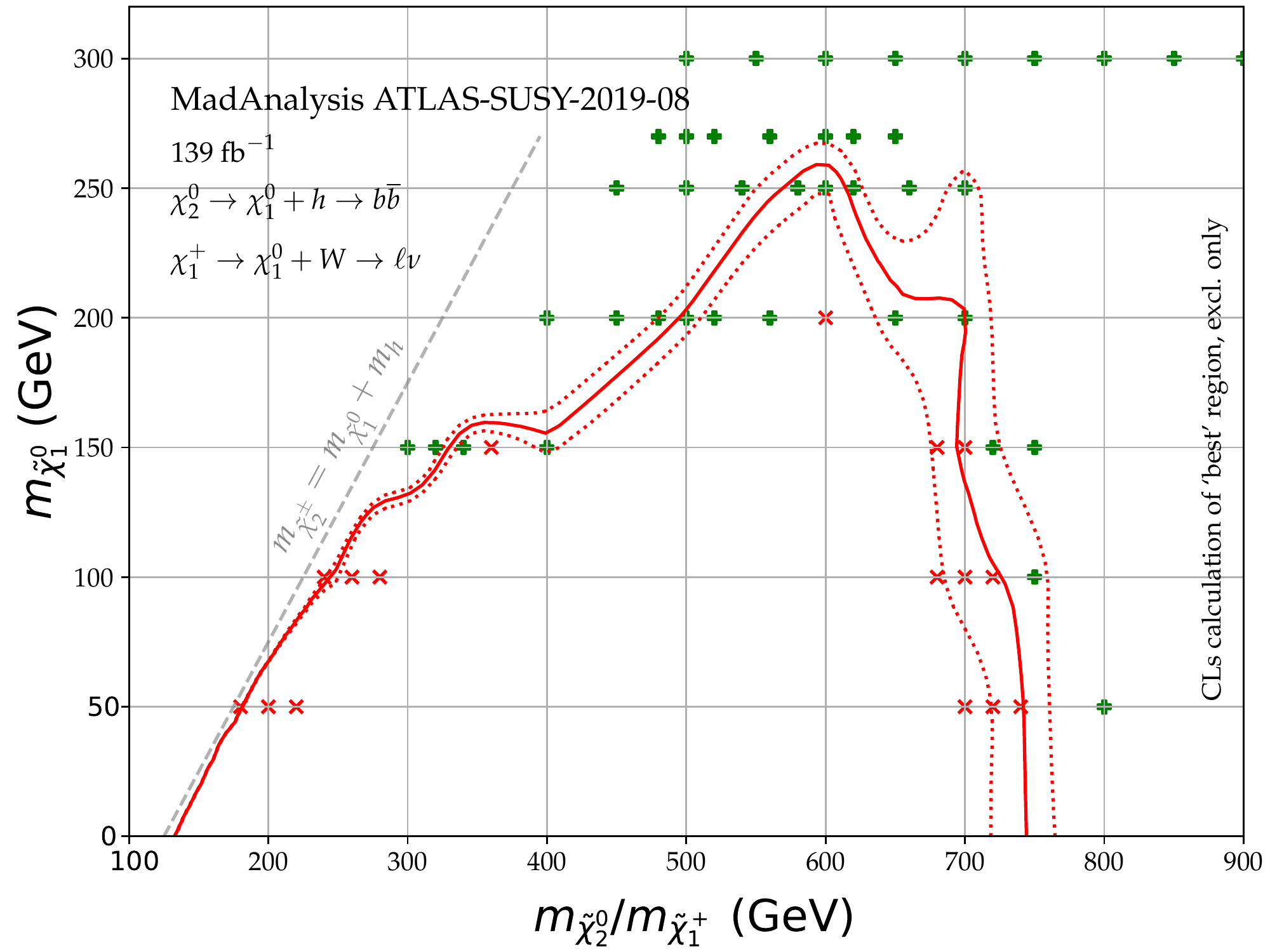}
  \caption{Comparison of experimental exclusion contour (solid red line; dotted lines are the $\pm 1 \sigma$ variation) provided in \HepData with the points simulated and tested with this analysis, using the standard \MA procedure for setting limits based on the ``best'' region. Points excluded at $1-\mathrm{CL}_s > 0.95$ are marked with red crosses; non-excluded points are shown as green plusses.}\label{FIG:Exclusion}
\end{figure}

\begin{figure}[]\centering
  \includegraphics[width=0.8\textwidth]{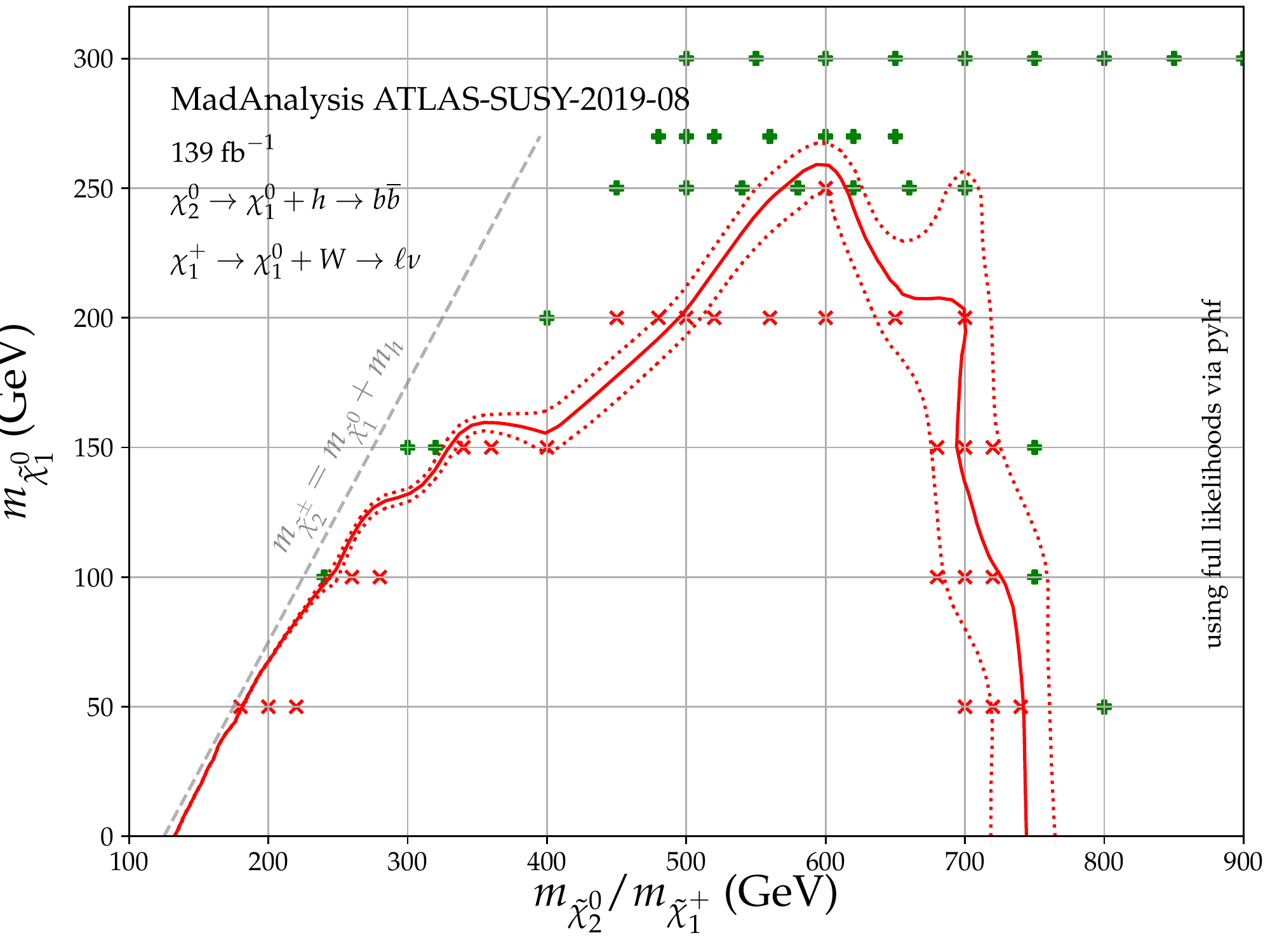}
  \caption{Comparison of experimental exclusion contour (solid red line; dotted lines are the $\pm 1 \sigma$ variation) provided in \HepData with the points simulated and tested with this analysis, with statistics calculated using {\tt pyhf} \cite{pyhf} and the provided background likelihood. Points excluded at $1-\mathrm{CL}_s > 0.95$ are marked with red crosses; non-excluded points are shown as green plusses.}\label{FIG:ExclusionPyhf}
\end{figure}

To make a final comparison of the recasting quality, I also present in figure~\ref{FIG:Exclusion} a reconstruction of the excluded region in the $m_{\tilde{\chi}_2^0}$--$m_{\tilde{\chi}_1^0}$ plane using the procedure outlined above, by simulating a selection of points at given masses marked in the plot, and compare to the contour from \cite{Aad:2019vvf}. A point is considered excluded if $1-\mathrm{CL}_s > 0.95 $, where $\mathrm{CL}_s$ is determined by the procedure in \cite{Read:2002hq} and implemented in \MA.
As per the default \MA procedure, the $\mathrm{CL}_s$ value is computed separately for each \emph{exclusion} signal region, \emph{excluding the discovery regions}\footnote{I investigated including the discovery regions, and found worse agreement with the experimental plot, including in particular non-excluded points at lower masses since the region LMdisc has an excess.} (as discussed in section \ref{SEC:CUTS}), using the data in table \ref{TAB:expresults}, and the limit is taken from the signal region which has the smallest \emph{expected} $95\%$ confidence-level limit on the cross-section (that is, treating the observed number of events as equal to the expected background) for regions where the efficiency of the signal is not zero. We see that the exclusion contour from \cite{Aad:2019vvf} is reasonably well reconstructed by the recasting presented here. 

On the other hand, for this analysis, full likelihoods are available in \HepData \cite{MyHepData}, and so I make use of them via a private code adapted from the approach in {\tt SModelS} \cite{Alguero:2020grj}. The background-only likelihood contains data for the exclusion signal regions, so I patch it with the expected number of events for each of these, and remove the ``other'' (CR/VR) regions, and compute the $\mathrm{CL}_s$ value with {\tt pyhf} \cite{pyhf}. The results are shown in figure \ref{FIG:ExclusionPyhf}, which shows a very good agreement with the experimental plot. Since a future update of \MA will include a separate implementation of this calculation and a more thorough investigation for this analysis, I do not provide my code with this analysis. However, we see from the two comparisons that the full likelihood calculation gives a much better agreement for the exclusion contour, increasing the reach in $m_{\tilde{\chi}_1^0}$ from $200$ GeV to $250$ GeV -- although the reach on the mass of $m_{\tilde{\chi}_1^{\pm}}/m_{\tilde{\chi}_2^{0}}$ is not much affected.

\subsection{Conclusions}

The recast described in this note, implemented in the \MA framework and using fast detector simulation through \Delphes with a custom card, can well reproduce the experimental cutflows and exclusion plot for a wino-like electroweakino decaying to a bino-like lightest neutralino. The code is available online from the \MA dataverse~\cite{DVN/BUN2UX_2020}, at \href{https://doi.org/10.14428/DVN/BUN2UX}{https://doi.org/10.14428/DVN/BUN2UX},  which also includes the custom \Delphes and \Pythia cards. The code described here can therefore be applied to other models/scenarios, as was done using an early version in \cite{Goodsell:2020lpx} in the Minimal Dirac Gaugino Model, with the caveat that only promptly decaying particles will be reliably constrained (due to the method employed to eliminate ``background'' leptons). I also identified several other areas for future investigation: improvements in modelling the $b$-tagging, jet reconstruction efficiency, isolation, JVT and the missing energy calculation, in order to accurately match recast analyses to recent ATLAS studies.

\subsection*{Acknowledgments}

MDG acknowledges support from the grant
\mbox{``HiggsAutomator''} of the Agence Nationale de la Recherche
(ANR) (ANR-15-CE31-0002). I thank Sabine Kraml, Sophie Williamson and Humberto Gonzales Reyes for collaboration on a related project using this analysis; and Jack Araz and Benjamin Fuks for helpful discussions and comments on the draft.

\setcounter{subsection}{0}
\renewcommand{\thesubsection}{Appendix~\arabic{section}.\Alph{subsection}}
\renewcommand{\theHsubsection}{Appendix~\arabic{section}.\Alph{subsection}}
\subsection{Lepton isolation}
\label{APP:isolation}

The baseline electrons are required to be \emph{Loose} (or \emph{FCLoose} -- ``Fixed Cut Loose''). The signal ones are \emph{tight}, and for $p_T< 200$ GeV, and \emph{additionally} \emph{FCHighPtCaloOnly} for higher $p_T$. These are described in \cite{Aad:2019tso} page 37:
\begin{align}
  FCLoose: \qquad E_T^{\rm cone 20}/p_T < 0.2, \qquad p_T^{\rm varcone 20}/p_T < 0.15, \nn\\
  Tight: \qquad E_T^{\rm cone 20}/p_T < 0.06, \qquad p_T^{\rm varcone 20}/p_T < 0.06,\nn\\
  FCHighPtCaloOnly: \qquad E_T^{\rm cone 20} < \mathrm{max}(0.015 \times p_T, 3.5 \mathrm{GeV}).
\end{align}
Baseline electrons have $ p_T > 7$ GeV, $\eta < 2.47$, no further cuts on these are imposed on the signal.

For the muons, they should be \emph{FCLoose}, described in \cite{Aad:2016jkr} page 17:
\begin{align}
p_T^{\rm varcone 30}/p_T^\mu < 0.15, \qquad E_T^{\rm topocone 20}/p_T^\mu < 0.30.
\end{align}
Baseline muons are \emph{medium} with $p_T > 6 $ GeV, $\eta < 2.7$; the detailed ID criteria for \emph{medium} muons are relevant only to the actual ATLAS experiment and are not given in the paper. 
Signal muons have $\eta < 2.5$. 

The quantities above are defined as:
\begin{itemize}
\item $p_T^{\rm varcone 30} $ is the scalar sum of the transverse momenta of he tracks wih $p_T > 1 $ GeV in a cone of size $\Delta R = \mathrm{min}(10\ \mathrm{GeV}/p_T^\mu,\Delta R_{\rm max})$, \emph{excluding the electron/muon track itself}, where $\Delta R_{\rm max} $ is 0.3 for muons, 0.2 for electrons.
\item $p_T^{\rm coneXX}$ is the same  but with a fixed cone.
\item $E_T^{\rm topocone 20} $ is the sum of the transverse energy in the ``topological clusters'' in a cone of size $\Delta R = 0.2$ around the muon, after subtracting the energy of the muon itself. I treat this as being the total transverse energy recovered in the given cone. 
\item $E_T^{\rm cone 20} $ is considered to be the same as $E_T^{\rm topocone 20} $ for this analysis. 
\end{itemize}

\subsection{Pythia settings}
\label{APP:PYTHIA}

The \Pythia card provided with this analysis gives the A14 tune \cite{TheATLAScollaboration:2014rfk} parameters as well as the filters to select $W$ decays to electrons/muons and Higgs decays to $b$-quarks. I reproduce them here:
\begin{lstlisting}
24:offIfAny=1 2 3 4 5 6 15 16
25:oneChannel=1 0.5876728 0 -5 5
SigmaProcess:alphaSvalue = 0.140 
SpaceShower:pT0Ref = 1.56
SpaceShower:pTmaxFudge = 0.91
SpaceShower:pTdampFudge = 1.05
SpaceShower:alphaSvalue = 0.127
TimeShower:alphaSvalue = 0.127
BeamRemnants:primordialKThard = 1.88
MultipartonInteractions:pT0Ref = 2.09
MultipartonInteractions:alphaSvalue = 0.126
!BeamRemnants:reconnectRange = 1.71
\end{lstlisting}
The final command appears to be incompatible with the MLM merging and so I comment it out.
\renewcommand{\thesubsection}{\arabic{section}.\arabic{subsection}}
\renewcommand{\theHsubsection}{\arabic{section}.\arabic{subsection}}

\cleardoublepage
\markboth{Malte Mrowietz, Sam Bein and Jory Sonneveld}{Implementation of the CMS-SUS-19-006 analysis}

\lstset { %
    language=C++,
    backgroundcolor=\color{black!5}, 
    basicstyle=\footnotesize,
}

\newcommand{\mht}{$H_{\text{T}}^{\text{miss}}~$}
\newcommand{\HT}{$H_{\text{T}}~$}

\section{Implementation of the CMS-SUS-19-006 analysis (supersymmetry with large hadronic activity and missing
   transverse energy; 137~fb$^{-1}$)}
  \vspace*{-.1cm}\footnotesize{\hspace{.5cm}By Malte Mrowietz, Sam Bein and Jory Sonneveld}
\label{sec:susyhad}


\subsection{Introduction}
Proton-proton collisions (events) that feature significant hadronic activity in combination with large missing transverse momentum $E_{T}^{\text{miss}}$ in the final state can act as a probe for a general class of beyond the Standard Model (BSM) models. In particular, models of $R$-parity conserving supersymmetry (SUSY) that feature TeV-scale squarks or gluinos often have these attributes as hallmark signal event characteristics. Therefore, the data analyzed in the all-hadronic multi-jet channel \cite{Sirunyan:2019ctn} provide an important constraint on generic dark matter models and strong-production SUSY. 

A \texttt{MadAnalysis 5}~\cite{Conte:2012fm,Conte:2014zja,Dumont:2014tja,Conte:2018vmg} implementation of the CMS-SUS-19-006 analysis of ref.~\cite{Sirunyan:2019ctn} has been carried out for the purpose of allowing for the reinterpretation of the results of this search in any new physics context~\cite{4DEJQM_2020}. This note provides supporting documentation for the implementation and details steps taken to validate the work using information made public by CMS. This information pertains to the efficiency and acceptance of signal events of benchmark points within the simplified models of gluino pair production with decays of $\tilde{g}\rightarrow q\bar{q}\tilde{\chi}_{1}^{0}$, $\tilde{g}\rightarrow t\bar{t}\tilde{\chi}_{1}^{0}$, and $\tilde{g}\rightarrow b\bar{b}\tilde{\chi}_{1}^{0}$, denoted T1qqqq, T1tttt, T1bbbb, T5qqqqVV, respectively, as well as squark pair production featuring $\tilde{q}\rightarrow q\tilde{\chi}_{1}^{0}$, $\tilde{q}\rightarrow b\tilde{\chi}_{1}^{0}$, and $\tilde{t}\rightarrow q\tilde{\chi}_{1}^{0}$, denoted T2qq, T2tt, and T2bb, respectively. 
Additionally, the T5qqqqVV simplified model is considered. Its signature is identical to the T1qqqq one, but with an intermediate electroweakino entering the decay chain. 

\subsection{Description of the analysis}
The analysis defines 174 signal regions (SRs) that target a variety of final states. The region definitions are based on requirements on the missing transverse hadronic activity \mht, the transverse hadronic activity \HT, jet and $b$-jet multiplicity. Here, the missing transverse hadronic momentum \mht is used as a proxy for the missing transverse energy $E_{T}^{\text{miss}}$. The lower multiplicity regions probe squark pair production models where large multiplicities are more sensitive to gluinos. Categories with large $b$-jet multiplicity help target scenarios with kinematically accessible third-generation squarks, while the 0-$b$ bins increase sensitivity to first and second generation squark models. Larger and smaller \mht and \HT regions respectively target compressed and uncompressed mass spectra.

\subsubsection{Object definitions}
The primary objects used in the CMS analysis are particle flow jets, obtained by a clustering of all reconstructed particles with trajectories pointing to the primary vertex using the anti-$k_{\text{T}}$~\cite{Cacciari:2008gp} jet algorithm with a cone size parameter of 0.4. Jets are required to have 
\begin{itemize}
\item $p_{\text{T}}>30$ GeV and 
\item $|\eta|<5.0$. 
\end{itemize}
Because particle flow jets are the basis of the calculation of $E_{\text{T}}^{\text{miss}}$, they are inclusive with respect to all reconstructed energy in an event.  To emulate this behavior in \texttt{Delphes}, we avoid the use of the \texttt{UniqueObjectFinder} module, and this is consistent with the detector card recommended as default in association with \texttt{MadAnalysis 5}. 

Detector smearing of hadron energy is kept as default, and the jet energy scale (JES) applied by default in \texttt{Delphes} has been removed. Additionally, the jet cone size parameter for the anti-$k_{\text{T}}$ algorithm used in \texttt{Delphes} has been changed from the default to R=0.4, as required in the paper. Finally, the cone size parameter for flavor assignment in the \texttt{Delphes} card has been changed to 0.4. The \texttt{MadAnalysis 5} interface to \texttt{Delphes} also removes jets which are identified as originating from $\tau$ lepton decays from the \texttt{Jet} collection. These jets are added back in to the jet collection in the implemented C++ analyzer code. 

Leptons are identified if they point to the primary vertex, are isolated, and satisfy
\begin{itemize}
\item $p_{\text{T}}>$10 GeV and 
\item $|\eta|<2.5$ (2.4) for electrons (muons)
\item $I<0.1 (0.2)$ for electrons (muons).
\end{itemize}
For the isolation $I$, we implement the so-called ``mini" relative isolation definition, which for a lepton candidate $i$, is given by 
\begin{itemize}
\item $(1/p_{Ti})\sum_{j\neq i}^{n} p_{Tj}<0.2$.
\end{itemize}
Here, the sum runs over all particles $j$ with a cone of variable radius around the candidate lepton.  The radius is given by 
\[
R^* = 
  \begin{cases} 
  0.2 & \text{: } p_{T} \leq 50 \text{ GeV} \\
  (10 \text{ GeV})/p_{T} & \text{: } 50 < p_{T} \leq 200 \text{ GeV} \\  
   0.05 & \text{: } p_{T} > 200 \text{ GeV}. \\
  \end{cases}
\]
\noindent 
Photons are identified if they have 
\begin{itemize}
\item $p_{\text{T}}>$100 GeV and 
\item $|\eta|<2.5$,
\end{itemize}
and are relatively isolated based on a fixed cone size of 0.3. Note that the isolation criterion for photons is somewhat simplified compared to the paper, since the CMS isolation is performed for each component's contribution to the $p_{T}$ sum, the components being: charged hadrons, neutral hadrons, and electromagnetic particles. The impact of the photon veto on signal efficiency is small or negligible for the interpreted models. The analysis also applies a veto based on the presence of isolated tracks which were not identified as a lepton, aimed at further suppressing backgrounds from $W$+Jets and $t\bar{t}$ processes. Slightly different object criteria are placed on isolated tracks attributed to electrons, muons, and pions, all together summarized as
\begin{itemize}
\item $p_{\text{T}}>$5 GeV,
\item $|\eta|<2.4$,
\item $m_{\text{T}}(\text{track},H_{\text{T}}^{\text{miss}})<100$ GeV, and
\item $I<0.2$ (0.1) for electron and muon (pion) tracks,
\end{itemize}
where $I$ is the relative isolation taken with respect to other tracks within a constant-size cone of radius 0.3 around the candidate track. 

\subsubsection{Event selection}
The baseline selection is as follows:
\begin{itemize}
\item $H_{\text{T}}^{\text{miss}} = |\vec{H}_{\text{T}}^{\text{miss}}| > 300$ GeV, where $\vec{H}_{\text{T}}^{\text{miss}}$ is the negative $\vec{p}_{T}$ sum of all selected jets;
\item $H_{\text{T}}>300$ GeV, where $H_{\text{T}}$ is the scalar sum of the $p_{\text{T}}$ of jets within $|\eta|<2.4$;
\item $H_{\text{T}}>H_{\text{T}}^{\text{miss}}$;
\item $n_{j}>1$, where $n_{j}$ is the number of jets within $|\eta|<2.4$;
\item $n_{e}=n_{\mu}=n_{\gamma}=n_{\text{iso}\,\text{tracks}}=0$;
\item $\Delta\phi(\vec{j}_{1,2,3,4}, \vec{H}_{\text{T}}^{\text{miss}})>0.5,0.5,0.3,0.3,$ where \{$j_{i}$\} is the $p_{\text{T}}$-ordered list of jets in a given event.
\end{itemize}
Our implementation accounts for cases in which long-lived charginos are reconstructed as muons, and thus trigger the muon event veto. This is done by treating any chargino with a decay length of $>3$ m to be a muon. 

\begin{figure}[t]
\centering
\includegraphics[width=.65\linewidth]{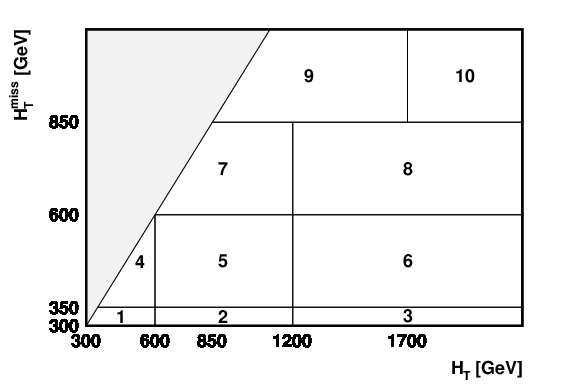}
\caption{Boundaries in the $H_{T}-H_{T}^{\text{miss}}$ plane which go into defining the final search bins \cite{Sirunyan:2019ctn}.}
\label{fig:htmhtboundaries}
\end{figure}

Events passing the baseline selection are further categorized into orthogonal signal regions defined by ranges of $H_{\text{T}}^{\text{miss}}$, $H_{\text{T}}$, $n_{j}$, and $n_{b}$. The boundaries in the $H_{T}-H_{T}^{\text{miss}}$ plane are shown in Fig. \ref{fig:htmhtboundaries}. Each region is further split into categories based on $n_{jets}$ of [2,3], [4,5], [6,8], [8,10], $>10$, and $n_{b-\text{tags}}=0,1,2,\geq3$, and the complete list of signal regions is given in Tables 3-7 of ref.~\cite{Sirunyan:2019ctn}. It is noted that the search bins correspond to the regions 1 and 4 in Fig. \ref{fig:htmhtboundaries} are dropped for N $\geq$ 8.

An alternative, smaller set of aggregate signal regions, totaling 12 in number, is also defined in Table 9 of ref.~\cite{Sirunyan:2019ctn}. The former SR's are mutually exclusive event categories and thus can be more safely used in a combination analysis, whereas the latter aggregate regions have significant overlaps in phase space. Each aggregate region has been defined in a way to give reasonably good sensitivity to a particular type of signal model. For example, Aggregate region 11, which requires $H_{T}$ and $H_{T}^{\text{miss}}>600$ GeV, $>5$ jets, and at least 1 $b$-tagged jet, should provide good sensitivity to models with top squark production, particularly in uncompressed mass regimes. By contrast, Bin 1 is more inclusive and may be most suitable for probing generic 1st generation squark or dark matter production, given that it is more inclusive and that it vetoes events with one or more $b$-jets. In some cases, the use of a likelihood based on a single aggregate bin, e.g., that giving the largest expected significance, can be a good choice for establishing constraints on a model. The the most sensitive approach is to use a combination of several or even all 174 bins in a likelihood. CMS has provided covariance and correlation matrixes for the regions, which can be utilized for this purpose. 

The provided analysis code produces an estimate of the signal acceptance calculation for all 174 signal regions, as well as the aggregated signal regions. 

\subsection{Validation}
We validate the implementation based on cut flow tables and distributions of kinematic observables provided by the analysis in its public webspace~\cite{1749379}. The needed event samples as well as the results of the validation are described in the following. 

\subsubsection{Event samples}
Simulated event samples have been prepared for a host of benchmark signal model scenarios, corresponding to the simplified models specified in the introduction. For each model (Tables \ref{cutflow:T1bbbb-1800-200}-\ref{yield:T2tt-950-100}) one compressed and one uncompressed scenario has been examined. The event generation has been carried out using \texttt{MadGraph5\_aMC@NLO} version 2.7.2\cite{Alwall:2014hca}, making use of the UFO\cite{Degrande:2011ua} file \texttt{MSSM\_SLHA2}\cite{Duhr:2011se} and relying on matrix elements including up to three additional partonic jet constituents. The parton distribution function (PDF) used is \texttt{NNPDF23\_nlo\_as\_0119}~\cite{Ball:2012cx} as implemented with  \texttt{LHAPDF}\cite{Buckley:2014ana}. Gluino and squark decays are implemented in Pythia8~\cite{Sjostrand:2007gs,Sjostrand:2014zea}, as well as parton showering and hadronization. Jet merging is implemented with an \texttt{xqcut} value of 30 GeV and \texttt{qcut} parameter values ranging from 68 to 171 according to the mother particle mass. For additional information, equivalent event samples have been generated and hadronized (full chain) using leading order (LO) \texttt{Pythia8}. In all cases the \texttt{Delphes} implementation with a lightly modified detector card has been used to simulate the response of the CMS detector to these events. The modifications include the inclusion of the CMS $b$-tagging efficiency parameterization for the \textit{deepCSV} flavor algorithm, provided in ref.~\cite{Sirunyan:2017ezt}. The medium working point efficiency has been used, as that is what is used by the analysis.  This parameterization has been modified by removing the quadratic term from the $p_{T}>250\,$GeV part and approximating with a constant efficiency for $p_{T}>1300\,$GeV. The main reason for this is that the parameterization is not defined for $p_{T}>1000$, and the functional form with the quadratic term gains a positive slope above 1000 GeV, which is not physical. The fact that the efficiency continues to decrease for high-$p_{T}$ is evidenced by our over-prediction of $n_{b}$ for uncompressed bottom squark models. The original, rounded to two significant digits, is 
\[
  \small
\epsilon = 
  \begin{cases} 
  (20,50]\text{:} & \hspace{-.2cm}.19+.021p_{T}-.00035p_{T}^{2}+2.8\cdot10^{-6}p_{T}^{3}-1.0\cdot10^{-8}p_{T}^4+1.5\cdot10^{-11}p_{T}^5\\
  (50,250]\text{:} & \hspace{-.2cm}.56\hspace{-.05cm}+\hspace{-.05cm}.0034p_{T}\hspace{-.05cm}-\hspace{-.05cm}3.3\cdot10^{-5}p_{T}^{2}+1.5\cdot10^{-7}p_{T}^{3}\hspace{-.05cm}-\hspace{-.05cm}3.6\cdot10^{-10}p_{T}^{4}\hspace{-.05cm}+\hspace{-.05cm}3.5\cdot10^{-13}p_{T}^{5} \\
  >250\text{:} & \hspace{-.2cm}.77-.00055p_{T}+2.9\cdot10^{-7}p_{T}^{2} .\\
  \end{cases}
\]
where the ranges are given for $p_{T}$ in units of GeV. We have dropped the last term, and seen that this change leads to a 15\% improvement in the highest and lowest $b$-tag mulitiplicity regions for the models T1bbbb and T2bb, and has minimal effect on other models.

\subsubsection{Comparison with the official results}
\begin{figure}[t]
	\includegraphics[width=0.49\linewidth]{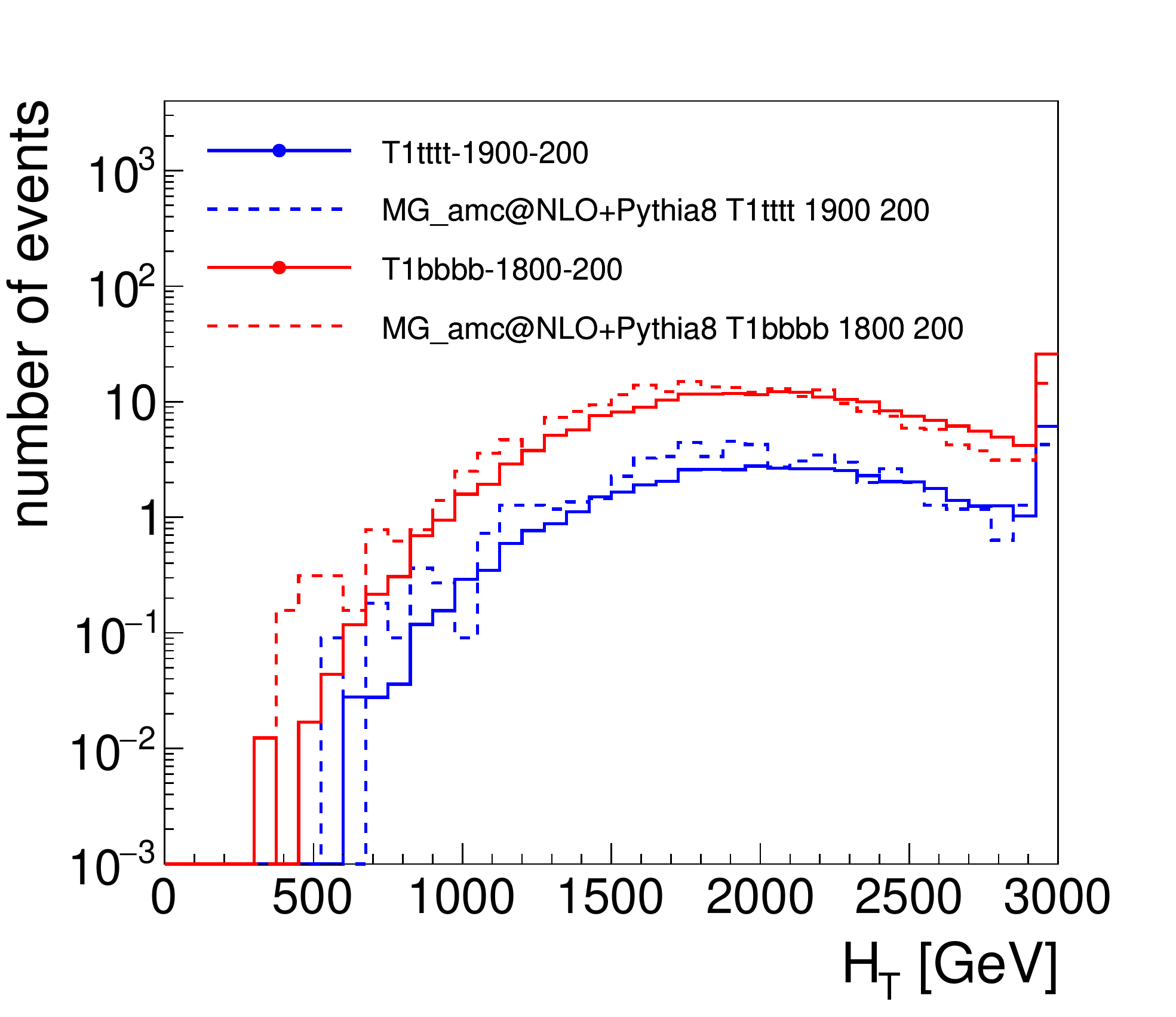}
	\includegraphics[width=0.49\linewidth]{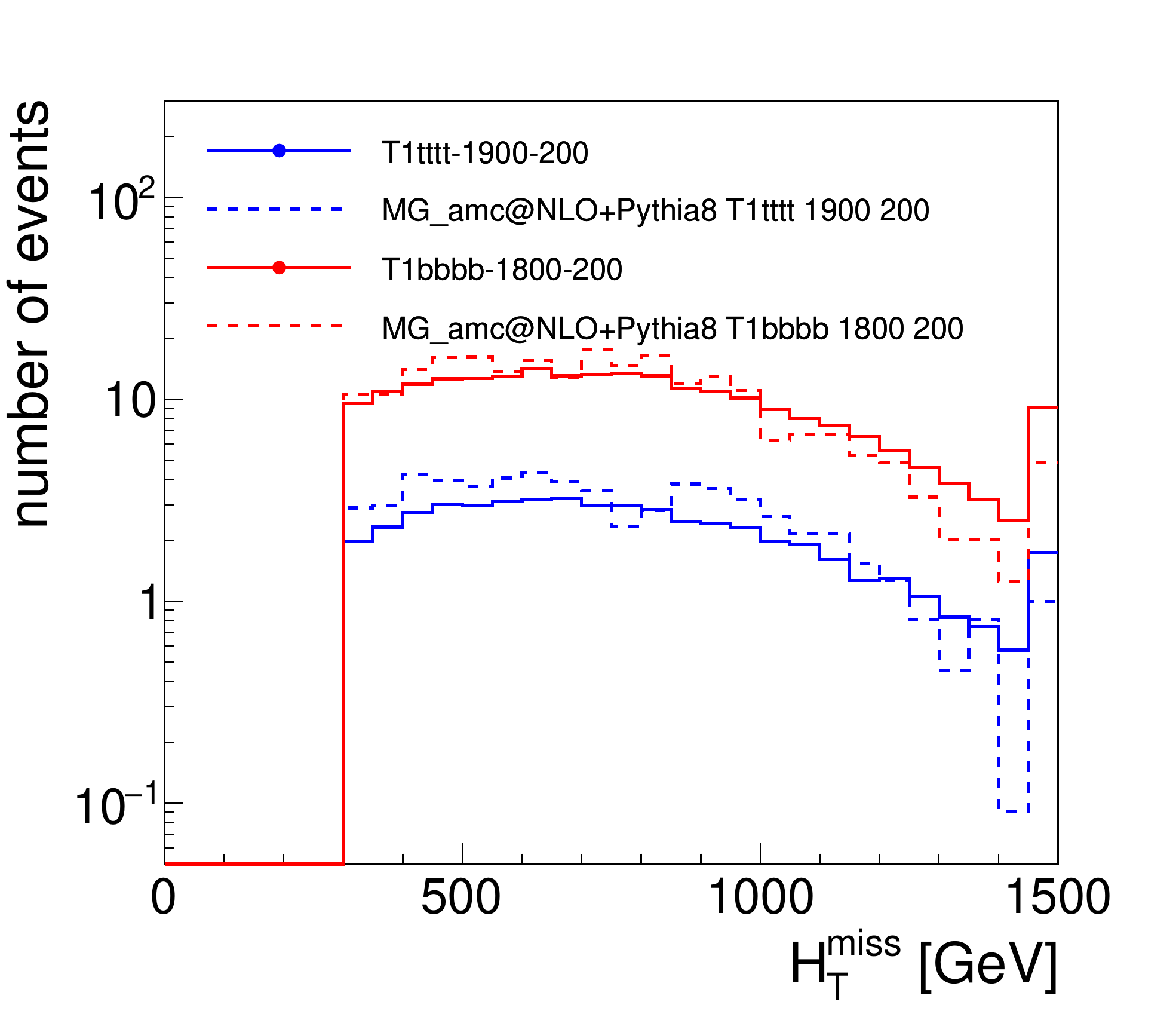}
	\includegraphics[width=0.49\linewidth]{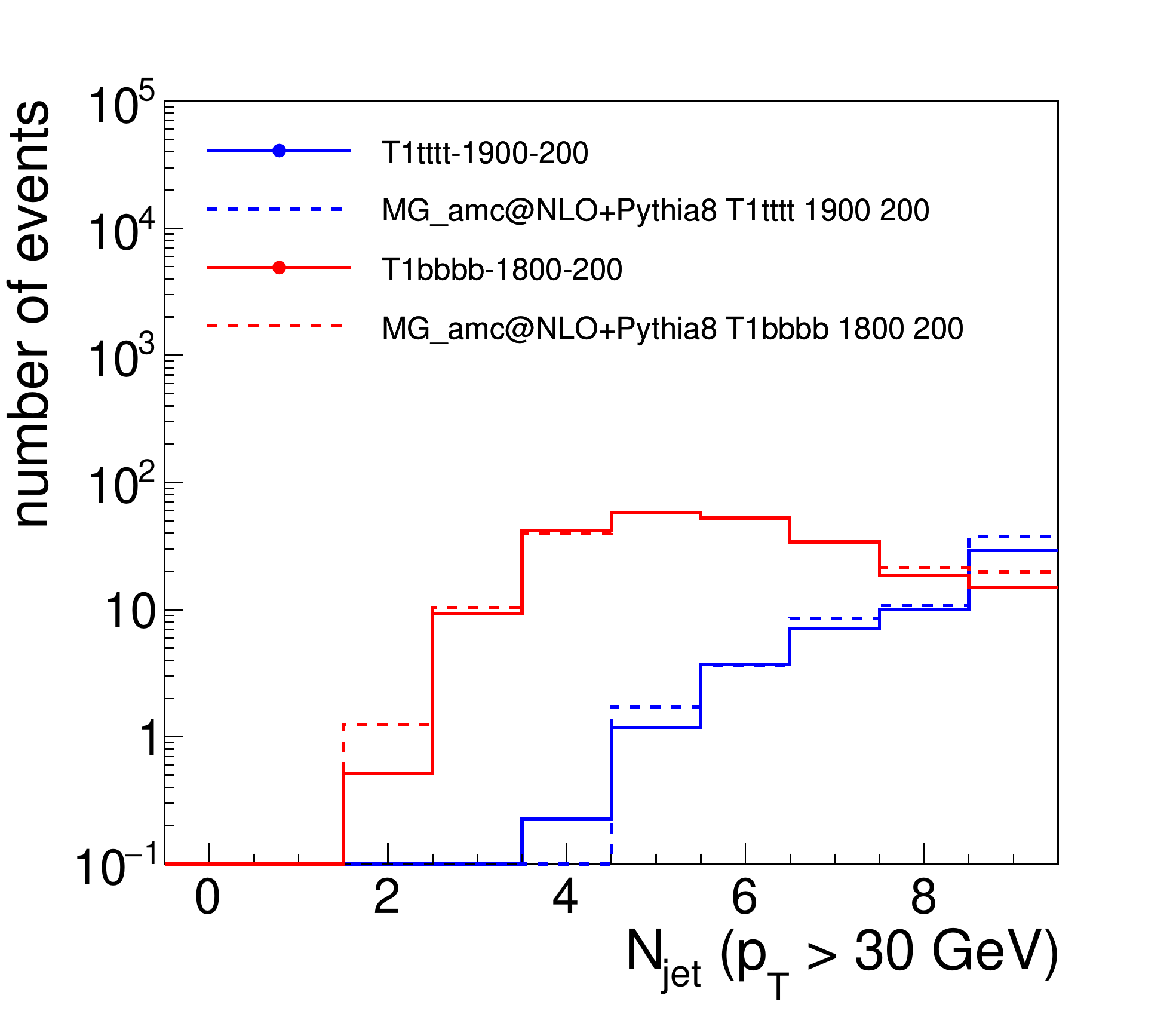}
	\includegraphics[width=0.49\linewidth]{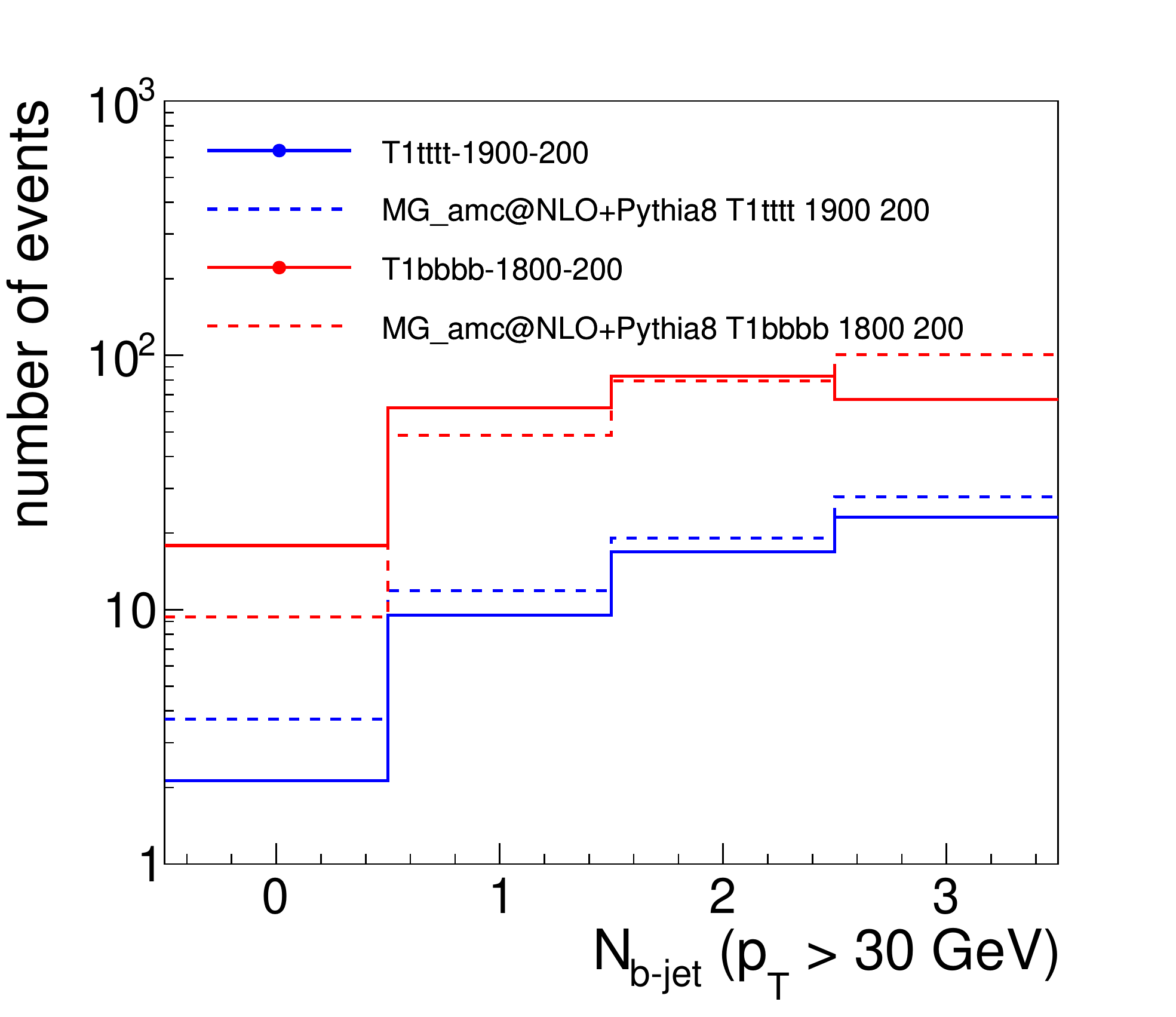}
	\vspace*{8pt}
	\caption{Comparison of kinematic distributions that define the signal regions for non-compressed gluino models between the \texttt{MadAnalysis 5} (dashed line) implementation and CMS (full line)\protect\label{fig:CMS-SUS-19-006_Figure-aux_011-a}}
\end{figure}

\begin{figure}[t]
	\includegraphics[width=0.49\linewidth]{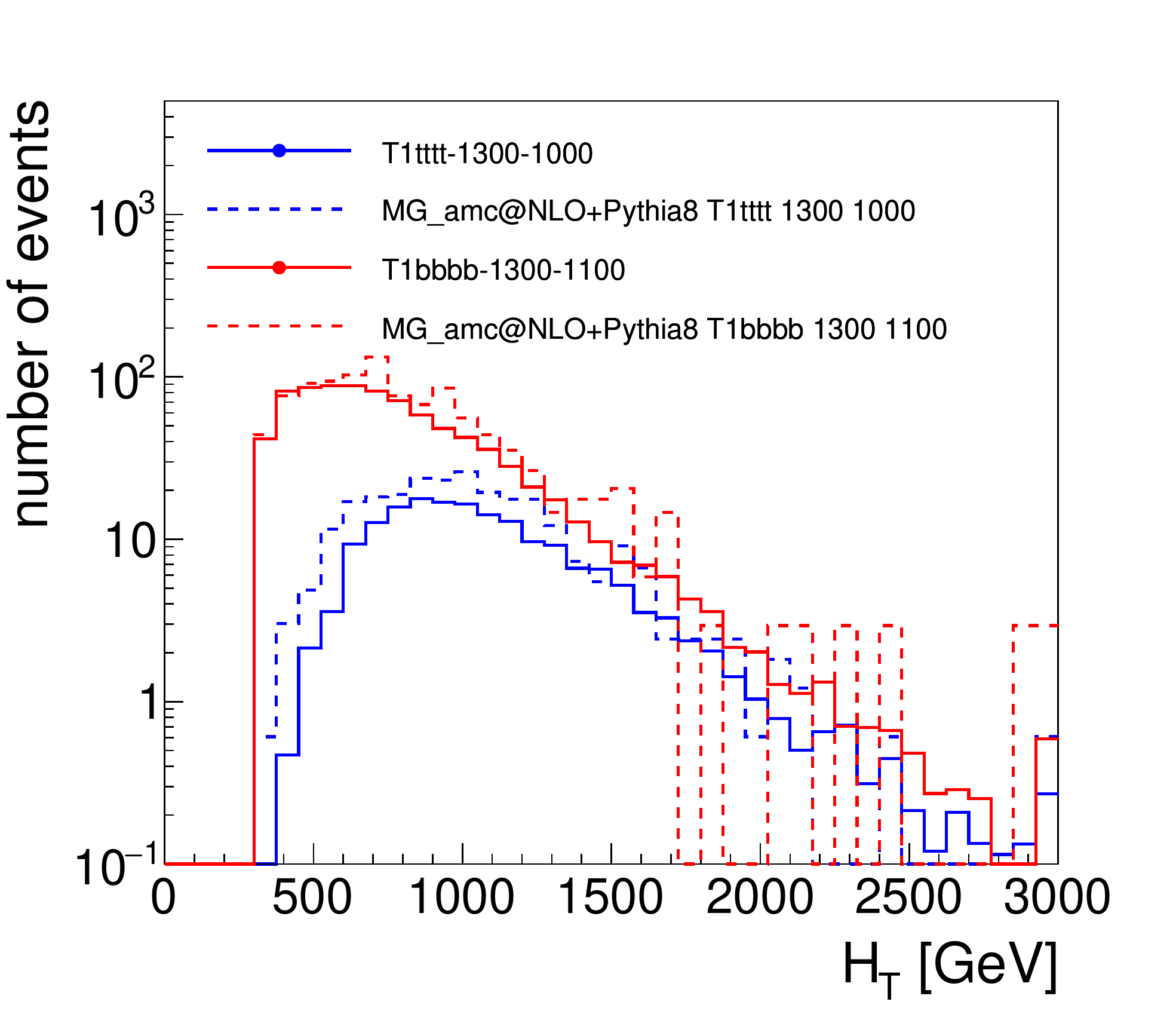}
	\includegraphics[width=0.49\linewidth]{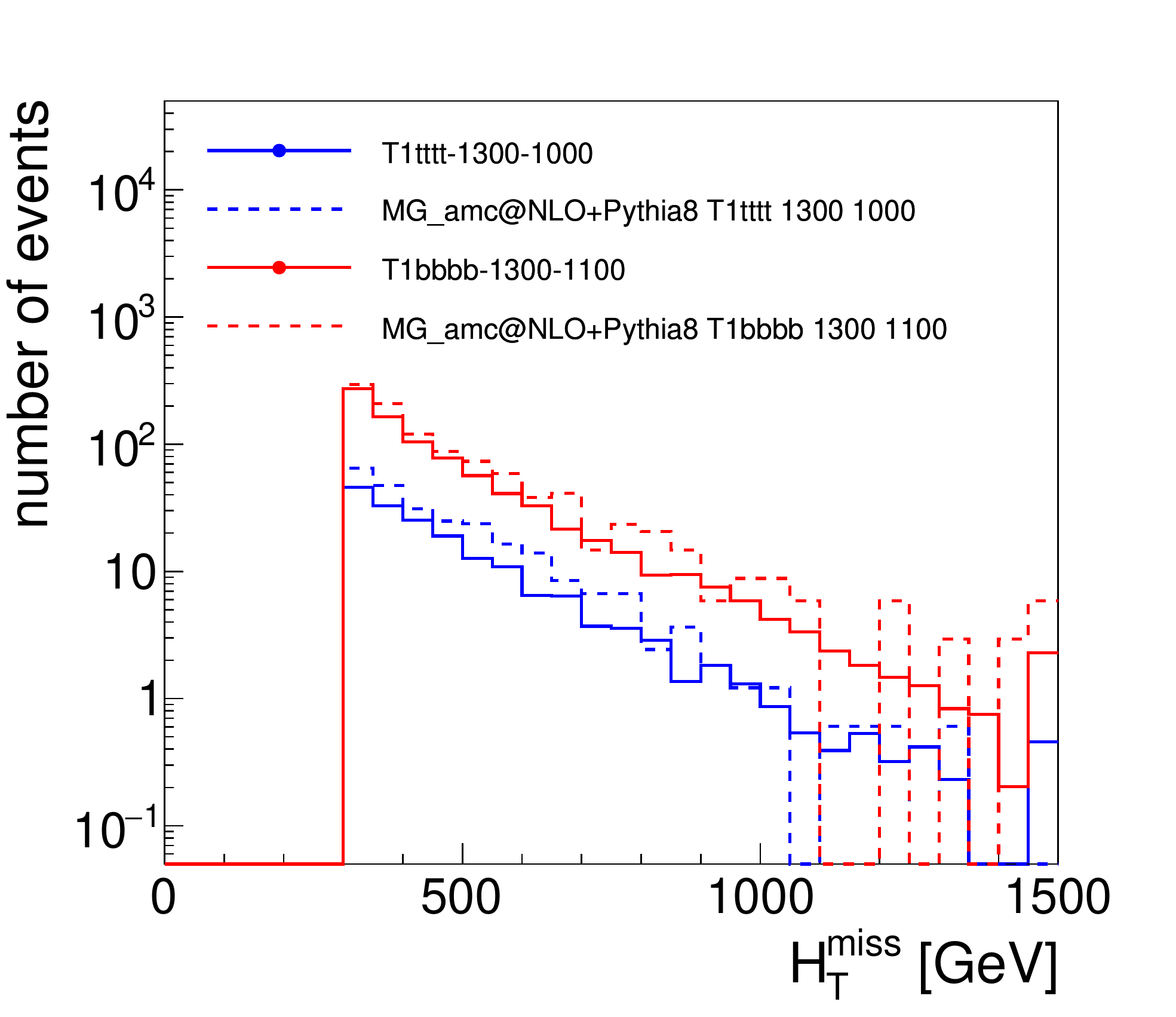}
	\includegraphics[width=0.49\linewidth]{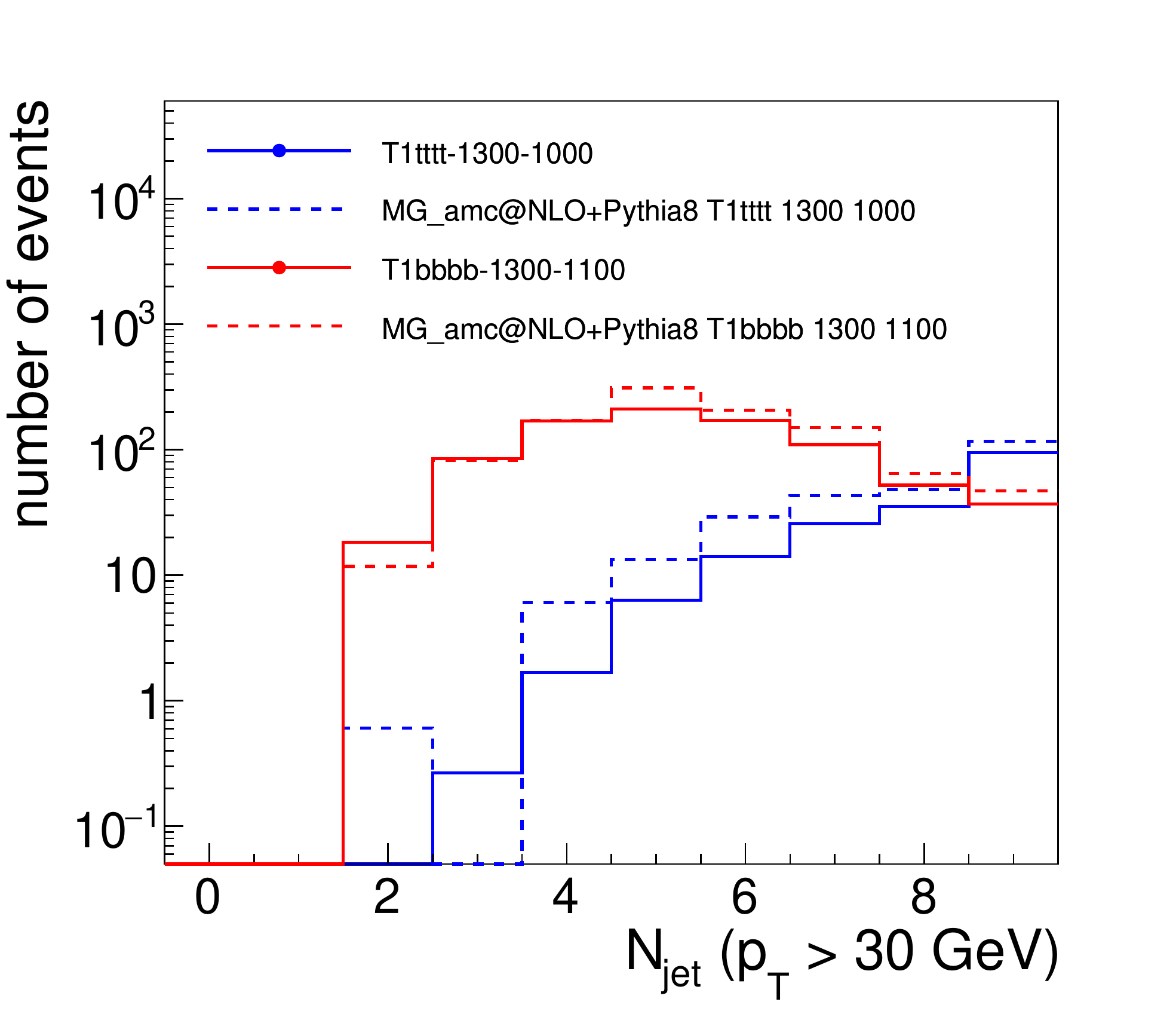}
	\includegraphics[width=0.49\linewidth]{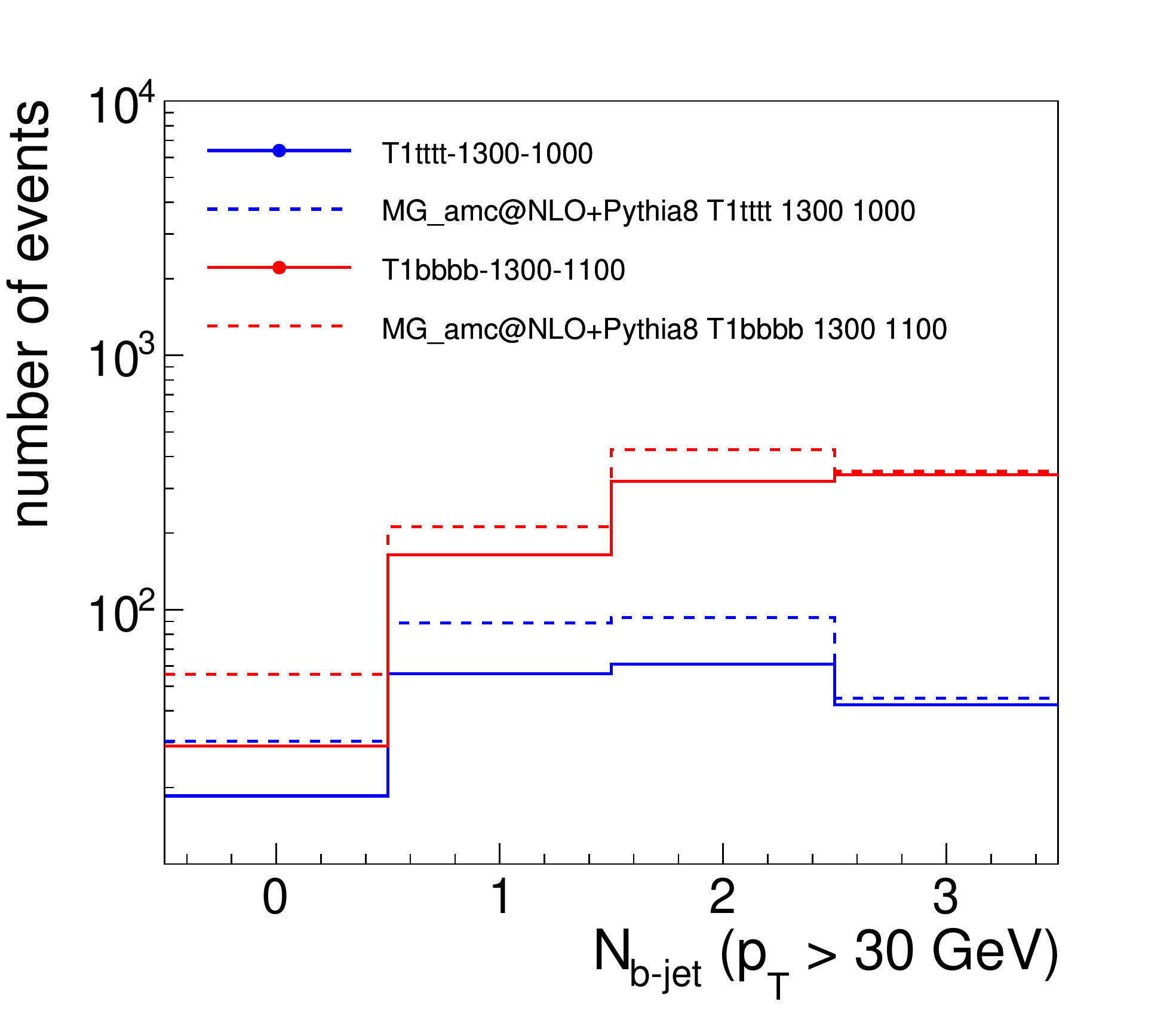}
	\vspace*{8pt}
	\caption{Comparison of kinematic distributions that define the signal regions for compressed gluino models between the \texttt{MadAnalysis 5} (dashed line) implementation and CMS (full line)\protect\label{fig:CMS-SUS-19-006_Figure-aux_012-a}}
\end{figure}

\begin{figure}[t]
	\includegraphics[width=0.49\linewidth]{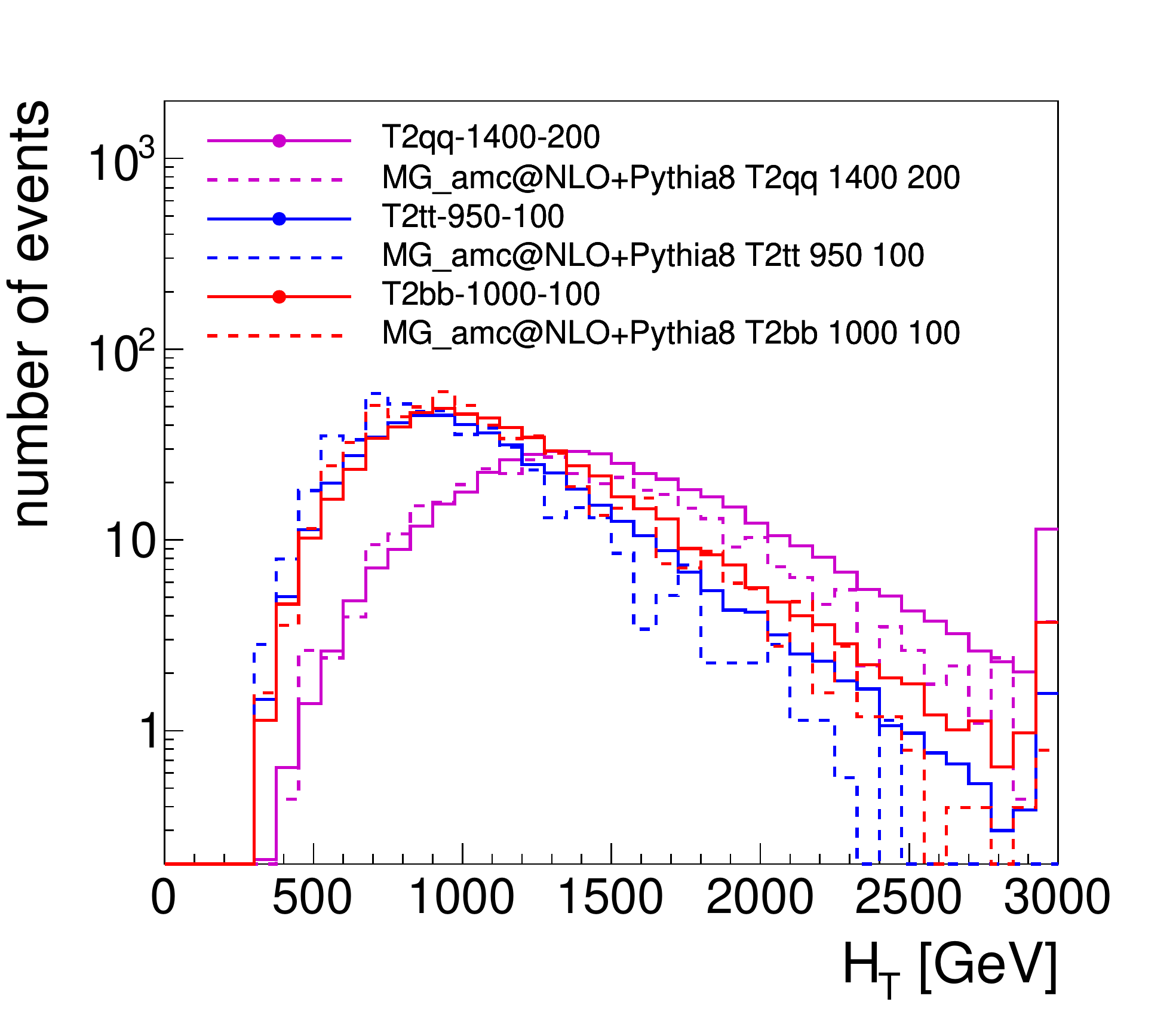}
	\includegraphics[width=0.49\linewidth]{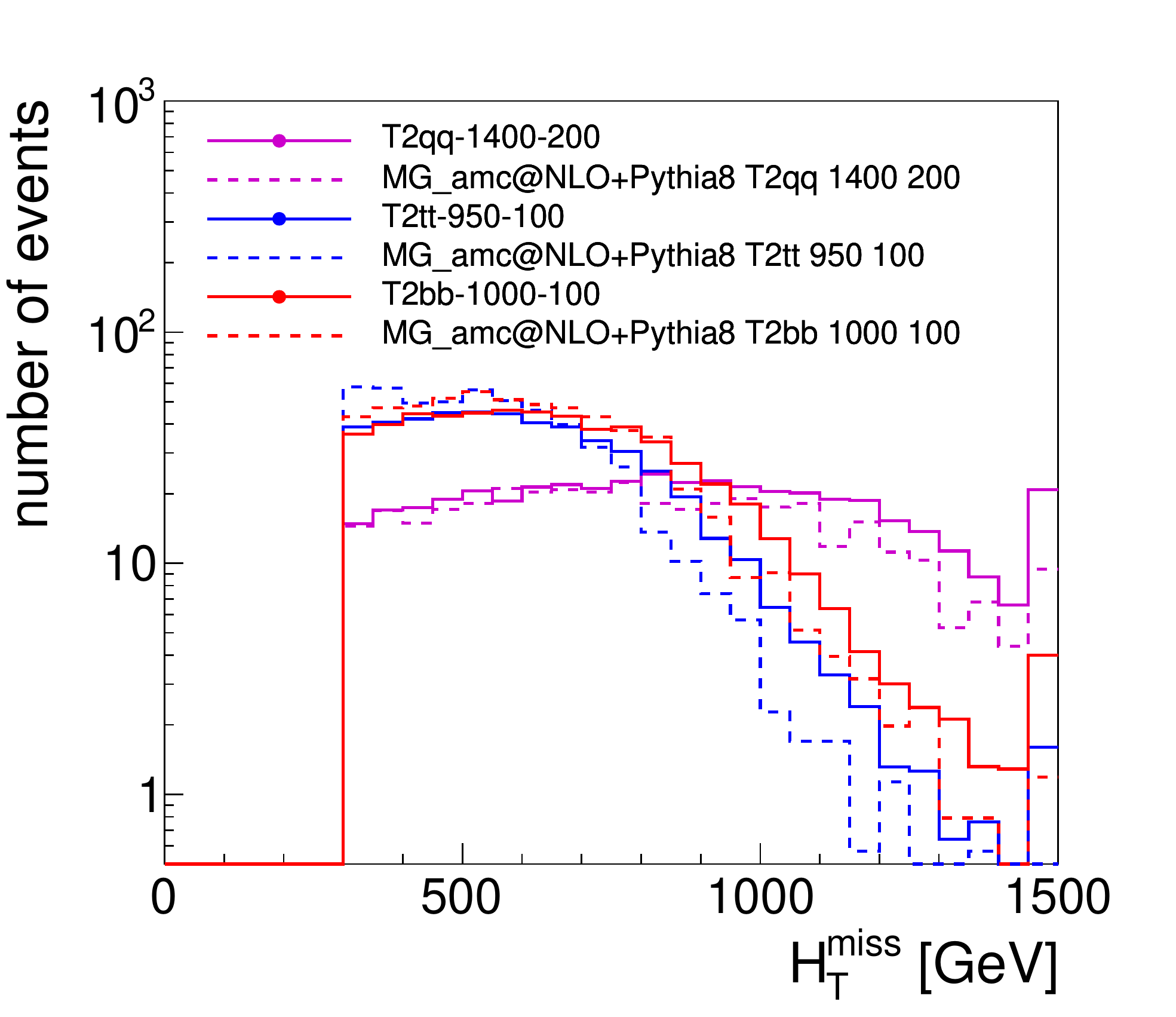}
	\includegraphics[width=0.49\linewidth]{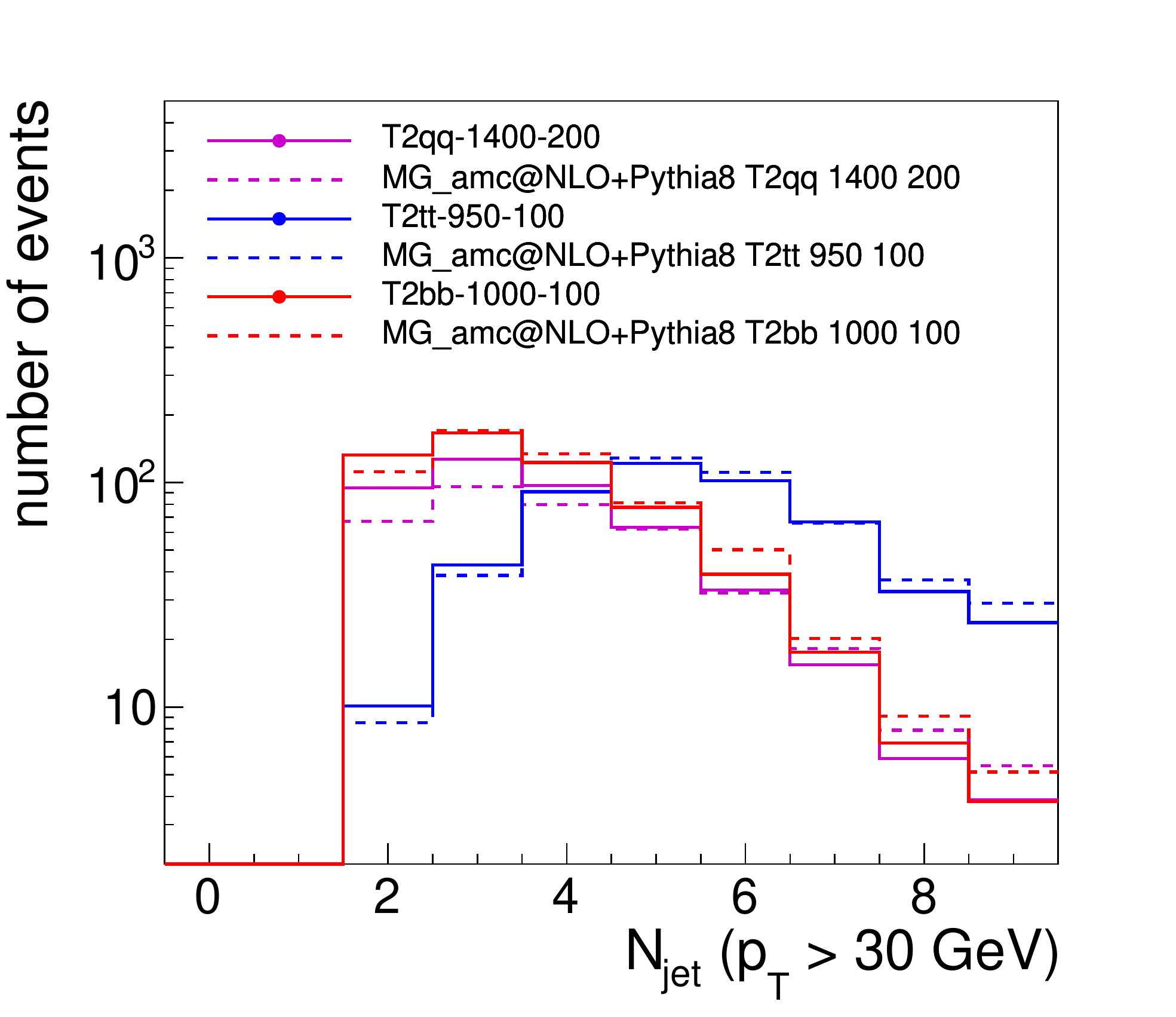}
	\includegraphics[width=0.49\linewidth]{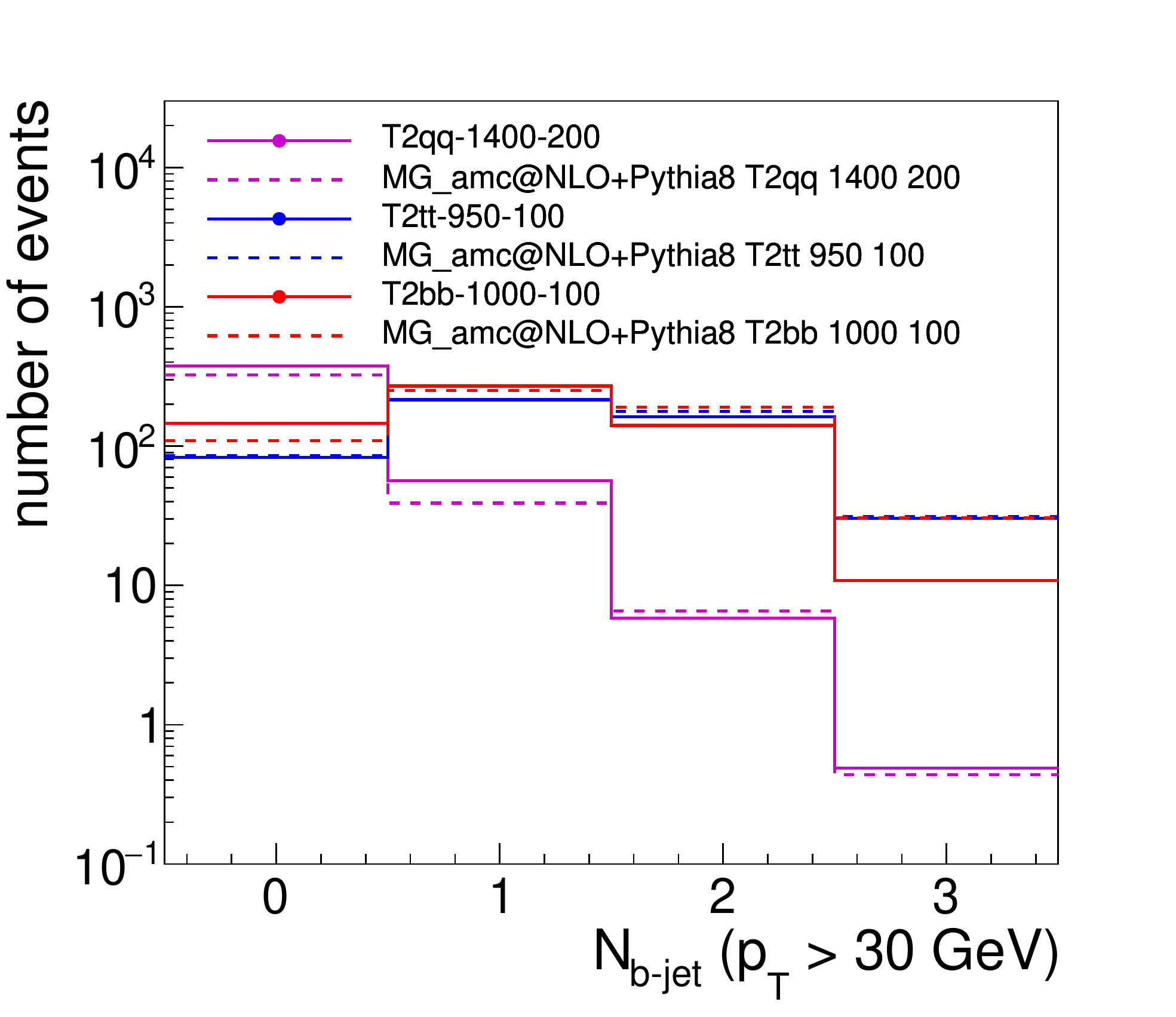}
	\vspace*{8pt}
	\caption{Comparison of kinematic distributions that define the signal regions for non-compressed squark models between the \texttt{MadAnalysis 5} (dashed line) implementation and CMS (full line)\protect\label{fig:CMS-SUS-19-006_Figure-aux_013-a}}
\end{figure}

\begin{figure}[t]
	\includegraphics[width=0.49\linewidth]{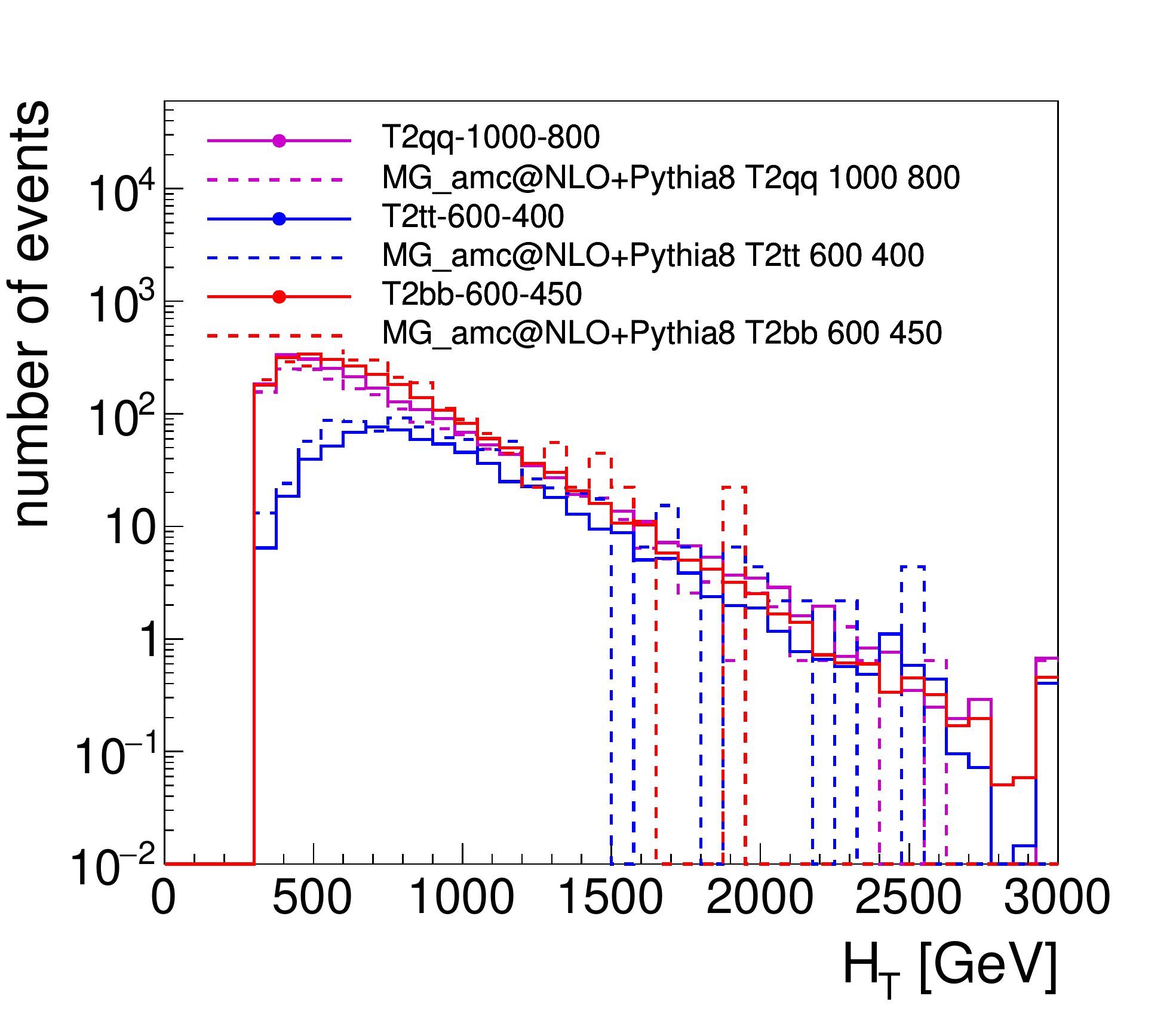}
	\includegraphics[width=0.49\linewidth]{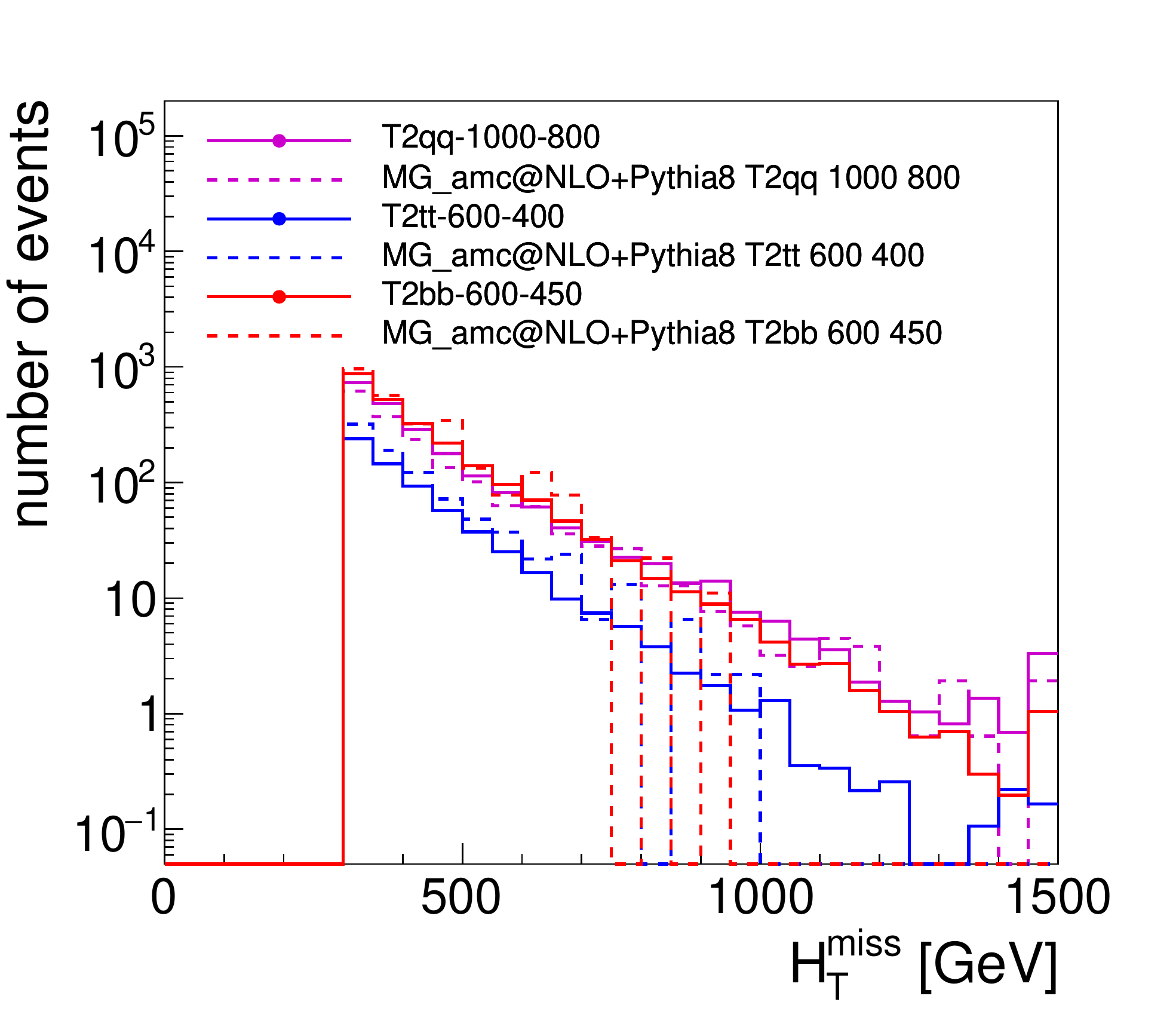}
	\includegraphics[width=0.49\linewidth]{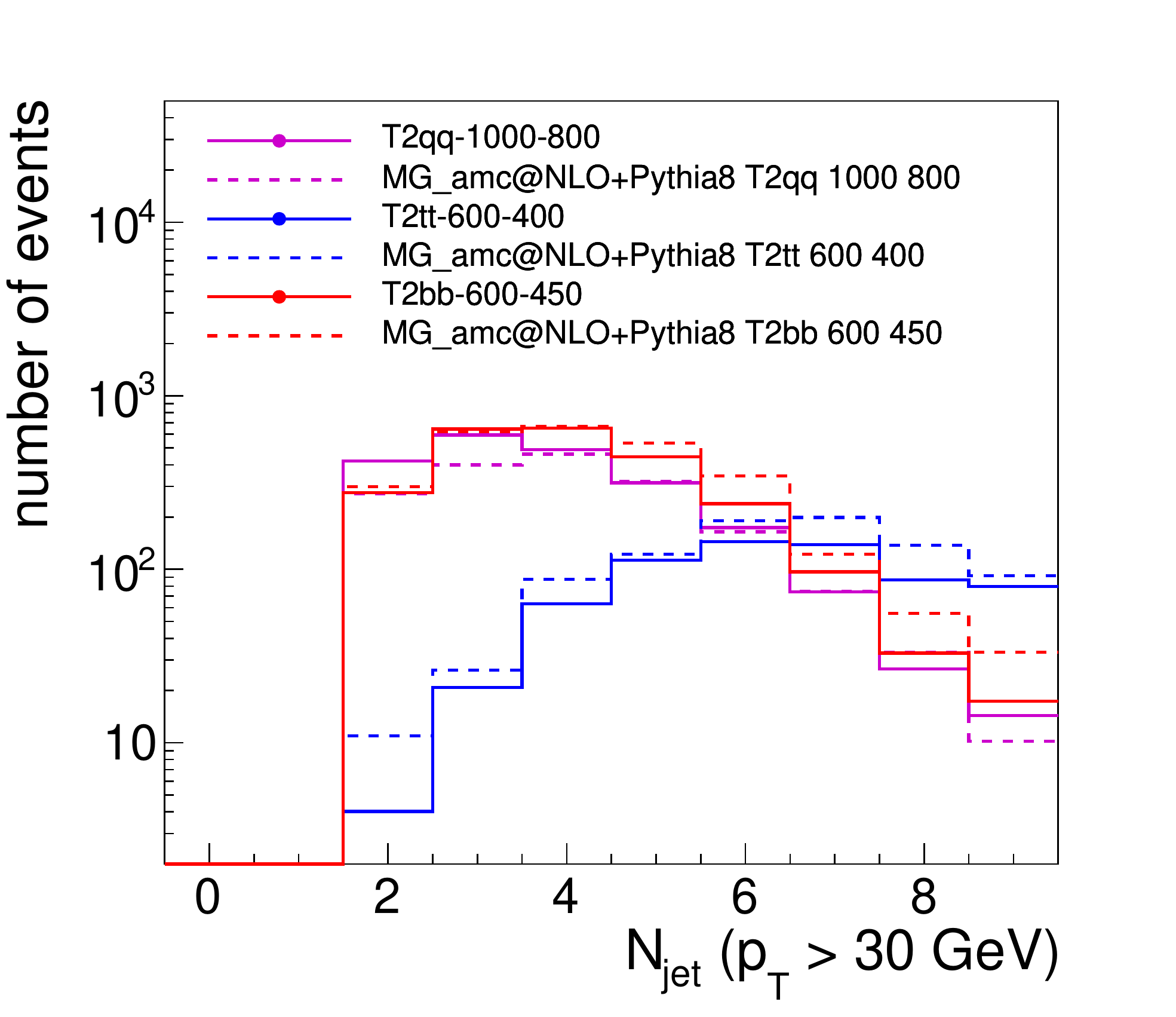}
	\includegraphics[width=0.49\linewidth]{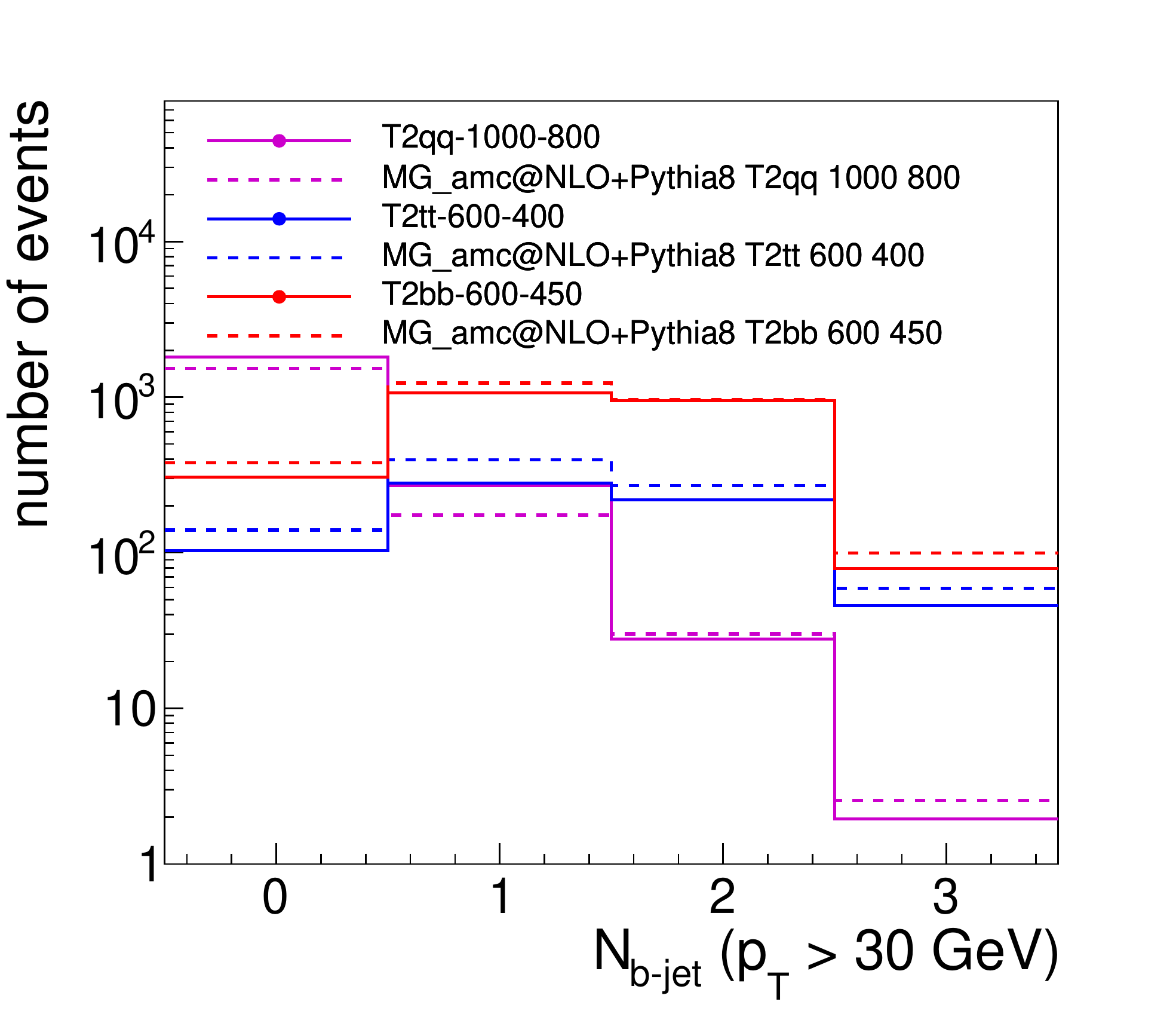}
	\vspace*{8pt}
	\caption{Comparison of kinematic distributions that define the signal regions for compressed squark models between the \texttt{MadAnalysis 5} (dashed line) implementation and CMS (full line)\protect\label{fig:CMS-SUS-19-006_Figure-aux_014-a}}
\end{figure}

This section compares results derived from the recast implementation with the official results from CMS. A table showing a comparison of the cut efficiencies is given for each model, along with the signal event count in each aggregate signal region. Additionally, distributions of kinematic observables used to define the signal regions are compared after the application of the baseline event selection and shown in Figures \ref{fig:CMS-SUS-19-006_Figure-aux_011-a}-\ref{fig:CMS-SUS-19-006_Figure-aux_014-a}. The degree of validation of the recast implementation is reflected in the comparison between the \texttt{MadGraph5\_aMC@NLO} and the CMS result.

A generally satisfactory agreement is seen between the recast implementation and official versions, with a few exceptions. Particularly, there is a logical inconsistency in the aggregate signal region counts for the model T1qqqq(1300,100), where the numbers indicate the mother particle and LSP mass in GeV.  We have reported these anomalies to the CMS team and they are working to fix it. However, we think this only impacts the validation material for these models, and does not undermine this implementation. 

Minor trends and disagreements come into the picture when considering models that produce heavy flavor jets. Particularly, models with real or virtual bottom squarks exhibit a slight bias toward larger $b$-tagged jet multiplicities, while the opposite is true for events with top squarks. This effect is most notable in the uncompressed mass regimes. We have conducted numerous tests to investigate the source of this discrepancy, including making adjustments to the {\tt Delphes} $b$-tagging efficiency, the jet energy scale, the way in which true $b$-jets are defined at the level of the generator, as well as changes to the PDF used in the {\tt MadGraph5\_aMC@NLO} generation. We find that the choice of PDF has the most pronounced impact on the distribution of $b$-tagged jet multiplicity than any other change, which lead to difference with respect to CMS. However, our final validation is based on the \texttt{LHAPDF} implementation consistent with that described by the analysis, and in the case of compressed third generation squark models, a slight over-prediction in $b$-tagged jet multiplicity is observed. We think this is due to inefficiencies that arise from excess transverse event activity, which are not captured by the efficiencies reported by CMS.

In a handful of cases, larger discrepancies of order 50\% appear in the count comparisons in the signal regions, but these are typically within the statistical uncertainties in the signal counts. In rare cases, there is no predicted value for the signal in a given aggregate signal region. This should not have an unwanted effect on limit setting because such bins are not typically sensitive to models with nearly negligible yield, and the inclusion of such bins will not impact the likelihood.

\clearpage
\begin{table}[t]
	\tbl{Pre-selection cutflow for the T1bbbb simplified models.}
	{\setlength\tabcolsep{3.0pt}
		\begin{tabular}{@{}cccccc||ccccc@{}} \toprule
			& T1bbbb &            &                     &                     &                     & T1bbbb &                      &                     &                     &                     \\
			& 1800-200 &            &                     &                     &                     & 1300-1100 &                      &                     &                     &                     \\\hline
			Cut & MA5        & CMS        & MA5                 & MA5                 & CMS                 & MA5        & CMS                  & MA5                 & MA5                 & CMS                 \\
			&            &            & diff [\%] & drop [\%] & drop [\%] &            &                      & diff [\%] & drop [\%] & drop [\%] \\ \colrule
			N$_{\text{jet}} \geq 2$\hphantom{0} & 100.0$\pm^{0.0}_{0.1}$& 100.0$\pm$0.5& 0.0 & 0.0 & 0.0 & 99.6$\pm^{0.1}_{0.2}$& 99.3$\pm$0.1& -0.3 & 0.4 & 0.7 \\ 
			H$_{\text{T}}>300$\hphantom{0} & 100.0$\pm^{0.0}_{0.1}$& 100.0$\pm$0.5& 0.0 & 0.0 & 0.0 & 74.8$\pm^{0.9}_{0.9}$& 74.8$\pm$0.5& 0.0 & 24.8 & 24.5 \\ 
			$\text{H}_{\text{T}}^{\text{miss}}>300$\hphantom{0} & 83.9$\pm^{0.8}_{0.8}$& 86.8$\pm$1.9& 3.34 & 16.1 & 13.2 & 20.0$\pm^{0.9}_{0.8}$& 19.9$\pm$0.5& -0.5 & 54.8 & 54.9 \\ 
			H$_{\text{T}}>\text{H}_{\text{T}}^{\text{miss}}$\hphantom{0} & 83.8$\pm^{0.8}_{0.8}$& 86.8$\pm$1.9& 3.46 & 0.1 & 0.0 & 19.7$\pm^{0.9}_{0.8}$& 19.5$\pm$0.5& -1.03 & 0.3 & 0.4 \\ 
			IsoMuons\hphantom{0} & 83.6$\pm^{0.8}_{0.8}$& 86.0$\pm$2.0& 2.79 & 0.2 & 0.8 & 19.7$\pm^{0.9}_{0.8}$& 19.2$\pm$0.5& -2.6 & 0.0 & 0.3 \\ 
			MuonTracks\hphantom{0} & 83.6$\pm^{0.8}_{0.8}$& 85.8$\pm$2.0& 2.56 & 0.0 & 0.2 & 19.5$\pm^{0.9}_{0.8}$& 18.9$\pm$0.5& -3.17 & 0.2 & 0.3 \\ 
			IsoElectrons\hphantom{0} & 83.5$\pm^{0.8}_{0.8}$& 85.3$\pm$2.0& 2.11 & 0.1 & 0.5 & 19.4$\pm^{0.9}_{0.8}$& 18.8$\pm$0.5& -3.19 & 0.1 & 0.1 \\ 
			ElectronTracks\hphantom{0} & 83.3$\pm^{0.8}_{0.8}$& 85.0$\pm$2.0& 2.0 & 0.2 & 0.3 & 19.2$\pm^{0.9}_{0.8}$& 18.4$\pm$0.5& -4.35 & 0.2 & 0.4 \\ 
			IsoTracks\hphantom{0} & 83.0$\pm^{0.8}_{0.8}$& 84.3$\pm$2.0& 1.54 & 0.3 & 0.7 & 19.1$\pm^{0.9}_{0.8}$& 18.2$\pm$0.5& -4.95 & 0.1 & 0.2 \\ 
			IsoPhotons\hphantom{0} & 82.3$\pm^{0.8}_{0.8}$& 81.5$\pm$2.1& -0.98 & 0.7 & 2.8 & 19.0$\pm^{0.9}_{0.8}$& 17.8$\pm$0.5& -6.74 & 0.1 & 0.4 \\ 
			$\Delta\Phi_{1}>0.5$\hphantom{0} & 80.5$\pm^{0.8}_{0.8}$& 80.0$\pm$2.2& -0.63 & 1.8 & 1.5 & 19.0$\pm^{0.9}_{0.8}$& 17.7$\pm$0.5& -7.34 & 0.0 & 0.1 \\ 
			$\Delta\Phi_{2}>0.5$\hphantom{0} & 74.1$\pm^{0.9}_{0.9}$& 71.8$\pm$2.4& -3.2 & 6.4 & 8.2 & 17.0$\pm^{0.8}_{0.8}$& 16.2$\pm$0.4& -4.94 & 2.0 & 1.5 \\ 
			$\Delta\Phi_{3}>0.3$\hphantom{0} & 68.1$\pm^{1.0}_{1.0}$& 66.6$\pm$2.5& -2.25 & 6.0 & 5.2 & 16.0$\pm^{0.8}_{0.8}$& 15.1$\pm$0.4& -5.96 & 1.0 & 1.1 \\ 
			$\Delta\Phi_{4}>0.3$\hphantom{0} & 62.4$\pm^{1.0}_{1.0}$& 61.1$\pm$2.6& -2.13 & 5.7 & 5.5 & 15.3$\pm^{0.8}_{0.8}$& 14.2$\pm$0.4& -7.75 & 0.7 & 0.9 \\ \botrule
		\end{tabular}\label{cutflow:T1bbbb-1800-200} }
\end{table}
\begin{table}[t]
	\tbl{Signal yield in the aggregated signal regions for the T1bbbb simplified models.}
	{\setlength\tabcolsep{3.0pt}\begin{tabular}{@{}ccccc|ccc||ccc@{}} \toprule
			& Agg.             &   SR            &              &                                        &T1bbbb-1800-200&       &                      & T1bbbb-1300-1100        &       &                     \\\hline
			bin    & N$_{\text{jet}}$&N$_{\text{bjet}}$&H$_{\text{T}}$&$\text{H}_{\text{T}}^{\text{miss}}$& MA5      & CMS   & MA5                  & MA5               & CMS   & MA5                  \\
			&                  &                 &[GeV]         &[GeV]                                   &yield     & yield &  diff [\%] & yield             & yield & diff [\%] \\ \colrule
			1&$\geq$ 2&0&$>$600&$>$600\hphantom{0} & 5.0$\pm$0.88& 11.88$\pm$0.15 & 57.91& 14.69$\pm$6.57& 4.85$\pm$0.19 & -202.89  \\ 
			2&$\geq$ 4&0&$>$1700&$>$850\hphantom{0} & 2.5$\pm$0.62& 4.95$\pm$0.1 & 49.49& 0.0$\pm$0.0& 0.32$\pm$0.03 & 100.0  \\ 
			3&$\geq$ 6&0&$>$600&$>$600\hphantom{0} & 1.72$\pm$0.52& 4.77$\pm$0.08 & 63.94& 5.87$\pm$4.15& 1.68$\pm$0.08 & -249.4  \\ 
			4&$\geq$ 8&$\leq$ 2&$>$600&$>$600\hphantom{0} & 3.44$\pm$0.73& 5.68$\pm$0.12 & 39.44& 14.69$\pm$6.57& 3.21$\pm$0.15 & -357.63  \\ 
			5&$\geq$ 10&$\leq$ 2&$>$1700&$>$850\hphantom{0} & 0.78$\pm$0.35& 0.43$\pm$0.03 & -81.4& 0.0$\pm$0.0& 0.1$\pm$0.02 & 100.0  \\ 
			6&$\geq$ 4&$\geq$ 2&$>$300&$>$300\hphantom{0} & 175.05$\pm$5.23& 143.99$\pm$0.82 & -21.57& 728.43$\pm$46.26& 589.39$\pm$4.54 & -23.59  \\ 
			7&$\geq$ 2&$\geq$ 2&$>$600&$>$600\hphantom{0} & 120.55$\pm$4.34& 100.49$\pm$0.69 & -19.96& 146.86$\pm$20.77& 105.43$\pm$1.68 & -39.3  \\ 
			8&$\geq$ 6&$\geq$ 2&$>$350&$>$350\hphantom{0} & 98.85$\pm$3.93& 79.3$\pm$0.6 & -24.65& 305.47$\pm$29.95& 236.98$\pm$2.48 & -28.9  \\ 
			9&$\geq$ 4&$\geq$ 2&$>$600&$>$600\hphantom{0} & 118.06$\pm$4.29& 97.83$\pm$0.67 & -20.68& 140.99$\pm$20.35& 102.58$\pm$1.65 & -37.44  \\ 
			10&$\geq$ 8&$\geq$ 3&$>$300&$>$300\hphantom{0} & 21.39$\pm$1.83& 12.97$\pm$0.26 & -64.92& 64.62$\pm$13.78& 51.73$\pm$1.23 & -24.92  \\ 
			11&$\geq$ 6&$\geq$ 1&$>$600&$>$600\hphantom{0} & 82.76$\pm$3.6& 74.87$\pm$0.57 & -10.54& 96.93$\pm$16.87& 76.83$\pm$1.33 & -26.16  \\ 
			12&$\geq$ 10&$\geq$ 3&$>$850&$>$850\hphantom{0} & 0.62$\pm$0.31& 0.86$\pm$0.06 & 27.91& 0.0$\pm$0.0& 1.07$\pm$0.16 & 100.0  \\ \botrule
		\end{tabular}\label{yield:T1bbbb-1800-200} }
\end{table}
\clearpage
\begin{table}[t]
	\tbl{Pre-selection cutflow for the T1tttt simplified models.}
	{\setlength\tabcolsep{3.0pt}
		\begin{tabular}{@{}cccccc||ccccc@{}} \toprule
			& T1tttt &            &                     &                     &                     & T1tttt &                      &                     &                     &                     \\
			& 1900-200 &            &                     &                     &                     & 1300-1000 &                      &                     &                     &                     \\\hline
			Cut & MA5        & CMS        & MA5                 & MA5                 & CMS                 & MA5        & CMS                  & MA5                 & MA5                 & CMS                 \\
			&            &            & diff [\%] & drop [\%] & drop [\%] &            &                      & diff [\%] & drop [\%] & drop [\%] \\ \colrule
			N$_{\text{jet}} \geq 2$\hphantom{0} & 100.0$\pm^{0.0}_{0.1}$& 100.0$\pm$0.8& 0.0 & 0.0 & 0.0 & 100.0$\pm^{0.0}_{0.0}$& 100.0$\pm$0.0& 0.0 & 0.0 & 0.0 \\ 
			H$_{\text{T}}>300$\hphantom{0} & 100.0$\pm^{0.0}_{0.1}$& 100.0$\pm$0.8& 0.0 & 0.0 & 0.0 & 87.7$\pm^{0.3}_{0.3}$& 90.1$\pm$0.4& 2.66 & 12.3 & 9.9 \\ 
			$\text{H}_{\text{T}}^{\text{miss}}>300$\hphantom{0} & 84.6$\pm^{0.7}_{0.8}$& 85.5$\pm$2.7& 1.05 & 15.4 & 14.5 & 14.6$\pm^{0.3}_{0.3}$& 13.8$\pm$0.4& -5.8 & 73.1 & 76.3 \\ 
			H$_{\text{T}}>\text{H}_{\text{T}}^{\text{miss}}$\hphantom{0} & 84.5$\pm^{0.7}_{0.8}$& 85.5$\pm$2.7& 1.17 & 0.1 & 0.0 & 14.5$\pm^{0.3}_{0.3}$& 13.8$\pm$0.4& -5.07 & 0.1 & 0.0 \\ 
			IsoMuons\hphantom{0} & 56.9$\pm^{1.0}_{1.0}$& 53.4$\pm$3.6& -6.55 & 27.6 & 32.1 & 9.2$\pm^{0.3}_{0.3}$& 8.8$\pm$0.3& -4.55 & 5.3 & 5.0 \\ 
			MuonTracks\hphantom{0} & 56.4$\pm^{1.0}_{1.0}$& 52.6$\pm$3.6& -7.22 & 0.5 & 0.8 & 9.0$\pm^{0.3}_{0.3}$& 8.5$\pm$0.3& -5.88 & 0.2 & 0.3 \\ 
			IsoElectrons\hphantom{0} & 38.0$\pm^{1.0}_{1.0}$& 34.2$\pm$3.4& -11.11 & 18.4 & 18.4 & 6.5$\pm^{0.2}_{0.2}$& 5.8$\pm$0.3& -12.07 & 2.5 & 2.7 \\ 
			ElectronTracks\hphantom{0} & 37.6$\pm^{1.0}_{1.0}$& 33.3$\pm$3.4& -12.91 & 0.4 & 0.9 & 6.3$\pm^{0.2}_{0.2}$& 5.4$\pm$0.3& -16.67 & 0.2 & 0.4 \\ 
			IsoTracks\hphantom{0} & 36.9$\pm^{1.0}_{1.0}$& 32.1$\pm$3.4& -14.95 & 0.7 & 1.2 & 5.8$\pm^{0.2}_{0.2}$& 5.0$\pm$0.3& -16.0 & 0.5 & 0.4 \\ 
			IsoPhotons\hphantom{0} & 36.4$\pm^{1.0}_{1.0}$& 30.3$\pm$3.3& -20.13 & 0.5 & 1.8 & 5.8$\pm^{0.2}_{0.2}$& 4.9$\pm$0.3& -18.37 & 0.0 & 0.1 \\ 
			$\Delta\Phi_{1}>0.5$\hphantom{0} & 35.6$\pm^{1.0}_{1.0}$& 29.5$\pm$3.3& -20.68 & 0.8 & 0.8 & 5.7$\pm^{0.2}_{0.2}$& 4.9$\pm$0.3& -16.33 & 0.1 & 0.0 \\ 
			$\Delta\Phi_{2}>0.5$\hphantom{0} & 32.2$\pm^{1.0}_{1.0}$& 26.5$\pm$3.2& -21.51 & 3.4 & 3.0 & 4.8$\pm^{0.2}_{0.2}$& 4.1$\pm$0.2& -17.07 & 0.9 & 0.8 \\ 
			$\Delta\Phi_{3}>0.3$\hphantom{0} & 30.0$\pm^{1.0}_{0.9}$& 24.8$\pm$3.2& -20.97 & 2.2 & 1.7 & 4.2$\pm^{0.2}_{0.2}$& 3.5$\pm$0.2& -20.0 & 0.6 & 0.6 \\ 
			$\Delta\Phi_{4}>0.3$\hphantom{0} & 28.1$\pm^{0.9}_{0.9}$& 23.1$\pm$3.1& -21.65 & 1.9 & 1.7 & 3.8$\pm^{0.2}_{0.2}$& 3.1$\pm$0.2& -22.58 & 0.4 & 0.4 \\ \botrule
		\end{tabular}\label{cutflow:T1tttt-1900-200} }
\end{table}
\begin{table}[t]
	\tbl{Signal yield in the aggregated signal regions for the T1tttt simplified models.}
	{\setlength\tabcolsep{3.0pt}\begin{tabular}{@{}ccccc|ccc||ccc@{}} \toprule
			& Agg.             &   SR            &              &                                        &T1tttt-1900-200&       &                      & T1tttt-1300-1000        &       &                     \\\hline
			bin    & N$_{\text{jet}}$&N$_{\text{bjet}}$&H$_{\text{T}}$&$\text{H}_{\text{T}}^{\text{miss}}$& MA5      & CMS   & MA5                  & MA5               & CMS   & MA5                  \\
			&                  &                 &[GeV]         &[GeV]                                   &yield     & yield &  diff [\%] & yield             & yield & diff [\%] \\ \colrule
			1&$\geq$ 2&0&$>$600&$>$600\hphantom{0} & 2.72$\pm$0.5& 1.43$\pm$0.02 & -90.21& 5.47$\pm$1.82& 2.86$\pm$0.17 & -91.26  \\ 
			2&$\geq$ 4&0&$>$1700&$>$850\hphantom{0} & 1.27$\pm$0.34& 0.68$\pm$0.02 & -86.76& 0.0$\pm$0.0& 0.43$\pm$0.07 & 100.0  \\ 
			3&$\geq$ 6&0&$>$600&$>$600\hphantom{0} & 2.63$\pm$0.49& 1.3$\pm$0.02 & -102.31& 5.47$\pm$1.82& 2.73$\pm$0.17 & -100.37  \\ 
			4&$\geq$ 8&$\leq$ 2&$>$600&$>$600\hphantom{0} & 7.43$\pm$0.82& 5.12$\pm$0.06 & -45.12& 14.58$\pm$2.98& 8.58$\pm$0.36 & -69.93  \\ 
			5&$\geq$ 10&$\leq$ 2&$>$1700&$>$850\hphantom{0} & 2.08$\pm$0.43& 0.98$\pm$0.02 & -112.24& 0.61$\pm$0.61& 0.77$\pm$0.09 & 20.78  \\ 
			6&$\geq$ 4&$\geq$ 2&$>$300&$>$300\hphantom{0} & 46.85$\pm$2.06& 39.89$\pm$0.26 & -17.45& 137.3$\pm$9.13& 102.74$\pm$1.45 & -33.64  \\ 
			7&$\geq$ 2&$\geq$ 2&$>$600&$>$600\hphantom{0} & 29.63$\pm$1.64& 27.13$\pm$0.21 & -9.21& 25.52$\pm$3.94& 18.52$\pm$0.61 & -37.8  \\ 
			8&$\geq$ 6&$\geq$ 2&$>$350&$>$350\hphantom{0} & 43.68$\pm$1.99& 37.52$\pm$0.25 & -16.42& 101.45$\pm$7.85& 74.45$\pm$1.23 & -36.27  \\ 
			9&$\geq$ 4&$\geq$ 2&$>$600&$>$600\hphantom{0} & 29.63$\pm$1.64& 27.13$\pm$0.21 & -9.21& 25.52$\pm$3.94& 18.52$\pm$0.61 & -37.8  \\ 
			10&$\geq$ 8&$\geq$ 3&$>$300&$>$300\hphantom{0} & 24.01$\pm$1.48& 19.32$\pm$0.2 & -24.28& 34.02$\pm$4.55& 36.26$\pm$0.89 & 6.18  \\ 
			11&$\geq$ 6&$\geq$ 1&$>$600&$>$600\hphantom{0} & 36.52$\pm$1.82& 32.65$\pm$0.23 & -11.85& 40.7$\pm$4.97& 27.2$\pm$0.73 & -49.63  \\ 
			12&$\geq$ 10&$\geq$ 3&$>$850&$>$850\hphantom{0} & 5.07$\pm$0.68& 3.69$\pm$0.09 & -37.4& 0.0$\pm$0.0& 1.26$\pm$0.17 & 100.0  \\ \botrule
		\end{tabular}\label{yield:T1tttt-1900-200} }
\end{table}
\clearpage

\clearpage
\begin{table}[t]
	\tbl{Pre-selection cutflow for the T2qq simplified models.}
	{\setlength\tabcolsep{3.0pt}
		\begin{tabular}{@{}cccccc||ccccc@{}} \toprule
			& T2qq &            &                     &                     &                     & T2qq &                      &                     &                     &                     \\
			& 1400-200 &            &                     &                     &                     & 1000-800 &                      &                     &                     &                     \\\hline
			Cut & MA5        & CMS        & MA5                 & MA5                 & CMS                 & MA5        & CMS                  & MA5                 & MA5                 & CMS                 \\
			&            &            & diff [\%] & drop [\%] & drop [\%] &            &                      & diff [\%] & drop [\%] & drop [\%] \\ \colrule
			N$_{\text{jet}} \geq 2$\hphantom{0} & 99.2$\pm^{0.2}_{0.2}$& 99.1$\pm$0.5& -0.1 & 0.8 & 0.9 & 97.9$\pm^{0.1}_{0.1}$& 97.8$\pm$0.2& -0.1 & 2.1 & 2.2 \\ 
			H$_{\text{T}}>300$\hphantom{0} & 98.9$\pm^{0.2}_{0.3}$& 98.9$\pm$0.6& 0.0 & 0.3 & 0.2 & 80.5$\pm^{0.4}_{0.4}$& 83.0$\pm$0.4& 3.01 & 17.4 & 14.8 \\ 
			$\text{H}_{\text{T}}^{\text{miss}}>300$\hphantom{0} & 86.6$\pm^{0.7}_{0.8}$& 88.1$\pm$1.4& 1.7 & 12.3 & 10.8 & 28.5$\pm^{0.4}_{0.4}$& 31.3$\pm$0.5& 8.95 & 52.0 & 51.7 \\ 
			H$_{\text{T}}>\text{H}_{\text{T}}^{\text{miss}}$\hphantom{0} & 85.7$\pm^{0.7}_{0.8}$& 86.8$\pm$1.5& 1.27 & 0.9 & 1.3 & 27.7$\pm^{0.4}_{0.4}$& 30.2$\pm$0.5& 8.28 & 0.8 & 1.1 \\ 
			IsoMuons\hphantom{0} & 85.6$\pm^{0.7}_{0.8}$& 86.7$\pm$1.5& 1.27 & 0.1 & 0.1 & 27.7$\pm^{0.4}_{0.4}$& 30.1$\pm$0.5& 7.97 & 0.0 & 0.1 \\ 
			MuonTracks\hphantom{0} & 85.6$\pm^{0.7}_{0.8}$& 86.7$\pm$1.5& 1.27 & 0.0 & 0.0 & 27.7$\pm^{0.4}_{0.4}$& 30.1$\pm$0.5& 7.97 & 0.0 & 0.0 \\ 
			IsoElectrons\hphantom{0} & 85.6$\pm^{0.7}_{0.8}$& 86.4$\pm$1.5& 0.93 & 0.0 & 0.3 & 27.6$\pm^{0.4}_{0.4}$& 30.0$\pm$0.5& 8.0 & 0.1 & 0.1 \\ 
			ElectronTracks\hphantom{0} & 85.6$\pm^{0.7}_{0.8}$& 86.2$\pm$1.5& 0.7 & 0.0 & 0.2 & 27.6$\pm^{0.4}_{0.4}$& 29.9$\pm$0.5& 7.69 & 0.0 & 0.1 \\ 
			IsoTracks\hphantom{0} & 85.3$\pm^{0.7}_{0.8}$& 85.6$\pm$1.5& 0.35 & 0.3 & 0.6 & 27.5$\pm^{0.4}_{0.4}$& 29.6$\pm$0.5& 7.09 & 0.1 & 0.3 \\ 
			IsoPhotons\hphantom{0} & 84.3$\pm^{0.8}_{0.8}$& 83.6$\pm$1.6& -0.84 & 1.0 & 2.0 & 27.3$\pm^{0.4}_{0.4}$& 28.8$\pm$0.5& 5.21 & 0.2 & 0.8 \\ 
			$\Delta\Phi_{1}>0.5$\hphantom{0} & 84.3$\pm^{0.8}_{0.8}$& 83.5$\pm$1.6& -0.96 & 0.0 & 0.1 & 27.3$\pm^{0.4}_{0.4}$& 28.8$\pm$0.5& 5.21 & 0.0 & 0.0 \\ 
			$\Delta\Phi_{2}>0.5$\hphantom{0} & 80.5$\pm^{0.8}_{0.9}$& 78.7$\pm$1.7& -2.29 & 3.8 & 4.8 & 25.9$\pm^{0.4}_{0.4}$& 27.1$\pm$0.5& 4.43 & 1.4 & 1.7 \\ 
			$\Delta\Phi_{3}>0.3$\hphantom{0} & 76.5$\pm^{0.9}_{0.9}$& 74.4$\pm$1.8& -2.82 & 4.0 & 4.3 & 25.0$\pm^{0.4}_{0.4}$& 26.0$\pm$0.5& 3.85 & 0.9 & 1.1 \\ 
			$\Delta\Phi_{4}>0.3$\hphantom{0} & 74.3$\pm^{0.9}_{1.0}$& 71.4$\pm$1.9& -4.06 & 2.2 & 3.0 & 24.2$\pm^{0.4}_{0.4}$& 25.2$\pm$0.5& 3.97 & 0.8 & 0.8 \\ \botrule
		\end{tabular}\label{cutflow:T2qq-1400-200} }
\end{table}
\begin{table}[t]
	\tbl{Signal yield in the aggregated signal regions for the T2qq simplified models.}
	{\setlength\tabcolsep{3.0pt}\begin{tabular}{@{}ccccc|ccc||ccc@{}} \toprule
			& Agg.             &   SR            &              &                                        &T2qq-1400-200&       &                      & T2qq-1000-800        &       &                     \\\hline
			bin    & N$_{\text{jet}}$&N$_{\text{bjet}}$&H$_{\text{T}}$&$\text{H}_{\text{T}}^{\text{miss}}$& MA5      & CMS   & MA5                  & MA5               & CMS   & MA5                  \\
			&                  &                 &[GeV]         &[GeV]                                   &yield     & yield &  diff [\%] & yield             & yield & diff [\%] \\ \colrule
			1&$\geq$ 2&0&$>$600&$>$600\hphantom{0} & 233.98$\pm$7.16& 285.27$\pm$2.44 & 17.98& 172.66$\pm$10.51& 188.65$\pm$4.54 & 8.48  \\ 
			2&$\geq$ 4&0&$>$1700&$>$850\hphantom{0} & 40.97$\pm$3.0& 35.93$\pm$0.81 & -14.03& 7.67$\pm$2.22& 10.1$\pm$0.96 & 24.06  \\ 
			3&$\geq$ 6&0&$>$600&$>$600\hphantom{0} & 38.56$\pm$2.91& 33.64$\pm$0.72 & -14.63& 45.4$\pm$5.39& 48.21$\pm$1.99 & 5.83  \\ 
			4&$\geq$ 8&$\leq$ 2&$>$600&$>$600\hphantom{0} & 8.76$\pm$1.39& 7.19$\pm$0.28 & -21.84& 10.23$\pm$2.56& 11.76$\pm$0.77 & 13.01  \\ 
			5&$\geq$ 10&$\leq$ 2&$>$1700&$>$850\hphantom{0} & 0.22$\pm$0.22& 0.4$\pm$0.06 & 45.0& 0.64$\pm$0.64& 0.34$\pm$0.11 & -88.24  \\ 
			6&$\geq$ 4&$\geq$ 2&$>$300&$>$300\hphantom{0} & 5.26$\pm$1.07& 5.17$\pm$0.15 & -1.74& 26.86$\pm$4.14& 23.23$\pm$0.61 & -15.63  \\ 
			7&$\geq$ 2&$\geq$ 2&$>$600&$>$600\hphantom{0} & 5.04$\pm$1.05& 4.55$\pm$0.12 & -10.77& 7.67$\pm$2.22& 4.57$\pm$0.26 & -67.83  \\ 
			8&$\geq$ 6&$\geq$ 2&$>$350&$>$350\hphantom{0} & 2.85$\pm$0.79& 2.38$\pm$0.1 & -19.75& 14.71$\pm$3.07& 8.68$\pm$0.39 & -69.47  \\ 
			9&$\geq$ 4&$\geq$ 2&$>$600&$>$600\hphantom{0} & 3.72$\pm$0.9& 3.72$\pm$0.12 & 0.0& 7.03$\pm$2.12& 4.23$\pm$0.25 & -66.19  \\ 
			10&$\geq$ 8&$\geq$ 3&$>$300&$>$300\hphantom{0} & 0.0$\pm$0.0& 0.12$\pm$0.03 & 100.0& 0.64$\pm$0.64& 0.36$\pm$0.07 & -77.78  \\ 
			11&$\geq$ 6&$\geq$ 1&$>$600&$>$600\hphantom{0} & 7.45$\pm$1.28& 10.6$\pm$0.24 & 29.72& 17.27$\pm$3.32& 14.65$\pm$0.62 & -17.88  \\ 
			12&$\geq$ 10&$\geq$ 3&$>$850&$>$850\hphantom{0} & 0.0$\pm$0.0& 0.02$\pm$0.01 & 100.0& 0.0$\pm$0.0& 0.0$\pm$0.0 & nan  \\ \botrule
		\end{tabular}\label{yield:T2qq-1400-200} }
\end{table}
\clearpage
\begin{table}[t]
	\tbl{Pre-selection cutflow for the T2bb simplified models.}
	{\setlength\tabcolsep{3.0pt}
		\begin{tabular}{@{}cccccc||ccccc@{}} \toprule
			& T2bb &            &                     &                     &                     & T2bb &                      &                     &                     &                     \\
			& 1000-100 &            &                     &                     &                     & 600-450 &                      &                     &                     &                     \\\hline
			Cut & MA5        & CMS        & MA5                 & MA5                 & CMS                 & MA5        & CMS                  & MA5                 & MA5                 & CMS                 \\
			&            &            & diff [\%] & drop [\%] & drop [\%] &            &                      & diff [\%] & drop [\%] & drop [\%] \\ \colrule
			N$_{\text{jet}} \geq 2$\hphantom{0} & 99.2$\pm^{0.2}_{0.2}$& 98.8$\pm$0.5& -0.4 & 0.8 & 1.2 & 95.2$\pm^{0.4}_{0.5}$& 95.4$\pm$0.1& 0.21 & 4.8 & 4.6 \\ 
			H$_{\text{T}}>300$\hphantom{0} & 98.8$\pm^{0.2}_{0.3}$& 98.3$\pm$0.5& -0.51 & 0.4 & 0.5 & 56.5$\pm^{1.0}_{1.0}$& 58.2$\pm$0.3& 2.92 & 38.7 & 37.2 \\ 
			$\text{H}_{\text{T}}^{\text{miss}}>300$\hphantom{0} & 78.2$\pm^{0.9}_{0.9}$& 79.6$\pm$1.4& 1.76 & 20.6 & 18.7 & 12.4$\pm^{0.7}_{0.7}$& 13.6$\pm$0.2& 8.82 & 44.1 & 44.6 \\ 
			H$_{\text{T}}>\text{H}_{\text{T}}^{\text{miss}}$\hphantom{0} & 77.3$\pm^{0.9}_{0.9}$& 78.2$\pm$1.4& 1.15 & 0.9 & 1.4 & 12.2$\pm^{0.7}_{0.7}$& 13.2$\pm$0.2& 7.58 & 0.2 & 0.4 \\ 
			IsoMuons\hphantom{0} & 77.2$\pm^{0.9}_{0.9}$& 77.9$\pm$1.4& 0.9 & 0.1 & 0.3 & 12.1$\pm^{0.7}_{0.7}$& 13.1$\pm$0.2& 7.63 & 0.1 & 0.1 \\ 
			MuonTracks\hphantom{0} & 77.2$\pm^{0.9}_{0.9}$& 77.8$\pm$1.4& 0.77 & 0.0 & 0.1 & 12.1$\pm^{0.7}_{0.7}$& 13.1$\pm$0.2& 7.63 & 0.0 & 0.0 \\ 
			IsoElectrons\hphantom{0} & 77.0$\pm^{0.9}_{0.9}$& 77.5$\pm$1.5& 0.65 & 0.2 & 0.3 & 12.1$\pm^{0.7}_{0.7}$& 13.0$\pm$0.2& 6.92 & 0.0 & 0.1 \\ 
			ElectronTracks\hphantom{0} & 76.7$\pm^{0.9}_{0.9}$& 77.2$\pm$1.5& 0.65 & 0.3 & 0.3 & 12.0$\pm^{0.7}_{0.7}$& 12.9$\pm$0.2& 6.98 & 0.1 & 0.1 \\ 
			IsoTracks\hphantom{0} & 76.4$\pm^{0.9}_{0.9}$& 76.8$\pm$1.5& 0.52 & 0.3 & 0.4 & 12.0$\pm^{0.7}_{0.7}$& 12.8$\pm$0.2& 6.25 & 0.0 & 0.1 \\ 
			IsoPhotons\hphantom{0} & 75.8$\pm^{0.9}_{0.9}$& 75.2$\pm$1.5& -0.8 & 0.6 & 1.6 & 11.9$\pm^{0.7}_{0.7}$& 12.5$\pm$0.2& 4.8 & 0.1 & 0.3 \\ 
			$\Delta\Phi_{1}>0.5$\hphantom{0} & 75.7$\pm^{0.9}_{0.9}$& 75.1$\pm$1.5& -0.8 & 0.1 & 0.1 & 11.9$\pm^{0.7}_{0.7}$& 12.5$\pm$0.2& 4.8 & 0.0 & 0.0 \\ 
			$\Delta\Phi_{2}>0.5$\hphantom{0} & 72.0$\pm^{1.0}_{1.0}$& 70.6$\pm$1.6& -1.98 & 3.7 & 4.5 & 10.8$\pm^{0.7}_{0.6}$& 11.3$\pm$0.2& 4.42 & 1.1 & 1.2 \\ 
			$\Delta\Phi_{3}>0.3$\hphantom{0} & 68.2$\pm^{1.0}_{1.0}$& 67.0$\pm$1.6& -1.79 & 3.8 & 3.6 & 10.4$\pm^{0.7}_{0.6}$& 10.7$\pm$0.2& 2.8 & 0.4 & 0.6 \\ 
			$\Delta\Phi_{4}>0.3$\hphantom{0} & 65.4$\pm^{1.0}_{1.0}$& 64.5$\pm$1.6& -1.4 & 2.8 & 2.5 & 10.0$\pm^{0.7}_{0.6}$& 10.2$\pm$0.2& 1.96 & 0.4 & 0.5 \\ \botrule
		\end{tabular}\label{cutflow:T2bb-1000-100} }
\end{table}
\begin{table}[t]
	\tbl{Signal yield in the aggregated signal regions for the T2bb simplified models.}
	{\setlength\tabcolsep{3.0pt}\begin{tabular}{@{}ccccc|ccc||ccc@{}} \toprule
			& Agg.             &   SR            &              &                                        &T2bb-1000-100&       &                      & T2bb-600-450        &       &                     \\\hline
			bin    & N$_{\text{jet}}$&N$_{\text{bjet}}$&H$_{\text{T}}$&$\text{H}_{\text{T}}^{\text{miss}}$& MA5      & CMS   & MA5                  & MA5               & CMS   & MA5                  \\
			&                  &                 &[GeV]         &[GeV]                                   &yield     & yield &  diff [\%] & yield             & yield & diff [\%] \\ \colrule
			1&$\geq$ 2&0&$>$600&$>$600\hphantom{0} & 52.59$\pm$4.56& 81.08$\pm$0.94 & 35.14& 0.0$\pm$0.0& 30.02$\pm$0.66 & 100.0  \\ 
			2&$\geq$ 4&0&$>$1700&$>$850\hphantom{0} & 4.75$\pm$1.37& 3.24$\pm$0.19 & -46.6& 0.0$\pm$0.0& 0.84$\pm$0.09 & 100.0  \\ 
			3&$\geq$ 6&0&$>$600&$>$600\hphantom{0} & 7.12$\pm$1.68& 7.38$\pm$0.22 & 3.52& 0.0$\pm$0.0& 6.0$\pm$0.21 & 100.0  \\ 
			4&$\geq$ 8&$\leq$ 2&$>$600&$>$600\hphantom{0} & 3.56$\pm$1.19& 3.38$\pm$0.16 & -5.33& 11.1$\pm$11.1& 4.93$\pm$0.27 & -125.15  \\ 
			5&$\geq$ 10&$\leq$ 2&$>$1700&$>$850\hphantom{0} & 0.0$\pm$0.0& 0.11$\pm$0.03 & 100.0& 0.0$\pm$0.0& 0.15$\pm$0.05 & 100.0  \\ 
			6&$\geq$ 4&$\geq$ 2&$>$300&$>$300\hphantom{0} & 133.26$\pm$7.26& 84.14$\pm$0.85 & -58.38& 843.84$\pm$96.79& 682.87$\pm$4.09 & -23.57  \\ 
			7&$\geq$ 2&$\geq$ 2&$>$600&$>$600\hphantom{0} & 102.42$\pm$6.36& 76.83$\pm$0.81 & -33.31& 99.93$\pm$33.31& 90.18$\pm$1.49 & -10.81  \\ 
			8&$\geq$ 6&$\geq$ 2&$>$350&$>$350\hphantom{0} & 35.19$\pm$3.73& 23.03$\pm$0.43 & -52.8& 288.68$\pm$56.61& 143.37$\pm$1.69 & -101.35  \\ 
			9&$\geq$ 4&$\geq$ 2&$>$600&$>$600\hphantom{0} & 67.22$\pm$5.16& 43.57$\pm$0.6 & -54.28& 88.82$\pm$31.4& 74.32$\pm$1.31 & -19.51  \\ 
			10&$\geq$ 8&$\geq$ 3&$>$300&$>$300\hphantom{0} & 1.58$\pm$0.79& 1.01$\pm$0.08 & -56.44& 22.21$\pm$15.7& 6.33$\pm$0.28 & -250.87  \\ 
			11&$\geq$ 6&$\geq$ 1&$>$600&$>$600\hphantom{0} & 33.61$\pm$3.65& 29.97$\pm$0.54 & -12.15& 88.82$\pm$31.4& 51.75$\pm$1.01 & -71.63  \\ 
			12&$\geq$ 10&$\geq$ 3&$>$850&$>$850\hphantom{0} & 0.0$\pm$0.0& 0.04$\pm$0.02 & 100.0& 0.0$\pm$0.0& 0.07$\pm$0.03 & 100.0  \\ \botrule
		\end{tabular}\label{yield:T2bb-1000-100} }
\end{table}
\clearpage
\begin{table}[t]
	\tbl{Pre-selection cutflow for the T2tt simplified models.}
	{\setlength\tabcolsep{3.0pt}
		\begin{tabular}{@{}cccccc||ccccc@{}} \toprule
			& T2tt &            &                     &                     &                     & T2tt &                      &                     &                     &                     \\
			& 950-100 &            &                     &                     &                     & 600-400 &                      &                     &                     &                     \\\hline
			Cut & MA5        & CMS        & MA5                 & MA5                 & CMS                 & MA5        & CMS                  & MA5                 & MA5                 & CMS                 \\
			&            &            & diff [\%] & drop [\%] & drop [\%] &            &                      & diff [\%] & drop [\%] & drop [\%] \\ \colrule
			N$_{\text{jet}} \geq 2$\hphantom{0} & 99.9$\pm^{0.1}_{0.1}$& 99.9$\pm$0.2& 0.0 & 0.1 & 0.1 & 99.2$\pm^{0.1}_{0.1}$& 99.6$\pm$0.0& 0.4 & 0.8 & 0.4 \\ 
			H$_{\text{T}}>300$\hphantom{0} & 97.9$\pm^{0.3}_{0.3}$& 98.7$\pm$0.4& 0.81 & 2.0 & 1.2 & 67.4$\pm^{0.4}_{0.4}$& 72.2$\pm$0.3& 6.65 & 31.8 & 27.4 \\ 
			$\text{H}_{\text{T}}^{\text{miss}}>300$\hphantom{0} & 68.9$\pm^{1.0}_{1.0}$& 74.5$\pm$1.2& 7.52 & 29.0 & 24.2 & 8.5$\pm^{0.3}_{0.3}$& 9.2$\pm$0.2& 7.61 & 58.9 & 63.0 \\ 
			H$_{\text{T}}>\text{H}_{\text{T}}^{\text{miss}}$\hphantom{0} & 68.3$\pm^{1.0}_{1.0}$& 73.6$\pm$1.3& 7.2 & 0.6 & 0.9 & 8.4$\pm^{0.3}_{0.3}$& 9.1$\pm$0.2& 7.69 & 0.1 & 0.1 \\ 
			IsoMuons\hphantom{0} & 56.4$\pm^{1.1}_{1.1}$& 58.7$\pm$1.4& 3.92 & 11.9 & 14.9 & 6.9$\pm^{0.2}_{0.2}$& 7.0$\pm$0.2& 1.43 & 1.5 & 2.1 \\ 
			MuonTracks\hphantom{0} & 55.8$\pm^{1.1}_{1.1}$& 58.2$\pm$1.4& 4.12 & 0.6 & 0.5 & 6.8$\pm^{0.2}_{0.2}$& 6.9$\pm$0.2& 1.45 & 0.1 & 0.1 \\ 
			IsoElectrons\hphantom{0} & 45.7$\pm^{1.1}_{1.1}$& 47.2$\pm$1.4& 3.18 & 10.1 & 11.0 & 5.4$\pm^{0.2}_{0.2}$& 5.4$\pm$0.1& 0.0 & 1.4 & 1.5 \\ 
			ElectronTracks\hphantom{0} & 45.5$\pm^{1.1}_{1.1}$& 46.4$\pm$1.4& 1.94 & 0.2 & 0.8 & 5.3$\pm^{0.2}_{0.2}$& 5.2$\pm$0.1& -1.92 & 0.1 & 0.2 \\ 
			IsoTracks\hphantom{0} & 45.3$\pm^{1.1}_{1.1}$& 45.5$\pm$1.4& 0.44 & 0.2 & 0.9 & 5.1$\pm^{0.2}_{0.2}$& 4.8$\pm$0.1& -6.25 & 0.2 & 0.4 \\ 
			IsoPhotons\hphantom{0} & 44.6$\pm^{1.1}_{1.1}$& 43.8$\pm$1.4& -1.83 & 0.7 & 1.7 & 5.1$\pm^{0.2}_{0.2}$& 4.7$\pm$0.1& -8.51 & 0.0 & 0.1 \\ 
			$\Delta\Phi_{1}>0.5$\hphantom{0} & 44.5$\pm^{1.1}_{1.1}$& 43.6$\pm$1.4& -2.06 & 0.1 & 0.2 & 5.1$\pm^{0.2}_{0.2}$& 4.7$\pm$0.1& -8.51 & 0.0 & 0.0 \\ 
			$\Delta\Phi_{2}>0.5$\hphantom{0} & 42.1$\pm^{1.1}_{1.1}$& 41.1$\pm$1.4& -2.43 & 2.4 & 2.5 & 4.3$\pm^{0.2}_{0.2}$& 3.9$\pm$0.1& -10.26 & 0.8 & 0.8 \\ 
			$\Delta\Phi_{3}>0.3$\hphantom{0} & 41.0$\pm^{1.1}_{1.0}$& 39.8$\pm$1.4& -3.02 & 1.1 & 1.3 & 3.7$\pm^{0.2}_{0.2}$& 3.4$\pm$0.1& -8.82 & 0.6 & 0.5 \\ 
			$\Delta\Phi_{4}>0.3$\hphantom{0} & 39.5$\pm^{1.1}_{1.0}$& 38.5$\pm$1.4& -2.6 & 1.5 & 1.3 & 3.2$\pm^{0.2}_{0.2}$& 3.0$\pm$0.1& -6.67 & 0.5 & 0.4 \\ \botrule
		\end{tabular}\label{cutflow:T2tt-950-100} }
\end{table}
\begin{table}[t]
	\tbl{Signal yield in the aggregated signal regions for the T2tt simplified models.}
	{\setlength\tabcolsep{3.0pt}\begin{tabular}{@{}ccccc|ccc||ccc@{}} \toprule
			& Agg.             &   SR            &              &                                        &T2tt-950-100&       &                      & T2tt-600-400        &       &                     \\\hline
			bin    & N$_{\text{jet}}$&N$_{\text{bjet}}$&H$_{\text{T}}$&$\text{H}_{\text{T}}^{\text{miss}}$& MA5      & CMS   & MA5                  & MA5               & CMS   & MA5                  \\
			&                  &                 &[GeV]         &[GeV]                                   &yield     & yield &  diff [\%] & yield             & yield & diff [\%] \\ \colrule
			1&$\geq$ 2&0&$>$600&$>$600\hphantom{0} & 32.91$\pm$4.32& 40.94$\pm$0.5 & 19.61& 8.75$\pm$4.37& 7.51$\pm$0.53 & -16.51  \\ 
			2&$\geq$ 4&0&$>$1700&$>$850\hphantom{0} & 0.57$\pm$0.57& 1.56$\pm$0.1 & 63.46& 0.0$\pm$0.0& 0.32$\pm$0.07 & 100.0  \\ 
			3&$\geq$ 6&0&$>$600&$>$600\hphantom{0} & 11.92$\pm$2.6& 13.38$\pm$0.23 & 10.91& 8.75$\pm$4.37& 4.53$\pm$0.36 & -93.16  \\ 
			4&$\geq$ 8&$\leq$ 2&$>$600&$>$600\hphantom{0} & 10.21$\pm$2.41& 12.77$\pm$0.27 & 20.05& 10.93$\pm$4.89& 8.71$\pm$0.54 & -25.49  \\ 
			5&$\geq$ 10&$\leq$ 2&$>$1700&$>$850\hphantom{0} & 0.0$\pm$0.0& 0.28$\pm$0.04 & 100.0& 0.0$\pm$0.0& 0.23$\pm$0.07 & 100.0  \\ 
			6&$\geq$ 4&$\geq$ 2&$>$300&$>$300\hphantom{0} & 197.47$\pm$10.59& 181.57$\pm$1.13 & -8.76& 325.77$\pm$26.69& 254.99$\pm$3.37 & -27.76  \\ 
			7&$\geq$ 2&$\geq$ 2&$>$600&$>$600\hphantom{0} & 76.04$\pm$6.57& 87.22$\pm$0.78 & 12.82& 26.24$\pm$7.57& 19.85$\pm$0.92 & -32.19  \\ 
			8&$\geq$ 6&$\geq$ 2&$>$350&$>$350\hphantom{0} & 97.6$\pm$7.44& 96.21$\pm$0.8 & -1.44& 161.79$\pm$18.81& 130.96$\pm$2.34 & -23.54  \\ 
			9&$\geq$ 4&$\geq$ 2&$>$600&$>$600\hphantom{0} & 71.5$\pm$6.37& 81.86$\pm$0.76 & 12.66& 26.24$\pm$7.57& 19.67$\pm$0.91 & -33.4  \\ 
			10&$\geq$ 8&$\geq$ 3&$>$300&$>$300\hphantom{0} & 9.08$\pm$2.27& 7.36$\pm$0.16 & -23.37& 28.42$\pm$7.88& 20.36$\pm$0.71 & -39.59  \\ 
			11&$\geq$ 6&$\geq$ 1&$>$600&$>$600\hphantom{0} & 72.63$\pm$6.42& 88.82$\pm$0.8 & 18.23& 59.03$\pm$11.36& 30.97$\pm$1.16 & -90.6  \\ 
			12&$\geq$ 10&$\geq$ 3&$>$850&$>$850\hphantom{0} & 0.0$\pm$0.0& 0.19$\pm$0.03 & 100.0& 0.0$\pm$0.0& 0.09$\pm$0.03 & 100.0  \\ \botrule
		\end{tabular}\label{yield:T2tt-950-100} }
\end{table}
\clearpage

\clearpage

\subsection{Conclusions}
We have presented a recast-ready implementation of \textit{Search for supersymmetry in proton-proton collisions at 13 TeV in final states with jets and missing transverse momentum} (CMS-SUS-19-006)~\cite{Sirunyan:2019ctn}. The implementation has been validated using a plethora of information made available by the CMS experiment/analysis team. We find that the accompanying code provides a good description of the signal acceptance of a wide range of simplified models of supersymmetry, where acceptances and distributions generally agree with official results to within 20\%, or else within statistical uncertainties. We note that in a few cases this general statement is made exception to and discuss the results and implications. We finally remind that it is most robust to use a combination of signal regions in any interpretation, since a signal event that wrongfully migrates from one bin of $n_{b}$ to another will be in this way be picked up by another signal region, and a comparable sensitivity will be retained. 

The {\sc MadAnalysis}~5 {\sc C++} recast code is available online for download from the MA5 dataverse (\href{https://doi.org/10.14428/DVN/4DEJQM}{https://doi.org/10.14428/DVN/4DEJQM})~\cite{4DEJQM_2020}, together with the corresponding validation material (Monte Carlo event generator cards).

\cleardoublepage
\renewcommand{\mspin}{{\sc MadSpin}}
\renewcommand{\mw}{{\sc MadWidth}}
\renewcommand{\ma}{{\sc MadAnalysis~5}}
\renewcommand{\py}{{\sc Pythia~8}}
\renewcommand{\del}{{\sc Delphes~3}}
\renewcommand{\fj}{{\sc FastJet}}
\renewcommand{\be}{\begin{equation}}
\renewcommand{\ee}{\end{equation}}
\def\bsp#1\esp{\begin{split}#1\end{split}}
\def\gev{\textrm{GeV}}

\markboth{Luc Darm\'e and Benjamin Fuks}{Implementation of the CMS-TOP-18-003 analysis}

\section{Implementation of the CMS-TOP-18-003 analysis (four top quarks with at least two leptons; 137~fb$^{-1}$)}
  \vspace*{-.1cm}\footnotesize{\hspace{.5cm}By Luc Darm\'e and Benjamin~Fuks}
\label{sec:4tops}


\subsection{Introduction}

In the Standard Model (SM), four top quark production at the LHC ($pp\to t
\bar{t}t\bar{t}$) mainly proceeds through pure QCD contributions and the
associated production of a top-antitop pair with a Higgs boson in an
$s$-channel-like topology. The total production cross section is predicted,
at the next-to-leading order accuracy in the strong coupling, to
be $\sigma_{4t}^{\rm SM} =
11.97^{+2.15}_{-2.51}~{\rm fb}$~\cite{Frederix:2017wme}. By virtue of the size
of the top Yukawa coupling, Higgs-boson exchange diagrams contribute
significantly.  Four-top probes, which have already been under deep scrutiny until
now~\cite{Sirunyan:2017roi,Aaboud:2018xpj,Sirunyan:2019wxt}, have been consequently expected to provide soon an alternative channel to measure the top Yukawa coupling and more generally to play a key role at the upcoming third run of the LHC.

The four-top channel is in addition expected to be important in the search
for new physics scenarios, as such a signature could be representative of a variety of new
physics scenarios featuring top-philic new scalar or vector particles~\cite{Battaglia:2010xq,Greiner:2014qna,Alvarez:2016nrz,Kim:2016plm,Fox:2018ldq}.
Such particles arise for instance in composite Higgs solutions to the hierarchy 
problem~\cite{Lillie:2007hd,Pomarol:2008bh,Zhou:2012dz,Cacciapaglia:2015eqa,Cacciapaglia:2020vyf}, in models featuring extended supersymmetry\cite{Fox:2002bu,Plehn:2008ae,Choi:2008ub,GoncalvesNetto:2012nt,Calvet:2012rk,Benakli:2014cia,Kotlarski:2016zhv,Darme:2018dvz}, in models derived from the minimal flavour violation principle\cite{Gerbush:2007fe,Hayreter:2017wra}, but also simply when the new physics interact with the SM via a scalar portal mechanism (for which the top Yukawa coupling dominates the interactions).
Furthermore, four-top production can often be associated with a significant production of missing energy (although in that case supersymmetry-driven searches are typically better suited~\cite{Aaboud:2017vwy}).
While we will not focus on this mass range in this note,  the case of a top-philic particle with mass below $m_t$ has been also considered in the literature~\cite{Alvarez:2019uxp}. 
Finally, the measurement of the top Yukawa itself can be used to indirectly probe new physics.

New physics contributions generally lead to an enhancement of four-top
production, such an enhancement featuring kinematical properties significantly distinct from the SM.
It is therefore crucial to be able to extract reasonable bounds on models
under consideration from SM four-top production searches, as well as to
study the properties of the corresponding new physics signal to design a better suited
analysis strategy fully dedicated to the quest for beyond standard model particles. The \ma\
platform\cite{Conte:2012fm,Conte:2014zja,Dumont:2014tja,Conte:2018vmg} is one
of the public software aiming at such an objective. It allows for the
derivation of predictions detailing how the different signal regions of a
given LHC analysis are populated by an arbitrary new physics signal. The analysis impact on the signal
properties can furthermore be estimated.

We present in this note the implementation of the latest CMS analysis targeting
the production of four top quarks in the Standard Model\cite{Sirunyan:2019wxt}
in the \ma~framework, briefly describing the analysis itself in
Sec.~\ref{sec:descr}. In Sec.~\ref{sec:valid}, we provide information on the
procedure that we have followed in order to validate our implementation, so
that any potential user can check how robust is our work and to which level any
phenomenological outcome should be trusted. In this context, we have
verified the compatibility between a SM four-top signal as obtained with our
implementation and the official results as reported by the CMS collaboration,
both for event counts in the different signal regions of the analysis and
various differential distributions. A practical recasting example is shown in
Sec.~\ref{sec:scalars} and a summary of our work is given in
Sec.~\ref{sec:conclusions}.

\subsection{Description of the analysis}
\label{sec:descr}
The production of four top quarks and their subsequent decay at the LHC typically leads to final states
featuring a large number of leptons and hard jets with an important heavy-flavour content. In particular, a pair of leptons carrying the
same electric charge typically arises from 10\% of the decays. In contrast to any
other channel, a same-sign di-lepton probe is known to enjoy a low SM
background, and is thus an excellent way to search for any new phenomenon.
In order to increase the signal efficiency, the analysis additionally
considers a final-state with more than two leptons, as the SM background is in
that case is also known to be reducible to a small enough level.

\subsubsection{Object definitions}
\label{sec:obj}

The CMS-TOP-18-003 analysis~\cite{Sirunyan:2019wxt} defines 14 signal regions
that differ in the details on the selection criteria on the leptons and jets
reconstructed in the events. The signal object candidates are required to
satisfy mild kinematics requirements and to be isolated. The latter
criterion is particularly important as the analysis targets the identification
of events featuring a large multiplicity of isolated jets and leptons.

The signal selection process considers leptons with properties fulfilling
\begin{align}
    p_T > 20 ~\gev ~\textrm{ and }~ |\eta|  < 2.5 ~\textrm{(electrons)~or} ~2.4 ~\textrm{(muons)} \ .
\end{align}
Jets are reconstructed using the anti-$k_T$ algorithm\cite{Cacciari:2008gp}
with a distance parameter $R=0.4$, and the analysis is restricted to jets
featuring
\begin{align}
    p_T > 40 ~\gev ~\textrm{ and }~ |\eta|  < 2.4 \ .
\end{align}
In addition, all jets that are overlapping with a lepton are discarded, the
overlap being defined by constraining the angular distance in the transverse
plane $\Delta R$ so that it is smaller than 0.4. The angular distance is defined
in a standard way, with
\begin{align}
    \Delta R \equiv \sqrt{(\eta_j - \eta_\ell)^2 + (\varphi_j - \varphi_\ell)^2} > 0.4\ ,
\end{align}
where $\eta_j$ ($\eta_\ell$) is the jet (lepton) pseudo-rapidity and $\varphi_j$
($\varphi_\ell$) is the corresponding azimuthal angle.

At the same time, lepton isolation requirements~\cite{Khachatryan:2016kod}
restrict the amount of hadronic activity around the leptons, this activity being
evaluated by including the contributions of all (isolated and non-isolated)
jets and by ignoring any $p_T$ requirement on the jets. Lepton isolation is
enforced by means of three variables: first the mini-isolation variable
defined as the scalar sum of the transverse momenta of all charged hadrons,
neutral hadrons and photons within a cone of radius depending on the lepton
$p_T$; then the ratio of the lepton $p_T$ to the one of the closest jet within a
$\Delta R=0.4$ distance; and finally the $p_T^{\rm rel}$ variable defined as the
transverse momentum of the lepton relative to the residual momentum of the
closest jet (within a $\Delta R=0.4$ angular distance from the lepton), after
having subtracted the lepton momentum.

Since the analysis requires typically many jets and $b$-tagged jets (as much as
at least four in one of the analysis signal regions, for instance), controlling
precisely the performance of the $b$-tagging algorithm is critical.
The considered CMS analysis relies a deep neural network algorithm,
named DeepCSV~\cite{Sirunyan:2017ezt}, with a medium working point. The
corresponding $b$-tagging efficiency $\mathcal{E}_{b|b}$ approximately reads
\begin{equation}
  \hspace{-.2cm} \mathcal{E}_{b|b}(p_T)   = \begin{cases}
   0.13 +0.028  ~p_T - 5.07 \!\cdot\! 10^{-4} ~p_T^2 +4.07 \!\cdot\! 10^{-6}   ~p_T^3
      -1.21  \!\cdot\! 10^{-8}  ~p_T^4 \\    
   \quad   \ \   \textrm{ for } ~25 ~\gev <  p_T < 115 ~\gev\ , \\[0.5em]
        0.65 + 0.00143  ~p_T  - 1.03  \!\cdot\! 10^{-5}  ~p_T^2  + 2.55  \!\cdot\! 10^{-8} p_T^3
        - 2.78  \!\cdot\! 10^{-11} \\ \quad \ \  \ p_T^4 + 1.11 \cdot10^{-14}   p_T^5
     \   \  ~ \textrm{ for }  115  ~\gev \leqslant p_T < 950 ~\gev\ , \\[0.5em]
    0.50  \ \  \textrm{ for }  ~ p_T \geqslant 950  ~\gev \ ,
      \end{cases}
\end{equation}
and is associated with the mistagging rate of a charmed jet
($\mathcal{E}_{b|c}$) and a light jet ($\mathcal{E}_{j|b}$) as a $b$-jet given
by
\be\bsp
  \mathcal{E}_{b|c}(p_T) = & \begin{cases}
     0.0571+ 0.00603~p_T -  1.74\!\cdot\! 10^{-4} ~p_T^2
      + 2.15 \!\cdot\! 10^{-6}~p_T^3
      - 1.20 \!\cdot\! 10^{-8}~p_T^4 \\ \quad
      + 2.50 \!\cdot\! 10^{-11}~p_T^5
      \  \textrm{ for } ~25 ~\gev <  p_T < 155 ~\gev\ , \\[0.5em]
     15.8 - 0.432~p_T +  4.87\!\cdot\! 10^{-3} ~p_T^2
      - 2.88 \!\cdot\! 10^{-5}~p_T^3
      + 9.43 \!\cdot\! 10^{-8}~p_T^4 \\ \quad
      - 1.62 \!\cdot\! 10^{-10}~p_T^5
      + 1.14 \!\cdot\! 10^{-13}~p_T^6
      \  \textrm{ for } ~155 ~\gev <  p_T < 318 ~\gev\ , \\[0.5em]
        0.119 - 0.000225~p_T  + 1.36 \!\cdot\! 10^{-6}~p_T^2 
        - 1.96  \!\cdot\! 10^{-9}~p_T^3
        + 7.38  \!\cdot\! 10^{-13}~p_T^4 \\ \quad
        + 1.11 \!\cdot\! 10^{-16}~  p_T^5
        \ \textrm{ for }  318  ~\gev \leqslant p_T < 950 ~\gev\ , \\[0.5em]
    0.14  \ \  \textrm{ for }  ~ p_T \geqslant 950  ~\gev \ ,
  \end{cases}\\
  \mathcal{E}_{b|j}(p_T) = & \begin{cases}
     0.0194 - 0.000344~p_T +  3.66\!\cdot\! 10^{-6} ~p_T^2
      - 1.43 \!\cdot\! 10^{-8}~p_T^3
      + 1.27 \!\cdot\! 10^{-11}~p_T^4 \\ \quad
      + 4.82 \!\cdot\! 10^{-14}~p_T^5 -8.56\!\cdot\! 10^{-17}~p_T^6
      \  \textrm{ for } ~25 ~\gev <  p_T < 360 ~\gev\ , \\[0.5em]
        1.26 - 0.0134~p_T  + 5.83 \!\cdot\! 10^{-5}~p_T^2 
        - 1.30  \!\cdot\! 10^{-7}~p_T^3
        + 1.57  \!\cdot\! 10^{-10}~p_T^4 \\ \quad
        - 9.79 \!\cdot\!10^{-14}~  p_T^5
        + 2.48 \!\cdot\!10^{-17}~  p_T^6
        \ \textrm{ for }  260  ~\gev \leqslant p_T < 950 ~\gev\ , \\[0.5em]
    0.035  \ \  \textrm{ for }  ~ p_T \geqslant 950  ~\gev \ .
  \end{cases}
\esp\ee
We have accordingly designed a customised \del~\cite{deFavereau:2013fsa} card,
which should be used for the simulation of the detector response associated with
our implementation (see below). The above performance corresponds to an
average tagging efficiency ranging from $50\%$ to $70\%$, for quite small
associated false positive rates.

In the CMS-TOP-18-003 analysis, signal $b$-jet candidates are selected by
enforcing their transverse momentum to satisfy
\begin{align}
    p_T > 25 ~\gev \ .
\end{align}

\subsubsection{Event selection}

Strong selection cuts are applied to unravel the signal from the
large background. One first requires event final states to exhibit the
presence of at least two jets ($N_j\geq2$) and two $b$-tagged jets ($N_b\geq2$),
and then constrains the sum of the transverse momenta of all reconstructed jets
to satisfy
\begin{align}
    H_T  = \sum_{i=1}^{N_j} > 300 ~\gev \ .
\label{eq:ht}\end{align}
As a sensible amount of missing transverse energy $p_T^{\rm{miss}}$ is expected to arise from the leptonic top-quark decays for the considered signal, we ask events to satisfy
\begin{align}
p_T^{\rm{miss}} > 50 ~\gev \ .
\end{align}
As usual $p_T^{\text{miss}}$ denotes the magnitude of the projection of the
negative sum of the momenta of all reconstructed candidates in the event on the
plane perpendicular to the beams.

\begin{table}[t]
  \setlength\tabcolsep{10pt}
  \tbl{Preselection cuts as defined in the CMS-TOP-18-003
    analysis\cite{Sirunyan:2019wxt}. We recall that the sum of the transverse
    momenta of all jets is given by $H_T$, as defined in Eq.~\eqref{eq:ht}.}
  {\begin{tabular}{@{}cccccc@{}} \toprule
    \multicolumn{6}{l}{\textbf{Basic kinematic requirements}} \\
      & Electrons & Muons & Jets & $b$-tagged jets &\\
      $p_T$ (GeV) & $>20$ &  $>20$ &  $>40$ &  $>25$ &\\
      $\eta$ (GeV) & $>2.5$ &  $>2.4$ &  $>2.4$ &  $>2.4$&\\
    \colrule
    \multicolumn{6}{l}{\textbf{Baseline selection}} \\
    Jets & \multicolumn{5}{|c}{$H_T > 300$ GeV,  $p_T^{\text{miss}} > 50$ GeV, at least two  jets and two $b$-tagged jets} \\
    Leptons& \multicolumn{5}{|c}{If same charge pair: $p_T(\ell_1) > 25$ GeV and $p_T(\ell_i) > 20$ GeV for $i\neq 1$} \\
Isolation & \multicolumn{5}{|c}{Jets and b-tagged jets $\Delta R > 0.4$ w.r.t the selected leptons} \\
    \colrule
    \multicolumn{6}{l}{\textbf{Further vetoes}} \\
    Vetoed & \multicolumn{5}{|p{10cm}}{Same sign electron pairs with pair mass below $12$ GeV} \\
    Vetoed & \multicolumn{5}{|p{10cm}}{Third lepton with $p_T > 5 (7)$ GeV for $e$ ($\mu$) forming an opposite-sign same-flavour pair with an invariant mass $m_{\text{OS}} < 12$ GeV or $m_{\text{OS}} \in [76,106]$~GeV} \\
  \botrule
  \end{tabular}\label{tab:preselection} }
\end{table}

One then restricts the kinematical properties of the leptons and enforces that
the leading lepton has a transverse momentum $p_T(\ell_1) > 25$~GeV and that
there exists a trailing lepton of the same electric charge with a $p_T(\ell_i) >
20$~GeV (with $i\neq 1$). In addition, events featuring more than two leptons
are allowed, provided that no other same-sign lepton pair can be formed with the
leading lepton.

Extra selections are imposed to reject the possibility
that a lepton pair originates from a hadronic resonance or from a $Z$-boson
decay. The invariant mass $m_{\ell\ell}$ of any electron pair and any
opposite-sign muon pair that can be formed from the leptonic content of the
event has to be larger than 12~GeV. Moreover, $m_{\ell\ell}$ has to lie outside
the $Z$-boson mass window in the case of an opposite-sign same-flavour pair
($m_{\ell\ell}\not\in [76, 106]$~GeV). Those preselection cuts are summarised in
Table~\ref{tab:preselection}.

\begin{table}[t]
  \tbl{Definition of the signal regions of the CMS-TOP-18-003 analysis, together
    with the expectation from SM $t\bar{t} t\bar{t}$ production as reported by
    the CMS collaboration (pre-fit results are shown)\cite{Sirunyan:2019wxt}.}
  {\begin{tabular}{@{}cccc|c@{}} \toprule
     $N_\ell$  & $N_b$ & $N_j$ & Region & $t\bar{t} t\bar{t}$ (SM - CMS)  \\
     2 & 2       & 6        & SR1 & $1.89\pm 1.14 $ \\
     2 & 2       & 7        & SR2 & $1.04\pm 0.57 $ \\
     2 & 2       & $\geq$8  & SR3 & $0.67\pm 0.38 $ \\[.2cm]
     2 & 3       & 5        & SR4 & $1.51\pm 0.85 $ \\
     2 & 3       & 6        & SR5 & $1.61\pm 0.90 $ \\
     2 & 3       & 7        & SR6 & $1.14\pm 0.66 $ \\
     2 & 3       & $\geq$8  & SR7 & $0.85\pm 0.47 $ \\[.2cm]
     2 & $\geq4$ & $\geq$5  & SR8 & $2.08\pm 1.23 $ \\
  \botrule\end{tabular}
  \renewcommand{\arraystretch}{1.1}
  \begin{tabular}{@{}cccc|c@{}} \toprule
     $N_\ell$  & $N_b$ & $N_j$ & Region & $t\bar{t} t\bar{t}$ (SM - CMS)  \\
     $\geq 3$ & 2        & 5        & SR9  & $0.66\pm 0.38 $ \\
     $\geq 3$ & 2        & 6        & SR10 & $0.33\pm 0.21 $ \\
     $\geq 3$ & 2        & $\geq$7  & SR11 & $0.22\pm 0.13 $ \\[.2cm]
     $\geq 3$ & $\geq 3$ & 4        & SR12 & $0.56\pm 0.32 $ \\
     $\geq 3$ & $\geq 3$ & 5        & SR13& $0.66\pm 0.38 $  \\
     $\geq 3$ & $\geq 3$ & $\geq$6  & SR14 & $0.76\pm 0.45 $ \\ \botrule
  \end{tabular}\label{sr} }
\end{table}

Once signal leptons, jets and $b$-tagged jets have been identified and selected,
the CMS analysis then splits all surviving events into 14 distinct signal
regions (SR), according to the number of leptons present in the event $N_\ell$,
as well as the number of $b$-jets $N_b$ and jets $N_j$.
This selection is summarised signal region by signal region, in Table~\ref{sr},
along with the predicted number of SM $t\bar{t} t\bar{t}$ events that is
expected for each SR. The selection cuts are very
stringent and typically retain only around $2\%$ of the cross section.

Very importantly, one should pay attention to how
the numbers of SM four-top events populating each signal region are reported by
the CMS collaboration. The final results are provided ``post-fit'', {\it i.e.}
after the cross section related to the four-top SM signal has been fitted so
that theory and measurement match. In order to recover proper predictions, one
needs to rescale the results by the theoretical cross section
$\sigma_{4t}^{\rm SM}$. The obtained numbers of events, referred to henceforth
as ``pre-fit'', are the values to be compared with our \ma\ predictions when
validation is at stake.

\subsection{Validation}
\label{sec:valid}
\subsubsection{Event generation}
In order to validate our implementation, we generate SM four-top signal
events at the next-to-leading-order (NLO) accuracy in the strong coupling,
convoluting NLO matrix elements with the NLO set of NNPDF3.0 parton
densities\cite{Ball:2014uwa} that is provided through the LHAPDF~6 library~\cite{Buckley:2014ana}. In our simulations, we set the factorisation and renormalisation scales to the average transverse mass of the final-state particles, and the corresponding scale variation uncertainties are obtained by varying this choice by a factor of two up and down. Parton density uncertainties are extracted using replica sets.

After including the top quark decays with
the \mspin\ package\cite{Artoisenet:2012st} (so that spin correlations
are retained) and \mw~\cite{Alwall:2014bza}, the hard-scattering fixed-order
results are matched with parton showers as described by
\py\cite{Sjostrand:2014zea} that further includes the simulation of
hadronisation effects. We finally model the response of the CMS detector with
\del\cite{deFavereau:2013fsa}, which internally relies on
\fj\cite{Cacciari:2011ma} for object reconstruction.

We have created our own
\del\ card for this analysis, in order to match accurately the lepton and jet
reconstruction efficiencies as required by the CMS-TOP-18-003 analysis and the
corresponding $b$-tagging performance~\cite{Sirunyan:2019wxt} described in
Section~\ref{sec:obj}.

Our validation relies on 2,500,000 simulated SM events, generated according to the
procedure described above. This leads to about 50,000 events passing all
selection cuts. Accordingly, this allows us to neglect the statistical
uncertainties with respect to the theoretical ones when validation histograms
and cutflows are extracted.

\subsubsection{Comparison with the official results}

\begin{figure}[t]
	\centering
	\includegraphics[width=0.45\textwidth]{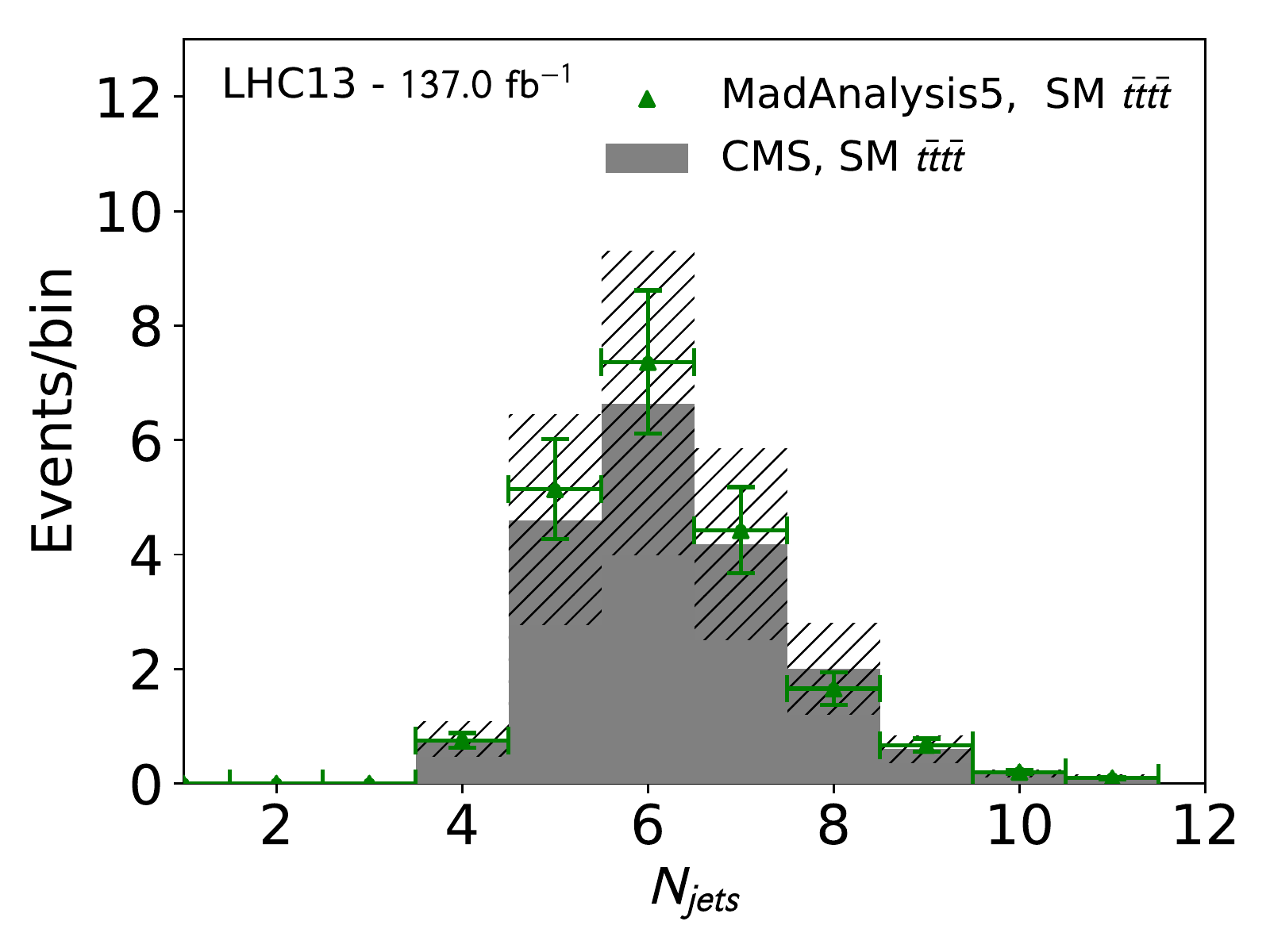}
	\includegraphics[width=0.45\textwidth]{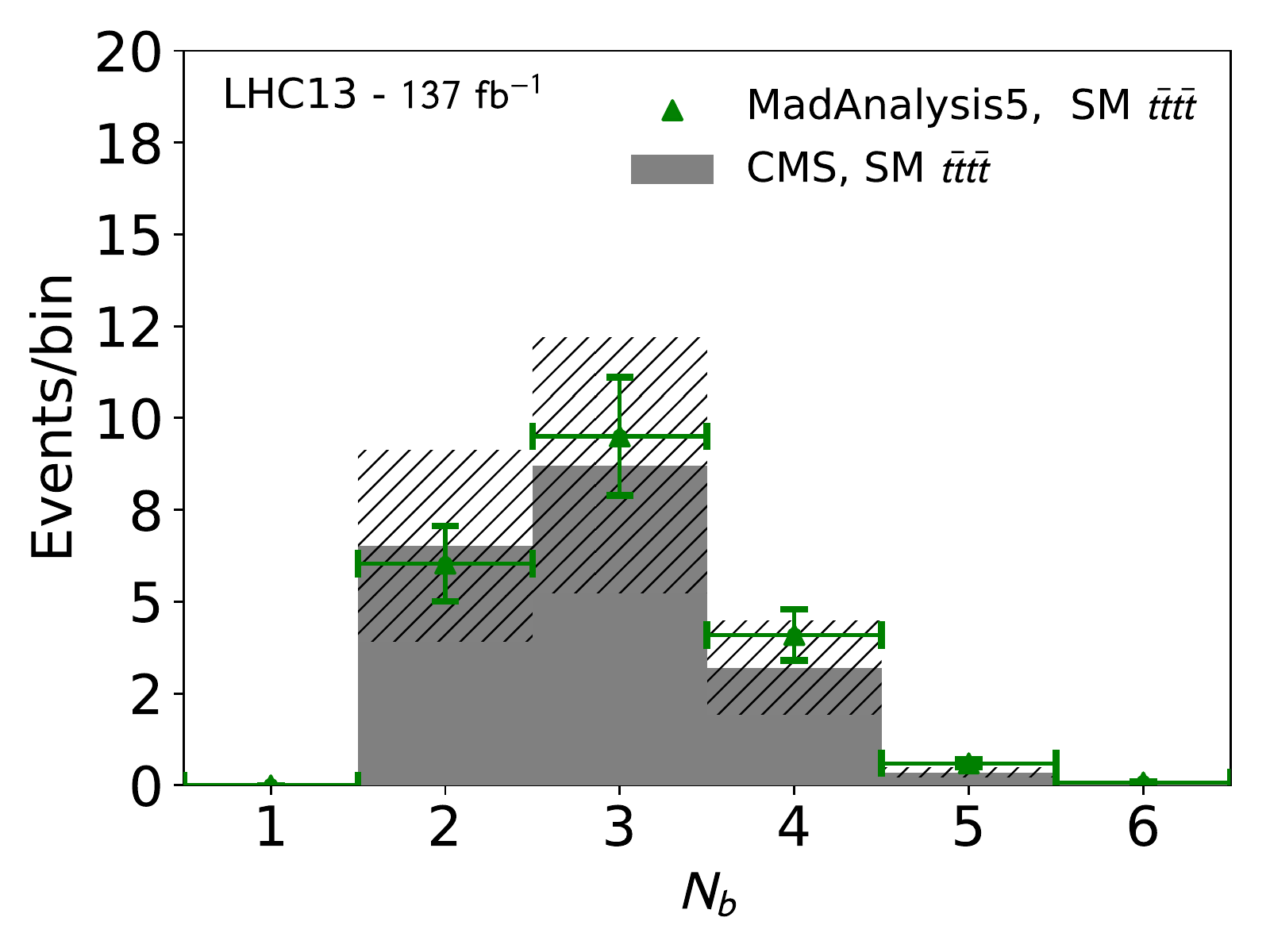}\\
	\includegraphics[width=0.45\textwidth]{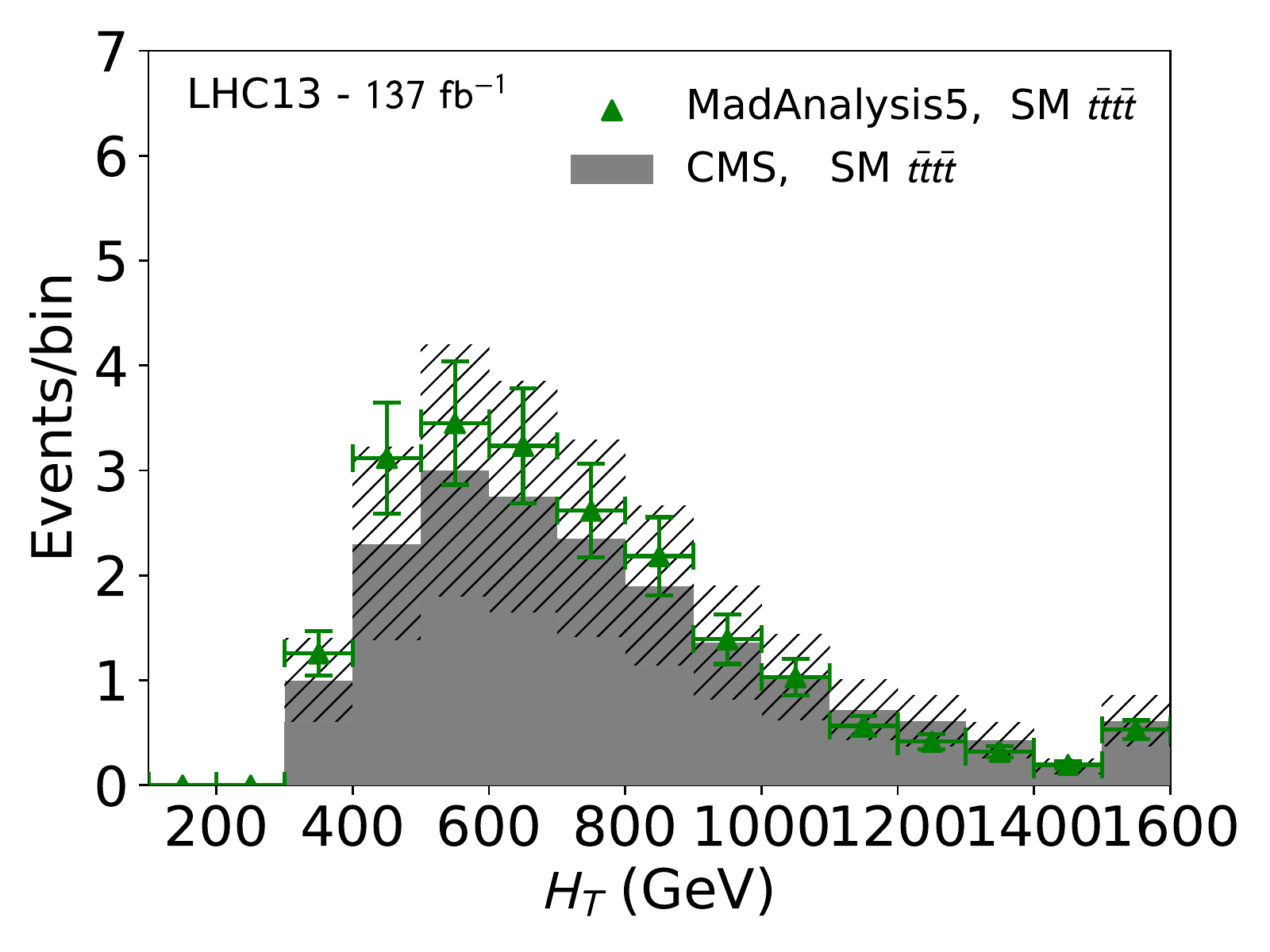}
	\includegraphics[width=0.5\textwidth]{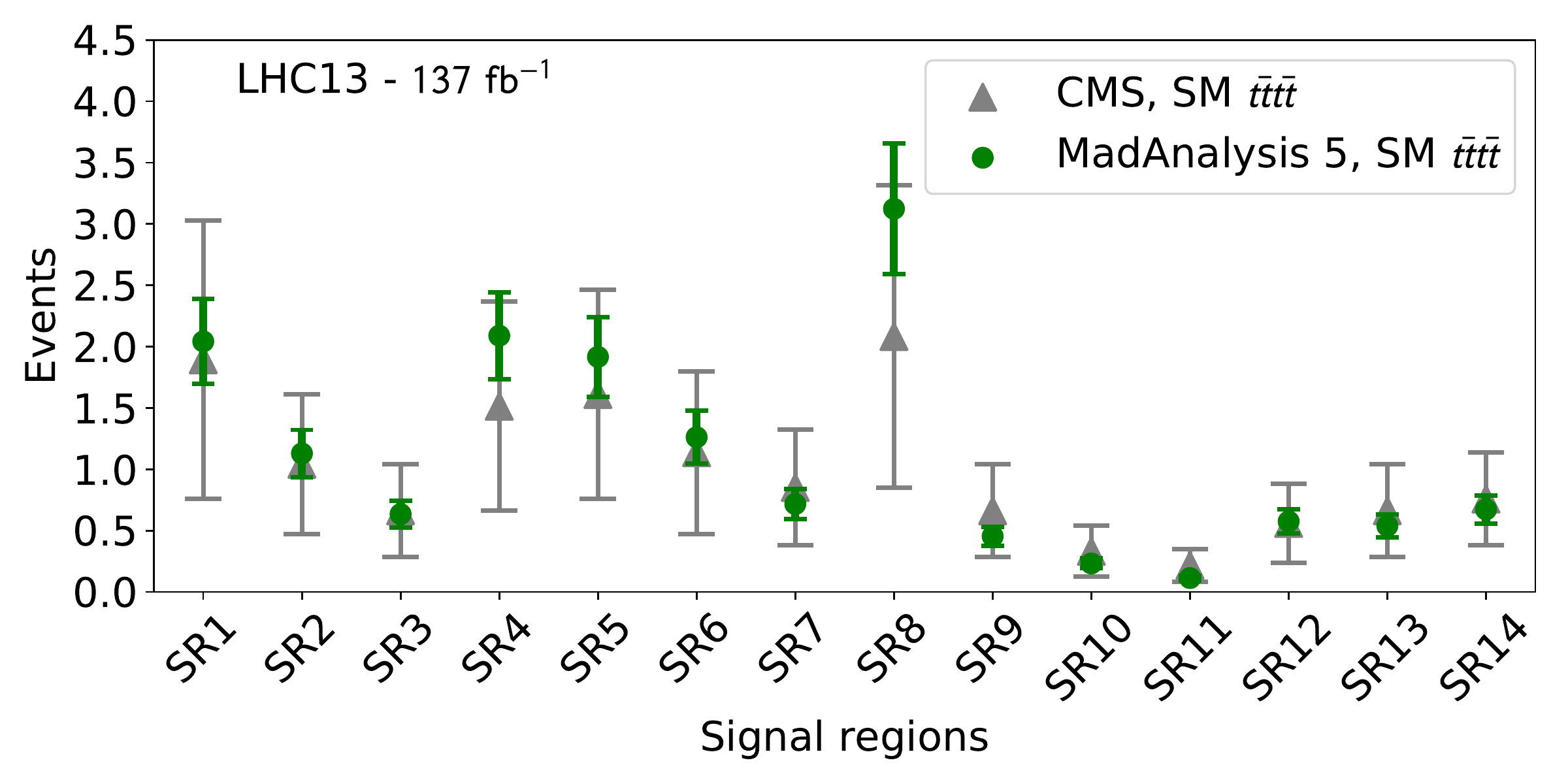}
  \caption{Validation figures of our implementation, in the \ma\ framework, of
		the CMS four-top analysis of Ref.~\cite{Sirunyan:2019wxt}. We compare \ma\
		predictions (green) with the CMS official results (dark grey) for the jet
		multiplicity (upper-left panel), $b$-jet multiplicity (upper-right panel)
		and $H_T$ (central-left panel)
		spectra, as well as for the event counts populating each signal region (lower
		right panel). The \ma\ predictions include theoretical uncertainties (green error
		bars) whilst the CMS numbers include both systematical and statistical errors
		(black dashed bands and light grey error bars in the lower panel).\protect\label{fig1}}
\end{figure}

We validate our implementation of the CMS-TOP-18-003 analysis by comparing
predictions obtained with our \ma\ implementation and the SM four-top events
generated following the above strategy. We show in Fig.~\ref{fig1} the result of
such a comparison for various differential distributions, and display
histograms representing the jet multiplicity $N_j$ (upper left), the $b$-jet
multiplicity $N_b$ (upper right) and the hadronic activity $H_T$ (lower left).
We find in all three cases a very good agreement, after accounting for the
errors, between the \ma\ predictions (green) and the CMS official
results\cite{Sirunyan:2019wxt} (grey).

Moreover, we also present the event yields in the different signal regions
(lower right) and compare again the \ma\ numbers (green) to the CMS results
(grey). A very good agreement is found, for all signal region.

We therefore consider our implementation as validated, so that it has been
added to the \ma\ Public Analysis Database (PAD) and is available from the \ma\ dataverse~\cite{OFAE1G_2020}.

\subsection{A practical example: top-philic scalars}
\label{sec:scalars}

As a simple illustrative example, we consider a simplified ($SU(2)$-violating) model in which a top-philic real scalar $S$ of mass $M_S$ interacts with the
Standard Model through the top-quark. The corresponding new physics Lagrangian
reads
\begin{align}
 \mathcal{L}_{s0} \supset \frac{1}{2} \partial_\mu S \partial^\mu S- \frac{1}{2} M_S^2 S^2 +
   y_0 \bar{t} t\ S + h.c\ ,
\end{align}
where $y_0$ denotes the new Yukawa coupling.
The main production mechanism of the four-top signal induced by new physics,
when $M_S > 2 m_t$ and $M_S$ is around or below the TeV-scale proceeds via
associated production,
\begin{align}
p p \to t \bar{t} S \to t \bar{t} t \bar{t} \ .
\end{align}
For lower scalar masses, the on-shell production of the scalar $S$ dominates,
implying that the cross section scales as $y_{0}^2$. On the contrary, for
higher mass, the off-shell contribution dominates instead, so that the cross
section scales as $y_{0}^4$. 

We present in Table~\ref{tab:limitCS} limits on the new physics signal cross
section that we derive with our \ma\ implementation. We consider two
scenarios in which $M_S = 600$ GeV and $M_S = 1000$ GeV respectively, and show
results for each SR of the analysis. We observe that the strongest limits
arise for the SR4, SR5 and SR8 region. The lack of sensitivity of the SR6 region
is associated with an observed large upward fluctuation of events in CMS data.
We moreover present  projections for the HL-LHC as derived with the machinery
introduced in Ref.~\cite{Araz:2019otb}.

\begin{table}[t]
  \setlength\tabcolsep{3.pt}
  \tbl{Observed and limits on the considered new physics signal production rate,
    as obtained with \ma\ and by using the implementation of the
    CMS-TOP-18-003 analysis~\cite{Sirunyan:2019wxt}. We consider two scenarios
    for which $M_S = 600$ GeV (upper row) and $M_S = 1000$ GeV (lower row). We
    moreover also show the projected limits at the HL-LHC.}
  {\begin{tabular}{@{}c|c c c | c c c c | c c c c| c c c @{}} \toprule
   				\rule{0pt}{3pt}  & SR1 & SR2 & SR3 & SR4 & SR5 & SR6 & SR7 & SR8 & SR9 & SR10 & SR11 & SR12 & SR13 & SR14 \\
     $M_S = 600 ~\rm{GeV}$ & & & & &&&&&&&& \\[0.3em]
     $~\sigma^{\rm{lim}}_{\rm{obs}}$ (fb) & $86$ & $70$ & $124$ & $55$ & $29$ & $97$ & $49$ & $\mathbf{23}$ & $169$ & $363$ & $979$ & $87$ & $95$ & $80$ \\
     $ ~\sigma^{\rm{lim}}_{\rm{HL-LHC}}$ (fb)  & $26$ & $29$ & $34$ & $19$ & $\mathbf{13}$ & $16$ & $19$ & $12$ & $58$ & $65$ & $351$ & $26$ & $23$ &  $21$  \\[0.4em]
     \hline				\rule{0pt}{14pt}
     $M_S = 1000 ~\rm{GeV}$ & & & & &&&&&&&& \\[0.3em]
      $~\sigma^{\rm{lim}}_{\rm{obs}}$ (fb) & $74$ & $49$ & $61$ & $50$ & $25$ & $74$ & $25$ & $\mathbf{17}$ & $125$ & $175$ & $189$ & $100$ & $78$ & $45$ \\
     $ ~\sigma^{\rm{lim}}_{\rm{HL-LHC}}$ (fb)  & $22$ & $20$ & $17$ & $17$ & $11$ & $12$ & $10$ & $\mathbf{9}$ & $43$ & $32$ & $68$ & $30$ & $18$ &  $13$ \\[0.4em]
  \hline\end{tabular}\label{tab:limitCS} }
\end{table}

\begin{figure}[t]
	\centering
	\includegraphics[width=0.75\textwidth]{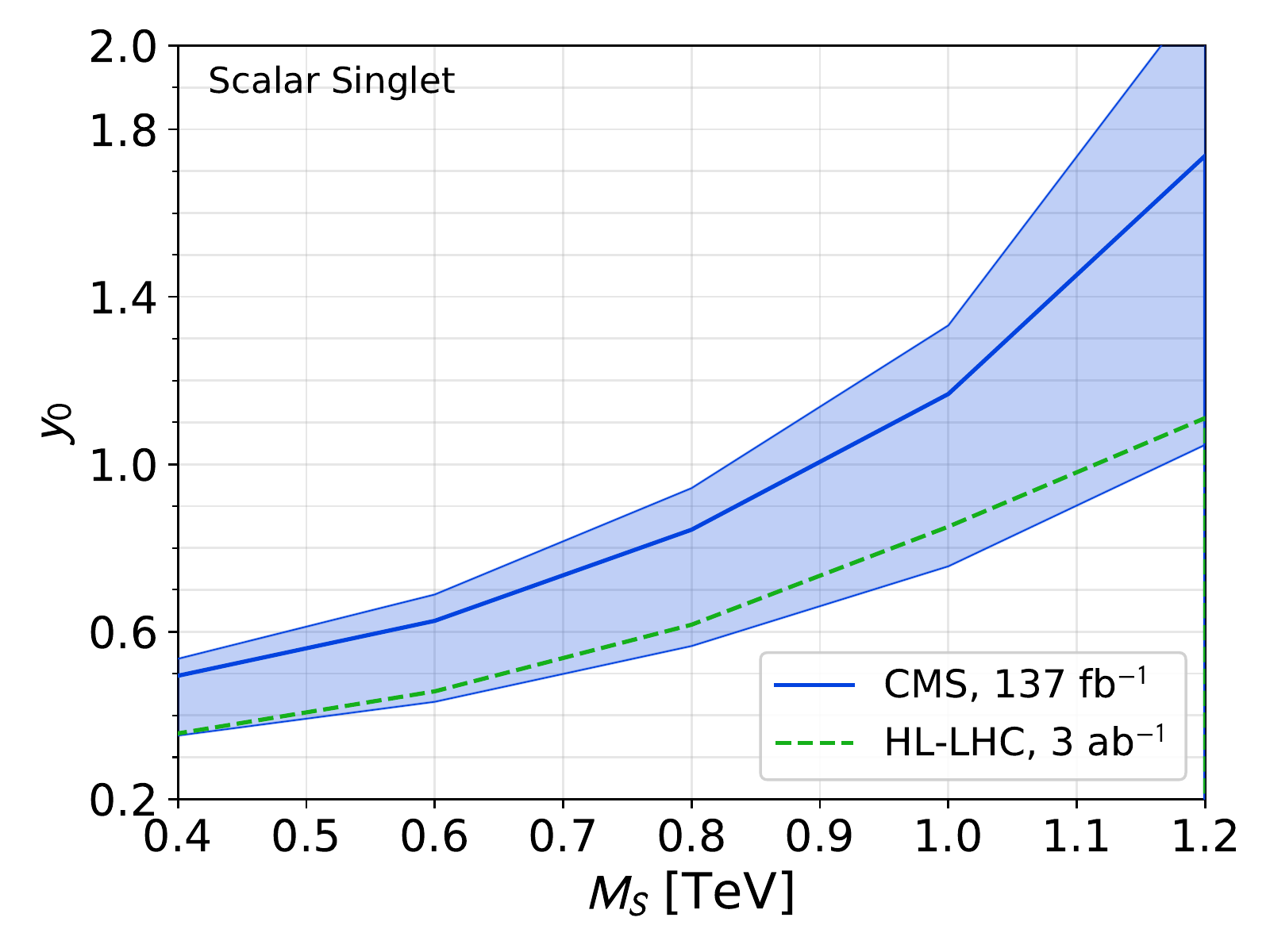}
  \caption{Limit on $y_0$ as a function of masses derived from the above procedure in \ma~ along with projections for the HL-LHC based on $3$ ab$^{-1}$ of data. The regions  above the curves are excluded, and the shaded areas show the associated theoretical uncertainties for both limits.\protect\label{fig2}}
\end{figure}

Next, we scan over the singlet mass in the $[400, 1200]$~GeV range, and
translate the limits on the cross section as a direct constraint on the $S$
coupling to the top quark. We remind that $t\bar tS$ associated production
typically dominates for such mass values. We present limits derived from the
CMS analysis under consideration (blue), as well as projections for the HL-LHC (green) in Fig.~\ref{fig2}. The regions above the blue and green thick line in the figure correspond to a 95\% confidence level exclusion by the CMS-TOP-18-003 analysis when considering the run~2 and HL-LHC luminosity respectively. In order to derive these limits, we use standard build-in features from \ma\ allowing for the calculation of the exclusion confidence level associated with a given signal. These are extensively documented in ref.~\cite{Conte:2018vmg}. The large error bars (corresponding to the shaded regions in the figure) are related to the significant theoretical uncertainties associated with our leading-order signal simulations. We refer to ref.~\cite{Darme:2020} for a more refined analysis.

At face value, the additional statistics provided by the HL-LHC imply a $50\%$
improvements in the limits. However, the new physics signal is expected to
strongly deviate from the SM background in various observables, such as the sum
of the transverse momenta of all reconstructed jets $H_T$. One can therefore
expect a significant improvement on these limits from a dedicated search
strategy, as already mentioned in Ref.~\cite{Darme:2018dvz}.

\subsection{Conclusions}
\label{sec:conclusions}
We have described in this work the implementation of the CMS-TOP-18-003
analysis in the \ma\ framework. Such an analysis can be used to target new
physics expected to show up in four-top events at LHC. We have validated our
work by comparing predictions relying on the Monte Carlo simulations of SM
four-top production. We have found an agreement with the CMS official results, well within their $1\sigma$ uncertainties. In particular, all the SR event counts agree with the CMS $t\bar t t\bar t$ projection within $~30\%$, as do the differential distributions in $H_T$, $N_{\rm jets}$ and $N_b$. Consequently, the present work can be considered as validated and used
without restriction to probe and test novel new physics models.

As an illustrative example of usage, we have reintepreted the CMS-TOP-18-003
analysis to extract bounds on a simplified top-philic scalar model, together
with their projection at the HL-LHC.

The {\sc MadAnalysis}~5 C++ recast code is available online from the MA5 dataverse (\href{https://doi.org/10.14428/DVN/OFAE1G}{https://doi.org/10.14428/DVN/OFAE1G})~\cite{OFAE1G_2020}.

\subsection*{Acknowledgments}
LD is supported by the INFN “Iniziativa Specifica” Theoretical Astroparticle Physics (TAsP-LNF).
\cleardoublepage

\markboth{B.~Fuks {\it et al.}}{Proceedings of the second MadAnalysis 5 workshop on LHC recasting in Korea}

\bibliographystyle{ws-mpla}
\bibliography{bibliography}
\end{document}